\documentclass[11pt,a4paper,twoside]{book}
\usepackage{titlesec}
\usepackage{fancyhdr}
\usepackage[utf8]{inputenc}
\usepackage[english]{babel}
\usepackage[autostyle,italian=guillemets]{csquotes}
\usepackage[style=numeric,citestyle=numeric-comp,backend=biber,sorting=none]{biblatex}
\usepackage[T1]{fontenc}
\usepackage{bm,bbold}
\usepackage{amsmath}
\usepackage{amsfonts}
\usepackage{amssymb}
\usepackage{empheq}
\usepackage{graphicx}
\usepackage{pdfpages}
\usepackage{subfigure}
\usepackage{booktabs}
\usepackage{slashed}
\usepackage[a4paper]{geometry}
\usepackage{microtype}
\usepackage[
breaklinks=true,colorlinks=true,
linkcolor=blue,urlcolor=blue,citecolor=blue,
bookmarks=true,bookmarksdepth=2,bookmarksopenlevel=2,bookmarksnumbered=true]{hyperref}

\titleformat{\chapter}[display]
  {\bfseries\Large}
  {\Huge\thechapter}
  {1ex}
  {\titlerule\vspace{1ex}\filright}
  [\vspace{1ex}\titlerule]
\DeclareSymbolFont{bbold}{U}{bbold}{m}{n}
\DeclareSymbolFontAlphabet{\mathbbold}{bbold}  
\newcommand{\f}{\frac}											
\newcommand{\s}{\sqrt}											
\newcommand{\de}{\partial}										
\renewcommand{\l}{\left}										
\renewcommand{\r}{\right}										
\renewcommand{\Re}{\text{Re}}									
\renewcommand{\Im}{\text{Im}}									
\newcommand{\mc}{\ensuremath{\mathcal}}        					
\newcommand{\h}{\widehat}                    	      			
\newcommand{\N}[1]{\mathop{:\!{#1}\!:}}							
\newcommand{\ped}[1]{_{\textup{#1}}}							
\newcommand{\apic}[1]{^{\textup{#1}}}							
\renewcommand{\mod}[1]{\left|#1\right|}							
\newcommand{\ket}[1]{|#1\rangle}								
\newcommand{\bra}[1]{\langle#1|}
\newcommand{\braket}[1]{\langle#1\rangle}
\newcommand{\acom}[2]{\{#1,\,#2\}}

\def\wT{{\widehat T}}
\def\wj{{\widehat j}}
\def\wJ{{\widehat J}}
\def\wK{{\widehat K}}
\def\wP{{\widehat P}}
\def\wQ{{\widehat Q}}

\def\wPsi{{\sf{\Psi}}}

\def\wrho{{\widehat{\rho}}}

\newcommand{\D}{\mathrm{d}}										
\newcommand{\di}{\mathrm{d}}									
\newcommand{\Dpi}{\mathcal{D}}									
\newcommand{\E}{\mathrm{e}}										
\newcommand{\e}{\mathrm{e}}										
\newcommand{\parz}{\mathcal{Z}}									

\newcommand{\I}{{\rm i}}										
\newcommand{\ii}{{\rm i}}										

\newcommand{\idmat}{\mathbbold{1}}								
\renewcommand{\vec}[1]{\ensuremath{\mathchoice					
                     {\mbox{\boldmath$\displaystyle\mathbf{#1}$}}
                     {\mbox{\boldmath$\textstyle\mathbf{#1}$}}
                     {\mbox{\boldmath$\scriptstyle\mathbf{#1}$}}
                     {\mbox{\boldmath$\scriptscriptstyle\mathbf{#1}$}}}}
\newcommand{\group}[1]{\relax\ifmmode\mathsf{#1}\else\textsf{#1}\fi}	
\newcommand{\tr}{{\text{tr}}}									
\newcommand{\res}{\text{Res}}									
\DeclareMathOperator*{\sumint}{
\mathchoice%
  {\ooalign{$\displaystyle\sum$\cr\hidewidth$\displaystyle\int$\hidewidth\cr}}
  {\ooalign{\raisebox{.14\height}{\scalebox{.7}{$\textstyle\sum$}}\cr\hidewidth$\textstyle\int$\hidewidth\cr}}
  {\ooalign{\raisebox{.2\height}{\scalebox{.6}{$\scriptstyle\sum$}}\cr$\scriptstyle\int$\cr}}
  {\ooalign{\raisebox{.2\height}{\scalebox{.6}{$\scriptstyle\sum$}}\cr$\scriptstyle\int$\cr}}
}
\let\oldphi\phi \let\phi\varphi \let\varphi\oldphi		
\newcommand{\mean}[1]{\langle #1 \rangle}						
\newcommand{\intmod}[1]{\!\int_0^\infty\!\!\!\!\D #1\,}			
\newcommand{\oraw}[1]{\overrightarrow{#1}}
\newcommand{\olaw}[1]{\overleftarrow{#1}}
\newcommand{\oper}{\mathcal{O}_{\alpha\dots\beta}}
\newcommand{\mycorr}[1]{\langle\langle#1\rangle\rangle}
\DeclareMathSymbol{\Gamma}{\mathalpha}{letters}{"00}
\DeclareMathSymbol{\Delta}{\mathalpha}{letters}{"01}
\DeclareMathSymbol{\Theta}{\mathalpha}{letters}{"02}
\DeclareMathSymbol{\Lambda}{\mathalpha}{letters}{"03}
\DeclareMathSymbol{\Xi}{\mathalpha}{letters}{"04}
\DeclareMathSymbol{\Pi}{\mathalpha}{letters}{"05}
\DeclareMathSymbol{\Sigma}{\mathalpha}{letters}{"06}
\DeclareMathSymbol{\Upsilon}{\mathalpha}{letters}{"07}
\DeclareMathSymbol{\Phi}{\mathalpha}{letters}{"08}
\DeclareMathSymbol{\Psi}{\mathalpha}{letters}{"09}
\DeclareMathSymbol{\Omega}{\mathalpha}{letters}{"0A}
\DeclareMathSymbol{\varGamma}{\mathalpha}{operators}{"00}
\DeclareMathSymbol{\varDelta}{\mathalpha}{operators}{"01}
\DeclareMathSymbol{\varTheta}{\mathalpha}{operators}{"02}
\DeclareMathSymbol{\varLambda}{\mathalpha}{operators}{"03}
\DeclareMathSymbol{\varXi}{\mathalpha}{operators}{"04}
\DeclareMathSymbol{\varPi}{\mathalpha}{operators}{"05}
\DeclareMathSymbol{\varSigma}{\mathalpha}{operators}{"06}
\DeclareMathSymbol{\varUpsilon}{\mathalpha}{operators}{"07}
\DeclareMathSymbol{\varPhi}{\mathalpha}{operators}{"08}
\DeclareMathSymbol{\varPsi}{\mathalpha}{operators}{"09}
\DeclareMathSymbol{\varOmega}{\mathalpha}{operators}{"0A}

\newcommand{\allmodesymb}[2]{\relax\ifmmode{\mathchoice
{\mbox{\fontsize{\tf@size}{\tf@size}#1{#2}}}
{\mbox{\fontsize{\tf@size}{\tf@size}#1{#2}}}
{\mbox{\fontsize{\sf@size}{\sf@size}#1{#2}}}
{\mbox{\fontsize{\ssf@size}{\ssf@size}#1{#2}}}}
\else
\mbox{#1{#2}}\fi}

\hypersetup{pdfinfo={Title={PhD Thesis},Author={Matteo Buzzegoli}}}
\graphicspath{{./Images/}}
\begin{document}
\frontmatter
\pagestyle{empty}
\begin{titlepage}
 \begin{flushleft}
  \includegraphics[width=.4\textwidth]{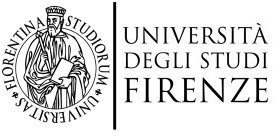}\newline
  \large{\textsc{Dipartimento di Fisica e Astronomia}}
 \end{flushleft}
\vspace{1cm}
\begin{center}
 \Large{\textsc{Dottorato di Ricerca in Fisica e Astronomia}}\\
 \Large{\textsc{Indirizzo Fisica - XXXII Ciclo}}\\
 \large{\textsc{Settore Scientifico Disciplinare: FIS/02}}\\[3cm]
 \huge{\textbf{Thermodynamic equilibrium of massless fermions with vorticity, chirality and magnetic field}}\\[1 cm]
 \huge{Matteo Buzzegoli}
\end{center}
\vspace{4cm}
\begin{minipage}{.5\textwidth}
	\begin{flushleft}
		 \large{\textbf{Supervisor:}\\ Prof. Francesco Becattini}
	\end{flushleft}
\end{minipage}%
\begin{minipage}{.5\textwidth}
	\begin{flushright}
		 \large{\textbf{Coordinator of PhD program:}\\ Prof. Raffaello D'Alessandro}
	\end{flushright}
\end{minipage}
\vfill
\begin{center}
\large{\textsc{Academics years 2016-2019}}
\end{center}
\end{titlepage}


\cleardoublepage
\pdfbookmark{Contents}{Contents}
\tableofcontents
\pagestyle{fancy}
\fancyhead{} 
\renewcommand{\chaptermark}[1]{\markboth{#1}{}}
\renewcommand{\sectionmark}[1]{\markright{\thesection.\ #1}}
\fancyhead[LE]{\thepage \hspace{1cm} \leftmark}
\fancyhead[RO]{\rightmark \hspace{1cm} \thepage}
\fancyfoot{} 
\renewcommand{\headrulewidth}{0pt}
\renewcommand{\footrulewidth}{0pt}
\chapter{Notation and conventions}
\label{ch:notazioni}
In this work we mostly use the \emph{natural unit} system in which $\hslash=c=G=k_B=1$.

The \emph{Minkowski metric} is defined by the tensor $\eta_{\mu\nu}=$diag$(1,-1,-1,-1)$
with greek indices running on four-vectors components $\mu=0,1,2,3$; instead we use latin indices for space coordinates only $i=1,2,3$ and we adopt
the Einstein index summation convention but only if at least one repeated index is up and the other is down.

The four-vectors and the tensors in general are indicated both with their components (e.g.~$\beta^\mu$) and their symbol (e.g.~$\beta$),
instead the tri-vectors are indicated with their component (e.g.~$k^i$) or with a bold letter (e.g.~$\vec{k}$).
The module of the four-vector is indicated as $\mod{\beta}$ or $\s{\beta^2}$. The inner product is generally indicated with ``$\,\cdot\,$'', the same symbol is used
for the contraction of two indices of two tensors between the most right of the first with the most left of the second (e.g. 
$(\alpha \cdot \varpi)_\mu = \alpha_\lambda \varpi^{\lambda\mu}$). We use the symbol ``$\wedge$'' for the wedge product.

The \textbf{sansserif} letters are used to indicate the element of a group (e.g. $\group{R} \in$ SO(3)).
If it is not explicitly marked, the stress-energy tensor $\h{T}_{\mu\nu}$ is symmetric.

We use  Weyl (chiral) basis for Gamma matrices $\gamma^\mu$; in imaginary time we denote Euclidean Gamma matrices with $\tilde{\gamma}_\mu$ and when it is
clear that we switched into imaginary formalism the tilde symbol is omitted; in curved spacetime we define the coordinate dependent Dirac gamma matrices 
as $\underline{\gamma}^\mu(x)\equiv e_a^{\,\,\mu}(x)\gamma^a$, where $e_a^{\,\,\mu}(x)$ is the vierbein field. 
\vfill
\newpage
The following special symbols and abbreviations are used throughout:
\begin{center} 
	\begin{tabular}{lp{0.45\textwidth}}
		$\I$ & imaginary unit;\\
		$\E$ & Euler's number;\\
		$*$ & complex conjugate;\\
		$\dagger$ or h.c. & Hermitian conjugate;\\
		$\bar{\,}$ & Dirac adjoint (e.g. $\bar{\psi}$);\\
		$\h{\;}$ & quantum operator (e.g. $\h{T}^{\mu\nu}$);\\
		$\hat{}$ & versor (e.g. $\hat{u}$);\\
		$\Re$ ($\Im$) & real (imaginary) part;\\
		$\tr,\,\log,\,\det$ & trace, natural logarithm, determinant;\\
		$\l[A,B\r]$ & commutator $AB-BA$;\\
		$\l\{A,B\r\}$ & anticommutator $AB+BA$;\\
		$\N{\,}$ & normal ordering;\\
		$a_{(\mu,\nu)}$ & symmetric sum on index $\f{1}{2}(a_{\mu,\nu}+a_{\nu,\mu})$;\\
		$\epsilon^{\mu\nu\rho\sigma}$ & Levi-Civita symbol such that $\epsilon^{0123}=+1$;\\
		$\f{\de}{\de x^\mu}$ or $\de_\mu$ & partial derivative;\\
		$g_{\mu\nu},\,g\equiv\det(g_{\mu\nu})$ & curved spacetime metric with inverse metric $g^{\mu\nu}$;\\
		$\Gamma^\lambda_{\,\mu\nu}=\Gamma^\lambda_{\,\nu\mu}=\f{1}{2}g^{\lambda\sigma}\l(g_{\sigma\nu,\mu}+g_{\mu\sigma,\nu}-g_{\mu\nu,\sigma}\r)$ & Christoffel connection;\\
		$\nabla_\mu$ & covariant derivative;\\
		$R^\lambda_{\,\tau\mu\nu}=\Gamma^\lambda_{\,\tau\mu,\nu}-\Gamma^\lambda_{\,\tau\nu,\mu}
		+\Gamma^\lambda_{\,\nu\sigma}\Gamma^\sigma_{\,\tau\mu}-\Gamma^\lambda_{\,\mu\sigma}\Gamma^\sigma_{\,\tau\nu}$ & Riemann tensor;\\
		$R_{\mu\nu}=R^\alpha_{\,\mu\alpha\nu}$ & Ricci tensor;\\
		$\simeq$ & approximately equal to;\\
		$\sim$ & order of magnitude estimate;\\
		$\approx$ & asymptotically approximate to;\\
		$\equiv$ & defined to be equal to;\\
		$\tau=\!\I\, t,\quad X\equiv(\tau,x^i)$ & Euclidean coordinates;\\
		$\int_X \equiv \int_0^\beta \D\tau \int_{\vec{x}},\quad \int_{\vec{x}}\equiv \int \D^d\vec{x}$ & integral  in Euclidean space-times
		where $\beta \equiv 1 / T$,  and $d=3$ is the space dimensionality;\\
		$K\equiv(k_n,k_i),\quad$ sometimes $\omega_n=k_n$ & Fourier modes in Matsubara formalism;\\
		$\sumint_K \equiv T\sum_{k_n}\int_{\vec{k}},\quad \int_{\vec{k}}\equiv \int \D^k\vec{k} / (2\pi)^d$ & Fourier-analysis in Matsubara formalism;\\ 
		$\sumint_{\{K\}} \equiv T\sum_{\{k_n\}}\int_{\vec{k}}$ & antiperiodic (or fermionic) summation ;\\
		$n\ped{F}(x)=\frac{1}{\E^{\beta x}+1}$ & Fermi-Dirac distribution function;\\
		$n\ped{B}(x)=\frac{1}{\E^{\beta x}-1}$ & Bose-Einstein distribution function;\\
		$\zeta(z)$ & Riemann zeta function;\\
		$\Gamma(z)$ & Euler gamma function;\\
		$D_n(z)$ & parabolic cylinder function.
	\end{tabular}
\end{center}

Other notation is introduced as needed.
\chapter{Introduction}
\label{ch:intro}

The quantum behavior of a physical system is usually relevant at microscopic lengths, while the
classical behavior is recovered at macroscopic scales. However, quantum properties may emerge even
at macroscopic scale, giving rise to phenomena not allowed in a classical world. Some notable
instances are superfluidity, superconductivity, ultra-cold atom gases and quantum Hall effects.
In the last decades, these macroscopic quantum phenomena have produced great advances both
in theoretical and in technological fields.

Macroscopic quantities are understood as emergent properties of a system and are obtained integrating out all
the detailed information about microscopic dynamics but retaining collective behavior. For example, the
temperature $T$ measures the mean kinetic energy of the system constituents, which is roughly
given by $k_B T$, where $k_B$ is the Boltzmann constant. While temperature gives the scale of macroscopic
quantities, the (reduced) Planck constant $\hbar$ sets the scale for quantum behavior. For instance,
consider a Bose-Einstein condensate, which is required for most of the aforementioned phenomena.
In that case, for a fluid with particle density $\rho$ and consisting of mass $m$ bosons,
we can associate a quantum energy $\hbar^2\rho^{2/3}/m$. Indeed, boson fluids show quantum behavior
(the condensate) when this energy is larger than the thermal energy $k_B T$. That is why all the
aforementioned phenomena require low-temperature environments to keep the quantum effects dominant.

Since most of the systems in nature are affected by rotation or by acceleration, such as the one caused by Earth
gravitational attraction, we may wonder if an interplay between angular velocity~$\omega$, or acceleration~$a$, and quantum
effects may produce macroscopic phenomena. Indicating with $c$ the speed of light in vacuum, this is likely to occur if
the quantum energy associated to angular velocity~$\hbar\, \omega$ or to acceleration~$\hbar\, a/c$
\footnote{Inside the core of the text it would be clear and motivated from where those constant combinations pop out.}
is larger than thermal energy $k_B T$, i.e. when the following ratios $w$ and $\alpha$ become relevant:
\begin{equation*}
w\equiv\frac{\hbar\, \omega}{ k_B T},\quad \alpha\equiv\frac{\hbar\, a}{c\, k_B T}.
\end{equation*}
When replacing some numbers related to ``common'' physical systems, we realize that $\alpha$ and $w$ are usually
frighteningly small: $\alpha,\,w\sim 10^{-20}$. However, nature and humans resourcefulness provided some
instances of matter under extreme conditions where those parameters play a significant role. 

With the Relativistic Heavy Ion Collider at Brookhaven National Laboratory and
with the Large Hadron Collider at CERN, ions of heavy nuclei (usually
Au and Pb) are collided between themselves at ultra-relativistic speed. As a result, a new state of nuclear matter,
dubbed the Quark-Gluon Plasma (QGP), is created~\cite{Heinz:2000bk}. The QGP phase occurs at extremely high
temperatures and densities when the
hadrons 
dissolve into an asymptotically free state of their elementary constituents: quarks and gluons.
The QGP phase produced in high energy nuclear collisions lasts for about 5~fm$/c\simeq 10^{-23}$~s. During that
time it expands and cools down until it once again forms hadron states, which are then detected by instruments.
Even the Universe, some picoseconds after the Big Bang and lasting for 10 microseconds, is thought to have
taken the form of quark-gluon plasma. This new phase of matter has become a great benchmark for phase transitions,
symmetries breaking, collective behavior and baryonic asymmetry inside the Standard Model of elementary particles,
especially for Quantum Chromo-Dynamics (QCD), the theory that describes nuclear interactions.


QCD possesses the remarkable property of asymptotic freedom, meaning that at short distances, the QCD running
coupling constant is small and quantities can be reliably evaluated in perturbation theory. However, at strong
coupling, we still lack a reliable theoretical method for performing analytical calculations. As a consequence,
despite QCD has well-known symmetries and established elementary constituents, its emergent behavior
remains a mystery to us. For instance, perturbation theory does not contain confinement of color, i.e. it does
not explain how the asymptotic states of QCD perturbation theory (colored quarks and gluons) transform into the
asymptotic states actually observed in experiments (the color-singlet hadrons). The confinement scale can be
estimated from the proton radius, roughly 1 fm, which corresponds to an energy $\Lambda$ of about 200 MeV.
Color confinement can be overcome rising the temperature above a critical value $T_c$; intuitively when it
reaches confinement energy $k_B T_c\sim \Lambda$ (lattice calculations indicate $k_B T_c=154\pm8$ MeV~\cite{Karsch:2000kv}),
then one obtains a gas composed by almost free quarks and gluons, the aforementioned QGP. For the same critical value,
the weakly broken chiral symmetry, responsible for the three light pions, gets restored.

Experiments already revealed and established remarkable properties of the QGP. Regarding collective behavior, it
is found that the QGP is indeed a fluid and what's more the one with the lowest viscosity ever observed~\cite{Heinz:2013th,Teaney:2009qa}.
Moreover, most of the collisions are not central and so produce matter with angular momentum of order $1000\hbar$.
This vorticity couples with quantum behavior with an energy associated to acceleration of $\hbar\, a/c\sim 10$ MeV
and to rotation of $\hbar\, \omega\sim 12$ MeV~\cite{STAR:2017ckg}. Compared to the thermal energy of the plasma
$k_B T\sim 200$ MeV, this means a ratios of $\alpha=0.05$ and $w=0.06$. In agreement with local
thermodynamic equilibrium predictions~\cite{Becattini:2007sr}, these values of local vorticity and acceleration are sufficient
to induce polarization on the particles produced by the plasma, as recent measures on Lambda particle polarization
have confirmed~\cite{STAR:2017ckg,Adam:2018ivw}. This, beside vesting QGP as the most vorticous fluid ever observed,
is the proof that quantum effects induced by rotation are displayed at a macroscopic level even at extremely
high temperatures.

In recent years, studies on the interplay of quantum effects with vorticity and magnetic field identified several
novel non-dissipative transport phenomena~\cite{Kharzeev:2015znc} similar to the effects that lead to particle
polarization. The phenomenology is even richer in systems possessing chiral fermions. Elementary particles are
classified according to the Lorentz group, which has two unitarily inequivalent spinor representations related
by parity transformation. A particle is said to have right or left chirality if it belongs to one or the other
of such representations. Or, in other words, a particle is chiral if it is possible to distinguish the particle from its
mirror image. The identification is easier for massless spin $1/2$ fermions because in this case chirality corresponds
to particle helicity, which is whether the particle spin is or is not aligned with the particle momentum.

Among those non-dissipative transport phenomena, the Chiral Magnetic Effect (CME)~\cite{Fukushima:2008xe} is the induced
electric current along an external magnetic field driven by chiral imbalance. Since both the magnetic field and the electric
current are odd quantities under time reversal transformation $\group{T}$, the CME conductivity must be $\group{T}$-even,
revealing that its occurrence does not involve any dissipative process. On the contrary, the magnetic field is even
under parity transformation $\group{P}$ but the electric current is $\group{P}$-odd. This is why to connect the
two we need a chiral medium: matter possessing an imbalance between right- and left-handed particles.

An instance of chiral matter is once again found in QGP, where chirality is thought to be generated in the topological
sector of QCD vacuum states. The compactness of SU(3) group, not addressed in perturbative approaches, allows
the existence of nontrivial topological solutions of gluon configurations. These solutions can possess chirality
which can be transferred to quarks through the chiral anomaly. Indeed, the Atiyah-Singer index theorem relates the
topological charge of the background gauge field configuration to the number of fermionic zero-modes of the Dirac
operator. Since quarks also possess electric charge, we can probe their topology using a magnetic field.
\enlargethispage{\baselineskip}

In heavy-ion collisions, together with the QGP, a magnetic field is generated with an initial magnitude
of $eB\sim 10 m_\pi^2\sim 10^{18}$ G~\cite{Skokov:2009qp,Deng:2012pc,Bloczynski:2012en}. Actually, it is the strongest magnetic field in
the present Universe, thousands of times stronger than those found on the surface of magnetars, where it may also
affect its cold dense nuclear matter. In QGP produced by heavy-ion collisions, we then have all the ingredients
for the generation of CME, and its observable manifestation would be the proof of the existence of topological
structure of QCD. CME is expected to cause a charge-dependent azimuthal asymmetry on the spectrum of produced
particles, which has already been observed by STAR Collaboration at RHIC~\cite{PhysRevLett.103.251601}. However,
mundane backgrounds unrelated to the CME may be able to explain the observed charge separation and dedicated
experiments with isobar collisions~\cite{Kharzeev:2019zgg} are currently ongoing to asses this possibility.
Moreover, contrary to angular velocity, which persists longer in the plasma due to angular momentum conservation,
the magnetic field has a very short lifetime,  about $\sim 10^{-23}$ s, and there are still significant theoretical
uncertainties due to its evolution. That is why recently many efforts have been made to develop a proper
chiral magnetohydrodynamics~\cite{Boyarsky:2015faa,DelZanna:2018cme,Hattori:2017usa} to run simulations in
which both the medium and the field are evolved dynamically.

Magnetic field has already been successfully used to probe topological properties, as in the case of (fractional)
quantum Hall effect in condensed matter physics. It is then not surprising that thanks to the discovery of chiral
materials (Dirac and Weyl semimetals), CME has received its experimental evidence in condensed matter~\cite{Li:2014bha}.
In those materials, chirality is pumped into the system with an external parallel magnetic and electric field by
leveraging chiral anomaly. This reinforces the idea that the CME coefficient is entirely dictated by chiral anomaly and, because
of that, is topologically protected. Indeed various computations in severals regime from free to strongly coupled
systems found the same conductivity for CME independently of dynamical details.

We refer to quantum anomalies when a classical symmetry of a theory is not compatible with quantum theory.
When massless quarks are addressed, QCD is invariant under chiral transformations and, for Noether's theorem, the axial current
is conserved. However, this symmetry is explicitly broken by the regularization of quantum processes in background
gauge fields~\cite{Adler:2004qt}. Among several consequences, chiral anomaly is responsible for the large mass of
the pseudoscalar $\eta'$ meson and for the decay of neutral pion into two photons. The salient feature of quantum
anomalies is that they bridge collective motion of particles with arbitrarily large momenta in the vacuum to the
finite momentum world~\cite{Gribov:1981ku}. That is how quantum anomalies can possibly affect macroscopic
phenomena, even though the latter involves a large number of quanta. Moreover, the strength of anomalies resides on
the non-renormalization theorem~\cite{Adler:2004qt}, stating that the anomaly is exact as an operator relation
at one loop.

Quantum anomalies have a primary role in the understanding (and in the discovery) of non-dissipative phenomena
of quantum many-body physics we are discussing. We already mentioned CME and its role in heavy-ion collisions;
in astrophysics transport phenomena induced by anomalies could explain the sudden acceleration during the birth
of neutron stars (neutron star kicks)~\cite{Kaminski:2014jda} and in cosmology the origin of primordial magnetic
fields~\cite{Giovannini:1997eg}. Chiral anomaly also induces macroscopic effects when coupling with vorticity.
For example, the Chiral Vortical Effect (CVE) is the generation of an electric current along the global
rotation in a chiral medium. This effect shares many properties with CME and has similar consequences on the
phenomenology of QGP, although, being CVE independent of charge, the two effects can be distinguished.

Connections and similarities between rotation and magnetic field have a long history, starting from noticing
that a charged particle under magnetic force undergoes into a circular motion, to the Barnett and Einstein-de
Hass effects. Spin-orbit coupling can transfer angular momentum to spin and vice versa. Therefore, in
ferromagnetic materials, mechanical rotation can induce magnetization (Barnett effect) and, due to angular
momentum conservation, also a variation on magnetization induces rotation (Einstein-de Haas effect). It has
been suggested that CVE could be interpreted as an extension of a relativistic Barnett effect,
see~\cite{Fukushima:2018grm}.

For the sake of understanding, it is important to make a clear distinction between CVE and another macroscopic
quantum phenomena: the Axial Vortical Effect (AVE), which is the induced axial current along the global rotation.
Contrary to electric current, the axial current has a tight connection with both chirality and spin. In
relativistic theories axial current is dual to spin tensor and in systems consisting of massless fermions
and without background gauge fields, the global chiral charge is defined as the integral of axial current. Other
substantial differences with CVE are that AVE does not require a chiral imbalance and that it is not entirely
topologically protected~\cite{Hou:2012xg}, questioning its possible proposed relation with mixed-axial
gravitational anomaly~\cite{Landsteiner:2011iq}.

In this work, we consider thermodynamic properties of systems composed by massless fermions with a chiral imbalance
and subject to global vorticity and external magnetic field. The systematic derivations of all constitutive
equations are carried out using the covariant operator approach~\cite{BecaBetaF} based on Zubarev method of
stationary non-equilibrium density operator~\cite{Zubarev1966,Zubarev1992,Becattini:2019dxo}.
With this procedure, the underlying theory has full-quantum properties and does not breaks down when a kinetic
description is no longer valid. Indeed, close to the critical temperature, the QGP constituents are strongly interacting
and the mean free path length is comparable to the quantum oscillations of the fields, hindering a kinetic
description in terms of quasiparticles. Moreover, the effects of global vorticity in this formalism arise
naturally from the covariance properties of the density matrix.

The arguments in the thesis are divided as follows.
An introduction to quantum relativistic statistical mechanics is provided in Chapter~\ref{chp:QRelStatMech}.
In particular, the key concept for this work is the global thermal equilibrium in the presence of vorticity,
which is given in Section~\ref{sec:GenGlobEquil}. The chiral anomaly is introduced in Section~\ref{sec:Symm},
where we also define a conserved current via the Chern-Simons current. A chiral conserved current is needed
to reach global equilibrium with chiral imbalance when dynamic or external gauge fields are considered.
The issue of a conserved axial current is discussed again in Section~\ref{sec:GlobEqEM} for external
electromagnetic field and in Section~\ref{sec:AVENon-Univ} for dynamical gauge fields.
Global equilibrium with both chiral imbalance and vorticity is considered in Section~\ref{sec:GenGlobEquil}.
In chapter~\ref{ch:ftft}, we review the finite temperature field theory methods for fermions and gauge fields.
In Section~\ref{sec:homogeq}, we address thermodynamics of massless fermions at finite density and finite chirality.

In chapter~\ref{ch:ChiralVort}, we discuss the global equilibrium of massless fermions in a chiral vorticous medium.
We provide thermodynamic constitutive equations at the second-order of thermal vorticity for relevant physical
quantities, such as the stress-energy tensor and both axial and electric currents. In Section~\ref{sec:CVFree},
we evaluate thermal coefficients up to second order in thermal vorticity for the case of free fermions.

In chapter~\ref{ch:ExternalB}, we consider global vorticous thermal equilibrium in presence of an external
electromagnetic field. First, in Section~\ref{sec:DiracInB}, we review the properties of fermions under
external electromagnetic field; in particular (Section~\ref{subsec:SymInFconst}) how translation
and Lorentz generators are affected. Then, in Section~\ref{sec:ThermoInB}, we provide exact solutions for the
thermodynamics of non-interacting chiral fermions subject to a constant magnetic field but with vanishing
thermal vorticity. Then, in Section~\ref{sec:GlobEqEM}, we discuss the general properties of thermal equilibrium
with both external electromagnetic field and thermal vorticity and we give constitutive equations as an expansion of
only thermal vorticity. In Section~\ref{subsec:Feffect}, we derive the effects of electromagnetic field on
first order thermal coefficients using the conservation equations of operators. At last, in Section~\ref{sec:AVENon-Univ},
we consider dynamic gauge fields and we show that the AVE receives radiative corrections.

\mainmatter
\chapter{Quantum relativistic statistical mechanics}
\label{chp:QRelStatMech}
In this work, we deal with global thermodynamic equilibrium of relativistic quantum systems.
In this chapter, we review how to derive macroscopic behavior at thermal equilibrium using
the methods of statistical mechanics in a full quantum relativistic framework.
We are using the method of non-equilibrium statistical operator introduced by Zubarev~\cite{Zubarev1966}
and later reworked by van Weert~\cite{vanWeert1982}, see~\cite{Becattini:2019dxo} for a
revision of original arguments. Zubarev's method provides a stationary statistical
operator, which is then well-suited to be used in relativistic quantum field theory in
Heisenberg representation. Therefore, this method can be used to describe systems
that reached local thermal equilibrium and to derive their hydrodynamic evolution.
These are the main reasons why recently Zubarev's method has been adopted in~\cite{BecaBetaF,Hayata2015}
and in sequent works to derive quantum effects on relativistic systems.

Before turning on relativistic system, we briefly summarize how global equilibrium is recovered
in quantum statistical mechanics using the principle of maximum information entropy.
Global thermodynamic equilibrium is described as a static statistical ensembles of systems which
are only defined by the specification of macroscopic variables. The \emph{statistical operator}
(also called \emph{density matrix}) $\h{\rho}$ provides the ``statistical mixture'' and
it is given by
\begin{equation*}
\h{\rho}=\sum_n p_n \ket{n}\bra{n},\quad \sum_n p_n=1,
\end{equation*}
where the $p_n$'s are the statistical weight of $n$th state and the $\{\ket{n}\}$'s form a
complete set. Starting from the statistical operator one calculates various quantities related
to thermal equilibrium, such as the partition function, the Green's functions and the Wigner
function. Average value of an observable related to the quantum operator $\h{O}$ is simply
obtained as
\begin{equation*}
\mean{\,\h{O}\,}=\tr (\h{\rho}\,\h{O} )=\sum_n p_n \bra{n}\h{O}\ket{n}.
\end{equation*}

At global equilibrium, the density matrix operator gives the stationary thermal states of the system; therefore, it can only depend on conserved quantities. In Newtonian physics, there exist seven first
integrals related to the conservation of energy, momentum and angular velocity. To these constants,
other conserved quantities can be added, such as the particle number, the baryon number and the
electric charge.

The maximum entropy principle provides a methods to build the density matrix appropriate to the
system we are considering. For instance, in the canonical ensemble only the Hamiltonian $\h{H}$ and
a conserve charge $\h{Q}$ are taken in account. In general, we can add any amount of conserved
quantities as long as the statistical operator remains time-independent. It is postulated~\cite{Balian}
that the equilibrium state maximizes the \emph{Von Neumann entropy} or \emph{information entropy}
\begin{equation*}
S[\,\h{\rho}\,]=-\tr(\h{\rho}\log\h{\rho})
\end{equation*}
under the following constraints
\begin{equation*}
\mean{\h{H}}=E=\text{constant},\quad\mean{\h{Q}}=Q=\text{constant},\quad \tr\left(\,\h{\rho}\,\right)=1.
\end{equation*}

In order to take into account the constraints, we use the method of Lagrange multipliers, which translates
our problem to the extremization of the free energy functional $F$ with respect to $\h\rho$:
\begin{equation*}
F_{\beta,\zeta,\lambda_0}[\,\h{\rho}\,]=-\tr(\h{\rho}\log\h{\rho})-\beta [\tr(\h{\rho}\,\h{H})-E]
+\zeta [\tr(\h{\rho}\,\h{Q})-Q]-\lambda_0 [\tr(\h{\rho})-1]
\end{equation*}
where $\beta$, $\zeta$ and $\lambda_0$ are respectively the Lagrange multiplier conjugate of energy $E$,
charge $Q$  and to normalization. The maximal solution~\cite{Balian} is the well known Gibbs distribution
\begin{equation*}
\h{\rho}=\frac{1}{\parz}\E^{-\beta \l( \h{H}-\mu\h{Q}\r)},\quad \parz=\tr\l[\E^{-\beta \l(\h{H}-\mu\h{Q}\r)}\r],
\end{equation*}
where we used the common notation for the chemical potential $\mu=\zeta\,T$ and for the inverse
temperature $\beta=1/T$. The functional $\parz$ is called \emph{partition function}.
In Heisenberg representation the density matrix does not evolve, instead in Schr\"{o}dinger
representation evolves with unitary operator $\h{U}(t,t_0)=\E^{-\I\h{H}(t-t_0)}$, where $\h{H}$ is
the hamiltonian of the system:
\begin{equation*}
\h{\rho}(t)=\h{U}(t,t_0)\h{\rho}(t_0)\h{U}^\dagger(t,t_0).
\end{equation*}

We now proceed to extend these basic ideas on how to describe local thermal equilibrium to relativistic
systems. As usual in relativity, energy and momentum must be dealt with in the same footing and in general,
all quantities must respect appropriate covariant properties. As a major consequence of relativity, finite
volume does not acquire an invariant definition and only local quantities have a real meaning. To address this
necessity, as the first step of next section, we introduce the Arnowitt-Deser-Misner (ADM) decomposition of
space-time and we review the symmetries and the conservation of a relativistic quantum system consisting
of fermions (with external forces). After that, we proceed to the construction of the covariant local
thermal equilibrium density operator.

\section{Space-time foliation}
\label{sec:Foliation}
\begin{figure}[htbp]
	\centering
	\includegraphics[height=6cm]{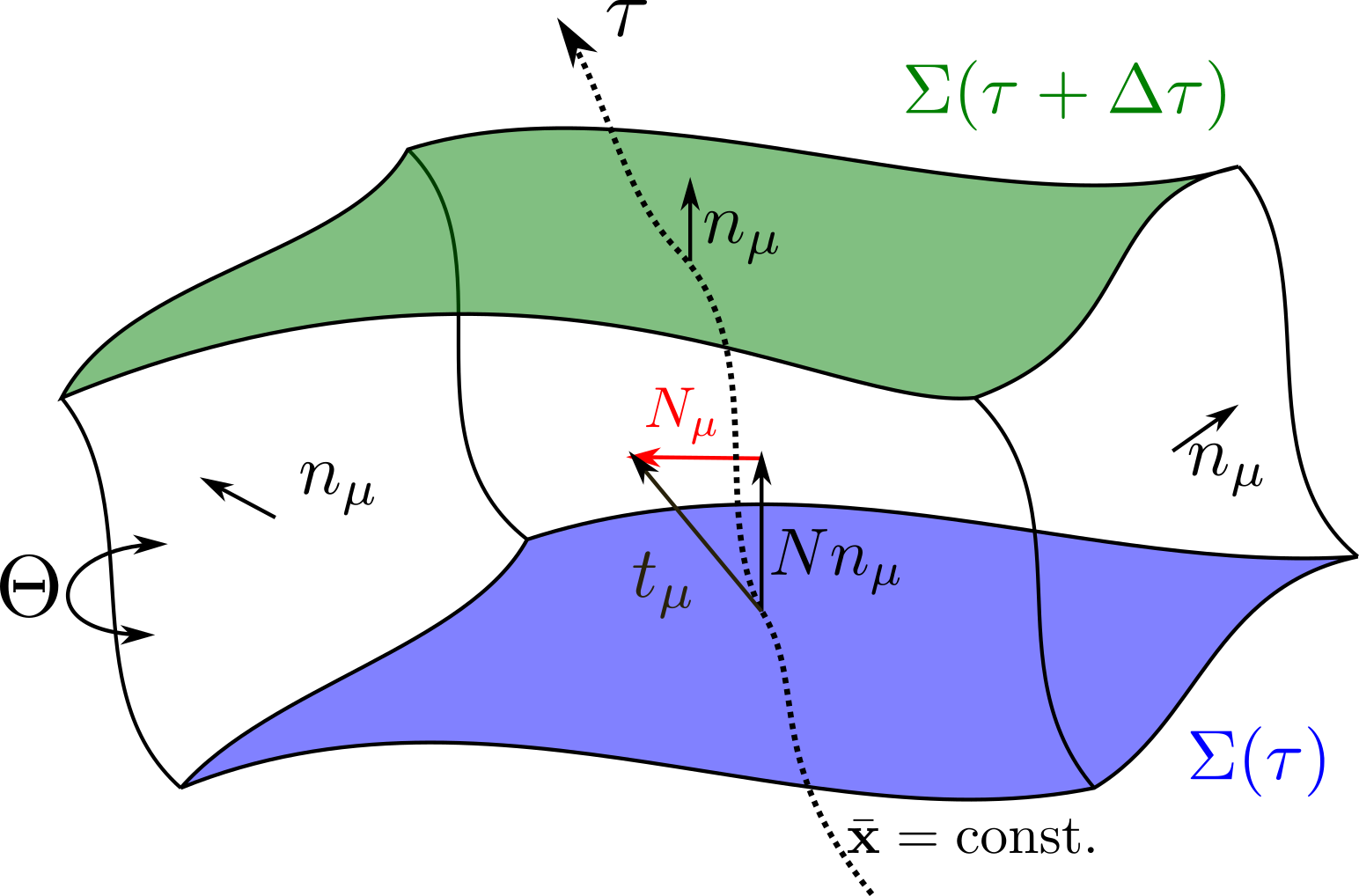}
	\caption{The Arnowitt-Deser-Misner (ADM) foliation of the spacetime. $\Sigma(\tau)$ is a space-like hyper-surface
		parametrized by $\tau(x)=$const. The normal unit vector to hyper-surfaces is denoted by $n^\mu$. The figure
		shows the decomposition of the time vector $t^\mu=\de_\tau x^\mu=N\, n^\mu+N^\mu$, with $N(x)$ the lapse function
		and $N^\mu$ the shift vector. The whole region represents the space-time volume $\Omega$ enclosed by 
		two space-like hyper-surface $\Sigma(\tau)$ and $\Sigma(\tau+\Delta\tau)$ and a time-like boundary $\Theta$.}
	\label{fig:foliation}
\end{figure}
In a relativistic system the symmetries of space-times are dictated by the Poincaré group, which has ten
generators. To each generator is associated a conserved quantity. The densities of those conserved quantities
are retrieved with a foliation of space-time, slicing it with space-like hyper-surfaces. In a generic curved
space-time with metric $g_{\mu\nu}$, we choose a function $\tau(x)$ as a time which parametrize the
space-like  hyper-surfaces $\Sigma(\tau)$. We indicate with $\bar{\vec{x}}=\bar{\vec{x}}(x)$ the space
coordinates on $\Sigma(\tau)$. We can now distinguish two future oriented time-like vectors $n_\mu$ and
$t_\mu$ (see Fig.~\ref{fig:foliation}). The unit vector normal to the space-like hyper-surfaces is
\begin{equation*}
n_\mu(x)=N(x)\de_\mu\tau(x)\quad\text{with}\quad N(x)\equiv(\de^\mu\tau(x)\de_\mu\tau(x)))^{-1/2},
\end{equation*}
where $n_\mu$ is normalized as $n_\mu n^\mu=1$ and $N>0$ is the lapse function. $t_\mu$ is the local time
direction in our coordinate system, which we decompose into the parts perpendicular and parallel to $n_\mu$:
\begin{equation*}
t^\mu(x)\equiv \de_\tau x^\mu (\tau, \bar{\vec{x}})=N n^\mu+N^\mu\quad\text{with}\quad n_\mu N^\mu=0,
\end{equation*}
$N^\mu$ is called the shift vector. The unit vector $n$ defines the world lines of observer, while $\tau$ in
general does not correspond to the proper time of comoving clock. An other significant distinction between
the two vectors, is that for Frobenius theorem $n$ must fulfill
\begin{equation*}
\epsilon_{\mu\nu\lambda\sigma}n^{\mu}(\partial^{\nu}n^{\lambda}-\partial^{\lambda}n^{\nu})=0
\end{equation*}
meaning that it is irrotational, while, on the contrary, $t$ can posses a vortical distribution.

Denoting with $\D\Sigma(\tau)=N^{-1}\sqrt{-g}\,\D^{d-1}\bar{\vec{x}}$ the volume element on
the hyper-surface at constant $\tau(x)$, we define the hyper-surface vector
$\D\Sigma_\mu=\D\Sigma(\tau) n_\mu$. Let $\h{j}^{\mu}(x)$  be a conserved current of the system,
at a given fixed value of the parametric time $\tau$, the total charge $\h{Q}$ is obtained
projecting the current along the normal direction $n$ and integrating over all
the hyper-surface $\Sigma(\tau)$: 
\begin{equation*}
\h{Q}=\int_{\Sigma(\tau)}\D\Sigma_\mu\,  \h{j}^{\mu}(x).
\end{equation*}   
The charge $\h{Q}$ do not depend on the specific value of $\tau$, indeed consider the difference of
the charge at time $\tau$ and at time $\tau+\Delta\tau$
\begin{equation*}
\Delta\h{Q}=\int_{\Sigma(\tau+\Delta\tau)}\D\Sigma_\mu\,  \h{j}^{\mu}(x)-\int_{\Sigma(\tau)}\D\Sigma_\mu\,  \h{j}^{\mu}(x).
\end{equation*}   
Call $\Omega$ the region enclosed by the two hyper-surfaces $\Sigma(\tau+\Delta\tau)$ and $\Sigma(\tau)$
and call $\Theta$  the time-like hyper-surface that surrounds $\Omega$ (see Fig.~\ref{fig:foliation}).
Then, the difference of charge is the difference between the current flux on the boundary
$\de\Omega$ of the whole region and on the surface $\Theta$:
\begin{equation*}
\Delta\h{Q}=\int_{\de\Omega}\D\Sigma_\mu\,  \h{j}^{\mu}(x)-\int_{\Theta}\D\Sigma_\mu\,  \h{j}^{\mu}(x).
\end{equation*}   
The second term could be sent to zero by a suitable boundary conditions and the first term can be written
by Gauss theorem  as the divergence of the current, which is vanishing:
\begin{equation*}
\Delta\h{Q}=\int_{\Omega}\D\Omega\nabla_\mu\,  \h{j}^{\mu}(x)=0.
\end{equation*}   
%

\section{Symmetries and conservation laws}
\label{sec:Symm}
Geometry being settled, we are now considering the microscopic description of matter. The most general
situation we are dealing in this work is a Dirac field $\psi(x)$ interacting with an external gauge field
$A^\mu(x)$ in a general curved  space-time background $g_{\mu\nu}$. The action related to a particle with
spin in curved space-time is not written in terms of the metric, the vierbein vector field $e^a_{\,\mu}(x)$
is used instead. It is defined by~\cite{BirrellDavies}:
\begin{equation*}
g_{\mu\nu}(x)=\eta_{ab}e^a_{\,\mu}(x)e^b_{\,\nu}(x),\quad e\equiv\det e^a_{\,\mu}=\sqrt{-g},
\end{equation*}
where the Greek letters $(\mu,\nu,\dots)$ denote curved space-times indices in the coordinate system
$(t,\vec{x})$, while Latin letter $(a,b,\dots)$ denote the local Lorentz indices. The action is then given
by the following form
\begin{equation}
\label{eq:GenAction}
S\left[\psi,\bar\psi;e^a_{\,\mu},A_\mu\right]=\int \D^d x \,e\,
\mc{L}\left( \psi(x),\bar\psi(x),D_\mu \psi(x) ,D_\mu\bar\psi(x) ; e^a_{\,\mu}(x),A_\mu(x) \right)
\end{equation}
where the covariant derivative acts on the Dirac field as follows:
\begin{equation*}
D_\mu\psi=\nabla_\mu-\I A_\mu=\de_\mu\psi+\frac{1}{2}\omega^{ab}_\mu\sigma_{ab}\psi-\I A_\mu,
\end{equation*}
where $\sigma_{ab}=\frac{1}{4}[\gamma_a,\gamma_b]$, $\gamma_a$ are the flat space time gamma matrices
satisfying $\{\gamma^a,\gamma^b\}=2\eta^{ab}$ and $\omega^{ab}_\mu$ is the spin connection
given by
\begin{equation*}
\omega^{ab}_\mu=\frac{1}{2}e^{a\nu}(\de_\mu e_\nu^b-\de_\nu e_\mu^b)-\frac{1}{2}e^{b\nu}(\de_\mu e_\nu^a-\de_\nu e_\mu^a)
+\frac{1}{2}e^{a\nu}e^{b\rho}(\de_\rho e_\nu^c-\de_\nu e_\rho^c)e_{c\mu}.
\end{equation*}
As a notable examples the Dirac Lagrangian in curved spacetime is written as
\begin{equation}
\label{eq:DiracLcurved}
\mc{L}=\frac{\I}{2}\bar\psi\left(\underline{\gamma}^\mu \oraw{D}_\mu -\olaw{D}_\mu\underline{\gamma}^\mu\right)\psi-m\bar\psi\psi,
\end{equation}
where we have introduced the coordinate dependent Dirac matrices $\underline{\gamma}^\mu(x)\equiv e_a^{\,\,\mu}(x)\gamma^a$.

Before moving on considering symmetries of the action, once the form of the action is known we can
define several quantities that would come in handy. The stress-energy tensor and the electric current are obtained
as variation of the action respect to the vierbien and to the guage field respectively:
\begin{equation*}
T^\mu_{\;a}\equiv \frac{1}{e}\frac{\delta S}{\delta e^{\;a}_\mu},\quad
J^\mu\equiv \frac{1}{e}\frac{\delta S}{\delta A_\mu}.
\end{equation*}
The stress-energy tensor with full curved space-times indices is simply derived from the previous one:
$T^\mu_{\;\nu}\equiv  e_\nu^{\;a}T^\mu_{\;a}$. For the free Dirac field of the Lagrangian~(\ref{eq:DiracLcurved})
the stress-energy tensor is
\begin{equation}
\label{eq:setfermion}
T^{\mu\nu}=\f{1}{4}\l[\bar{\psi}\gamma^\mu(D^\nu\psi) +\bar{\psi}\gamma^\nu(D^\mu\psi)
-(D^\nu\bar{\psi})\gamma^\mu\psi-(D^\mu\bar{\psi})\gamma^\nu\psi\r].
\end{equation}
Moreover, the field strength tensor of gauge field is defined as $F_{\mu\nu}\equiv \de_\mu A_\nu-\de_\nu A_\mu$.

Consider first the internal symmetries of the action. We have explicitly chosen the form in Eq.~(\ref{eq:GenAction})
with a gauged covariant derivative such that the action posses gauge invariance under the U(1) transformation
related to the field $A_\mu$. Upon a transformation with an infinitesimal parameter $\alpha(x)$, assumed to
be vanishing on the boundary, the fields change into
\begin{equation*}
\delta_\alpha A_\mu=\de_\mu \alpha,\quad \delta_\alpha\psi=\I q\alpha\psi,\quad \delta_\alpha\bar\psi=-\I q\alpha\bar\psi
\end{equation*}
with $q=\pm1$ or $0$ the elementary charge of Dirac field. The variation of the action respect to this
transformation is vanishing because we supposed it invariant. However, we could explicitly write the form
of the variation. Dirac field gives a contribution proportional to the equations of motions $\delta S/\delta \psi=0$
and $\delta S/\delta \bar\psi=0$, therefore the action variation is written as
\begin{equation*}
\begin{split}
\delta_\alpha S =&\int \D^d x\, e\,J^\mu \delta_\alpha A_\mu=\int \D^d x\, e\,J^\mu \de_\mu\alpha\\
=&\int \D^d x\, e \nabla_\mu \l(J^\mu\alpha\r)-\int \D^d x\, e\,\alpha\,\nabla_\mu J^\mu.
\end{split}
\end{equation*}
The first term vanishes because $\alpha$ is vanishing on the boundary and since the action variation
is zero for any function $\alpha$, this implies that the current $J^\mu$ must be a conserved current:
\begin{equation}
\label{eq:conscurr}
\nabla_\mu J^\mu=0.
\end{equation}

The action may posses several internal symmetries, for instance related to electric or baryon charge
conservation. One peculiar case of such charges is chiral charge, also called axial charge. In the
case of massless field ($m=0$) the Dirac Lagrangian~(\ref{eq:DiracLcurved}) is invariant under the unitary
``chiral'' transformation defined by:
\begin{equation*}
\psi\to\psi'=\E^{\I\alpha\gamma^5}\psi,
\end{equation*}
where $\gamma^5$ is obtained from flatspace gamma matrices $\gamma^5=\I\gamma^0\gamma^1\gamma^2\gamma^3$.
This correspond to the infinitesimal change of the fields:
\begin{equation*}
\delta_\alpha\psi=\I \alpha\gamma^5 \psi,\quad \delta_\alpha\bar\psi=\I \alpha\bar\psi\gamma^5 .
\end{equation*}
The explicit variation of the action with Lagrangian (\ref{eq:DiracLcurved}) is
\begin{equation*}
\delta_\alpha S=-\int \D^d x\,e\left[\left(\bar\psi \underline{\gamma}^\mu\gamma^5\psi\right)\de_\mu\alpha+2\I m\bar\psi\gamma^5\psi\,\alpha\right],
\end{equation*}
which upon part integration of the $\de_\mu\alpha$ term gives the conservation equation
\begin{equation}
\label{eq:consAxialcurr}
\nabla_\mu J\ped{A}^\mu=2\I m\bar\psi\gamma^5\psi,
\end{equation}
where we defined
\begin{equation}
\label{eq:AxialCurr}
J\ped{A}^\mu\equiv \bar\psi\underline{\gamma}^\mu\gamma^5\psi
\end{equation}
the axial (or chiral) current. The relation (\ref{eq:consAxialcurr}) can also be obtained
from equation of motions starting from the current definition (\ref{eq:AxialCurr}). What is
special about axial current is that, contrary to electric current, the conservation equation
holds only at classical level and may be altered when quantum processes are considered.
When fermions interacts with some gauge fields~\cite{Adler:2004qt} or in presence of 
space-time curvature~\cite{Parker:2009uva}, quantum processes generate axial current
and the net charge is not conserved. The current is said to be anomalous and, in $d=4$
dimensions, the axial current divergence is
\begin{equation}
\label{eq:ChiralAnomaly}
\nabla_\mu J^\mu\ped{A}=2\I m\bar\psi\gamma^5\psi-\frac{1}{8}\epsilon^{\mu\nu\lambda\sigma}\frac{q^2}{2\pi^2}F_{\mu\nu}F_{\lambda\sigma}
+\frac{1}{8}\epsilon^{\mu\nu\lambda\sigma}\frac{1}{48\pi^2}R^\alpha_{\,\beta\mu\nu}R^\beta_{\,\alpha\lambda\sigma}.
\end{equation}
In this work, we mostly consider Dirac fermions with zero mass and in flat space-time, therefore the
only source of axial non conservation can only come from gauge fields. When gauge fields 
are external electromagnetic field, there exist configurations of electric and
magnetic fields such that the second term in Eq.~(\ref{eq:ChiralAnomaly}) is vanishing
and the chiral charge is indeed conserved. Replacing the definitions of strength tensor
and of electric $E$ and magnetic $B$ field, we find that term is written as
\begin{equation*}
\frac{1}{2}\epsilon^{\mu\nu\lambda\sigma}F_{\mu\nu}F_{\lambda\sigma}=4B\cdot E.
\end{equation*}
We therefore see that this term is vanishing when the electric and magnetic fields are orthogonal one to an other
or simply when one of the two is vanishing. If, on the other hand, this term is not vanishing or
we are dealing with dynamical gauge fields, the right hand side of Eq. (\ref{eq:ChiralAnomaly}) at zero mass
can be written as the divergence of the so-called Chern-Simons current~\cite{Flachi:2017vlp}:
\begin{equation*}
\begin{split}
K^\mu=&\frac{q^2}{8\pi^2}\epsilon^{\mu\nu\rho\sigma}A_\nu F_{\rho\sigma}
	-\frac{1}{8}\epsilon^{\mu\nu\rho\lambda}\frac{1}{48\pi^2}\Gamma^\alpha_{\,\nu\beta}\left(\de_\rho \Gamma^\beta_{\alpha\lambda}
	+\frac{2}{3}\Gamma^\beta_{\,\rho\sigma}\Gamma^\sigma_{\,\alpha\lambda}\right),\\
\nabla^\mu K_\mu =& \frac{1}{8}\epsilon^{\mu\nu\lambda\sigma}\frac{q^2}{2\pi^2}F_{\mu\nu}F_{\lambda\sigma}
	-\frac{1}{8}\epsilon^{\mu\nu\lambda\sigma}\frac{1}{48\pi^2}R^\alpha_{\,\beta\mu\nu}R^\beta_{\,\alpha\lambda\sigma}
\end{split}
\end{equation*}
where $\Gamma$'s are the Christoffel symbols and the integral of Chern-Simons current can be embodied
in the definition of a gauge-invariant overall conserved axial charge including both fermion and gauge
contribution:
\begin{equation*}
J^\mu\ped{CS}=J^\mu\ped{A}+K^\mu,\quad
\de_\mu J^\mu\ped{CS}=0.
\end{equation*}

Moving to other symmetries, we also want our action to be compatible with relativity principles, therefore we ask that our description is
unaffected by the local inertial frame chosen.  The action is then invariant under Lorentz transformations:
\begin{equation*}
\delta_\Omega\, e^{\;a}_\mu=\Omega^a_{\;b}(x)e^{\;b}_\mu,\quad
\delta_\Omega\psi=-\frac{\I}{2}\Omega^{ab}(x)\Sigma_{ab}\psi,\quad
\delta_\Omega\bar\psi=\frac{\I}{2}\Omega^{ab}(x)\bar\psi\Sigma_{ab},
\end{equation*}
where $\Omega_{ab}=-\Omega_{ba}$ denotes an infinitesimal local rotation angle and $\Sigma_{ab}$ are
the generators of Lorentz group. Again, taking advantage of the Eq.s of motions, the action variation is
given by
\begin{equation*}
\begin{split}
\delta_\Omega\, S =&\int \D^d x\, e\,T^\mu_{\;a}\,\, \delta_\Omega\, e^{\;a}_\mu=-\int \D^d x\, e\,T^{ab} \Omega_{ab}\\
=&-\int \D^d x\, e\frac{1}{2}\l(T^{ab}-T^{ba}\r)\Omega_{ab}.
\end{split}
\end{equation*}
Therefore, Lorentz invariance requires that the stress-energy-tensor is symmetric under the exchange of local
Lorentz indices:
\begin{equation}
\label{eq:setsym}
T^{ab}-T^{ba}=0.
\end{equation}

At last, the action must be invariant under diffeomorphism transformations. Consider then the general
infinitesimal coordinate transformation
\begin{equation*}
x^\mu \to x'^\mu=x^\mu-\xi^\mu(x)
\end{equation*}
where $\xi$ is chosen such that it is vanishing on the boundary region. Under this transformation,
the variation of the fields are given by the Lie derivative along $\xi^\mu$
\begin{gather*}
\mc{L}_\xi e^a_{\,\mu}\equiv e'^a_{\,\mu}(x)-e^a_{\,\mu}(x)=\xi^\nu\nabla_\nu e^a_{\,\mu}+e^a_{\,\nu}\nabla_\mu\xi^\nu\\
\mc{L}_\xi A_\mu\equiv A'_\mu(x)-A_\mu(x)=\xi^\nu\nabla_\nu A_\mu+A_\nu\nabla_\mu\xi^\nu\\
\mc{L}_\xi\psi\equiv\psi'(x)-\psi(x)=\xi^\nu\de_\nu\psi\\
\mc{L}_\xi\bar\psi\equiv\bar\psi'(x)-\bar\psi(x)=\xi^\nu\de_\nu\bar\psi.
\end{gather*}
Consequently, the action transforms as follows
\begin{equation*}
\begin{split}
\delta_\xi S=&\int\D^d x\, e\l[ T^\mu_{\;a}\mc{L}_\xi e^{\;a}_\mu +J^\mu \mc{L}_\xi A_\mu\r]\\
=&\int\D^d x\, e\l[T^\mu_{\;a}\l(\xi^\nu\nabla_\nu e^{\;a}_\mu+e^{\;a}_\nu\nabla_\mu\xi^\nu \r)+J^\mu\l(\xi^\nu\nabla_\nu A_\mu+A_\nu \nabla_\mu\xi^\nu\r) \r]\\
=&-\int\D^d x\, e\l[\l(\nabla_\mu T^\mu_{\;\nu}-F_{\nu\lambda}J^\lambda\r)\xi^\nu \r]    +\int\D^d x\, e\, T_{ab}\,\omega^{ab}_\nu\,\xi^\nu\\
&+\int\D^d x\, e\nabla_\mu\l[\l(T^\mu_{\;\nu}+J^\mu A_\nu\r)\xi^\nu \r]-\int\D^d x\, e\,\xi^\nu A_\nu\nabla_\mu J^\mu,
\end{split}
\end{equation*}
where we used the identity~\cite{BirrellDavies}
\begin{equation*}
D_\mu e^{\;a}_\nu=\nabla_\mu e^{\;a}_\nu +\omega^{\;a}_{\mu\;b}e^{\;b}_\nu=0.
\end{equation*}

The last term in the action variation vanishes because $J$ is a conserved current, the third term because
$\xi$ is vanishing on the boundary and the second term because the stress-energy tensor is symmetric,
while spin-connection is anti-symmetric. Therefore, diffeomorphism invariance impose the following
conservation Eq.s for the stress-energy tensor:
\begin{equation}
\label{eq:consset}
\nabla_\mu T^\mu_{\;\nu}=F_{\nu\lambda}J^\lambda.
\end{equation}
If $J$ is an electric current, then $F_{\nu\lambda}J^\lambda$ is the external Lorentz force which drives
the energy and the momentum of the system. It is trivial to notice that in absence of external gauge (force)
the stress-energy tensor is conserved.

\section{Local equilibrium}
\label{sec:LocEquil}
In the previous sections we provided a space-time foliation in order to access local covariant
quantities and we derived the conserved quantities related to the invariance of coordinate, Lorentz
and U(1) transformations. We can now proceed to define the covariant local equilibrium density operator.

Choose a foliation of space-time and suppose that the system in consideration thermalize faster than
the evolution of ``time'' $\tau$ in which we are interested. Then, at each step of evolution $\D\tau$,
the system is at local thermal equilibrium and the macroscopic behavior of the system is described by
a stress-energy density $T_{\mu\nu}(x)$ and a current density $j_\mu(x)$ lying on the space-like
hyper-surface $\Sigma(\tau)$. Now, we want to describe the system based on the density operator which
lives on $\Sigma(\tau)$. As in the non-relativistic case, the local density operator at local
equilibrium $\h\rho\ped{LE}$ is defined as the operator which maximizes the entropy $S=-\tr(\h\rho\log\h\rho)$.
As for the constraints, since we want to reproduce the thermodynamics on the hyper-surface,
we impose that the mean values of the stress-energy tensor and of the current on $\Sigma(\tau)$
corresponds to the actual values of the densities $T_{\mu\nu}(x)$ and $j_\mu(x)$. To obtain these
densities, we project the stress–energy tensor and the current mean values onto the normalized vector
perpendicular to $\Sigma$:
\begin{equation}
\label{eq:ConsRel}
\begin{split}
n_\mu(x)\tr(\h{\rho}\,\,\h{T}^{\mu\nu}(x))=&\,n_\mu(x)\mean{\h{T}^{\mu\nu}(x)}\equiv n_\mu(x)T^{\mu\nu}(x),\\
n_\mu(x)\tr(\h{\rho}\,\,\h{j}^{\mu}(x))=&\,n_\mu(x)\mean{\h{j}^\mu(x)}\equiv n_\mu(x)j^\mu(x).
\end{split}
\end{equation}
In addition to the energy, momentum, and charge densities, one should include the angular momentum density,
but if the stress–energy tensor is the Belinfante, this further constraint is redundant and can be disregarded~\cite{Becattini:2018duy}.
Therefore, from now on, the stress-energy tensor is considered to be symmetric.

At any fixed $\tau$, the maximization of entropy is obtained introducing the Lagrange multiplier
functions $\beta^{\nu}(x)$ and $\zeta(x)$ and finding the extreme with respect to $\h{\rho}$ for the
functional $F$:
\begin{equation*}
\begin{split}
F[\,\h\rho\,]=&-\tr\left(\,\h{\rho}\log\h{\rho}\,\right)+\\
&+\int_{\Sigma(\tau)} \D\Sigma \; n_{\mu}(x)\left[\left(\braket{\h{T}^{\mu\nu}(x)}-T^{\mu\nu}(x)\right)\beta_{\nu}(x)
-\left(\braket{\h{j}^{\mu}(x)}-j^{\mu}(x) \right)\zeta(x)\right].
\end{split}
\end{equation*}  
The maximum solution $\h{\rho}\ped{LE}$ gives the \emph{Local Equilibrium Density Operator} (LEDO)~\cite{BecaBetaF,Hayata2015}:
\begin{equation}
\label{eq:LEDO}
\h{\rho}\ped{LE}=\dfrac{1}{\parz\ped{LE}}\exp\left[-\int_{\Sigma(\tau)}\D\Sigma \; n_{\mu}(x)\left(\h{T}^{\mu\nu}(x)\beta_{\nu}(x)
-\h{j}^{\mu}(x)\zeta(x)\right)\right],
\end{equation}
where $\mc{Z}\ped{LE}$ is chosen such that the density operator is normalized, i.e. $\tr(\,\h\rho\ped{LE})=1$.
The physical meaning of the four-vector $\beta$ and the scalar $\zeta$ are respectively those of the temperature
four-vector and the ratio between local chemical potential and temperature. They are obtained
as solution of the constraints (\ref{eq:ConsRel}) with $\h\rho=\h\rho\ped{LE}$:
\begin{equation*}
n_\mu(x) T^{\mu\nu}\ped{LE}[\beta,\zeta,n]= n_\mu(x)T^{\mu\nu}(x),\quad
n_\mu(x)j^\mu\ped{LE}[\beta,\zeta,n]= n_\mu(x)j^\mu(x).
\end{equation*}
These equations identify a time-like four-vector $\beta$ which in turn can be used as hydrodynamic frame
by choosing the unitary fluid four-velocity $u$ as
\begin{equation*}
u=\frac{\beta}{\sqrt{\beta^2}}=T\beta,
\end{equation*}
referred as $\beta$-frame~\cite{BecaBetaF} or thermodynamic frame~\cite{Jensen:2012jh}.

It should be noted that $\h\rho\ped{LE}$ is not the true density operator. Indeed,
the true density operator in Heisenberg picture is time-independent, while the
operator in Eq. (\ref{eq:LEDO}) is explicitly dependent on time through the time
dependence of the operators. The true density operator can be derived from  $\h\rho\ped{LE}$
assuming that at a certain time $\tau_0$ the system is at local thermal equilibrium.
This means that at that particular time the true density operator coincides with the
LEDO. Then the true statistical operator is obtained evolving the system in $\tau$~\cite{BecaBetaF,Becattini:2019dxo}. 

We are instead interested in the system when it reaches the global thermal equilibrium.
This is obtained when the LEDO is time-independent and do not depend on the parametrization
chosen for the foliation.

\section{General global equilibrium}
\label{sec:GenGlobEquil}
We want to make the LEDO~(\ref{eq:LEDO}) independent of time parametrization and consequently
stationary. There exist peculiar configurations of the $\beta$ and $\zeta$ fields which make
the exponent of Eq.~(\ref{eq:LEDO}) divergent free~\cite{BecCov}. As we are now proving, this is sufficient
to keep the statistical operator independent of the foliation.

Choose any foliation and consider two space-like hyper-surfaces at different time: $\Sigma(\tau)$
and $\Sigma(\tau+\Delta\tau)$. As in figure~\ref{fig:foliation}, we call $\Omega$ the region of
space-time enclosed by those hyper-surfaces and by $\Theta$ the time-like part of the boundary $\de\Omega$.
We integrate the four-divergence of the local operator in the exponent of the LEDO~(\ref{eq:LEDO})
over the region $\Omega$; the integral can be turned by Gauss theorem into the flux through the boundary:
\begin{equation*}
\int_{\Omega}\D \Omega\; \nabla_{\mu}\left(\h{T}^{\mu\nu}(x)\beta_{\nu}(x)-\h{j}^{\mu}(x)\zeta(x) \right)
=\int_{\de\Omega}\D\Sigma_\mu \left(\h{T}^{\mu\nu}(x)\beta_{\nu}(x) -\h{j}^{\mu}(x)\zeta(x) \right).
\end{equation*}
The whole flux can be decomposed into the sum of the single fluxes over each boundary surface of region $\Omega$:
\begin{equation*}
\begin{split}
\int_{\de\Omega}&\D\Sigma_\mu \left(\h{T}^{\mu\nu}(x)\beta_{\nu}(x)-\h{j}^{\mu}(x)\zeta(x) \right)=
\int_{\Theta}\D\Sigma_\mu \left(\h{T}^{\mu\nu}(x)\beta_{\nu}(x)-\h{j}^{\mu}(x)\zeta(x) \right)\\
&+\int_{\Sigma(\tau+\Delta\tau)}\!\!\!\!\D\Sigma_\mu \left(\h{T}^{\mu\nu}(x)\beta_{\nu}(x)-\h{j}^{\mu}(x)\zeta(x) \right)
-\int_{\Sigma(\tau)}\!\!\D\Sigma_\mu\left(\h{T}^{\mu\nu}(x)\beta_{\nu}(x)-\h{j}^{\mu}(x)\zeta(x) \right).
\end{split}
\end{equation*}
The first term is vanishing for a suitable boundary condition of the fields $\beta$ and $\zeta$.
As found in section~\ref{sec:Foliation} for a charge of a system, the shift by time evolution of the local
operator in the exponent is given by the integral of a divergence:
\begin{equation*}
\begin{split}
\int_{\Omega}&\D \Omega\; \nabla_{\mu}\left(\h{T}^{\mu\nu}(x)\beta_{\nu}(x)-\h{j}^{\mu}(x)\zeta(x) \right)=
+\int_{\Sigma(\tau+\Delta\tau)}\D\Sigma_\mu \left(\h{T}^{\mu\nu}(x)\beta_{\nu}(x)-\h{j}^{\mu}(x)\zeta(x) \right)\\
&-\int_{\Sigma(\tau)}\D\Sigma_\mu\left(\h{T}^{\mu\nu}(x)\beta_{\nu}(x)-\h{j}^{\mu}(x)\zeta(x) \right).
\end{split}
\end{equation*}
It follows that the vanishing of the integrand on the l.h.s is a sufficient condition for the exponent of
the LEDO to be time independent. Once it is time independent, it does not matter at which time $\tau$ are
evaluating your local operator and by extensions it does not matter which time functions $\tau(x)$ you
picked for space-time foliation. In other words, it is also independent on the foliation chosen.

The specific configurations of the $\beta$ and $\zeta$ fields that make the four-divergence vanishing
are obtained taking advantage of the conservation equation of stress-energy tensor~(\ref{eq:consset})
and current~(\ref{eq:conscurr}). In the following derivation, we are also including a conserved axial
current $\h{j}\ped{A}$ with a corresponding Lagrange multiplier $\zeta\ped{A}(x)$. As discussed in
Sec.~\ref{sec:Symm}, for a massless Dirac field, even in the presence of anomalies, we can always define
a conserved current $\h{j}\ped{CS}$ related to chiralty by means of Chern-Simons current. Therefore, when we are
considering massless fermions affected by external or dynamic gauge fields, we always implicitly refer to global
equilibrium with chiral imbalance related to conserved current $\h{j}\ped{CS}$. In the case of massive field,
or more generally when a conserved axial current can not be defined, global thermal equilibrium with a
non-vanishing Lagrange multiplier can not be reached because the axial current will make the statistical
operator dependent on time. In that case, we can simply set $\zeta\ped{A}=0$ so that the following expressions
and arguments remain valid. To enforce a vanishing four-divergence the $\beta$, $\zeta$ and $\zeta\ped{A}$ field
must satisfy 
\begin{equation*}
\begin{split}
0=&\nabla_{\mu}\left(\h{T}^{\mu\nu}(x)\beta_{\nu}(x)-\h{j}^{\mu}(x)\zeta(x)-\h{j}\ped{A}^{\mu}(x)\zeta\ped{A}(x) \right)\\
=& \beta_\nu(x)\nabla_\mu \h T^{\mu\nu}(x) +\h T^{\mu\nu}(x) \nabla_\mu \beta_\nu(x)\\
&-\h j^\mu(x)\nabla_\mu \zeta(x)-\zeta(x)\,\nabla_\mu \h j^\mu(x)-\h j\ped{A}^\mu(x)\nabla_\mu \zeta\ped{A}(x)-\zeta\ped{A}(x)\,\nabla_\mu \h j\ped{A}^\mu(x)\\
=&\h T^{\mu\nu}(x) \nabla_\mu \beta_\nu(x)-\left[\nabla^\mu \zeta(x)-F^{\nu\mu}(x)\beta_\nu(x)\right]\h j_\mu(x)-\nabla_\mu \zeta\ped{A}(x)\h j^\mu\ped{A}(x)=0
\end{split}
\end{equation*}
Then, using the fact that stress-energy tensor is symmetric, we see that the
previous quantity is vanishing if the $\beta$, $\zeta$ and $\zeta\ped{A}$ fields satisfy:
\begin{equation*}
\nabla_{\mu}\beta_{\nu}(x)+\nabla_{\nu}\beta_{\mu}(x)=0,\qquad \nabla^\mu \zeta(x)=F^{\nu\mu}(x)\beta_\nu(x)
,\qquad \nabla^\mu \zeta\ped{A}(x)=0.
\end{equation*}
The first equation impose that $\beta$ is a Killing vector field. The second equation has a simple physical
interpretation. In the beta frame, the fluid velocity $u$ is the unit four vector directed along the four 
inverse-temperature $\beta$. Starting from that time-like vector we can define the comoving electric
and magnetic fields: 
\begin{equation}
\label{eq:ComvEandB}
E^\mu=F^{\mu\lambda}u_\lambda,\quad B^\mu=\frac{1}{2}\epsilon^{\mu\nu\rho\sigma}F_{\nu\rho}u_\sigma;\quad
F^{\mu\nu}=E^\mu u^\nu-E^\nu u^\mu-\epsilon^{\mu\nu\rho\sigma}B_\rho u_\sigma.
\end{equation}
Indicating with $T(x)$ the temperature $1/T(x)=\sqrt{\beta^2(x)}$, the second condition can be written as
\begin{equation*}
T(x)\nabla^\mu \left(\frac{\mu(x)}{T(x)}\right)-E^\mu(x)=0,
\end{equation*}
meaning that a system subject to an external electric field will develop a gradient of $\mu/T$ in order
to compensate the applied field and ensure that the equilibrium is maintained. It is also worth noticing that
even in presence of a comoving magnetic field if the comoving electric field is vanishing, equilibrium
is reached only with a constant $\zeta$ filed. The condition on $\zeta\ped{A}$ simply states that
the ratio between chiral chemical potential and temperature $\zeta\ped{A}=\mu\ped{A}(x)/T(x)$ must
be constant.

In flat space-time, the general solution for the Killing equation $\de_\mu\beta_\nu+\de_\mu\beta_\mu=0$
is given in terms of a constant vector $b_{\nu}$ and a constant anti-symmetric tensor $\varpi_{\mu\nu}$:
\begin{equation}
\label{eq:betakilling}
\beta_{\mu}(x)=b_{\mu}+\varpi_{\mu\nu}x^{\nu}.
\end{equation}
The anti-symmetric tensor $\varpi_{\mu\nu}$ is dubbed \emph{thermal vorticity} and can be expressed
by means of Eq.~(\ref{eq:betakilling}) as an exterior derivative of the $\beta$ field:
\begin{equation*}
\varpi_{\mu\nu}=-\frac{1}{2}\left(\de_\mu\beta_\nu-\de_\nu\beta_\mu\right).
\end{equation*}
Since in this chapter we are introducing global equilibrium with vorticity, in order to not
overburden the chapter, we now deal with a vanishing electromagnetic field.
The global equilibrium with both vorticity and external electromagnetic
field is discussed in Sec.~\ref{sec:GlobEqEM}. Consider then a system without an external gauge
field. In this case, at global equilibrium the $\zeta$ field must be a constant. Inserting the
configurations (\ref{eq:betakilling}), $\zeta=$const. and $\zeta\ped{A}=$const. into the
LEDO~(\ref{eq:LEDO}), we find the general global equilibrium density operator:
\begin{equation*}
\h{\rho}\ped{GE}=\dfrac{1}{\parz\ped{GE}}\exp\left[-\int_{\Sigma(\tau)}\D\Sigma \; n_{\mu}(x)\left(\h{T}^{\mu\nu}(x)\left(b_{\nu}+\varpi_{\nu\rho}x^{\rho}\right)
-\h{j}^{\mu}(x)\zeta-\h{j}\ped{A}^{\mu}(x)\zeta\ped{A}\right)\right].
\end{equation*}
The quantities $b$, $\varpi$, $\zeta$ and $\zeta\ped{A}$ are constant and can be taken outside the integral.
The integration in the exponential reproduce the Poincaré generators, which are conserved quantities.
The term linear in $b$ reproduces the four-momentum of the system which is the generator of translation
and is given by:
\begin{equation*}
\h{P}^{\nu}=\displaystyle\int_{\Sigma(\tau)}\D\Sigma_{\mu}(x) \h{T}^{\mu\nu}(x).
\end{equation*}
Using the antisymmetry of $\varpi$, the term in vorticity is written as 
\begin{equation*}
\h{T}^{\mu\nu}\varpi_{\nu\rho}x^{\rho}=\frac{1}{2}\left(\h{T}^{\mu\nu}x^{\rho}-\h{T}^{\mu\rho}x^{\nu}\right)\varpi_{\nu\rho}
\end{equation*}
and upon integration, it reproduces the Lorentz generators:
\begin{equation}
\label{eq:LorenzGenerators}
\h{J}^{\mu\nu}=\int_\Sigma \D\Sigma_\lambda(x^\mu\h{T}^{\lambda\nu}-x^\nu\h{T}^{\lambda\mu}).
\end{equation}
The term proportional to $\zeta$ ($\zeta\ped{A}$) gives the electric (axial) charge:
\begin{equation}
\label{eq:ConsCharges}
\h{Q}=\int_{\Sigma(\tau)}\D\Sigma_{\mu}(x) \h{j}^{\mu}(x),\quad \h{Q}\ped{A}=\int_{\Sigma(\tau)}\D\Sigma_{\mu}(x) \h{j}\ped{A}^{\mu}(x).
\end{equation}
The whole global equilibrium density operator is therefore written in terms of conserved quantities
of the system:
\begin{equation}
\label{eq:GEDO}
\h{\rho}\ped{GE}=\dfrac{1}{\parz\ped{GE}}\exp\left[-b_\mu\h{P}^\mu+\frac{1}{2}\varpi:\h{J}+\zeta\h{Q}+\zeta\ped{A}\h{Q}\ped{A}\right].
\end{equation}
This general global equilibrium can be regarded both as a local equilibrium with a specific local
temperature $\beta(x)$ or as a true global equilibrium with time independent entropy. Excluding the
parameters $\zeta$ and $\zeta\ped{A}$ which correspond to internal symmetries, the density operator depends on 10
parameters, corresponding to the 4 components of $b_{\mu}$ and the 6 independent components of
the anti-symmetric tensor $\varpi_{\mu\nu}$. This is the maximum number of first integrals
for a relativistic system, each one corresponding to a generator of the Poincaré group.

Given a fixed point in space-time $x$, we can write the general equilibrium density operator in terms
of the inverse temperature evaluated at that point. First, taking advantage of Poincaré algebra
identities~\cite{WeinField}, we evaluate the angular momentum operator at $x$ by using the
translational operator $\h {\sf T}(x)=\exp[\,\I x\cdot\h{P}\,]$:
\begin{equation}\label{jshift}
\begin{split}
\h J^{\mu\nu}_x \equiv&  \h {\sf T}(x) \h J^{\mu\nu} \h {\sf T}^{-1}(x)=\h J^{\mu\nu}-x^{\mu}\h P^{\nu}+x^{\nu}\h P^{\mu}\\
&=\int_\Sigma \D \Sigma_{\lambda} \left[ (y^\mu -x^{\mu})\h T^{\lambda\nu}(y)-(y^\nu - x^{\nu})\h T^{\lambda\mu}(y)\right].
\end{split}
\end{equation}
With this definition and writing $b$ in terms of $\beta(x)$ thanks to Eq. (\ref{eq:betakilling}), the density
matrix takes the form
\begin{equation}
\label{eq:GEDObeta}
\h{\rho}\ped{GE}=\dfrac{1}{\parz\ped{GE}}\exp\left[-\beta_\mu(x)\h{P}^\mu+\frac{1}{2}\varpi:\h{J}_x+\zeta\h{Q}+\zeta\ped{A}\h{Q}\ped{A}\right].
\end{equation}
This form will be useful for evaluating mean values of observables around the point in $x$ as a perturbative
expansion in thermal vorticity. 

The statistical operator (\ref{eq:GEDO}) describes a stationary global equilibrium that in presence of vorticity and it
is not homogeneous. Indeed, the four-temperature $\beta$ in Eq. (\ref{eq:betakilling}) depends linearly on the coordinate $x$.
However for a vanishing thermal vorticity $\varpi$ the inverse temperature is a constant $\beta=b$  and the usual
homogeneous global thermal equilibrium is recovered:
\begin{equation*}
\h{\rho}=\dfrac{1}{\parz}\exp\left[-\beta_\mu\h{P}^\mu+\zeta\h{Q}+\zeta\ped{A}\h{Q}\ped{A}\right].
\end{equation*}
Moreover, specializing the thermal vorticity into the values $b_{\mu}=(1/T,\mathbf{0})$ and
$\varpi_{\mu\nu}=\omega/T(\delta^{1}_{\mu}\delta^2_{\nu}-\delta^2_{\mu}\delta^1_{\nu})$ we recover
the covariant form of the statistical operator used for rotating systems in~\cite{LandauStat}:
\begin{equation*}
\h{\rho}_\omega=\dfrac{1}{\parz_\omega}\exp\left(- \h{H}/T+\omega \h{J}_z/T+\mu\h{Q}/T\right),
\end{equation*}
where $\h{J}_z$ is the angular momentum of the system along the $z$ axis. This operator has been
used to address quantum effects in relativistic matter under rotation~\cite{Vilenkin:1980zv,BecTin2011}.
Similarly, indicating with $\h{K}_z$ the generator of a Lorentz boost along the $z$ axis, the
parameter choice
\begin{equation*}
b_\mu=(1/T,0,0,0),\quad \varpi_{\mu\nu}=(a/T)(g_{0\mu}g_{3\nu}-g_{3\mu}g_{0\nu})
\end{equation*}
reproduces an ensemble representing the equilibrium of a relativistic fluid with constant comoving
acceleration $a$ along the $z$ direction~\cite{Becattini:2017ljh}:
\begin{equation*}
\h{\rho}_a=\frac{1}{\parz_a}\exp\l(-\h{H}/T+a\h{K}_z/T\r).
\end{equation*}
%

\section{Thermal vorticity and Lie transported basis}
\label{sec:ThermalVort}
Here we show that the anti-symmetric tensor $\varpi$, coming from the Killing solution
of inverse temperature, has indeed the physical meaning of fluid acceleration and rotation.
If the $\beta$ four-vector (\ref{eq:betakilling}) is a time-like vector, then we
can choose the $\beta$-frame as hydrodynamic frame~\cite{BecCov,Van:2013sma}. 
The unitary four-vector fluid velocity $u$ is therefore identified with the direction of $\beta$:
\begin{equation*}
u^\mu(x)=\frac{\beta^\mu(x)}{\sqrt{\beta_\rho(x)\beta^\rho(x)}}.
\end{equation*}
As long as we are considering physical observables in a region where the coordinate $x$ are
such that $\beta(x)$ is a time-like vector, this choice is perfectly legit and free of issues.

We can decompose the thermal vorticity into two space-like vector fields, each having three
independent components, by projecting along the time-like fluid velocity $u$.
This is analogous to the decomposition of the also anti-symmetric electromagnetic strength tensor
$F$ into an electric and magnetic part. For the thermal vorticity $\varpi$, the decomposition
\begin{equation*}
\varpi^{\mu\nu}=\epsilon^{\mu\nu\rho\sigma}w_\rho u_\sigma+\alpha^\mu u^\nu-\alpha^\nu u^\mu
\end{equation*}
defines the four-vectors $\alpha$ and $w$ which are explicitly written inverting the previous relation: 
\begin{equation}
\label{eq:accvort}
\alpha^\mu(x)\equiv\varpi^{\mu\nu}u_\nu,\quad w^\mu(x)\equiv-\frac{1}{2}\epsilon^{\mu\nu\rho\sigma}\varpi_{\nu\rho}u_\sigma.
\end{equation}
Notice that $w$ has a minus sign compared to comoving magnetic field in Eq.~(\ref{eq:ComvEandB}).
The vectors $\alpha$ and $w$ depends on the coordinate and are space-like and orthogonal to $u$.
All the quantity $u,\,\varpi,\,\alpha,\,w$ are dimensionless. A physical interpretation is easily
assigned to the vectors $\alpha$ and $w$ if we express them as derivative of the fluid velocity.

First, notice that, at global equilibrium, the thermal vorticity is just the derivative of $\beta$
four-vector: $\varpi_{\mu\nu}=\de_\nu\beta_\mu$. Then, it easily realized that comoving temperature
along the flow lines does not changes:
\begin{equation*}
u_\nu \de^\nu \s{\beta^2}=\frac{u_\nu\beta^\rho\de^\nu\beta_\rho}{\s{\beta^2}}=
\frac{u_\nu\beta^\rho\varpi^\nu_\rho}{\s{\beta^2}}= u^\rho\alpha_\rho=0.
\end{equation*}
The $\alpha$ vector is then written as
\begin{equation*}
\begin{split}
\alpha^\mu=&\varpi^{\mu\nu}u_\nu=u_\nu\de^\nu\beta^\mu=u^\mu u_\nu\de^\nu \s{\beta^2}+\s{\beta^2}u_\nu \de^\nu u^\mu\\
=&\s{\beta^2}u_\nu\de^\nu u^\mu=\frac{1}{T}a^\mu,
\end{split}
\end{equation*}
where $T=1/\s{\beta^2}$ is the proper temperature and $a^\mu=u_\nu\de^\nu u^\mu$ is the acceleration
field of the fluid. Therefore, the vector $\alpha$ is the local acceleration of the fluid divided
by the temperature. The four-vector $w$ is instead expressed as the local rotation of the fluid
$\omega$, sometimes called local vorticity,
\begin{equation*}
\omega^\mu\equiv\frac{1}{2}\epsilon^{\mu\nu\rho\sigma}u_\sigma \de_\nu u_\rho.
\end{equation*}
From the definition of $w$ we indeed recover $\omega$
\begin{equation*}
\begin{split}
w^\mu&=-\frac{1}{2}\epsilon^{\mu\nu\rho\sigma}\varpi_{\nu\rho}u_\sigma=\frac{1}{2}\epsilon^{\mu\nu\rho\sigma}\de_\nu\beta_\rho u_\sigma\\
&=\frac{1}{2}\epsilon^{\mu\nu\rho\sigma}\s{\beta^2}u_\sigma \de_\nu u_\rho=\frac{1}{T}\omega^\mu,
\end{split}
\end{equation*}
and we see that $w$ is the local rotation divided by comoving temperature.

It also proves useful to define the projector into orthogonal space of fluid velocity:
\begin{equation*}
\Delta_{\mu\nu}\equiv g_{\mu\nu}-u_\mu u_\nu,
\end{equation*}
and the four vector $\gamma$ orthogonal to the other ones: $u,\alpha,w$
\begin{equation}
\label{eq:transversedir}
\gamma^\mu=\epsilon^{\mu\nu\rho\sigma}w_\nu\alpha_\rho u_\sigma=(\alpha\cdot \varpi)_\lambda \Delta^{\lambda\mu}.
\end{equation}
The $\varpi$ decomposition therefore defines a tetrad $\{u,\alpha,w,\gamma\}$ that can be used as a basis.
It must be noticed, however, that the tetrad is neither unitary nor orthonormal, indeed in general we
have $\alpha\cdot w\neq 0$.

To asses the order of magnitude of thermal vorticity in several physical systems, we express the
previous four-vectors in the local rest frame where local acceleration is $\vec{a}$ and local
angular velocity is $\vec{\omega}$:
\begin{equation*}
\alpha=\l(0,\f{\vec{a}}{T}\r),\quad w=\l(0,\f{\vec{\omega}}{T}\r),\quad \gamma=\l(0,\f{\vec{a}\wedge\vec{\omega}}{T}\r)
\end{equation*}
and hence restoring the physical constants
\begin{equation*}
\mod{\alpha}=\frac{\hslash \mod{\vec{a}}}{c\, k_B\, T},\quad \mod{w}=\frac{\hslash \mod{\vec{\omega}}}{k_B\, T},\quad
\mod{\gamma}=\frac{\hslash^2 \mod{\vec{a}\wedge\vec{\omega}}}{c\, k_B^2\, T^2}.
\end{equation*}
At human scales, an estimate of these parameters are obtained replacing room temperature $T=300$K,
an acceleration equal to  earth's gravitational acceleration and an angular velocity of $1 Hz$.
In this situation the modulus of $\alpha$ and $w$ are
\begin{equation*}
\mod{\alpha}\simeq 8\times 10^{-22},\quad \mod{w}\simeq 2\times 10^{-14}.
\end{equation*}
Therefore, we aspect the corrections from $\alpha$ and $w$ to be negligible for most common
systems in nature. However, in quark-gluon plasma produced by heavy ion collisions measures
of thermal vorticity indicate that it posses the values $\mod{\alpha}=0.05$ and $\mod{w}=0.06$
which are sufficient to induce observable effects~\cite{STAR:2017ckg,Adam:2018ivw}.
Nevertheless, the value of $\alpha$ and $w$ remains significantly smaller than 1, which validate
the use of perturbative expansion on thermal vorticity adopted in the following sections.

Now that we have defined the vectors $\alpha$ and $w$, we can write the global thermal equilibrium
statistical operator with explicit dependence of angular momenta and boost generators.
This is accomplished decomposing the Lorentz generator $\h{J}^{\mu\nu}$, which is also
antisymmetric, in the same fashion as the thermal vorticity:
\begin{equation*}
\h{J}^{\mu\nu}=u^\mu\h{K}^\nu-u^\nu\h{K}^\mu-u_\rho\epsilon^{\rho\mu\nu\sigma}\h{J}_\sigma;
\end{equation*}
as before this operation define the following conserved operators
\begin{equation}
\label{eq:boostrot}
\h{K}^\mu=u_\lambda\h{J}^{\lambda\mu},\quad
\h{J}^\mu=\f{1}{2}\epsilon^{\alpha\beta\gamma\mu}u_\alpha\h{J}_{\beta\gamma}.
\end{equation}
These are easily recognized as the local boost generators $\h{K}^\mu$ and as the local
angular momenta $\h{J}^\mu$. The contraction with thermal vorticity reproduce the vectors
related to acceleration and rotation
\begin{equation*}
\f{1}{2}\varpi_{\mu\nu}\h{J}^{\mu\nu}=-\alpha^\rho\h{K}_\rho-w^\rho\h{J}_\rho
\end{equation*}
and the statistical operator (\ref{eq:GEDO}) has the form
\begin{equation*}
\h{\rho}\ped{GE}=\dfrac{1}{\parz\ped{GE}}\exp\left[-b_\mu\h{P}^\mu-\alpha^\rho\h{K}_\rho-w^\rho\h{J}_\rho+\zeta\h{Q}+\zeta\ped{A}\h{Q}\ped{A}\right].
\end{equation*}

To conclude this section, we analyze the tetrad $\{u,\alpha,w,\gamma\}$ and when it really
forms a basis. More precisely, we are now building a tetrad from the $\beta$ field which is a
Lie transported basis along $\beta$, i.e. a basis composed by four-vectors that are
Lie transported along $\beta$. We remind that the Lie derivative of a four-vector
$V_\mu$ along another four-vector $\beta$ is defined as:
\begin{equation*}
\mc{L}_\beta\left(V_\mu\right)=\beta^\lambda \de_\lambda V_\mu +V_\lambda\de_\mu\beta^\lambda.
\end{equation*}
We are interested in the case of non-vanishing vorticity, where the $\beta$ four-vector has
the Killing form (\ref{eq:betakilling}). While it is known that the eigenvectors of a symmetric
non-singular matrix form a basis; the same statement is not true for an anti-symmetric
matrix such as $\varpi$. However, it is still possible to build a basis if we distinguish between
two cases: rotational and irrotational.

We start building the basis from the four vector $\beta$, given in terms of the constant
time-like four-vector $b^\mu$, the constant anti-symmetric tensor $\varpi$ and the
coordinate vector $x$:
\begin{equation*}
\beta^\mu=b^\mu+\varpi^{\mu\nu}x_\nu,\qquad \varpi^{\mu\nu}=\left(\begin{matrix} 0 & -a_1 & -a_2 & -a_3\\
a_1 & 0 & \Omega_3 & -\Omega_2\\ a_2 & -\Omega_3 & 0 & \Omega_1\\ a_3 & \Omega_2 & -\Omega_1 & 0 \end{matrix}\right),
\end{equation*}
with $a_1,a_2,a_3,\Omega_1,\Omega_2,\Omega_3$ real constants. To simplify the notation we can
choose a reference frame such that $b^\mu=(1,\vec{0})$. Then, the first vector that compose
the basis is $\beta$ itself (or its direction $u$), for which we know that $\mathcal{L}_\beta(u^\mu)= 0$.
As second vector, we can always choose the acceleration vector
\begin{equation*}
\alpha^\mu=\varpi^{\mu\nu}u_\nu.
\end{equation*}
As long as $\varpi$ is not vanishing, the vector $\alpha$ has at least one component and it
is well defined. This vector is both orthogonal to $u$ and has a vanishing Lie derivative
along $\beta$, $\mathcal{L}_\beta(\alpha^\mu)= 0$. We can then always pick $\alpha$ as
second vector of the Lie transported basis.

Then we consider the rotation vector $w$
\begin{equation*}
w^\mu=-\frac{1}{2}\epsilon^{\mu\nu\rho\sigma}\varpi_{\nu\rho}u_\sigma.
\end{equation*}
We can not always pick $w$ as a basis vector because it is vanishing if all the three constant
$\Omega_1,\Omega_2,\Omega_3$ are zero. We refer to the latter case as the \emph{irrotational}
case. On the other hand, if at least one of $\Omega_1,\Omega_2,\Omega_3$ is non-vanishing
we are in the \emph{rotational} case and $w$ is different from zero.

For the rotational case, we can pick $w$, which is Lie transported along $\beta$ and orthogonal
to $u$, albeit not to $\alpha$. Once we have three vectors, we can find a fourth vector that
is orthogonal to all the previous ones using Levi-Civita tensor:
\begin{equation*}
\gamma^\mu=\epsilon^{\mu\nu\rho\sigma}w_\nu\alpha_\rho u_\sigma.
\end{equation*}
Thanks to Leibniz rule, it follows from the definition that $\mathcal{L}_\beta(\gamma^\mu)= 0$.
As last step, since four vectors only form a basis if they are linearly independent,
we check that  $u,\alpha,w,\gamma$ are actually linear independent by solving the equations
\begin{equation*}
c_1\,u+c_2\,\alpha+c_3\,w+c_4\,\gamma=0
\end{equation*}
for $c_1,c_2,c_3,c_4$. We find that, unless $\Omega_1,\Omega_2,\Omega_3$ are all vanishing,
the solution is indeed $c_1=c_2=c_3=c_4=0$ and thus they are linear independent. The
four-vectors $u,\alpha,w,\gamma$ form a Lie transported basis in the rotational case.
With a Lie transported basis it is easy to give a four-vector 
that is Lie transported, indeed all the vectors $V$
\begin{equation*}
V^\mu=c_1(x)\,u^\mu+c_2(x)\,\alpha^\mu+c_3(x)\,w^\mu+c_4(x)\,\gamma^\mu
\end{equation*}
have vanishing Lie derivative along $\beta$ if
$\mathcal{L}_\beta(c_1(x))=\mathcal{L}_\beta(c_2(x))=\mathcal{L}_\beta(c_3(x))=\mathcal{L}_\beta(c_4(x))= 0$.

In the irrotational case ($\Omega_1=\Omega_2=\Omega_3=0$), there is not a special choice of vectors
that forms a Lie transported basis. Here, we provide one possible choice and we check that it satisfy
all the requirements. We introduce the vectors $V\ped{Irr}$ and $\gamma\ped{Irr}$
\begin{equation*}
\begin{split}
V\ped{Irr}^\mu=&\left(0,\, \frac{-a_2}{\sqrt{a_1^2+a_2^2}},\,\frac{a_1}{\sqrt{a_1^2+a_2^2}} ,\,0 \right),\\
\gamma\ped{Irr}^\mu=&\sqrt{\frac{a_1^2}{(a_1^2+a_2^2)(a_1^2+a_2^2+a_3^2)}}\left(0,\,-a_3,\,-\frac{a_2 a_3}{a_1},\frac{a_1^2+a_2^2}{a_1} \right).
\end{split}
\end{equation*}
It is straightforward to check that they are unitary $V\ped{Irr}\cdot V\ped{Irr}=\gamma\ped{Irr}\cdot\gamma\ped{Irr} =-1$
and that they are orthogonal to $u$ and $\alpha$ and between themselves. By direct computation we can also verify that the vectors
$V\ped{Irr}$ and $\gamma\ped{Irr}$ have vanishing Lie derivative along $\beta$ and together with $u$ and $\alpha$ forms
a linear independent set of four-vectors. Therefore $\{u,\alpha,V\ped{Irr},\gamma\ped{Irr}\}$ is an orthogonal Lie transported
basis. It is important to stress-out that in the rotational case the vectors $V\ped{Irr}$ and $\gamma\ped{Irr}$
loose both the properties of orthogonality with $u$ and $\alpha$ and the vanishing of Lie derivative along $\beta$;
only the previous rotational basis is a Lie transported basis in the rotational case. Back to the irrotational case,
a four-vector $V$ can be decomposed in
\begin{equation*}
V^\mu=c_1(x)\,u^\mu+c_2(x)\,\alpha^\mu+c_3(x)\,V\ped{Irr}^\mu+c_4(x)\,\gamma\ped{Irr}^\mu
\end{equation*}
and it is Lie transported along $\beta$ if
$\mathcal{L}_\beta(c_1(x))=\mathcal{L}_\beta(c_2(x))=\mathcal{L}_\beta(c_3(x))=\mathcal{L}_\beta(c_4(x))= 0$.

\section{Expansion on thermal vorticity}
\label{sec:VortExpan}
The main reason we have derived the statistical operator of a system at thermal equilibrium is that it connects
a local operator $\h{O}$ with its thermal mean value. Indeed, the mean value of a local operator $\h{O}$
is obtained tracing it with the statistical operator:
\begin{equation*}
\mean{\,\h{O}(x)\,}\ped{GE}=\tr(\h{\rho}\ped{GE}\,\h{O}(x))\ped{ren},
\end{equation*}
where the subscript indicates that non-physical divergences must be subtracted with a proper renormalization
procedure; for instance in the case of free fields it is sufficient to subtract the vacuum expectation values. 

One significant difference between generalized global equilibrium and global equilibrium without vorticity
is the lack of translational invariance of the former. Using the identity in~(\ref{jshift}), we can indeed
show that the generalized global statistical operator~(\ref{eq:GEDObeta}) transforms under translation
as following:
\begin{equation*}
\h {\sf T}(a)\, \h{\rho}\ped{GE} \h {\sf T}^{-1}(a)
=\dfrac{1}{\parz\ped{GE}}\exp\left[-\beta_\mu(x-a)\h{P}^\mu+\frac{1}{2}\varpi:\h{J}_x+\zeta\h{Q}+\zeta\ped{A}\h{Q}\ped{A}\right].
\end{equation*}
This is a consequence of the temperature not being homogeneous even if the system is at thermal equilibrium;
indeed, the temperature depends on the point according to Eq. (\ref{eq:betakilling}). This situation is exactly
what happens at thermal equilibrium in the case of gravitation, where the temperature measured by a local
observer depends on the gravitational potential where the measure is made. In that case, temperature follows
the Tolman-Ehrenfest law~\cite{Tolman:1930ona}, which states that for a static metric with a time-like Killing
vector  $\xi$ the product $T\cdot \sqrt{g_{\mu\nu}\xi^\mu\xi^\nu}$ remains constant. The temperature we
provided for a system with vorticity satisfy the Tolman-Ehrenfest law.

From the translated statistical operator we can still write the mean value of the operator $\h{O}$ at the point $x$
as a mean value of the operator at the point $x=0$:
\begin{equation*}
\begin{split}
\mean{\,\h{O}(x)\,}\ped{GE}=&\tr\left(\h{\rho}\ped{GE}\,\h{O}(x)\right)\\
=&\tr\left(\h {\sf T}^{-1}(x)\,\h{\rho}\ped{GE}\h {\sf T}(x)\h {\sf T}^{-1}(x)\,\h{O}(x)\h {\sf T}(x)\right)\\
=&\tr\left(\h {\sf T}^{-1}(x)\,\h{\rho}\ped{GE}\h {\sf T}(x)\,\h{O}(0)\right)\\
=&\dfrac{1}{\parz\ped{GE}}\tr\left(\exp\left[-\beta_\mu(x)\h{P}^\mu+\frac{1}{2}\varpi:\h{J}+\zeta\h{Q}+\zeta\ped{A}\h{Q}\ped{A}\right]\,\h{O}(0)\right).
\end{split}
\end{equation*}
From this relation, we see that, at small thermal vorticity, the leading term of the mean value of the
local operator at the point $x$ is equivalent to the thermal expectation values for homogeneous thermal equilibrium
with four-temperature given at point $x$. Since the solutions of thermal states for a homogeneous canonical system
are known, and thermal vorticity assumes only small values, we can derive the effects of thermal vorticity on mean
values of local operators adopting a perturbative series in thermal vorticity and expressing each term as a mean value
made with homogeneous statistical operator.

To develop an expansion on thermal perturbation we factorize the statistical operator into a homogeneous part and
into a part that contains thermal vorticity. We adopt the factorization adopted in~\cite{vanWeert1982}, which
explicitly preserve time-order products and which enables a straightforward use of imaginary-time thermal field
theory techniques. We consider the statistical operator in the form (\ref{eq:GEDObeta}) and we define the
operator $\h{A}$ and $\h{B}$ as:
\begin{equation*}
\h{\rho}\ped{GE}=\dfrac{1}{\parz\ped{GE}}\exp\left[\h{A}+\h{B}\right],\quad
\h{A}\equiv-\beta_\mu(x)\h{P}^\mu+\zeta\h{Q}+\zeta\ped{A}\h{Q}\ped{A},\quad\h{B}\equiv\frac{1}{2}\varpi:\h{J}_x.
\end{equation*}
Then, even if $\h{A}$ and $\h{B}$ do not commute, the statistical operator can be factorized as
\begin{equation*}
\h{\rho}\ped{GE}=\dfrac{1}{\parz\ped{GE}}\E^{\h{A}+\h{B}}=\dfrac{1}{\parz\ped{GE}}\E^{\h{A}}\,\Big[1+\sum_{n=1}^\infty\h{B}_n\Big],
\end{equation*}
where we denoted
\begin{gather*}
\h{B}_n\equiv \int_0^1\!\!\D\lambda_1\int_0^{\lambda_1}\!\!\D\lambda_2\dots\int_0^{\lambda_{n-1}}\!\!\!\D\lambda_n\,\h{B}(\lambda_1)\dots\h{B}(\lambda_n),\\
\h{B}(\lambda)\equiv\E^{-\lambda \h{A}}\,\h{B}\,\E^{\lambda \h{A}}.
\end{gather*}
The angular momentum $\h{J}^{\mu\nu}$ commutes with the electric and the axial charge, therefore we can see
the quantity $\h{B}(\lambda)$ as the translation of $\h{B}$ along $\beta$, except that the translation is
made of an imaginary quantity. We already know how a translation transforms the angular momentum, see (\ref{jshift}),
we can then write $\h{B}(\lambda)$ as
\begin{equation*}
\begin{split}
\h{B}(\lambda)&=\f{1}{2}\varpi_{\mu\nu}\E^{\lambda \beta(x)\cdot\h{P}}\,\h{J}^{\mu\nu}_x\,\E^{-\lambda \beta(x)\cdot\h{P}}\\
&=\f{1}{2}\varpi_{\mu\nu}\h{\group{T}}^{-1}\big(\I\lambda\beta(x)\big)\,\,\h{J}^{\mu\nu}_x\,\h{\group{T}}\big(\I\lambda\beta(x)\big)\\
&=\f{1}{2}\varpi_{\mu\nu}\,\h{J}^{\mu\nu}_{x-\I\lambda\beta}.
\end{split}
\end{equation*}
The integrals in $B_n$ can be rearranged such that they are given by the path-ordered products on $\lambda$.
path-ordered products sort the ordinary products according to the values of $\lambda$ and they are defined by
\begin{equation*}
T_\lambda \big(\h{O}_1(\lambda_1) \h{O}_2(\lambda_2) \cdots \h{O}_N(\lambda_N)\big)
\equiv \h{O}_{p_1}(\lambda_{p_1}) \h{O}_{p_2}(\lambda_{p_2}) \cdots \h{O}_{p_N}(\lambda_{p_N})
\end{equation*}
with $p$ the permutation that orders $\lambda$ by value:
\begin{align*}
&p \mathrel{:} \{1, 2, \dots, N\} \to \{1, 2, \dots, N\}\\
&\lambda_{p_1} \leq \lambda_{p_2} \leq \cdots \leq \lambda_{p_N}.
\end{align*}
Explicitly, the integrals are turned into
\begin{equation*}
\h{B}_n=\f{1}{n!}\int_0^1\!\!\D\lambda_1\dots\int_0^1\!\!\!\D\lambda_n\,T_\lambda\big(\h{B}(\lambda_1)\dots \h{B}(\lambda_n)\big)
\end{equation*}
and the statistical operator becomes
\begin{equation*}
\h{\rho}\ped{GE}=\f{1}{\parz\ped{GE}}\exp\big[-\beta(x)\cdot \h{P}+\xi\,\h{Q}\, \big]
\,T_\lambda\Bigl[\exp\Big( \int_0^1\!\!\D \lambda \,\f{1}{2}\varpi_{\mu\nu}\,\h{J}^{\mu\nu}(x-\I\lambda\beta)\Big)\Bigr].
\end{equation*}

Now, we indicate with a subscript $\beta$ the partition function and the mean value made with homogeneous statistical operator
\begin{equation}\label{leadorder}
\mean{\h O\,}_{\beta(x)}\equiv\frac{\tr\left[ \exp\left(-\beta(x)\cdot \wP+\zeta\wQ+\zeta\ped{A}\wQ\ped{A}\right)\h O\,\right]}{\tr\left[ \exp\left(-\beta(x)\cdot\wP+\zeta\wQ+\zeta\ped{A}\wQ\ped{A}\right)\right]}
=\f{1}{\parz_\beta}\tr\left[ \E^{\h A}\,\h O\,\right],
\end{equation}
and we remind that the connected correlators are defined as following
\begin{equation*}
\begin{split}
\mean{\h O}_{c}&=\mean{\h O},\\
\mean{\h J \h O}_{c}&= \mean{\h J\h O}-\mean{\h J}\mean{\h O},\\
\mean{\h J_1\h J_2\h O}_{c}&= \mean{\h J_1\h J_2\h O}-\mean{\h J_1}\mean{\h J_2 \h O}_{c}-
\mean{\h J_2}\mean{\h J_1 \h O}_{c}-\mean{\h O}\mean{\h J_1 \h J_2}_{c}-\mean{J_1}\mean{J_2}\mean{\h O},
\end{split}
\end{equation*}
and similarly for higher numbers of products.

We can use the derived product expansion to express thermal quantities at generalized
global equilibrium as expansion in thermal vorticity where thermal averages are made with homogeneous statistical
operator. As first example, consider the expansion on thermal vorticity of the logarithm of the partition function;
it is given by
\begin{equation*}
\begin{split}
\log\parz\ped{GE}=&\log \left(\tr\left[ \E^{\h A+\h B}\right]\right)=\log \left(\frac{\parz_\beta}{\parz_\beta}\tr\left[ \E^{\h A+\h B}\right]\right)\\
=&\log \parz_\beta +\log\left(1+\sum_{n=1}^\infty\mean{\h{B}_n}_{\beta(x)}\right)
\end{split}
\end{equation*}
and the partition function itself is
\begin{equation*}
\parz\ped{GE}=\E^{\log\parz\ped{GE}}= \parz_\beta \left(1+\sum_{n=1}^\infty\mean{\h{B}_n}_{\beta(x)}\right).
\end{equation*}
Then, expanding the logarithm in Taylor series, we recover the connected correlators
\begin{equation*}
\begin{split}
\log\parz\ped{GE}=&\log \parz_\beta +\sum_{r=1}^{\infty}\f{(-1)^{r+1}}{r}\left[\sum_{n=1}^{\infty}\mean{\h{B}_n}_{\beta(x)}\right]^r
=\log \parz_\beta +\sum_{N=1}^{\infty}\f{\mean{\h{B}_N}_{\beta(x),c}}{N!}.
\end{split}
\end{equation*}
Replacing all the definitions, the partition function is explicitly given by the series
\begin{equation*}
\log \parz\ped{GE}=\log\parz_{\beta} +\sum_{N=1}^\infty \f{1}{2^N N!}\prod_{n=1}^N\l[\int_0^1\!\!\!\D\lambda_n\varpi_{\mu_n\nu_n}\r]\,
\mean{T_\lambda\Big(\prod_{n=1}^N \h{J}^{\mu_n\nu_n}_{x-\I\lambda_n\beta}\Big)}\ped{$\beta(x)$,c},
\end{equation*}
or changing the integration variables into $\tau_n=\lambda_n\mod{\beta}$
\begin{equation*}
\log \parz\ped{GE}\!=\!\log\parz_{\beta}\!+\!\sum_{N=1}^\infty \f{1}{2^N N!}\prod_{n=1}^N\l[\f{1}{\mod{\beta}}\int_0^{\mod{\beta}}\!\!\!\!\D \tau_n\varpi_{\mu_n\nu_n}\r]\,
\braket{T_\tau\Big(\prod_{n=1}^N \h{J}^{\mu_n\nu_n}_{x-\I \tau_n u}\Big)}\ped{$\beta(x)$,c}.
\end{equation*}

The same argument is applied to find an expansion of the mean value of a local operator $\h{O}$:
\begin{equation*}
\braket{\h{O}(x)}\ped{GE}=\f{1}{\parz\ped{GE}}\tr\Big[\E^{\h{A}+\h{B}}\h{O}(x)\Big];
\end{equation*}
using the partition functional expansion obtained before and expanding the operator exponent, we obtain
\begin{equation*}
\begin{split}
\braket{\h{O}(x)}\ped{GE}&=\f{1}{\parz\ped{GE}}\tr\Big[\E^{\h{A}}\big(1+\sum_{n=1}^\infty\h{B}_n\big)\h{O}(x)\Big]\\
&=\f{1}{1+\sum_{n=2}^\infty\mean{\h{B}_n}_{\beta(x)}}\f{1}{\parz_\beta}\tr\Big[\E^{\h{A}}\big(1+\sum_{n=1}^\infty\h{B}_n\big)\h{O}(x)\Big]\\
&=\f{\mean{\h{O}(x)}_{\beta(x)}+\sum_{n=1}^\infty\mean{\h{B}_n\h{O}(x)}_{\beta(x)}}{1+\sum_{n=2}^\infty\mean{\h{B}_n}_{\beta(x)}}.
\end{split}
\end{equation*}
To complete the series we expand the denominator with the Taylor series $(1+x)^{-1}=\sum_n (-1)^n x^n$ and we recover
again the connected correlators:
\begin{equation*}
\braket{\h{O}(x)}\ped{GE}=\mean{\h{O}(x)}_{\beta(x)}+\sum_{N=1}^\infty\f{\mean{\h{B}_N\h{O}(x)}\ped{c}}{N!}.
\end{equation*}
The explicit form of the series is
\begin{equation}
\label{eq:MeanLocalO}
\mean{\h{O}(x)}\ped{GE}=\sum_{N=0}^\infty \f{1}{2^N N!}\prod_{n=1}^N\l[\f{1}{\mod{\beta}}\int_0^{\mod{\beta}}\!\!\!\D \tau_n\varpi_{\mu_n\nu_n}\r]\,
\mean{T_\tau\Big(\prod_{n=1}^N \h{J}^{\mu_n\nu_n}_{x-\I \tau_n u}\,\h{O}(x)\Big)}\ped{$\beta(x)$,c}.
\end{equation}
Moreover, by taking advantage of translation properties, we can remove the $x$ dependence inside the argument of the mean value:
\begin{equation*}
\mean{\h{O}(x)}\ped{GE}=\sum_{N=0}^\infty \f{1}{2^N N!}\prod_{n=1}^N\l[\f{1}{\mod{\beta}}\int_0^{\mod{\beta}}\!\!\!\D \tau_n\varpi_{\mu_n\nu_n}\r]\,
\mean{T_\tau\Big(\prod_{n=1}^N \h{J}^{\mu_n\nu_n}_{-\I \tau_n u}\,\h{O}(0)\Big)}\ped{$\beta(x)$,c}.
\end{equation*}
At the second order in thermal vorticity, we obtain~\cite{Buzzegoli:2017cqy}
\begin{equation}
\label{eq:MeanOSecOrd}
\begin{split}
\mean {\h O(x)}=&\mean{\h O(0)}_{\beta(x)}+\frac{\varpi_{\mu\nu}}{2|\beta|}\int_0^{|\beta|}\di \tau 
\mean{{\rm T}_\tau\left(\h J^{\mu\nu}_{-\ii \tau u}\h O(0)\right)}_{\beta(x),c}
\\& +\frac{\varpi_{\mu\nu}\varpi_{\rho\sigma}}{8|\beta|^2} \int_0^{|\beta|}\di \tau_1\di \tau_2
\mean{{\rm T}_\tau\left(\h J^{\mu\nu}_{-\ii \tau_1 u} \h J^{\rho\sigma}_{-\ii \tau_2 u}\h O(0)\right)}_{\beta(x),c}+\mathcal{O}(\varpi^3).
\end{split}
\end{equation}

Making use of the decomposition with the tetrad $\{u,\alpha,w,\gamma\}$, the contraction of angular momentum with thermal
vorticity is written in terms of comoving boost and rotation generators:
\begin{align*}
\frac{1}{2}\varpi_{\mu\nu}\h{J}^{\mu\nu}&=-\alpha^\rho\h{K}_\rho-w^\rho\h{J}_\rho\\
\varpi_{\mu\nu}\varpi_{\rho\sigma}\h{J}^{\mu\nu}\h{J}^{\rho\sigma}&=
2\alpha^\mu\alpha^\nu \{\h{K}_\mu,\,\h{K}_\nu\}+2w^\mu w^\nu \{\h{J}_\mu,\,\h{J}_\nu\}
+4\alpha^\mu w^\nu \{\h{K}_\mu,\,\h{J}_\nu\}.
\end{align*}
The mean value expansion is then decomposed into
\begin{equation*}
\begin{split}
\mean {\h O(x)}=&\,\mean{\h O(0)}_{\beta(x)}
-\frac{\alpha_\rho}{|\beta|}\int_0^{|\beta|}\di \tau\, \mean{{\rm T}_\tau\left(\h K^{\rho}_{-\ii \tau u} \h O(0)\right)}_{\beta(x),c}\\
&-\frac{w_\rho}{|\beta|} \int_0^{|\beta|}\di \tau\, \mean{{\rm T}_\tau \left(\h J^{\rho}_{-\ii \tau u} \h O(0)\right)}_{\beta(x),c}\\
&+\frac{\alpha_\rho\alpha_\sigma}{2|\beta|^2}\int_0^{|\beta|}\di \tau_1\di \tau_2
\,\mean{{\rm T}_\tau \left(\h K^{\rho}_{-\ii \tau_1 u } \h K^{\sigma}_{- \ii \tau_2 u} \h O(0)\right)}_{\beta(x),c}\\
&+\frac{w_\rho w_\sigma}{2|\beta|^2}\int_0^{|\beta|}\di \tau_1\di \tau_2 \,\mean{{\rm T}_\tau
	\left(\h J^{\rho}_{-\ii \tau_1 u } \h J^{\sigma}_{-\ii \tau_2 u} \h O(0)\right)}_{\beta(x),c}\\
&+\frac{\alpha_\rho w_\sigma}{2|\beta|^2}\int_0^{|\beta|}\di \tau_1\di \tau_2
\,\mean{{\rm T}_\tau \left(\{\h K^\rho_{-\ii \tau_1 u},\h J^{\sigma}_{-\ii \tau_2 u}\}\h O(0)\right)}_{\beta(x),c}
+\mathcal{O}(\varpi^3)
\end{split}
\end{equation*}
or, after defining the correlators
\begin{equation}
\begin{split}\label{mycorr}
\mycorr{\h K^{\rho_1}\cdots \h K^{\rho_n} \h J^{\sigma_1}\cdots \h J^{\sigma_m} \h O} \equiv&
\int_0^{|\beta|} \frac{\di\tau_1\cdots\di\tau_{n+m}}{|\beta|^{n+m}}\times\\
&\times\mean{{\rm T}_\tau\left(\h K^{\rho_1}_{-\ii \tau_1 u}\cdots\h K^{\rho_n}_{-\ii \tau_n u}
	\h J^{\sigma_1}_{-\ii \tau_{n+1} u} \cdots\h J^{\sigma_m}_{-\ii \tau_{n+m} u} \h O(0)\right)}_{\beta(x),c}
\end{split}
\end{equation}
the expansion in thermal vorticity of the mean value of a local operator $\widehat O(x)$
can be written as:
\begin{equation}
\begin{split} \label{meanvalueoper}
\mean {\h O(x)}=&\mean{\h O(0)}_{\beta(x)}-\alpha_\rho \mycorr{\,\h K^\rho \h O\,}-w_\rho \mycorr{\,\h J^\rho \h O\,}
+\frac{\alpha_\rho\alpha_\sigma}{2}\mycorr{\,\h K^\rho \h K^\sigma \h O\,}\\
&+\frac{w_\rho w_\sigma}{2} \mycorr{\,\h J^\rho \h J^\sigma \h O\,} +\frac{\alpha_\rho w_\sigma}{2}\mycorr{\,\{\h K^\rho,\h J^\sigma\}\h O\,}+\mathcal{O}(\varpi^3).
\end{split}
\end{equation}
Note that the above expansion applies to any local operator $\h O$, whether it is a scalar or
a component of a tensor of any rank.

Once a specific form of the local operator is chosen, we can further simplify the expression~(\ref{meanvalueoper})
taking advantage of rotational invariance of homogeneous statistical operator. This is
achieved taking into account the symmetries of the operator and selecting only the relevant components in the decomposition.
For instance, in the case of homogeneous equilibrium without vorticity the stress-energy tensor mean value is decomposed into
the so-called ideal form, which only contains two thermodynamic scalar quantities: pressure $p$ and energy density $\rho$~\cite{DeGroot:1980dk}:
\begin{equation*}
\mean{\,\h{T}_{\mu\nu}\,}_\beta=\rho\,u_\mu u_\nu-p\,\Delta_{\mu\nu}
\end{equation*}
where pressure and energy density are obtained as
\begin{equation*}
\rho\equiv u^\mu u^\nu\mean{\,\h{T}_{\mu\nu}\,}_\beta,\quad p=-\frac{1}{3}\Delta^{\mu\nu}\mean{\,\h{T}_{\mu\nu}\,}_\beta.
\end{equation*}
In the more general case of equilibrium with vorticity, the decomposition contains more terms than the ideal case
because the symmetries allow the thermal expectation value to be dependent on scalars and tensors built with
the tetrad $\{u,\alpha,w,\gamma\}$. In general, the global equilibrium mean value of an operator
$\h{O}_{\mu\cdots\nu}(x)$ is decomposed in
\begin{equation*}
\mean {\h O_{\mu\cdots\nu}(x)}=\sum_l C^{(l)} V^{(l)}_{\mu\cdots\nu},
\end{equation*}
where $C^{(l)}$ are a finite number of thermodynamic functions obtained averaging with homogeneous statistical
operator and not depending on thermal vorticity and $V^{(l)}_{\mu\cdots\nu}$ is an appropriate tensor built with
the tetrad $\{u,\alpha,w,\gamma\}$. For instance, the stress-energy tensor for parity even fluid at second
order in thermal vorticity is decomposed as~\cite{BecGro,Buzzegoli:2017cqy}:
\begin{equation*}
\begin{split}
\mean{\wT^{\mu\nu}}=&(\rho-\alpha^2 U_\alpha -w^2 U_w)u^\mu u^\nu -(p-\alpha^2D_\alpha-w^2D_w)\Delta^{\mu\nu}\\
&+A\,\alpha^\mu\alpha^\nu+Ww^\mu w^\nu+G \left(u^\mu\gamma^\nu+ u^\nu\gamma^\mu\right)+\mathcal{O}(\varpi^3).
\end{split}
\end{equation*}
For details on how this decomposition is made we refer to~\cite{BecGro} and to an explicit example in
Sec.~\ref{sec:SpinTens}. At each new term appearing in this way corresponds a new thermodynamical functions $C^{(l)}$
which represents a quantum correction induced by rotation if the term is coupled with vector $w$, or to acceleration if
it couples with $\alpha$.  In the example above $U_\alpha,\,D_\alpha$ and $A$ are thermodynamic coefficients representing
the response of acceleration, $U_w,\,D_w$ and $W$ are the response of rotation and $G$ is the response to their combined effect.
Every thermodynamic function $C^{(l)}$ introduced with this decomposition is a Lorentz scalar and therefore it is always
possible to evaluate it in the rest frame of thermal bath. For instance, one of the aforementioned
thermal coefficient of stress-energy tensor is
\begin{equation*}
U_\alpha=\frac{1}{2}\int_0^{|\beta|=1/T} \frac{\di\tau_1 \di\tau_2}{|\beta|^2}
\mean{{\rm T}_\tau\left(\h K^3_{-\ii \tau_1} \h K^3_{-\ii \tau_2} \wT^{00}\right)}_{T,c},
\end{equation*}
where we denoted the mean values in rest frame with a subscript $T$, that is: 
\begin{equation*}
\mean{\h O(x)}_T\equiv\frac{\tr\left[ \exp\left(-\h H/T +\zeta\wQ+\zeta\ped{A}\wQ\ped{A}\right)\h O(0)\right]}{\tr\left[ \exp\left(-\h H/T +\zeta\wQ+\zeta\ped{A}\wQ\ped{A}\right)\right]},
\end{equation*}
and $\h H$ is the Hamiltonian. In the rest frame, by construction the four-temperature direction $u$ has only the
time component, consequently the translation on the argument of the Lorentz generators is only a time shift.
Thus, imaginary time and euclidean space in the rest frame is well defined and is particularly convenient because the time-ordered features of the correlators are automatically satisfied. In the next chapter, we review thermal field theory in
imaginary-time, which we use to evaluate the before mentioned thermodynamic quantities.

\subsection{Comparison with Kubo Formulae}
\label{sec:KuboFormulae}
Thermal coefficients and, more generally, transport coefficients are usually given as
Kubo Formulae. The previous terms on the expansion~(\ref{meanvalueoper}) are Kubo
Formulae in the broad sense, as they are obtained as a linear response. We gave these
thermal coefficients as spatial and imaginary-time integrals of euclidean correlators,
see Eq.~(\ref{mycorr}). However, Kubo formulae are more commonly provided as Green
functions in the Fourier space. We show here how to turn first order correlators
of the form~(\ref{mycorr}) into Green function relations. This procedure, also used
in Ref.~\cite{Huang:2011dc}, can be also extended to higher orders.

Consider the first order term of the thermal vorticity expansion~(\ref{eq:MeanOSecOrd}).
Suppose the operator $\h{O}$ is a scalar, then we can evaluate it in the rest frame
of thermal bath, in which $u=(1,\vec{0})$. Therefore, the first term reads:
\begin{equation*}
\mean{\,\h{O}\,}^{(1)} = \varpi_{\rho\sigma}\int_0^{|\beta|} \frac{\di\tau}{2|\beta|}
	\mean{\, \h J^{\rho\sigma}_{-\I\tau} \h O(0)\, }_{T,c}.
\end{equation*}
We then replace the Lorentz generators $\h{J}^{\rho\sigma}$ with their definition~(\ref{eq:LorenzGenerators})
in terms of symmetric stress-energy tensor. Once this is done, we translate the stress-energy tensor
operator in imaginary time as required by the correlator, see Eq.~(\ref{jshift}). After that, we find:
\begin{equation*}
\mean{\,\h{O}\,}^{(1)} = \varpi_{\rho\sigma}\int_0^{|\beta|} \frac{\di\tau}{|\beta|}\int\D^3 x\, x^\rho\,
	\mean{\, \h T^{0\sigma}(-\I\tau,\vec{x}) \h O(0)\, }_{T,c}.
\end{equation*}
Remind that, since we are dealing with connected correlators, it is fair to assume that
\begin{equation*}
\lim_{t\to-\infty} 	\mean{\, \h T^{0\sigma}(t-\I\tau,\vec{x}) \h O(0)\, }_{T,c}=0,
\end{equation*}
taking advantage of which, we can write the correlator inserting also a real time integration:
\begin{equation*}
\begin{split}
\mean{\,\h{O}\,}^{(1)} =& \varpi_{\rho\sigma}\int_0^{|\beta|} \frac{\di\tau}{|\beta|}\int\D^3 x\, x^\rho\,
	\left[\mean{\, \h T^{0\sigma}(-\I\tau,\vec{x}) \h O(0)\,}_T-\mean{\,\h T^{0\sigma}(-\I\tau,\vec{x})}_T \mean{ \h O(0)\,}_T\right]\\
=& \varpi_{\rho\sigma}\int_0^{|\beta|} \frac{\di\tau}{|\beta|}\int\D^3 x x^\rho\int_{-\infty}^0 \D t \frac{\D}{\D t}
	\Big[\mean{\, \h T^{0\sigma}(t-\I\tau,\vec{x}) \h O(0)\,}_T \\
&\hphantom{\varpi_{\rho\sigma}\int_0^{|\beta|} \frac{\di\tau}{|\beta|}\int\D^3 x x^\rho\int_{-\infty}^0 \D t \frac{\D}{\D t}\Big[}
	-\mean{\,\h T^{0\sigma}(t-\I\tau,\vec{x})}_T \mean{ \h O(0)\,}_T\Big].
\end{split}
\end{equation*}
Notice that we can replace the derivative on $t$ with derivative on $\tau$:
\begin{equation*}
\frac{\D}{\D t}=\I \frac{\D}{\D \tau}.
\end{equation*}
Then, we can easily integrate over $\tau$ and, using the Kubo-Martin-Schwinger relation
\begin{equation*}
\mean{\, \h T^{0\sigma}(t-\I\beta,\vec{x}) \h O(0)\,}_T=\mean{\, \h O(0) \h T^{0\sigma}(t,\vec{x})\,}_T,
\end{equation*}
we obtain
\begin{equation*}
\begin{split}
\mean{\,\h{O}\,}^{(1)} =& -\frac{\I\varpi_{\rho\sigma}}{|\beta|}\int_{-\infty}^0\D t\int\D^3 x\, x^\rho\,
	\mean{\, \left[\h T^{0\sigma}(t,\vec{x}), \h O(0)\right]\, }_{T,c}.
\end{split}
\end{equation*}
We can now replace the spectral function~\cite{Laine:2016hma} in Fourier space
\begin{equation*}
\begin{split}
\rho^{AB}(p_0,\vec{p})\equiv& \int\D^4 x\, \E^{\I p\cdot x}\mean{\left[\h{A}(t,\vec{x}),\h{B}(0) \right]}_T,\\
\mean{\left[\h{A}(t,\vec{x}),\h{B}(0) \right]}_T=&\int\frac{\D^4 p}{(2\pi)^4}\E^{-\I p\cdot x}\rho^{AB}(p_0,\vec{p}),
\end{split}
\end{equation*}
in the previous equation:
\begin{equation*}
\begin{split}
\mean{\,\h{O}\,}^{(1)} =& -\frac{\I\varpi_{\rho\sigma}}{|\beta|}\int_{-\infty}^\infty\D t\,\theta(-t)\int\D^3 x\, x^\rho\,
	\int\frac{\D^4 p}{(2\pi)^4}\E^{-\I p\cdot x}\rho^{T^{0\sigma}O}(p_0,\vec{p}).
\end{split}
\end{equation*}
Then, using the integral representation of the Heaviside theta
\begin{equation*}
\theta(-t)=\I\int_{-\infty}^\infty \frac{\D\omega}{2\pi}\frac{\E^{\I\omega t}}{\omega+\I 0^+}
\end{equation*}
we write
\begin{equation*}
\mean{\,\h{O}\,}^{(1)} = \frac{\varpi_{\rho\sigma}}{|\beta|} \int_{-\infty}^\infty\frac{\D\omega}{2\pi}\int\frac{\D^4 p}{(2\pi)^4}
	\frac{\rho^{T^{0\sigma}O}(p_0,\vec{p})}{\omega+\I 0^+} \int\D^4 x\, x^\rho \E^{-\I p\cdot x+\I\omega t}.
\end{equation*}
Noticing that
\begin{equation*}
x^\rho \E^{-\I p\cdot x+\I\omega t}=\I \frac{\de}{\de p_\rho} \E^{-\I p\cdot x+\I\omega t}
\end{equation*}
and integrating by parts, we obtain
\begin{equation*}
\begin{split}
\mean{\,\h{O}\,}^{(1)} =& -\frac{\I\varpi_{\rho\sigma}}{|\beta|} \int_{-\infty}^\infty\frac{\D\omega}{2\pi}\int\frac{\D^4 p}{(2\pi)^4}
	\frac{1}{\omega+\I 0^+}\frac{\de}{\de p_\rho}\rho^{T^{0\sigma}O}(p_0,\vec{p}) \int\D^4 x\, \E^{-\I p\cdot x+\I\omega t}\\
=& -\frac{\I\varpi_{\rho\sigma}}{|\beta|} \int_{-\infty}^\infty\frac{\D\omega}{2\pi}\frac{1}{\omega+\I 0^+}\frac{\de}{\de p_\rho}\rho^{T^{0\sigma}O}(\omega,\vec{p})\Big|_{\vec{p}=0}.
\end{split}
\end{equation*}
Making use of
\begin{equation*}
\frac{1}{\omega+\I 0^+}=\text{PV}\left(\frac{1}{\omega}\right)-\I\pi\delta(\omega)
\end{equation*}
with PV the principal value,
and assuming that the spectral functions is real, we obtain
\begin{equation*}
\mean{\,\h{O}\,}^{(1)} = -\frac{\varpi_{\rho\sigma}}{2|\beta|} \lim_{\omega\to 0}\lim_{\vec{p}\to 0}\frac{\de}{\de p_\rho}\rho^{T^{0\sigma}O}(\omega,\vec{p}).
\end{equation*}
The spectral function is the imaginary part of retarded Green function~\cite{Laine:2016hma},
then we have the connection between first order correlators and Kubo Formule with
retarded Green functions:
\begin{equation}
\label{eq:KuboFormula}
\mean{\,\h{O}\,}^{(1)} = -\frac{\varpi_{\rho\sigma}}{2|\beta|} \lim_{\omega\to 0}\lim_{\vec{p}\to 0}\frac{\de}{\de p_\rho}\Im G\ped{R}^{T^{0\sigma}O}(\omega,\vec{p}).
\end{equation}
\chapter{Finite temperature filed theory}
\label{ch:ftft}
In this chapter, we review the finite temperature and finite density field theory techniques~\cite{KapustaGale,Bellac,Laine:2016hma}
and we extend them to include the effects of a conserved axial charge in the statistical operator.
Finite temperature field theory methods allow to write thermal expectation values as expectation
values in ordinary quantum field theory. Once this connection is accomplished, we are able to use
similar tools used in quantum field theory, such as Green functions and Feynman diagrams for
perturbative calculations. At the end of this chapter (Sec.~\ref{sec:homogeq}), we use the methods
described here to derive essential thermodynamic functions, such as thermodynamic potential, energy,
pressure, electric and axial charge density of a free Dirac gas at homogeneous equilibrium
with a chiral imbalance.

In the previous chapter, we proved that effects of thermal vorticity are given in terms of
Lorentz scalar functions, each one of these obtained with a thermal expectation value
made with the statistical operator of homogeneous thermal equilibrium:
\begin{equation}
\label{eq:rho_homogeq}
\h{\rho}\ped{h}=\dfrac{1}{\parz\ped{h}}\E^{-\beta_\mu\h{P}^\mu+\zeta\h{Q}+\zeta\ped{A}\h{Q}\ped{A}},
\end{equation}
with $\beta$ a constant four-vector. Since the thermal quantities are Lorenz scalars, it is convenient
to evaluate them in the rest frame of thermal bath, where the inverse four-temperature assumes
the form $\beta_\mu=\beta(1,\vec{0})$ with $\beta=1/T$, and the statistical operator is simply given by:
\begin{equation*}
\h{\rho}\ped{h}=\dfrac{1}{\parz\ped{h}}\E^{-\beta\l(\h{H}-\mu\h{Q}-\mu\ped{A}\h{Q}\ped{A}\r)}
\end{equation*}
and the partition function is
\begin{equation}
\label{eq:parzhomo}
\parz\ped{h}=\tr\left[\e^{-\beta\l(\h{H}-\mu\h{Q}-\mu\ped{A}\h{Q}\ped{A}\r)}\right].
\end{equation}
Several important thermal quantities can be derived directly from the partition function,
for example in the infinite-volume $V$ limit, pressure, electric and axial charge, entropy,
free energy and grand thermodynamic potential are given by:
\begin{gather*}
p=\f{\de(T\log\parz)}{\de V},\quad
Q=\f{\de(T\log\parz)}{\de \mu},\quad
Q\ped{A}=\f{\de(T\log\parz)}{\de \mu\ped{A}},\\
S=\f{\de(T\log\parz)}{\de T},\quad
E=-PV+TS+\mu\, Q+\mu\ped{A} Q\ped{A},\quad
\Omega=-\frac{T}{V}\log\parz.
\end{gather*}
Other quantities, such as the ones that gives vorticity corrections, can not be obtained as
derivative of the homogeneous partition function. However, partition functions in statistical mechanics have
the same role of generating functions in quantum field theory. Indeed, the starting point to
develop a finite temperature filed theory is a path integral evaluation of partition function.

\section{Path integral for Dirac field}
\label{sec:PathIntFermion}
Path integral in both quantum field theory and in finite temperature field theory is a well-known
subject and we refer to textbooks for details and reference~\cite{KapustaGale,Bellac,Laine:2016hma}.
Here we retrace the basic steps on how path integral formulation of statistical mechanics is obtained
with the aim of setting the notation, providing and motivating the modifications brought by the
conserved axial charge inside the statistical operator. We begin by reminding the Lagrangian density
of a free Dirac field~(\ref{eq:DiracLcurved}) in flat space-time:
\begin{equation}
\mc{L}=\f{\I}{2} [ \bar{\psi}\gamma^\mu (\de_\mu \psi) - (\de_\mu\bar{\psi})\gamma^\mu\psi]-m\,\bar{\psi}\psi.
\end{equation}
We already discussed symmetries of this theory in Sec.~\ref{sec:Symm}. The conserved currents in flat-space time
are simply given by\footnote{We do not remove the mass term in the Lagrangian to keep track of it and to provide
	an expression valid in the limit of vanishing $\mu\ped{A}$ even though it must not be considered if we want
	a conserved axial current.}:
\begin{gather*}
\h{j}^\mu=\h{\bar{\psi}}\gamma^\mu\h\psi,\quad \h{Q}=\int_{\vec{x}}\h{j}^{\,0}(x)\\
\h{j}^\mu\ped{A}=\h{\bar{\psi}}\gamma^\mu\gamma^5\h\psi,\quad \h{Q}\ped{A}=\int_{\vec{x}}\h{j}^{\,0}\ped{A}(x),
\end{gather*}
where we have introduced the symbol
\begin{equation*}
\int_{\vec{x}}\equiv\int\D\vec{x}.
\end{equation*}
From the Lagrangian, the conjugate momenta of the Dirac fields is
\begin{equation*}
\pi=\f{\de\mc{L}}{\de(\de_0\psi)}=\f{\I}{2}\psi^\dagger,\quad \pi^\dagger=\f{\de\mc{L}}{\de(\de_0\psi^\dagger)}=\f{\I}{2}\psi,
\end{equation*}
and by Legendre transformation we obtain the Hamiltonian density
\begin{equation*}
\mc{H}=-\f{\I}{2} [ \bar{\psi}\gamma^k (\de_k \psi) - (\de_k\bar{\psi})\gamma^k\psi]+m\,\bar{\psi}\psi.
\end{equation*}
Fermionic quantization imposes anti-commutating relations on canonical variables:
\begin{equation*}
\begin{split}
\left\{\h{\psi}_a(\vec{x},t),\,\h{\psi}^\dagger_b(\vec{y},t)\right\}=& \delta_{ab}\delta(\vec{x}-\vec{y}),\\
\left\{\h{\psi}_a(\vec{x},t),\,\h{\psi}_b(\vec{y},t)\right\}=& \left\{\h{\psi}^\dagger_a(\vec{x},t),\,\h{\psi}^\dagger_b(\vec{y},t)\right\}=0.
\end{split}
\end{equation*}
The Hamiltonian is now a time independent functional of the field operator and its conjugate momentum
and it is given by the spatial integral of Hamiltonian density
\begin{equation*}
\h{H}=\int_{\vec{x}}\mc{H}\left(\h{\psi},\,\h\pi\right).
\end{equation*}
The Hamiltonian controls the time evolution of the system through the evolution operator $\exp(-\I\h{H}t)$.
Indicating with $\h{\psi}(\vec{x},0)$ the Schr\"{o}dinger-picture field at time $t=0$ and with $\h{\pi}(\vec{x},0)$
its conjugate momentum operator, the eigenvectors  $\ket{\psi}$ and $\ket{\pi}$, defined by
\begin{equation*}
\h{\psi}(\vec{x},0)\ket{\psi}=\psi(\vec{x})\ket{\psi},\quad
\h{\pi}(\vec{x},0)\ket{\pi}=\pi(\vec{x})\ket{\pi},
\end{equation*}
form a complete orthogonal basis of the field operators $\h{\psi}$ and $\h{\pi}$.
If the system is in the state $\ket{\psi}$ at the time $t=0$, it then evolves to the states $\E^{-\I\h{H}\Delta t}\ket{\psi}$
after an interval of time $\Delta t$.

Consider a system of free Dirac particles at global thermal equilibrium with an imbalance of electric and chiral current
described by the statistical operator in Eq.~(\ref{eq:rho_homogeq}). The associated partition function~(\ref{eq:parzhomo})
is given by the trace of a quantum operator, therefore it is obtained as the following sum over all states represented
by the basis $\ket{\psi}$:
\begin{equation*}
\parz\ped{h}=\tr\l[ \E^{-\beta (\h{H}-\mu \h{Q}-\mu\ped{A} \h{Q}\ped{A})}\r]
=\sum_a \int \D \psi_a \bra{\psi_a} \E^{-\beta (\h{H}-\mu \h{Q}-\mu\ped{A} \h{Q}\ped{A})} \ket{\psi_a}.
\end{equation*}
In this expression, we can still use the exponential of the Hamiltonian as it was the evolution operator,
we just have to turn the inverse temperature $\beta$ into an imaginary time interval $\beta\equiv\I t_f$.
The Gibbs operator $\exp(-\beta\h{H})$ is then formally equivalent to the time evolution operator $\exp(-\I t_f\h{H})$.
This operation is analogous to performing a Wick rotation and to switch to the imaginary time $\tau\equiv\I t$.
In this way, the partition function reads
\begin{equation*}
\parz\ped{h}=\sum_a \int \D \psi_a \bra{\psi_a} \E^{-\I t_f (\h{H}-\mu \h{Q}-\mu\ped{A} \h{Q}\ped{A})} \ket{\psi_a}.
\end{equation*}
In this form, the partition function is similar to the generating functional of quantum field theory but the time
here is fictional and limited to the interval $[0,t_f]$, where the upper limit is bounded by the inverse-temperature.

Before addressing the transition amplitude inside the sum over the states, we give some important properties
of thermal fields. We define the two-point thermal Green function for the fermionic field as
\begin{equation*}
G(x,y)\equiv \mean{T_\tau\left\{\h\psi(x)\h\psi(y)\right\}}_\beta
=\tr\left[\h\rho\ped{h}\,T_\tau\left\{\h\psi(x)\h\psi(y)\right\}\right]
\end{equation*}
where $T_\tau$ denotes the time-$\tau$ ordered-product 
\begin{equation*}
T_\tau\left\{\h\psi(\tau_1)\h\psi(\tau_2) \right\}=\h\psi(\tau_1)\h\psi(\tau_2)\theta(\tau_1-\tau_2)-
\h\psi(\tau_2)\h\psi(\tau_1)\theta(\tau_2-\tau_1)
\end{equation*}
with $\theta$ the Heaviside theta function. Since $\h\rho\ped{h}$ is invariant under translation transformation
we can show that the thermal Green function is only depending on the relative distance of the two field:
$G(x,y)=G(x-y)$ and we can simply indicate $G=G(\tau,\vec{x})$. From translation invariance it also
follows that the thermal Green function is anti-periodic on time translation of period $\beta$~\cite{KapustaGale}:
\begin{equation*}
G(\tau,\vec{x})=-G(\tau-\beta,\vec{x})
\end{equation*}
which in turn also implies $\h\psi(\vec{x},0)=-\h\psi(\vec{x},\beta)$. Therefore, in evaluating
the partition function, we require that at the time $t_f$ the system must return to the starting state at
$t=0$ and in particular that the fermionic field is anti-periodic in imaginary time $\beta$.

The path integral formulation of partition function is then obtained dividing the amplitude
$\bra{\psi_a} \E^{-\I t_f (\h{H}-\mu \h{Q}-\mu\ped{A} \h{Q}\ped{A})} \ket{\psi_a}$ into $N$
equal time interval $\epsilon$ covering the full period $[0,t\ped{f}]$. At each step we insert
a complete set of state and we send $N$ to infinity. The transition amplitude is then written as
\begin{equation*}
\begin{split}
\bra{\psi_a} \E^{-\I t\ped{f}(\h{H}-\mu \h{Q}-\mu\ped{A} \h{Q}\ped{A})} \ket{\psi_a} =& \lim_{N\to\infty}\int
\l( \prod_{i=1}^N \f{\D \pi_i}{2\pi} \D \psi_i \r)\times\\
&\times\braket{\psi_a | \pi_N}\braket{\pi_N | \E^{-\I(\h{H}-\mu\h{Q}-\mu\ped{A}\h{Q}\ped{A})\epsilon} | \psi_N}\dots \braket{\psi_1 | \psi_a}.
\end{split}
\end{equation*}
The electric charge can be included inside a new definition of the state
\begin{equation*}
\ket{\Psi_i(\tau)}\equiv \e^{\mu\h{Q}\tau}\ket{\psi_i}=\e^{\ii\mu\h{Q}t} \ket{\psi_i}.
\end{equation*}
The action of the electric charge on the states is simply given by $\h{Q}\ket{\psi}=\ket{\psi}$ and $\h{Q}\ket{\bar{\psi}}=-\ket{\bar{\psi}}$,
therefore the relations between the evolved new and old states are
\begin{equation*}
\ket{\Psi(\tau)}=\e^{\mu\tau}\ket{\psi},\quad \ket{\bar{\Psi}(\tau)}=\e^{-\mu\tau}\ket{\bar{\psi}}.
\end{equation*}
The axial charge $\h{Q}\ped{A}$ can not be included in the same way in a state because the states $\ket{\psi}$ and $\ket{\bar{\psi}}$ are
not eigenstates of $\h{Q}\ped{A}$. After evaluating the transition amplitude, we obtain the path integral form of partition function
\begin{equation*}
\parz= C \int_{\Psi(\beta,\vec{x})=-\Psi(0,\vec{x})} \mathcal{D} \bar{\Psi}\,\mathcal{D} \Psi \,\, \exp \left\{ - S\ped{E}(\Psi,\bar{\Psi},\mu\ped{A}) \right\}
\end{equation*}
where $C$ is a constant that can be ignored and $S\ped{E}$ is the Euclidean thermal action defined by
\begin{equation*}
S\ped{E}(\Psi,\bar{\Psi},\mu\ped{A})=\int_0^\beta\di\tau\int_{\vec{x}}\mathcal{L}\ped{E}(\Psi,\bar{\Psi},\mu\ped{A}).
\end{equation*}
The Euclidean Lagrangian, denoted by $\mathcal{L}\ped{E}$, is obtained from regular Lagrangian with the correspondence
$\mathcal{L}\ped{E}(X)=-\mathcal{L}(\tau=\ii t,\vec{x},\mu\ped{A})$ where we introduced the symbol $X=(\tau,\vec{x})$ and we denote
\begin{equation*}
\int_X\equiv\int_0^\beta\di\tau\int_{\vec{x}}.
\end{equation*}
The Eucldean Lagrangian is easily written noticing that $\I\de_0=\I\de_t=-\de_\tau$ and introducing the so called \emph{Euclidean Dirac matrices} through
\begin{equation*}
\tilde{\gamma}_0 \equiv\gamma^0,\quad\tilde{\gamma}_k\equiv-\I\gamma^k.
\end{equation*}
According to traditional gamma matrix algebra, Euclidean gamma matrices satisfy
\begin{equation*}
\acom{\tilde{\gamma}_\mu}{\tilde{\gamma}_\nu}= 2\delta_{\mu\nu},\quad\text{and}\quad
\tilde{\gamma}_\mu^\dagger=\tilde{\gamma}_\mu .
\end{equation*}
Thereby, the Euclidean Lagrangian for the free Dirac field at finite axial density is written as
\begin{equation*}
\mathcal{L}\ped{E}(X)=\frac{1}{2}\left[\bar{\psi}\tilde{\gamma}_\mu(\de_\mu\psi)-(\de_\mu\bar{\psi})\tilde{\gamma}_\mu\psi\right]
-\mu\ped{A} \bar{\psi}\tilde{\gamma}_0\gamma^5\psi+m\bar{\psi}\psi.
\end{equation*}
From now on, to simplify notation, we drop tildes from the euclidean $\tilde{\gamma}_\mu$'s and two identical repeated indices down imply
the use of Euclidean metric and Euclidean matrices.

It is usually more convenient to work in momentum space. We already noticed that fermionic field are antiperiodic in the imaginary time
period $\beta$. The antiperiodicity implies that momenta $\omega_n$ related to imaginary time are discrete and they are known as
\emph{fermionic Matsubara frequencies}:
\begin{equation*}
\omega_n=2\pi T\l(n+\f{1}{2}\r),\quad n \in\mathbb{Z}.
\end{equation*}
The fields in momentum space are then written as
\begin{equation}
\label{eq:DiracMomenta}
\psi(X)=\sumint_{\{P\}}\E^{\I P^+\cdot X}\tilde{\psi}(P),\quad \bar{\psi}(X)=\sumint_{\{P\}}\E^{-\I P^+\cdot X}\tilde{\bar{\psi}}(P),
\end{equation}
with the introduction fo the symbols
\begin{equation*}
P^+=(\omega_n+\I\mu,\vec{p}),\quad \sumint_{\{P\}}\equiv \int \frac{\D^3 p}{(2\pi)^3}T\sum_{\{\omega_n\}},
\end{equation*}
where the curly brackets ($\{\omega_n\}$) are included to remind that the Matsubara frequencies are of fermionic nature
and they stand for the sum of $n$ from $-\infty$ to $\infty$. The additional imaginary shift of momentum time component
proportional to chemical potential is the result of the electric charge that we previously included inside the definition
of the field states $\ket{\Psi}$. To account for discrete Matsubara frequencies, we define a modified delta function:
\begin{equation*}
\bar{\delta}(P-Q)\!=\!\int_X\!\! \E^{\I(P+Q)\cdot X}\!=\!\!\int\!\! \D\vec{x}\,\E^{\I\vec{x}(\vec{p}-\vec{q})}\!\!\int_0^\beta\!\! \D\tau\,\E^{\I\tau(p_n-q_n)}
\!=\!\beta\delta_{p_n+q_n,0}\,(2\pi)^3 \delta^{(3)}(\vec{p}+\vec{q}).
\end{equation*}
Using the introduced notation, we express the Euclidean action in terms of momentum modes of fields:
\begin{equation*}
\begin{split}
S\ped{E}&=\int_X \frac{1}{2}\left[\bar{\psi}(X)\gamma_\mu\de_\mu\psi(X)-(\de_\mu\bar{\psi}(X))\gamma_\mu\psi(X)\right]
-\mu\ped{A}\bar{\psi}(X)\gamma_0\gamma_5\psi(X)+m\bar{\psi}(X)\psi(X)\\
&=\int_X \sumint_{\{P\}}\sumint_{\{Q\}}\e^{\ii(P^+-Q^+)\cdot X}\tilde{\bar{\psi}}(Q)\left[\ii\gamma_\mu\frac{P_\mu^+ + Q_\mu^+}{2} +m-\mu\ped{A}\gamma_0\gamma_5\right]\tilde{\psi}(P)\\
&=\sumint_{\{P\}}\sumint_{\{Q\}}\bar{\delta}(P-Q)\tilde{\bar{\psi}}(Q)\left[\ii\gamma_\mu\frac{P_\mu^+ + Q_\mu^+}{2} +m-\mu\ped{A}\gamma_0\gamma_5\right]\tilde{\psi}(P)\\
&=\sumint_{\{P\}}\tilde{\bar{\psi}}(P)\left[\ii\slashed{P}^+ +m - \mu\ped{A}\gamma_0\gamma_5\right]\tilde{\psi}(P).
\end{split}
\end{equation*}
In the path integral formulation of partition function, we can switch from coordinate space to momentum space just by changing integration variables.
As usual, the measure is modified according to the determinant of the transformation:
\begin{equation*}
\Dpi \Psi = \mod{\det \l[ \f{\delta\Psi(X)}{\delta\tilde{\Psi}(P)}\r]}\Dpi \tilde{\Psi}.
\end{equation*}
However, this change is purely kinematic and the determinant can be absorbed inside the factor $C$ defining a new unknown coefficient $C'$:
\begin{equation*}
C'\equiv C \, \mod{\det \l[ \f{\delta\Psi(X)}{\delta\tilde{\Psi}(P)}\r]}\;\mod{\det \l[ \f{\delta\bar\Psi(X)}{\delta{\tilde{\bar{\Psi}}}(P)}\r]}.
\end{equation*}
We thus obtained the partition function path integral in terms of Fourier modes:
\begin{equation}
\label{eq:parzFermion}
\parz= C' \int \mathcal{D} \tilde{\bar{\Psi}}\,\mathcal{D} \tilde\Psi \,\,
\exp \left\{ - \sumint_{\{P\}}\tilde{\bar{\psi}}(P)\left[\ii\slashed{P}^+ +m - \mu\ped{A}\gamma_0\gamma_5\right]\tilde{\psi}(P) \right\}.
\end{equation}
%

\section{Fermionic propagator at finite density and chirality}
\label{sec:FermiProp}
With the path integral completely set up, it is  straightforward to evaluate the thermal propagator
\begin{equation*}
\mean{\psi_a(X) \bar\psi_b(Y)}_T=\tr\left[\h\rho\ped{h}\,T_\tau\left\{\h\psi_a(X)\h{\bar\psi}_b(Y)\right\}\right],
\end{equation*}
where $a$ and $b$ are spinorial indices. Notice that we denoted the thermal expectation value with the subscript $T$
instead of the subscript $\beta$ to remind that we are carrying out our quantities in a specific reference system, the one where the
thermal bath is at rest: $u=(1,\vec{0})$. As first step, we expand the fermionic propagator in momentum modes:
\begin{equation*}
\mean{\psi_a(X) \bar\psi_b(Y)}_T=\sum_{\{P\}}\sum_{\{Q\}} T_\tau\left\{\E^{\I P^+\cdot X}\E^{-\I Q^+\cdot Y}\right\}\mean{\tilde{\psi}_a(P) \bar{\tilde\psi}_b(Q)}_T.
\end{equation*}
Then, we can use the path integral formulation to obtain the thermal propagator of momentum modes,
which is given by the correlation of the two modes weighted with the Euclidean action:
\begin{equation*}
\mean{\tilde{\psi}_a(P)\bar{\tilde\psi}_b(Q)}_T=\frac{\int \Dpi \tilde\Psi \,\Dpi \tilde{\bar{\Psi}} \; \exp \l( - S\ped{E}\r) \tilde{\psi}_a(P)\bar{\tilde\psi}_b(Q)}{\int\Dpi \tilde\Psi \,\Dpi \tilde{\bar{\Psi}}\; \exp \l( - S\ped{E}\r)}.
\end{equation*}
The integral is a well-known result of Grassmann variables $\theta$~\cite{KapustaGale}:
\begin{equation*}
\frac{\int \left(\prod_i \di\theta_i^\dagger\di\theta_i\right)\theta_k\theta_l\exp(-\theta^\dagger A\theta)}{\int \left(\prod_i \di\theta_i^\dagger\di\theta_i\right)\,\exp(-\theta^\dagger A\theta)}=(A^{-1})_{kl}
\end{equation*}
Therefore, the fermionic propagator in momentum space is
\begin{equation*}
\mean{\tilde{\psi}_a(P)\bar{\tilde\psi}_b(Q)}_T=\bar{\delta}(P-Q)[\ii\slashed{P}^+ +m -\mu\ped{A}\gamma_0\gamma_5]^{-1}_{ab}.
\end{equation*}
At this point, we just need to find the inverse matrix $S(P)_{ab}$:
\begin{equation*}
S(P)_{ab}\equiv\mc{M}(P)^{-1}\equiv[\ii\slashed{P}^+ +m -\mu\ped{A}\gamma_0\gamma_5]^{-1}_{ab}.
\end{equation*}
When there is no axial chemical potential, the propagator is the usual free propagator:
\begin{equation*}
S(P)_{ab}=\frac{\left(-\ii\slashed{P}^+ + m\right)_{ab}}{{P^+}^2+m^2}.
\end{equation*}
For the case $m=0$ and $\mu\ped{A}\neq 0$, we first show that the matrix is invertible. Notice that $\ii\slashed{P}$ and $\gamma_0\gamma_5$
are anti-Hermitian, so the eigenvalues of $\mathcal{M}=\ii\slashed{P}^+ -\mu\ped{A}\gamma_0\gamma_5$ are of the form $\ii\lambda_n$,
where $\lambda_n$ are reals numbers. Because $\gamma_5$ anticommutes with $\mathcal{M}$, all eigenvalues come in pairs, which means that
if $\ii\lambda_n$ is an eigenvalue, then $-\ii\lambda_n$ is also an eigenvalue. The determinant is the product of all eigenvalues and is
therefore given by $\det \mathcal{M}=\prod_n\lambda_n^2$. This proof that the determinant is real and positive semidefinite and hence
it exist the inverse matrix $\mathcal{M}^{-1}$. The inverse matrix $\mathcal{M}^{-1}$ is
\begin{equation*}
\begin{split}
S(P)=&\left[\frac{1}{(P^0+\ii\mu-\ii\mu\ped{A})^2+\vec{P}^2}\frac{1-\gamma_5}{2}\right.+\\
&+\left.\frac{1}{(P^0+\ii\mu+\ii\mu\ped{A})^2+\vec{P}^2}\frac{1+\gamma_5}{2}\right](-\ii\slashed{P}^+ -\mu\ped{A}\gamma_0\gamma_5).
\end{split}
\end{equation*}
This can be proved by explicit checking the identity $\mc{M}(P)S(P)=1$. We first define some auxiliary quantities
\begin{gather*}
P^+(\pm\mu\ped{A})\equiv(P_0+\ii\mu\pm\ii\mu\ped{A},\vec{p}),\quad P^+(\pm\mu\ped{A})^2=(P_0+\ii\mu\pm\ii\mu\ped{A})^2+\vec{p}^2,\\
I_1\equiv\frac{1}{P^+(-\mu\ped{A})^2}\frac{1-\gamma_5}{2}(-\ii\slashed{P}^+-\mu\ped{A}\gamma_0\gamma_5),\\
I_2\equiv\frac{1}{P^+(+\mu\ped{A})^2}\frac{1+\gamma_5}{2}(-\ii\slashed{P}^+-\mu\ped{A}\gamma_0\gamma_5),
\end{gather*}
from which the matrix $S$ is written as $S(P)=I_1+I_2$. Taking advantage of the following gamma matrix identities
\begin{gather*}
\gamma_0^2=1,\quad\gamma_5^2=1,\quad(\gamma_0\gamma_5)^2=-1,\quad
\{\gamma_\mu,\gamma_\nu\}=2\delta_{\mu,\nu},\quad\{\gamma_\mu,\gamma_5\}=0,\\
\gamma_5\frac{1+\gamma_5}{2}=\frac{1+\gamma_5}{2},\quad\gamma_5\frac{1-\gamma_5}{2}=\frac{\gamma_5-1}{2},\\
\end{gather*}
we find that for $I_1$ we have 
\begin{equation*}
\begin{split}
[\ii\slashed{P}^+-\mu\ped{A}\gamma_0\gamma_5]I_1 P^+(-\mu\ped{A})^2 &=\frac{1+\gamma_5}{2}\left(P^{+2}-\ii\mu\ped{A} P_\mu^+\gamma_\mu\gamma_0\gamma_5+\ii\mu\ped{A} P_\mu^+\gamma_0\gamma_5\gamma_\mu-\mu\ped{A}^2\right)\\
&=\frac{1+\gamma_5}{2}\left[P^{+2}-\ii\mu\ped{A} P_\mu^+\left(\gamma_\mu\gamma_0+\gamma_0\gamma_\mu\right)\gamma_5-\mu\ped{A}^2\right]\\
&=\frac{1+\gamma_5}{2}\left[P^{+2}-\ii\mu\ped{A} P_\mu^+\left(\gamma_\mu\gamma_0+2\delta_{\mu,0}-\gamma_\mu\gamma_0\right)\gamma_5-\mu\ped{A}^2\right]\\
&=\frac{1+\gamma_5}{2}\left[P^{+2}-2\ii\mu\ped{A} P_0^+\gamma_5-\mu\ped{A}^2\right]\\
&=\frac{1+\gamma_5}{2}\left[P^{+2}-2\ii\mu\ped{A} P_0^+-\mu\ped{A}^2\right]=\frac{1+\gamma_5}{2}P^+(-\mu\ped{A})^2.
\end{split}
\end{equation*}
Similarly, for $I_2$ we have
\begin{equation*}
\begin{split}
[\ii\slashed{P}^+-\mu\ped{A}\gamma_0\gamma_5]I_2 P^+(+\mu\ped{A})^2 &=\frac{1-\gamma_5}{2}\left(P^{+2}-\ii\mu\ped{A} P_\mu^+\gamma_\mu\gamma_0\gamma_5+\ii\mu\ped{A} P_\mu^+\gamma_0\gamma_5\gamma_\mu-\mu\ped{A}^2\right)\\
&=\frac{1-\gamma_5}{2}\left[P^{+2}-2\ii\mu\ped{A} P_0^+\gamma_5-\mu\ped{A}^2\right]\\
&=\frac{1-\gamma_5}{2}\left[P^{+2}+2\ii\mu\ped{A} P_0^+-\mu\ped{A}^2\right]=\frac{1-\gamma_5}{2}P^+(+\mu\ped{A})^2.
\end{split}
\end{equation*}
Together they give the needed identity:
\begin{equation*}
[\ii\slashed{P}^+-\mu\ped{A}\gamma_0\gamma_5][I_1+I_2]=\frac{1+\gamma_5}{2}+\frac{1-\gamma_5}{2}=1.
\end{equation*}
The matrix $S(P)$ can be written in compact notation using right and left chemical potential and the chiral projector.
The chiral projector transforms a spinor into its right or left chirality components and it is defined by:
\begin{equation*}
\mathbb{P}\ped{R}=\frac{1+\gamma_5}{2},\quad\mathbb{P}\ped{L}=\frac{1-\gamma_5}{2};
\end{equation*}
while right and left chemical potential are defined by a linear combination of electric and axial chemical potential:
\begin{equation*}
\mu\ped{R}\equiv\mu+\mu\ped{A},\quad\mu\ped{L}\equiv\mu-\mu\ped{A}.
\end{equation*}
In terms of right and left chemical potentials, we define the right or left charged momenta by
\begin{equation*}
P\ped{R/L}^\pm\equiv(\omega_n\pm\I \mu\ped{R/L},\vec{p})
\end{equation*}
and the matrix $S(P)$ can be written as
\begin{equation*}
S(P)=\mathbb{P}\ped{R}\frac{-\I\slashed{P}^+\ped{R}}{{P^+\ped{R}}^2}+ \mathbb{P}\ped{L}\frac{-\I\slashed{P}^+\ped{L}}{{P^+\ped{L}}^2}
\equiv \sum_{\chi}\mathbb{P}_\chi\frac{-\I\slashed{P}^+_\chi}{{P^+_\chi}^2}
\end{equation*}
where the sum is on the chirality $\chi=$R,L which correspond to $\chi=+1,-1$ respectively.
To conclude, the thermal propagator in momentum space is given by
\begin{equation*}
\mean{\psi_a(P)\bar{\psi}_b(Q)}_T=\bar{\delta}(P-Q)\sum_{\chi}\left(\mathbb{P}_\chi\frac{-\I\slashed{P}^+_\chi}{{P^+_\chi}^2}\right)_{ab}
\end{equation*}
or in the coordinate space
\begin{equation*}
\mean{\psi_a(X)\bar{\psi}_b(Y)}_T=\sumint_{\{P\}}\sum_{\chi}\e^{\ii P^+_\chi\cdot(X-Y)}\left(\mathbb{P}_\chi\frac{-\I\slashed{P}^+_\chi}{{P^+_\chi}^2}\right)_{ab}.
\end{equation*}
We can formally put together the expression of the propagator for massive Dirac field in a non-chiral medium and a massless Dirac field in a chiral medium
with
\begin{equation}
\label{eq:propFermCoord}
\mean{\psi_a(X)\bar{\psi}_b(Y)}_T=\sumint_{\{P\}}\sum_{\chi}\e^{\ii P^+_\chi\cdot(X-Y)}\left(\mathbb{P}_\chi\frac{-\I\slashed{P}^+_\chi+m}{{P^+_\chi}^2+m^2}\right)_{ab}.
\end{equation}
%

\section{Path integral for gauge field}
In Sec.~\ref{sec:Symm} we have seen that conservation of electric current requires a gauge invariant theory. Gauge
invariance for fermionic action is obtained through the covariant derivative that introduces a coupling between
fermionic field and a four-vector bosonic gauge field. This coupling describes quantum electrodynamics and quantum
chromodynamics when the gauge fields considered are respectively Abelian and non-Abelian gauge fields. In both cases
also the dynamical description of the gauge fields must be added in the action of the theory and must be properly
quantized. The dynamical part of a  $SU(N_c)$ non-Abelian gauge field has Lagrangian
\begin{equation*}
\mc{L}=-\frac{1}{4}F^a_{\mu\nu}F^{a\mu\nu},\quad
F^a_{\mu\nu}=\de_\mu A^a_\nu-\de_\nu A^a_\mu+gf^{abc}A^b_\mu A^c_\nu,
\end{equation*}
where $g$ is the gauge coupling.  The gauge strength tensor can also be written in terms of the covariant derivative in
the adjoint representation $\mc{D}^{ac}_\mu$
\begin{equation*}
\mc{D}^{ac}_\mu=\de_\mu \delta^{ac}+gf^{abc} A^b_\mu,\quad
F^a_{\mu\nu}=\de_\mu A^a_\nu-\mc{D}^{ac}_\nu A^c_\mu=\mc{D}^{ac}_\mu A^c_\nu-\de_\nu A^a_\mu.
\end{equation*}
Calling $T^a$  the Hermitean generators of $SU(N_c)$, which satisfy the algebra $[T^a,T^b]=\I f^{abc}T^c$ and normalized
to $\tr[T^a,T^b]=\delta^{ab}/2$, we define $U\equiv\exp(\I g\alpha^a(x)T^a)$. The Lagrangian is invariant under the gauge
transformation:
\begin{equation*}
\begin{split}
A_\mu\to& A_\mu'={A'_\mu}^a T^a=UA_\mu U^{-1}+\frac{\I}{g}U\de_\mu U^{-1},\\
A_\mu^a\to&{A'_\mu}^a=A_\mu^a+\mc{D}^{ac}_\mu \alpha^c+\mc{O}(\alpha^2).
\end{split}
\end{equation*}

Although physical quantities are gauge invariant, we need to break gauge invariance to carry on the quantization of the gauge field.
The connection with physical states is to be addressed after we provide a quantization in a specific gauge. We choose a gauge
fixing such that it provides
\begin{equation*}
A_0^a\equiv 0,
\end{equation*}
and we treat the spatial components $A_i^a$ as canonical variables. This choice corresponds to a ``soft'' breaking of
gauge invariance, since time-independent gauge transformation are still allowed. The Lagrangian then becomes
\begin{equation*}
\mc{L}=\frac{1}{2}\de_0 A_i^a\de_0 A^a_i-\frac{1}{4}F^a_{ij}F^a_{ij},
\end{equation*}
from which we derive the canonical conjugate momentum
\begin{equation*}
E^a_i\equiv\frac{\de\mc{L}}{\de(\de_0A_i^a)}=\de_0 A^a_i,
\end{equation*}
and consequently the Hamiltonian density expressed with canonical fields and their conjugate momentum is
\begin{equation*}
\mc{H}=E^a_i\de_0A^a_i-\mc{L}=\frac{1}{2}E^a_i E^a_i+\frac{1}{4}F^q_{ij}F^a_{ij}.
\end{equation*}
Then, quantization of gauge field is achieved promoting the canonical variables to operator such that they satisfy
the equal time commutation relation:
\begin{equation*}
[\h{A}^a_i(t,\vec{x}),\h{E}^b_j(t,\vec{y})]=\I\delta^{ab}\delta_{ij}\delta(\vec{x}-\vec{y}).
\end{equation*}
The Hamiltonian of the system is now written in terms of the field operators:
\begin{equation*}
\h{H}=\int_{\vec{x}}\left(\frac{1}{2}\h{E}^a_i \h{E}^a_i+\frac{1}{4}\h{F}^a_{ij}\h{F}^a_{ij}\right).
\end{equation*}
The electric charge and axial charge operator in the statistical operator involve only fermionic fields
and we can ignore them completely when we deal with gauge fields.

We previously chose a specific gauge without providing an identification of physical states. We now need to identify physical
states that are gauge invariant. To that purpose, we introduce the operator $\h{U}$ which parametrize the time-independent
gauge transformations:
\begin{equation*}
\h{G}^a\equiv\h{\mc{D}}^{ab}_i\h{E}_i^b,\quad \h{U}\equiv\exp\left\{-\I\int_{\vec{x}}\alpha^a(\vec{x})\h{G}^a(\vec{x})\right\}.
\end{equation*}
It can be shown that $\h{G}^a$ commutes with the Hamiltonian and that indeed the operator $\h{U}$ transforms eigenstates $A_i^a$ of the field
operator $\h{A}^a_i$ into eigenstates ${A'_i}^a$ of the gauge transformed operator ${\h{A}^{\prime a}_i}$. In this way physical, i.e. gauge
invariant, states ``$\ket{\text{phys}}$'' are identified as those who satisfy $\h{U}^{-1}\ket{\text{phys}}=\ket{\text{phys}}$, which
formalize gauge invariance into an equation for the state. Expanding the operators $\h{U}$ at first order in gauge transformation parameter
$\alpha$ we see that physical states correspond to states satisfying the condition
\begin{equation*}
\h{G}^a\ket{\text{phys}}=0.
\end{equation*}
Since $\h{H}$ and $\h{G}$ commutes, it exist a vector basis of Hilbert space $\ket{E_q,q}$, which is composed by simultaneous
eigenvectors of the Hamiltonian and the operator $\h{G}$: $\h{H}\ket{E_q,q}=E_q\ket{E_q,q}$ and $\h{G}^a\ket{E_q,q}=q\ket{E_q,q}$.
It follows that only the eigenvectors with vanishing eigeinvalue $q$ are physical states. This observation allows us to evaluate
the physical partition function expanding the trace in $\ket{E_q,q}$ basis and selecting only the states with vanishing $q$:
\begin{equation*}
\parz\ped{phys}=\sum_{E_0}\bra{E_0,0}\E^{\beta E_0}\ket{E_0,0}=\sum_{E_q,q}\bra{E_q,q}\delta_{q,0}\E^{\beta E_q}\ket{E_q,q}
=\tr\left[\delta_{\h{G}^a,\h{0}}\,\,\E^{-\beta \h{H}}\right].
\end{equation*}
As done for the fermionic field we can turn to the imaginary time and divide the interval in $N$ pieces. In the $N\to\infty$ limit,
after the evaluation of transition amplitude and the integration on conjugate momenta, we eventually arrive at the path integral
representation of partition function~\cite{Laine:2016hma}:
\begin{equation}
\label{eq:parzGauge1}
\parz\ped{phys}=C\int\Dpi A_o^a\int_{A^a_i(\beta,\vec{x})=A^a_i(0,\vec{x})}\Dpi A^a_i \exp\left\{-\int_X \mc{L}_E\right\},
\end{equation}
where, being the gauge field a bosonic field, the boundary condition are periodic and the Eucldean Lagrangian is
\begin{equation*}
\mc{L}_E=\frac{1}{4}F^a_{\mu\nu}F^a_{\mu\nu},\quad F^a_{0i}=\de_\tau A^a_i-\mc{D}^{ab}_i A_0^b.
\end{equation*}

The previous expression for partition function is gauge invariant but it is not suitable for perturbation theory as the
Euclidean Lagrangian contains a non-invertible matrix as quadratic term and we can not define a propagator. This problem
is overcome breaking again the gauge invariance. Consider a generic function of integration variables $G^a$. An insertion
of a term of the type
\begin{equation*}
\prod_{X,Y,a,b}\delta(G^a)\det\left|\frac{\delta G^a(X)}{\delta\alpha^b(Y)}\right|
\end{equation*}
inside the integral of Eq.~(\ref{eq:parzGauge1}) does not change results for gauge independent quantities. To see that,
we divide the integration variables according to the gauge fixing $G^a$, splitting the gauge integration variables into gauge
fields $\bar{A}_\mu$, not connected with transformation $G^a$, and those connected with it, parametrized with $\alpha$. We then
find
\begin{equation*}
\begin{split}
\int\Dpi A_\mu &\delta(G^a)\det\left|\frac{\delta G^a}{\delta\alpha^b}\right|\E^{-\int_X \mc{L}_E(A_\mu)}=
\int\Dpi \bar{A}_\mu\int\Dpi \alpha^b \delta(G^a)\det\left|\frac{\delta G^a}{\delta\alpha^b}\right|\E^{-\int_X \mc{L}_E(\bar{A}_\mu)}\\
=&\int\Dpi \bar{A}_\mu\int\Dpi G^a \delta(G^a)\E^{-\int_X \mc{L}_E(\bar{A}_\mu)}=\int\Dpi \bar{A}_\mu \E^{-\int_X \mc{L}_E(\bar{A}_\mu)},
\end{split}
\end{equation*}
where we took advantage of Lagrangian gauge invariance. We can then exploit the fact that integration is independent on the choice
of $G^a$ and we replace $\delta(G^a)$ with $\delta(G^a-f^a)$, where $f^a$ is a general function independent of $A_\mu^a$. 
Then we can average over the $f^a$’s with a Gaussian weight:
\begin{equation*}
\int\Dpi f^a \delta(G^a-f^a)\exp\left(-\frac{1}{2\xi}\int_X f^a\,f^a\right)=\exp\left(-\frac{1}{2\xi}\int_X G^a G^a\right),
\end{equation*}
with $\xi$ a real parameter. Instead, the determinant $\det|\delta G^a/\delta\alpha^b|$ can be written with auxiliary fields
$c$ taking advantage of the identity
\begin{equation*}
\det\left|\frac{\delta G^a}{\delta\alpha^b}\right|=\int\Dpi \bar{c}\Dpi c\exp\left(-\bar{c}^a\frac{\delta G^a}{\delta\alpha^b}c^b \right).
\end{equation*}
The fields $c$ are known as Faddeev-Popov ghosts and despite being Grassmann variables they must satisfy periodic boundary conditions
because of the bosonic nature of $G^a$. In conclusion, the path integral form of partition function for gauge fields is
\begin{equation}
\label{eq:parzGauge}
\begin{split}
\parz\ped{phys}=&C\int\Dpi A_o^a\int\ped{Periodic}\Dpi A^a_i \int\ped{Periodic}\Dpi \bar{c}^a \Dpi c^a \times\\
&\times\exp\left\{-\int_X \left[\frac{1}{4}F^a_{\mu\nu} F^a_{\mu\nu}+\frac{1}{2\xi}G^a G^a+\bar{c}^a\frac{\delta G^a}{\delta \alpha^b}c^b\right] \right\}.
\end{split}
\end{equation}
In this thesis, we are using the \emph{covariant gauge} $G^a\equiv -\de_\mu A_\mu^a$ for which we have
\begin{gather*}
\frac{1}{2\xi}G^a G^a=\frac{1}{2\xi}\de_\mu A_\mu^a \de_\nu A_\nu^a,\\
\frac{\delta G^a}{\delta \alpha^b}=\olaw{\de}_\mu\frac{\delta A_\mu^a}{\delta\alpha^b}=\olaw{\de}_\mu\left(\oraw{\de}_\mu \delta^{ab}+g f^{abc}A_\mu^c\right),\\
\bar{c}^a\frac{\delta G^a}{\delta \alpha^b}c^b=\de_\mu\bar{c}^a\de_\mu c^a+g f^{abc}\de_\mu\bar{c}^a A_\mu^b c^c.
\end{gather*}

We can now find the thermal propagator. First, we expand the fields in momentum modes. Both the gauge field and the ghost field
are periodic in imaginary-time translation $\beta$ and consequently the time component modes are given by the discrete \emph{bosonic
	Matsubara frequencies}:
\begin{equation*}
\omega_n=2\pi T\,n,\quad n \in\mathbb{Z}.
\end{equation*}
The expansion in momenta modes are then written as
\begin{equation*}
A^a_\mu(X)=\sumint_P \E^{\I P\cdot X}\tilde{A}_\mu^a(P),\quad c^a(X)=\sumint_P \E^{\I P\cdot X}\tilde{c}^a(P).
\end{equation*}
In covariant gauge, the quadratic part of Euclidean action inside the partition function expressed in momentum modes is:
\begin{equation*}
\begin{split}
\int_X\frac{1}{4}F^a_{\mu\nu}F^a_{\mu\nu}=&\sumint_{P,Q}\bar{\delta}(P+Q)\frac{1}{2}\tilde{A}_\mu(P)\tilde{A}_\nu(Q)\left[\delta_{\mu\nu}P^2-P_\mu P_\nu\right]
+\text{non-quadratic terms},\\
\int_X\frac{1}{2\xi}G^a G^a=&\sumint_{P,Q}\bar{\delta}(P+Q)\frac{1}{2\xi}P_\mu P_\nu\tilde{A}_\mu(P)\tilde{A}_\nu(Q),\\
\int_X \bar{c}^a\frac{\delta G^a}{\delta \alpha^b}c^b=&\sumint_{P,Q}\bar{\delta}(P-Q)P^2\tilde{\bar{c}}^a(P)\tilde{c}^a(Q)+\text{non-quadratic terms},\\
S\ped{E}(\tilde{A}^a_\mu,\tilde{\bar{c}}^a,\tilde{c}^a)=&\sumint_{P,Q}\bar{\delta}(P+Q)\frac{1}{2}\tilde{A}_\mu(P)\tilde{A}_\nu(Q)
\left[\delta_{\mu\nu}P^2-\left(1-\frac{1}{\xi}\right)P_\mu P_\nu\right]\\
&+\sumint_{P,Q}\bar{\delta}(P-Q)P^2\tilde{\bar{c}}^a(P)\tilde{c}^a(Q)+\text{non-quadratic terms}.
\end{split}
\end{equation*}
The thermal propagators are obtained as inverse matrix of the quadratic part. For gauge field we obtain
\begin{equation*}
\mean{\tilde{A}^a_\mu(P)\tilde{A}^a_\nu(Q)}_T=\bar{\delta}(P+Q)\delta^{ab}\frac{1}{P^2}\left[\delta_{\mu\nu}-(1-\xi)\frac{P_\mu P_\nu}{P^2}\right]
\end{equation*}
and for ghost field we have:
\begin{equation*}
\mean{\tilde{\bar{c}}^a(P)\tilde{c}^b(Q)}_T=\bar{\delta}(P-Q)\delta^{ab}\frac{1}{P^2}.
\end{equation*}
In coordinate space the gauge thermal propagator is
\begin{equation*}
\mean{ A^a_\mu(X)A^b_\nu(Y)}_T= \sumint_{P}\E^{\I P(X-Y)}\delta^{ab} \frac{G_{\mu\nu}(P)}{P^2}
\end{equation*}
where:
\begin{equation}
\label{propden}
G_{\mu\nu}(P) = \delta_{\mu\nu}-(1-\xi)\frac{P_\mu P_\nu}{P^2}.
\end{equation}
%

\section{Homogeneous thermal equilibrium with axial charge}
\label{sec:homogeq}
In this section, we carry out the calculations for thermal mean values of stress-energy tensor, electric current, axial current and
thermodynamical potential for a free gas of fermions at thermal equilibrium with a conserved axial charge. This serves to
illustrate the methods used in the other parts of the thesis and as a reference point for the other cases considered.

\subsection{Thermodynamic potential}
\label{subsec:thermpot}
We start by the thermodynamic potential $\Omega$ , for which we remind the definition:
\begin{equation*}
\Omega\equiv \lim_{V\to \infty} -\frac{T}{V}\log\parz\ped{h}.
\end{equation*}
In Sec.~\ref{sec:PathIntFermion} we derived the path integral functional for the fermionic part of the partition function~(\ref{eq:parzFermion}),
which we report here:
\begin{equation*}
\parz\ped{h}= C' \int \mathcal{D} \tilde{\bar{\Psi}}\,\mathcal{D} \tilde\Psi \,\,
\exp \left\{ -\sumint_{\{P\}} \tilde{\bar{\psi}}(P)  S\ped{F}^{-1}(P) \tilde\Psi(P) \right\}
\end{equation*}
with $S\ped{F}^{-1}(P)$ the inverse propagator matrix. We can explicitly write the spinorial component of the inverse propagator as
$2\times 2$ block matrix using the chiral representation of Eucldean gamma matrices:
\begin{equation*}
S\ped{F}^{-1}(P)\equiv\I\slashed{P}^+ +m-\gamma_0\gamma^5\mu\ped{A}=\left(\begin{array}{cc}
m & \I(\omega_n+\I\mu)+\sigma_i p_i-\mu\ped{A}\\
\I(\omega_n+\I\mu)-\sigma_i p_i+\mu\ped{A} & m
\end{array}\right),
\end{equation*}
where the $\sigma_i$ are the Pauli matrices. As mentioned in the sections above, the partition function is an integral over Grassmann
variables with a quadratic exponential weight. The result of integration is simply the determinant of the inverse propagator~\cite{KapustaGale}.
The determinant of the spinorial components are carried out using determinant properties of block matrices:
\begin{equation*}
\parz=\det\left|S\ped{F}^{-1}(P)\right|=\det\left|\begin{array}{cc}A & B\\ C & D \end{array}\right|
=\det\left|AD-BD^{-1}CD\right|.
\end{equation*}
The multiplication of the matrices is straightforward and we find:
\begin{equation*}
\parz=\det\left|\left({P^+}^2+m^2+\mu\ped{A}^2\right)\idmat_{2\times 2}-2\sigma_i p_i\mu\ped{A}\right|
=\det\left|\left({P^+}^2+m^2+\mu\ped{A}^2\right)^2-4 \vec{p}^2\mu\ped{A}^2\right|.
\end{equation*}
The functional determinant is found as the product of all the eigenvalues of the matrix which simply results in the product over
the momentum modes $P=\{\omega_n,\vec{p}\}$:
\begin{equation*}
\parz=\prod_{P}\left[\left({P^+}^2+m^2+\mu\ped{A}^2\right)^2-4 \vec{p}^2\mu\ped{A}^2\right].
\end{equation*}
The thermodynamic potential is defined as the logarithm of the partition function, therefore the previous products become
sums of the logarithms:
\begin{equation*}
\log \parz=\sum_{P}\log\left[\left({P^+}^2+m^2+\mu\ped{A}^2\right)^2-4 \vec{p}^2\mu\ped{A}^2\right].
\end{equation*}
At last, the thermodynamic potential $\Omega$ is obtained in the infinite volume as $\Omega=-T\log\parz/V$ and it is therefore given by:
\begin{equation*}
\Omega=\lim_{V\to\infty}-\frac{T}{V}\sum_{P}\log\left[\left({P^+}^2+m^2+\mu\ped{A}^2\right)^2-4 \vec{p}^2\mu\ped{A}^2\right].
\end{equation*}
The large volume limit exactly reproduces the sums on momenta we introduced for the Fourier transform of fermionic fields in Eq. (\ref{eq:DiracMomenta}).
Using that same notation, the thermodynamic potential is written as:
\begin{equation*}
\begin{split}
\Omega=&-\sumint_{\{P\}}\log\left[\left({P^+}^2+m^2+\mu\ped{A}^2\right)^2-4 \vec{p}^2\mu\ped{A}^2\right]\\
=&-\int\frac{\D^3 p}{(2\pi)^3}T\sum_{\{\omega_n\}}\log\left[\left({P^+}^2+m^2+\mu\ped{A}^2\right)^2-4 \vec{p}^2\mu\ped{A}^2\right].
\end{split}
\end{equation*}
This expression would acquire a clearer physical interpretation if we perform the sum on Matsubara frequencies. Since the sums of logarithmic functions
are hardly addressed we perform the sum with the help of an auxiliary function. First, we define the function
\begin{equation*}
j(y)\equiv T\sum_{\{\omega_n\}}\log\left[\left({P^+}^2+m^2+\mu\ped{A}^2\right)^2-y^2\right];
\end{equation*}
the thermodynamic potential is obtained from $j(y)$ by the equation
\begin{equation*}
\Omega=-\int\frac{\D^3 p}{(2\pi)^3}j(y_0),
\end{equation*}
where we denoted $y_0\equiv 2\mu\ped{A}|\vec{p}|$. Then, from $j(y)$ we derive an additional function $i(y)$:
\begin{equation*}
i(y)\equiv \frac{1}{2y}\frac{\D}{\D y}j(y)=T\sum_{\{\omega_n\}}\frac{1}{\left[(\omega_n+\I\mu)^2+\vec{p}^2+m^2+\mu\ped{A}^2\right]^2-y^2}.
\end{equation*}
It is easier to perform the sum on $i(y)$. Once we have the closed form of $i(y)$ we obtain $j(y)$ inverting their relation, i.e. integrating $i(y)$.

We now proceed to evaluate the sum inside $i(y)$. Using standard techniques (see Appendix~\ref{sec:thermalsums}) we transform the sum on $\omega_n$ to a complex
integral:
\begin{equation*}
i(y)=T\sum_{\{\omega_n\}} f(\omega_n)=\int_{-\infty-\I 0^+}^{\infty-\I 0^+}\frac{\D z}{2\pi}\left\{f(z)[1-n\ped{F}(\I z)]-f(-z)n\ped{F}(\I z) \right\}
\end{equation*}
where $n\ped{F}(x)$ is the Fermi-Dirac distribution function and the function $f$ is
\begin{equation*}
f(z)=[(z+\I(E_+ +\mu))(z+\I(E_-+\mu))(z-\I(E_+-\mu))(z-\I(E_--\mu)]^{-1}
\end{equation*}
and we defined
\begin{equation*}
E_\pm^2=E^2\pm y=\vec{p}^2+m^2+\mu\ped{A}^2\pm y=(|\vec{p}|^2\pm\mu\ped{A})^2+m^2.
\end{equation*}
The complex integral is computed with residue theorem closing the path in the lower half-plane.
In that region\footnote{Here we considered the case of a massive field where $\mu,\mu\ped{A}<m$, hence $E_\pm>\mu$ for every value of
	$\vec{p},\mu,\mu\ped{A}$. For a vanishing mass several cases can be distinguished when computing the complex integral, however the final
	result coincides with the massive case in the $m\to 0$ limit.} $f(z)$ has two poles in $z=-\I(E_\pm+\mu)$ and $f(-z)$ has two poles in
$z=-\I(E_\pm-\mu)$, while $n\ped{F}(\I z)$ has poles only on the real axis. Therefore the residues theorem states that the result of the
integral is the sum of the following residues:
\begin{equation*}
\begin{split}
i(y)=&\sum_{s=\pm}\left\{-\I\res[f(z)]_{z=-\I(E_s+\mu)}(1-n\ped{F}(E_s+\mu))\right.\\
&\left.+\I\res[f(-z)]_{z=-\I(E_s-\mu)}n\ped{F}(E_S-\mu)\right\}.
\end{split}
\end{equation*}
The residues are easily computed:
\begin{gather*}
\I \res\left(f(z)\right)_{z=-\I(E_+ +\mu)}=\frac{1}{4y\,E_+},\quad \I \res\left(f(z)\right)_{z=-\I(E_-+\mu)}=-\frac{1}{4y\,E_-},\\
\I \res\left(f(-z)\right)_{z=-\I(E_+-\mu)}=\frac{1}{4y\,E_+},\quad \I \res\left(f(-z)\right)_{z=-\I(E_--\mu)}=-\frac{1}{4y\,E_-},
\end{gather*}
and in total the function $i(y)$ is
\begin{equation*}
i(y)=\frac{1}{4y}\left[\frac{1}{E_-}-\frac{1}{E_+}+\frac{n\ped{F}(E_+-\mu)}{E_+}-\frac{n\ped{F}(E_- -\mu)}{E_-}+\frac{n\ped{F}(E_++\mu)}{E_+}-\frac{n\ped{F}(E_-+\mu)}{E_-}\right].
\end{equation*}
The function $j(y)$ is obtained with the integration:
\begin{equation*}
j(y)=\int^{y_0} 2y\, i(y)\D y.
\end{equation*}
Noticing that
\begin{equation*}
\int^{y_0}\frac{\D y}{2\sqrt{E^2\pm y}}=\pm\sqrt{E^2\pm y_0}, \quad
\frac{\D}{\D y}\log\left(1+\E^{-\beta(E_\pm \pm\mu)}\right)=\frac{\mp\beta}{2E_\pm}n\ped{F}(E_\pm \pm\mu)
\end{equation*}
we can verify that the integration gives
\begin{equation*}
\begin{split}
j(y_0)=&-E_- -E_+-\frac{1}{\beta}\log\left(1+\E^{-\beta(E_-+\mu)}\right)-\frac{1}{\beta}\log\left(1+\E^{-\beta(E_++\mu)}\right)\\
&-\frac{1}{\beta}\log\left(1+\E^{-\beta(E_+-\mu)}\right)-\frac{1}{\beta}\log\left(1+\E^{-\beta(E_--\mu)}\right).
\end{split}
\end{equation*}
Finally the thermodynamic potential is
\begin{equation*}
\Omega=\int\frac{\D^3 p}{(2\pi)^3}\sum_{s=\pm}\left\{E_s +\sum_{\pm}T\log\left(1+\E^{-\beta(E_s\pm\mu)}\right)\right\}
\end{equation*}
with
\begin{equation*}
E^2_\pm=\vec{p}^2+m^2+\mu\ped{A}^2\pm2\mu\ped{A}|\vec{p}|^2.
\end{equation*}
From the thermodynamic potential we can derive electric charge density and axial charge density by means of a simple derivative:
\begin{gather*}
n\ped{c}=-\f{\de\Omega}{\de \mu},\quad
n\ped{A}=-\f{\de\Omega}{\de \mu\ped{A}},
\end{gather*}
but we are instead calculating them by the evaluation of thermal expectation values of the electric and axial current operators.
The same is done for the stress-energy tensor in the next section in order to illustrate the methods used for vorticous thermal
function in the next chapters.

\subsection{Point-splitting procedure}
\label{subsec:pointsplit}
Here we review how the mean value of local observables are obtained with the point-splitting procedure. Let $\h{O}(x)$ be a local
operator in Schr\"{o}dinger-representation and let $\h{O}(x)$ be Hermitian and quadratic in the fermionic field, such as
$\bar{\psi}\psi,\, \de_\mu\bar{\psi}\de_\nu\psi$ and so on. Many operators of physical interest satisfy these properties and it is
certainly the case for stress-energy tensor, electric and axial currents. Since we chosen the imaginary time representation of
thermal functions we first transform the operator in the Euclidean space. Consider the operator $\h{O}(0,\vec{x})$ at the time
$t=0$, then the operator in the imaginary time is obtained with the imaginary time-translation:
\begin{equation}
\label{eq:toimaginary}
\h{O}(X)=\h{O}(\tau,\vec{x})=\h {\sf T}((-\I\tau,\vec{0}))\h{O}(0,\vec{x}) \h {\sf T}^{-1}((-\I\tau,\vec{0})).
\end{equation}
Now, since it is quadratic, the operators can be written as
\begin{equation*}
\h{O}(X)=\h{\bar{\psi}}(X)\h{\mc{O}}\,\h\psi(X),
\end{equation*}
where $\h{\mc{O}}$ is a linear operator, such as a derivative or a multiplication by a scalar, who act on either $\bar\psi$ and $\psi$ field.
The point-splitting procedure prescribes to take the two fields $\bar{\psi}$ and $\psi$ in two different point $X_1$ and $X_2$ and to
split the action of the linear operator $\h{\mc{O}}$ in the two separated fields in a way that the splitting is symmetric and that the original
operator is recovered when the two points are again pinched together, i.e. when  $X_1,X_2\to X$. The operator is then written as a linear
operator $\mc{O}$ acting on the fields in two separate points:
\begin{equation*}
\h{O}(X)=\lim_{X_1,X_2\to X}\sum_{ab}\mathcal{O}(X_1,X_2)_{ab}\,\bar{\psi}_a(X_1)\psi_b(X_2),
\end{equation*}
where $a,b$ are spinor indices and from now on we omit the symbol `` $\,\hat{\,\,}\,$ '' in the operators.

The previous expression has the advantage that thermal expectation values are immediately given as the operator $\mc{O}$ acting
on thermal propagator of the fields that we already have derived:
\begin{equation*}
\begin{split}
\mean{\h{O}(X)}_\beta=&\lim_{X_1,X_2\to X}\sum_{ab}\mathcal{O}(X_1,X_2)_{ab}\,\mean{\bar{\psi}_a(X_1)\psi_b(X_2)}_\beta\\
=&\lim_{X_1,X_2\to X}\sum_{ab}-\mathcal{O}(X_1,X_2)_{ab}\,\mean{\psi_b(X_2)\bar{\psi}_a(X_1)}_\beta,
\end{split}
\end{equation*}
where in the last step we took advantage of fermionic properties to invert the order of the fields; in this way the sum on
spinor index forms the trace of the product of the two operators. Then, the thermal expectation value simplifies further when expanding
the fields propagator in momentum modes since the operator $\mc{O}$ acts only on the coordinates that are now located only at the exponential.
In the reference frame where the thermal bath is at rest, the fermion propagator is given by Eq.~(\ref{eq:propFermCoord}), replacing its
expression the local operator thermal expectation value becomes
\begin{equation*}
\begin{split}
\mean{\h{O}(X)}_T=&\lim_{X_1,X_2\to X}\sum_{ab}\sumint_{\{P\}}\sum_{\chi}-\left[\mathcal{O}(X_1,X_2)\e^{-\ii P^+_\chi\cdot(X_1-X_2)}\right]_{ab}\,
\left(\mathbb{P}_\chi\frac{-\I\slashed{P}^+_\chi}{{P^+_\chi}^2}\right)_{ba}\\
=&\lim_{X_1,X_2\to X}\sumint_{\{P\}}\sum_{\chi}-\tr\left[\left(\mathcal{O}(X_1,X_2)\e^{-\ii P^+_\chi\cdot(X_1-X_2)}\right)
\left(\mathbb{P}_\chi\frac{-\I\slashed{P}^+_\chi}{{P^+_\chi}^2}\right)\right].
\end{split}
\end{equation*}

We use this method in next sections for the mean values of the stress-energy tensor, the electric current
and the axial current with the homogeneous statistical operator (\ref{eq:rho_homogeq}). Then, the same
method is used in later chapters to evaluate vorticity corrections to the same observables.

\subsection{Stress-energy tensor thermal expectation value}
\label{subsec:stressenergygrandcanon}
The stress-energy tensor can be derived in curved space-time as the variation of Dirac action respect to the metric, or
equivalently by Eq.~(\ref{eq:setfermion}); then turning again in flat space-time we obtain the symmetric stress-energy tensor:
\begin{equation*}
\h{T}_{\mu\nu}(x)=\f{1}{4}\l\{ \l[\h{\bar{\psi}}\gamma_\mu\de_\nu\h\psi-(\de_\nu\h{\bar{\psi}})\gamma_\mu\h\psi\r]
+ \l[\h{\bar{\psi}}\gamma_\nu\de_\mu\h\psi-(\de_\mu\h{\bar{\psi}})\gamma_\nu\h\psi \r]\r\}.
\end{equation*}
More commonly, the stress-energy tensor is derived in flat space-time from the Dirac Lagrangian as the Noether current
resulting from translational invariance; in that case we would obtain the canonical stress-energy tensor:
\begin{equation*}
\h{T}_{\mu\nu}(x)=\f{1}{2} \l[\h{\bar{\psi}}\gamma_\mu\de_\nu\h\psi-(\de_\nu\h{\bar{\psi}})\gamma_\mu\h\psi\r].
\end{equation*}
The obvious main difference is that only the symmetric stress-energy tensor is symmetric under the exchange of Lorentz index $\mu\leftrightarrow\nu$.
However, they both provide the same Noether charge $\h{P}$:
\begin{equation*}
\h{P}^\mu=\int_\Sigma \D\Sigma_\lambda \h{T}^{\lambda\mu}.
\end{equation*}
Indeed, they both belong to a class of stress-energy tensors related one to another by a so-called pseudo-gauge transformation that leaves
the generator of translations unchanged. The pseudo-gauge transformation is given in terms of a rank three tensor field $\h{\Phi}$
antisymmetric in the last two indices known as superpotential:
\begin{equation*}
\h{T}'_{\mu\nu}=\h{T}_{\mu\nu}+\frac{1}{2}\nabla^\lambda\l(\h{\Phi}_{\lambda,\mu\nu}-\h{\Phi}_{\mu,\lambda\nu}-\h{\Phi}_{\nu,\lambda\mu}\r).
\end{equation*}
As mentioned in Sec.~(\ref{sec:LocEquil}), when representing the statistical operator of a system at thermal equilibrium, either including or
excluding thermal vorticity, we always chose the symmetric stress-energy tensor. On the other hand, this choice does not hinder the possibility
to consider the mean value of other stress-energy tensors like the canonical one. As we are now showing the mean value for the two operators
are identical in the case of homogeneous equilibrium while they provide different results in presence of vorticity~\cite{BecTin2012,Buzzegoli:2017cqy,Buzzegoli:2018wpy}.

Using the transformation in~(\ref{eq:toimaginary}), we build the Dirac field stress-energy tensor in Euclidean space-time.
Both the canonical and symmetric stress-energy tensor can be written in a single expression
\begin{equation}
\label{eq:seEuclidFerm}
\h{T}_{\mu\nu}(X)=\f{(\I)^{n_\mu+n_\nu}}{4}\l\{(2-\xi) \l[\h{\bar{\psi}}\gamma_\mu\de_\nu\h\psi-(\de_\nu\h{\bar{\psi}})\gamma_\mu\h\psi\r]
+ \xi \l[\h{\bar{\psi}}\gamma_\nu\de_\mu\h\psi-(\de_\mu\h{\bar{\psi}})\gamma_\nu\h\psi \r]\r\}
\end{equation}
where $n_\mu = \delta_{0,\alpha}$ and for $\xi=0$ we recover the canonical non-symmetric tensor, while for $\xi=1$ we recover the symmetric one.
The point-splitting procedure applied to the previous stress-energy tensor gives
\begin{equation*}
\h{T}_{\mu\nu}(X)=\lim_{X_1 X_2\to X}\sum_{a,b} \Dpi_{\mu\nu}(\de_{X_1},\de_{X_2})_{ab}\h{\bar{\psi}}_a(X_1)\h\psi_b(X_2)
\end{equation*}
where
\begin{equation*}
\Dpi_{\mu\nu}(X,Y)_{ab}=\f{(\I)^{n_\mu+n_\nu}}{4}\l[ (2-\xi){\gamma_\mu}_{ab}(Y-X)_\nu +\xi {\gamma_\nu}_{ab}(Y-X)_\mu \r].
\end{equation*}
It is not necessary to evaluate the thermal mean value for all the components of the tensor $\mean{\h{T}_{\mu\nu}}_\beta$,
but we can reduce to consider only two scalar functions. Since the statistical operator $\rho_h$ of
Eq.~(\ref{eq:rho_homogeq}) is invariant under translations and rotations, the stress-energy tensor mean value takes the ideal form:
\begin{equation*}
\mean{\h{T}_{\mu\nu}(X)}_\beta=\rho\,u_\mu u_\nu -p\,\Delta_{\mu\nu}
\end{equation*}
where $\rho$ is the energy density and $p$ is the pressure. Since the two thermodynamic functions are Lorentz invariant they can
be evaluated without loss of generality in the local rest frame of thermal bath. In that frame the two thermodynamic functions
are obtained with the thermal expectation values of the following components:
\begin{equation*}
\rho = \mean{\h{T}_{00}}_T\, ,\quad  p =\sum_i \f{\mean{\h{T}_{ii}}_T}{3}.
\end{equation*}

Following the point-splitting procedure, the thermal expectation value of stress-energy tensor at the rest frame is obtained from
\begin{equation*}
\mean{\h{T}_{\mu\nu}}_T=\lim_{X_1 X_2\to X} -\tr\left[\Dpi_{\mu\nu}(\de_{X_1},\de_{X_2})\mean{\psi(X_2)\bar{\psi}(X_1)}_T\right].
\end{equation*}
The operator $\Dpi$ acts on the Dirac thermal propagator only through derivatives, which from Eq.~(\ref{eq:propFermCoord}) are given by
\begin{gather*}
\de_{X_2} \mean{\psi(X_2)\bar{\psi}(X_1)}_T= \sumint_{\{P\}}\sum_{\chi}\e^{\ii P^+_\chi\cdot(X_2-X_1)}\I P^+_\chi\left(\mathbb{P}_\chi\frac{-\I\slashed{P}^+_\chi+m}{{P^+_\chi}^2+m^2}\right)\\
\de_{X_1} \mean{\psi(X_2)\bar{\psi}(X_1)}_T= \sumint_{\{P\}}\sum_{\chi}\e^{\ii P^+_\chi\cdot(X_2-X_1)}(-\I P^+_\chi)\left(\mathbb{P}_\chi\frac{-\I\slashed{P}^+_\chi+m}{{P^+_\chi}^2+m^2}\right).
\end{gather*}
The coordinate dependence is all gathered in the exponential which goes to one after we perform the limit $X_1,X_2\to X$, hence the results is homogeneous.
We are then left with
\begin{equation*}
\mean{\h{T}_{\mu\nu}(X)}_T =-\f{\I^{n_\mu+n_\nu+1}}{2}\sumint_{\{P\}}\sum_{\chi}\tr\left[\left((2-\xi)\gamma_\mu {P^+_\chi}_\nu+\xi\gamma_\nu {P^+_\chi}_\mu\right)
\left(\mathbb{P}_\chi\frac{-\I\slashed{P}^+_\chi+m}{{P^+_\chi}^2+m^2}\right)\right].
\end{equation*}
The trace on gamma matrices gives
\begin{equation*}
\tr\l[ \gamma_\mu \mathbb{P}_\chi\l(-\I\slashed{P}+m\r)\r]=\frac{1}{2}\tr\l[ \gamma_\mu \l(-\I\gamma_\lambda P_\lambda +m\r)\r]
+\frac{\chi}{2}\tr\l[ \gamma_\mu \gamma^5\l(-\I\gamma_\lambda P_\lambda+m\r)\r]=-2\I P_\mu.
\end{equation*}
We replace the trace and we simplify the terms:
\begin{equation*}
\begin{split}
\mean{\h{T}_{\mu\nu}(X)}_T &= -\f{\I^{n_\mu+n_\nu}}{2}\sumint_{\{P\}}\sum_\chi \f{2[(2-\xi){P^+_\chi}_\mu {P^+_\chi}_\nu+\xi {P^+_\chi}_\nu {P^+_\chi}_\mu]}{{P^+_\chi}^2+m^2}\\
&= -2\,\I^{n_\mu+n_\nu}\sumint_{\{P\}}\sum_\chi \f{{P^+_\chi}_\mu {P^+_\chi}_\nu}{{P^+_\chi}^2+m^2}\\
&= -2\,\I^{n_\mu+n_\nu}\int\frac{\D^3 p}{(2\pi)^3}\sum_\chi\,T\sum_{\{\omega_n\}} \f{{P^+_\chi}_\mu {P^+_\chi}_\nu}{{P^+_\chi}^2+m^2}.
\end{split}
\end{equation*}
Note that the result does not depend on $\xi$, meaning that it does not depend whether we choose canonical or symmetric stress-energy tensor.
We do not have to compute all the components but just the energy density and pressure. 

Let's start with the energy; in the rest frame it is given by the time-time component
\begin{equation*}
\begin{split}
\rho&= -2(\I)^2 \int\frac{\D^3 p}{(2\pi)^3}\sum_\chi T\sum_{\{\omega_n\}}\f{(\omega_n+\I \mu_\chi)^2}{{P^+_\chi}^2+m^2}\\
&=2\int\frac{\D^3 p}{(2\pi)^3}\sum_\chi T\sum_{\{\omega_n\}}\f{(\omega_n+\I \mu_\chi)^2}{(\omega_n+\I\mu)^2+E^2}
=2\int\frac{\D^3 p}{(2\pi)^3}\sum_\chi T\sum_{\{\omega_n\}} f(\omega_n)
\end{split}
\end{equation*}
where we have denoted the energy $E^2=\vec{p}^2+m^2$ and $f$ the integrand function. Then, using~(\ref{eq:thermalsumrule}) for the
Matsubara frequencies sum, we obtain
\begin{equation*}
\rho=2\int\frac{\D^3 p}{(2\pi)^3}\sum_\chi\l[\int_{-\infty}^{+\infty} \f{\D z}{2\pi} f(z)-
\int_{-\infty-\I 0^+}^{+\infty-\I 0^+} \f{\D z}{2\pi} [ f(z) + f(-z) ] n\ped{F}(\I z)\r].
\end{equation*}
The integral in $z$ can be solved with the residue theorem closing the contour in the lower half-plane.
Since for the relativistic chemical potential yields $0<\mu_\chi<m<E$, $f(z)$ has only a pole in the lower half-plane at $z=-\I(E+\mu_\chi)$; the 
value at the residue is
\begin{equation*}
\begin{split}
\res&\l(f(z)n\ped{F}(\I z)\r)_{z=-\I(E+\mu_\chi)}=\\
&=\lim_{z\to-\I(E+\mu_\chi)}\f{\I(z+\I\mu_\chi)^2 n\ped{F}(\I z)}{2E}\l[\f{1}{z+\I(\mu_\chi+E)}-\f{1}{z+\I(\mu_\chi-E)}\r](z+\I(E+\mu_\chi))\\
&=\f{\I n\ped{F}(E+\mu_\chi)(-\I E)^2}{2E}=-\f{\I E}{2}n\ped{F}(E+\mu_\chi).
\end{split}
\end{equation*}
Instead $f(-z)$ has a pole in $z=-\I(E-\mu_\chi)$ giving
\begin{equation*}
\begin{split}
\res&\l(f(-z)n\ped{F}(\I z)\r)_{z=-\I(E-\mu_\chi)}=\\
&=\lim_{z\to-\I(E-\mu_\chi)}\f{\I(\I\mu_\chi-z)^2 n\ped{F}(\I z)}{2E}\l[\f{1}{z+\I(E-\mu_\chi)}-\f{1}{z-\I(\mu_\chi+E)}\r](z+\I(E-\mu_\chi))\\
&=-\f{\I E}{2}n\ped{F}(E-\mu_\chi).
\end{split}
\end{equation*}
In total, the energy density is
\begin{equation*}
\begin{split}
\rho=&2\int\frac{\D^3 p}{(2\pi)^3}\sum_\chi\f{-2\pi\I}{2\pi}\l[\res\l[f(z)-f(z)n\ped{F}(\I z)\r]_{z=-\I(E+\mu_\chi)}+\r.\\
&\l.-\res\l[f(-z)n\ped{F}(\I z)\r]_{z=-\I(E-\mu_\chi)}\r]\\
=&2\int\frac{\D^3 p}{(2\pi)^3}\sum_\chi E\l[n\ped{F}(E-\mu_\chi)+n\ped{F}(E+\mu_\chi)-1\r].
\end{split}
\end{equation*}
The last term of the integrand does not depend on temperature and is therefore canceled when we subtract the vacuum expectation value.
The renormalized energy density value is then
\begin{equation}
\label{eq:energydensity}
\rho=\f{1}{2\pi^2}\intmod{p}\vec{p}^2 E\l[n\ped{F}(E-\mu\ped{R})+n\ped{F}(E+\mu\ped{R})+n\ped{F}(E-\mu\ped{L})+n\ped{F}(E+\mu\ped{L})\r].
\end{equation}
The energy density is obtained from the time-time component of the stress energy tensor which is a parity even operator and it also not
affected by charge conjugation transformations. This properties are reflected in the expression~(\ref{eq:energydensity}) which is even
on both the exchanges $\mu\to -\mu$ and  $\mu\ped{A}\to -\mu\ped{A}$, which corresponds to $\{\mu\ped{R}\to-\mu\ped{L},\,\mu\ped{L}\to-\mu\ped{R}\}$
and $\{\mu\ped{R}\to\mu\ped{L},\,\mu\ped{L}\to\mu\ped{R}\}$ respectively. For a massless field the integral can be carried out exactly
(see Appendix \ref{sec:masslessmomint}) and it gives:
\begin{equation*}
\rho= 3 p=\frac{30 \pi^2 \left(\zeta^2+\zeta\ped{A}^2\right)+15 \left(\zeta^4+6 \zeta^2 \zeta\ped{A}^2+\zeta\ped{A}^4\right)+7 \pi^4}{60 \pi^2 |\beta|^4},
\end{equation*}
where we replaced the chemical potential with $\zeta\ped{R/L}=|\beta|\mu\ped{R/L}$.
Instead, for a massive field and vanishing axial chemical potential we have
\begin{equation*}
\rho=\f{1}{\pi^2}\intmod{p}\vec{p}^2 E\l[n\ped{F}(E-\mu)+n\ped{F}(E+\mu)\r].
\end{equation*}

Similarly, the pressure is obtain by the following combinations of stress-energy tensor components
\begin{equation*}
p=\f{1}{3}\sum_{i=1}^3\mean{\h{T}_{ii}}_T=-\f{2}{3}\sumint_{\{P\}}\sum_\chi\f{\vec{P}^2}{{P^+_\chi}^2+m^2}
=-\f{2}{3}\int\frac{\D^3 p}{(2\pi)^3}\sum_\chi T\sum_{\{\omega_n\}} \f{\vec{p}^2}{(\omega_n+\I\mu_\chi)^2+E^2}.
\end{equation*}
Again we compute the sum on Matsubara frequencies with a complex integral
\begin{equation*}
\begin{split}
\int_{-\infty}^{+\infty} \f{\D z}{2\pi} \f{1}{(z+\I\mu_\chi)^2+E^2}&- \int_{-\infty-\I 0^+}^{+\infty-\I 0^+} \f{\D z}{2\pi} \l[ \f{1}{(z+\I\mu_\chi)^2+E^2}
+ \f{1}{(z-\I\mu_\chi)^2+E^2}\r] n\ped{F}(\I z)=\\
&=\f{1}{2E}-\f{1}{E}\l(n\ped{F}(E-\mu_\chi)+n\ped{F}(E+\mu_\chi)\r)
\end{split}
\end{equation*}
that, after renormalization, provides
\begin{equation}
\label{eq:pressure}
p=\f{1}{6\pi^2} \int\D p\, \f{\vec{p}^4}{E} \l[n\ped{F}(E-\mu\ped{R})+n\ped{F}(E+\mu\ped{R})+n\ped{F}(E-\mu\ped{L})+n\ped{F}(E+\mu\ped{L})\r].
\end{equation}
The pressure is invariant under parity and charge transformation as it was the energy density. For massless field, in accordance with conformal
invariance, the pressure is one third the energy $p=\rho/3$. For a massive field with vanishing axial chemical potential the pressure is
\begin{equation*}
p=\f{1}{3\pi^2} \int\D p\, \f{\vec{p}^4}{E} \l[n\ped{F}(E-\mu)+n\ped{F}(E+\mu)\r].
\end{equation*}
%

\subsection{Electric and axial current thermal expectation value}
\label{subsec:homogcurrents}
Consider now the mean value of electric and axial current. In the homogeneous thermal equilibrium, we only have one physical macroscopic four-vector,
the fluid velocity $u$. Therefore, electric and axial current can only by directed along that velocity:
\begin{equation*}
\mean{\,\h{j}^\mu\,}_\beta=n\ped{c}\,u^\mu,\quad\mean{\,\h{j}\ped{A}^\mu\,}_\beta=n\ped{A}\,u^\mu,
\end{equation*}
where $n\ped{c}$ and $n\ped{A}$ are two scalar thermodynamic functions representing electric charge density and axial charge density.
Since we have expressed the Dirac thermal propagator in terms of right and left chemical potential, it is more convenient to evaluate
the mean values for right and left currents, which are defined similarly to the chemical potentials as follow:
\begin{equation*}
\h{j}\apic{ R}_\mu=\frac{\h{j}_\mu+\h{j}\apic{A}_\mu}{2},\quad\h{j}\apic{ L}_\mu=\frac{\h{j}_\mu-\h{j}\apic{A}_\mu}{2}.
\end{equation*}
Using the point-splitting method, the two currents in the euclidean space read
\begin{equation*}
\begin{split}
\h{j}\apic{ R}_\mu(X)=&(-\ii)^{1-n_\mu}\h{\bar{\psi}}(X)\gamma_\mu\frac{1+\gamma_5}{2}\h{\psi}(X)\\
=&\lim_{X_1,X_2\to X}\sum_{a,b}(-\ii)^{1-n_\mu}\left(\gamma_\mu\frac{1+\gamma_5}{2}\right)_{ab}\h{\bar{\psi}}_a(X)\h{\psi}_b(X)
\end{split}
\end{equation*}
and
\begin{equation*}
\begin{split}
\h{j}\apic{ L}_\mu(X)=&(-\ii)^{1-n_\mu}\h{\bar{\psi}}(X)\gamma_\mu\frac{1-\gamma_5}{2}\h{\psi}(X)\\
=&\lim_{X_1,X_2\to X}\sum_{a,b}(-\ii)^{1-n_\mu}\left(\gamma_\mu\frac{1-\gamma_5}{2}\right)_{ab}\h{\bar{\psi}}_a(X)\h{\psi}_b(X).
\end{split}
\end{equation*}
Consider the left current; its mean value at rest frame is obtained using the propagator~(\ref{eq:propFermCoord})
and proceeding as previous section we obtain
\begin{equation*}
\begin{split}
\mean{\h{j}\apic{ L}_\mu(X)}_T &=\lim_{X_1,X_2\to X}-(-\ii)^{1-n_\mu}
\tr\left[\left(\gamma_\mu\frac{1-\gamma_5}{2}\right)\mean{\psi(X_2)\bar{\psi}(X_1)}_T\right]\\
&=\lim_{X_1,X_2\to X}-(-\ii)^{1-n_\mu}\sumint_{\{P\}}\sum_\chi\e^{\ii P^+_\chi\cdot (X_1-X_2)}
\tr\l[\left(\gamma_\mu\frac{1-\gamma_5}{2}\right)\left(\mathbb{P}_\chi\frac{-\I\slashed{P}^+_\chi+m}{{P^+_\chi}^2+m^2}\right)\r]\\
&=-(-\ii)^{1-n_\mu}\sumint_{\{P\}}\sum_\chi
\tr\l[\left(\gamma_\mu\frac{1-\gamma_5}{2}\right)\left(\mathbb{P}_\chi\frac{-\I\slashed{P}^+_\chi+m}{{P^+_\chi}^2+m^2}\right)\r].
\end{split}
\end{equation*}
It is no surprise that the non vanishing contribution of the trace comes only from the left projector, indeed we find
\begin{gather*}
\tr\left[\gamma_\mu\frac{1-\gamma_5}{2}\frac{1+\gamma_5}{2}\l(-\I\gamma_\lambda{P^+\ped{R}}_\lambda+m\r)\right]=0,\\
\tr\left[\gamma_\mu\frac{1-\gamma_5}{2} \frac{1-\gamma_5}{2} \l(-\I\gamma_\lambda{P^+\ped{L}}_\lambda+m\r) \right]
=\tr\left[\frac{1+\gamma_5}{2}\gamma_\mu\l(-\I\gamma_\lambda{P^+\ped{L}}_\lambda+m\r)\right]=-2\,\I\, {P^+\ped{L}}_\mu.
\end{gather*}
Hence, the mean value is
\begin{equation*}
\begin{split}
\mean{\h{j}\apic{ L}_\mu(X)}_T &=\ii(-\ii)^{1-n_\mu}\sumint_{\{P\}}\frac{2{P^+\ped{L}}_\mu}{{P^+\ped{L}}^2+m^2}\\
&=\ii(-\ii)^{1-n_\mu}\int\frac{\D^3 p}{(2\pi)^3} T\sum_{\{\omega_n\}}\frac{2{P^+\ped{L}}_\mu}{(\omega_n+\ii\mu\ped{L})^2+\vec{p}^2+m^2}\\
&=2\,\ii\,\delta_{\mu,0}\int\frac{\D^3 p}{(2\pi)^3}\sigma\ped{L}(\beta,\vec{p})=\delta_{\mu,0}\,\,n\ped{L},
\end{split}
\end{equation*}
where we identified the thermodynamic function $n\ped{L}$ which is the left-handed particle density.
The sum over frequencies is done with the correspondence in Eq.~(\ref{eq:thermalsumrule}). Introducing $E^2=\vec{p}^2+m^2$, after subtraction
of vacuum expectation value, we find
\begin{equation*}
\sigma\ped{L}(\beta,\vec{p})=\frac{1}{\beta}\sum_{\{\omega_n\}}\frac{{P^+\ped{L}}_\mu}{{P^+\ped{L}}^2}
=\frac{\ii}{2}\left[ n\ped{F}(E+\mu\ped{L})-n\ped{F}(E-\mu\ped{L})\right]
\end{equation*}
giving the final result
\begin{equation*}
n\ped{L}=\frac{1}{2\pi^2}\int_0^\infty \D p\,\, \vec{p}^2 \left[ n\ped{F}(E-\mu\ped{L})-n\ped{F}(E+\mu\ped{L})\right].
\end{equation*}
The same can be done for right current, giving
\begin{equation*}
\mean{\h{j}\apic{ R}_\mu(X)}_T=\delta_{\mu,0}\,\,n\ped{R},\quad
n\ped{R}=\frac{1}{2\pi^2}\int_0^\infty \D p\,\, \vec{p}^2 \left[ n\ped{F}(E-\mu\ped{R})-n\ped{F}(E+\mu\ped{R})\right].
\end{equation*}
From these results it is straightforward to obtain the axial and vectorial current
\begin{equation*}
\mean{\h{j}_\mu}_\beta=\mean{\h{j}\apic{ R}_\mu}_\beta+\mean{\h{j}\apic{ L}_\mu}_\beta=n\ped{c}\,u_\mu,\quad
\mean{\h{j}\apic{A}_\mu}_\beta=\mean{\h{j}\apic{ R}_\mu}_\beta-\mean{\h{j}\apic{ L}_\mu}_\beta=n\ped{A}\,u_\mu.
\end{equation*}
with
\begin{gather*}
n\ped{c}=\frac{1}{2\pi^2}\int_0^\infty \D p\,\, \vec{p}^2 \left[ n\ped{F}(E-\mu\ped{R})-n\ped{F}(E+\mu\ped{R})
+n\ped{F}(E-\mu\ped{L})-n\ped{F}(E+\mu\ped{L})\right]\\
n\ped{A}=\frac{1}{2\pi^2}\int_0^\infty \D p\,\, \vec{p}^2 \left[ n\ped{F}(E-\mu\ped{R})-n\ped{F}(E+\mu\ped{R})
-n\ped{F}(E-\mu\ped{L})+n\ped{F}(E+\mu\ped{L})\right].
\end{gather*}
The operator $\h{j}_0$ is odd by charge conjugation but even for parity transformation and indeed the function $n\ped{c}$
is odd if we flip the sign of electric chemical potential and is even if we flip the axial chemical potential sign.
The opposite scenario occurs for the axial current. The momentum integral can be done exactly for $m=0$
(see Appendix \ref{sec:masslessmomint}) and denoting $\zeta\ped{R/L}=\beta\mu\ped{R/L}$ we have
\begin{equation*}
n\ped{R/L}=\frac{\zeta\ped{R/L}}{6\pi^2|\beta|^3}\l(\pi^2+\zeta\ped{R/L}^2\r).
\end{equation*}
From previous equation, it immediately follows that
\begin{gather*}
n\ped{c}=\frac{\zeta}{3\pi^2|\beta|^3}\l(\pi^2+\zeta^2+3\zeta\ped{A}^2\r),\\
n\ped{A}=\frac{\zeta\ped{A}}{3\pi^2|\beta|^3}\l(\pi^2+3\zeta^2+\zeta\ped{A}^2\r).
\end{gather*}
For massive field and vanishing chemical potential axial, the charge density is vanishing and charge density becomes

\begin{equation}
\label{eq:ncfreefield}
n\ped{c}=\frac{1}{\pi^2}\int_0^\infty \D p\,\, \vec{p}^2 \left[ n\ped{F}(E-\mu)-n\ped{F}(E+\mu)\right].
\end{equation}
\chapter{Chiral fermions under vorticity}
\label{ch:ChiralVort}
Physics of relativistic chiral matter under the effects of vorticity is relevant
in several applications, such as the investigation of nuclear matter properties through
relativistic heavy ion collisions~\cite{Kharzeev:2015znc}, the study of Weyl semi-metals~\cite{Li:2014bha}
in condensed matter and in the evolution of neutron stars~\cite{Kaminski:2014jda}.
In the last decade, much efforts have been dedicated to the realization and development
of several different theoretical frameworks and approaches to the description of transport
phenomena in chiral relativistic matter. Most of the studies of chiral and vorticous fluids were aimed at
determining both thermodynamic equilibrium and non-equilibrium coefficients (either dissipative
or non-dissipative  according to the classification of ref.~\cite{Haehl:2015pja}) within different
theoretical frameworks:
with Wigner function~\cite{Gao:2012ix,Becattini:2013fla,Prokhorov:2017atp,,Huang:2018wdl,Prokhorov:2018qhq,Prokhorov:2018bql,Prokhorov:2019cik,Weickgenannt:2019dks,Gao:2019znl,Hattori:2019ahi};
with finite temperature field theory~\cite{Golkar:2012kb,Ambrus:2014uqa,Ambrus:2015lfr,Chowdhury:2015pba,Ambrus:2017opa};
withe hydrostatic partition function method and extensions~\cite{Banerjee:2012iz,Jensen:2013vta,Jensen:2013kka,Golkar:2015oxw,Kovtun:2018dvd};
within relativistic hydrodynamics~\cite{Son:2009tf,Neiman:2010zi,Hattori:2019lfp};
with effective field theory~\cite{Glorioso:2017lcn,Huang:2017pqe};
with kinetic theory~\cite{Pu:2010as,Stephanov:2012ki,Son:2012zy,,Manuel:2013zaa,Hidaka:2017auj,Mueller:2017lzw,Mueller:2017arw,Huang:2017tsq,Liu:2018xip};
with the fluid/gravity correspondence~\cite{Erdmenger:2008rm,Banerjee:2008th,Torabian:2009qk,Landsteiner:2011iq,Amado:2011zx,Landsteiner:2012kd};
and with Zubarev non-equilibrium statistical operator~\cite{BecGro,Hayata:2015lga,Hongo:2016mqm,Buzzegoli:2017cqy,Hongo:2019rbd}.

In this chapter, based on~\cite{Buzzegoli:2018wpy}, we use the latter method, introduced in Chapter~\ref{chp:QRelStatMech},
to analyze thermodynamic properties of vorticous chiral matter and we identify thermodynamic coefficients at
second order in thermal vorticity. In Sec.~\ref{sec:CVFree} we evaluate those coefficients for free massless Dirac field.

\section{Chiral thermodynamic coefficients}
Consider a system at thermal equilibrium with thermal vorticity and a conserved axial charge.
Thermal states are described by the generalized global equilibrium statistical operator of Eq.~(\ref{eq:GEDObeta}):
\begin{equation*}
\h{\rho}\ped{GE}=\dfrac{1}{\parz\ped{GE}}\exp\left[-\beta_\mu(x)\h{P}^\mu+\frac{1}{2}\varpi:\h{J}_x+\zeta\h{Q}+\zeta\ped{A}\h{Q}\ped{A}\right].
\end{equation*}
Note that, in this chapter, we consider the axial current to be conserved. This is indeed the case for a free
massless field, which we discuss in Sec.~\ref{sec:CVFree}. The consequences of having a non conserved axial current,
because of the particle mass, are discussed in Sec.~\ref{sec:CurrentsDec}.
In Sec.~\ref{sec:VortExpan} we showed how thermal expectation values of local observables made with the previous
statistical operator can be expanded in thermal vorticity using linear response theory. The expansion
up to second order~(\ref{meanvalueoper}) is reported here again:
\begin{equation*}
\begin{split}
\mean {\h O(x)}=&\mean{\h O(x)}_{\beta(x)}-\alpha_\rho \mycorr{\,\h K^\rho \h O\,}-w_\rho \mycorr{\,\h J^\rho \h O\,}
+\frac{\alpha_\rho\alpha_\sigma}{2}\mycorr{\,\h K^\rho \h K^\sigma \h O\,}\\
&+\frac{w_\rho w_\sigma}{2} \mycorr{\,\h J^\rho \h J^\sigma \h O\,} +\frac{\alpha_\rho w_\sigma}{2}\mycorr{\,\{\h K^\rho,\h J^\sigma\}\h O\,}+\mathcal{O}(\varpi^3).
\end{split}
\end{equation*}
Starting from this expansion, we can identify several thermodynamic coefficients related to the operator $\h{O}$~\cite{Buzzegoli:2017cqy}.
A general procedure to identify those thermal coefficients is give in Sec.~\ref{sec:VortExpan}.
Among those thermal coefficients, in the case of equilibrium with axial charge, new thermodynamic coefficients
arise as consequence of the breaking of parity symmetry inside the statistical operator.
We refer to these as chiral thermodynamic coefficients, which we are now defining more clearly.

Each term of the thermal vorticity expansion involves a connected correlator between
the observables and the Lorentz group generators and it is given by~(\ref{mycorr}):
\begin{equation}
\label{mycorr_repeat}
\begin{split}
\mycorr{\h K^{\rho_1}\cdots \h K^{\rho_n} \h J^{\sigma_1}\cdots \h J^{\sigma_m} \h O} \equiv&
\int_0^{|\beta|} \frac{\di\tau_1\cdots\di\tau_{n+m}}{|\beta|^{n+m}}\times\\
&\times\mean{{\rm T}_\tau\left(\h K^{\rho_1}_{-\ii \tau_1 u}\cdots\h K^{\rho_n}_{-\ii \tau_n u}
	\h J^{\sigma_1}_{-\ii \tau_{n+1} u} \cdots\h J^{\sigma_m}_{-\ii \tau_{n+m} u} \h O(0)\right)}_{\beta(x),c}.
\end{split}
\end{equation}
Since this correlator is the result of the expansion on thermal vorticity, the connected thermal expectation
value $\mean{\cdots}_{\beta(x),c}$ is not calculated with the statistical operator containing the thermal
vorticity $\wrho\ped{GE}$ but with the homogeneous statistical operator $\wrho\ped{h}$~(\ref{leadorder})
\begin{equation}\label{hdensop}
\wrho\ped{h}=\frac{1}{{\mc Z}\ped{h}} \exp\left(-\beta(x)\cdot \wP+\zeta\wQ+\zeta\ped{A}\wQ\ped{A}\right).
\end{equation}
This operator is symmetric under: translations, rotations in the hyperplane perpendicular
to $\beta(x)$, and time-reversal $\group{T}$ in the $\beta(x)$ rest frame. However, as $\wQ$ 
is odd under charge conjugation $\group{C}$ and $\wQ_A$ is odd under parity $\group{P}$ (see Table~\ref{tab:TransfProp}), $\wrho\ped{h}$ is not invariant under $\group{C}$ and $\group{P}$. 

The symmetries of the density operator have direct consequences on the value of the aforementioned 
correlators and, more generally, on mean values of operators. For instance, time-reversal invariance 
of $\wrho\ped{h}$ implies that any mean value of $\mean{\h O(x)}_{\beta(x)}$, such as the ones
in~(\ref{mycorr_repeat}), vanishes if the operator $\h O(x)$ is odd under time-reversal:
$$
\group{T}\h O\group{T}=-\h O .
$$
Similarly, rotational invariance implies that correlators in~(\ref{mycorr_repeat}) will be non-vanishing
only if the involved operators transform like the same irreducible vectors of the same representations 
of the rotation group. To study the impact of symmetries on the correlators~(\ref{mycorr_repeat}) it 
is convenient to indicate with the operator $\h A$ the products of
$$
\h A \equiv \h K^{\rho_1}\cdots \h K^{\rho_n} \h J^{\sigma_1}\cdots \h J^{\sigma_m}
$$
and to decompose both $\h A$ and $\h O$ in irreducible components under rotation: scalar, vector,
symmetric traceless tensor, etc. The only non-vanishing correlators will be those matching corresponding
components of the same kind {\em and} with the same transformation property under time-reversal.

In the homogeneous equilibrium density operator (\ref{hdensop}), if the Hamiltionian is parity invariant
the term $\mu_A \wQ_A$ is the only responsible for parity breaking. For this
reason, all correlators~(\ref{mycorr_repeat}) in which $\h A$ and $\h O$ transform with different 
sign under parity will be proportional to odd powers of $\mu_A$, because:
$$
\tr (\,\wrho_h\, \h A\, \h O\,) = \tr (\,\group{P}\, \wrho_h\, \h A\, \h O\, \group{P}^{-1}) = 
\tr (\,\wrho_h(-\mu_A) \group{P} \h A\, \h O\, \group{P}^{-1}) = - \tr (\,\wrho_h(-\mu_A) \h A\, \h O\,).
$$
These correlators will be henceforth dubbed as \emph{chiral correlators}, and the corresponding
coefficients chiral coefficients (see Table~\ref{tab:TransfProp}). 

\begin{table}[htb]
	\caption{Transformation properties under discrete transformation for various operators in the rest frame of four-temperature.
		$\group{P}$ is parity, $\group{T}$ is time-reversal and $\group{C}$ is charge conjugation transformation.
		$\wT^{\mu\nu}$ is the stress-energy tensor, $\wj\ped{V}^\mu$ is an (electric) vector current,  $\wj\ped{A}^\mu$ is an axial current,
		$\phi$ is a scalar and $\phi\ped{A}$ is a pseudo-scalar,  $\wK^\mu$ and $\wJ^\mu$ are the Lorentz generators,
		$\wQ$ and $\wQ\ped{A}$ are vector and axial charge respectively.}
	\label{tab:TransfProp}
	\[\begin{array}{lccc|cccccc|cccc|cc}
	& \wT^{00} & \wT^{0i} & \wT^{ij} & \wj^0\ped{V} & \h{\vec{j}}\ped{V} & \wj^0\ped{A} & \h{\vec{j}}\ped{A} & \h\phi & \h\phi\ped{A} &
	\h{\vec{K}} & \h{\vec{J}} & (\wK\wK, \wJ\wJ) & \wK\wJ & \wQ & \wQ\ped{A}\\  
	\hline
	\group{P} & + & - & + & + & - & - & + & + & - & - & + & + & - & + & - \\
	\group{T} & + & - & + & + & - & + & - & + & - & + & - & + & - & + & + \\
	\group{C} & + & + & + & - & - & + & + & + & + & + & + & + & + & - & + \\
	\end{array}\]	
\end{table}

As explained in Sec.~\ref{sec:VortExpan}, the coefficients can be expressed as thermal expectation values
in the rest frame, i.e. the frame where $\beta^\mu=(1/T(x),\vec{0})$. This coefficients are Lorentz scalar
and they are function of $T$, $\zeta = \mu/T$ and $\zeta_A = \mu_A/T$. If we denote the mean values in
rest frame with a subscript $T$, that is: 
\begin{equation*}
\mean{\h O(x)}_T\equiv\frac{\tr\left[ \exp\left(-\h H/T +\zeta\wQ+\zeta\ped{A}\wQ\ped{A}\right)\h O(0)\right]}{\tr\left[ \exp\left(-\h H/T +\zeta\wQ+\zeta\ped{A}\wQ\ped{A}\right)\right]},
\end{equation*}
where $\h H$ is the Hamiltonian, then the correlators~(\ref{mycorr_repeat}) in the rest frame are written as:
\begin{equation}
\begin{split}\label{mycorrrest}
\mycorr{\h K^{\rho_1}\cdots \h K^{\rho_n} \h J^{\sigma_1}\cdots \h J^{\sigma_m} \h O}_T \equiv&
\int_0^{|\beta|=1/T} \frac{\di\tau_1\cdots\di\tau_{n+m}}{|\beta|^{n+m}}\times\\
&\times\mean{{\rm T}_\tau\left(\h K^{\rho_1}_{-\ii \tau_1}\cdots\h K^{\rho_n}_{-\ii \tau_n}
	\h J^{\sigma_1}_{-\ii \tau_{n+1}} \cdots\h J^{\sigma_m}_{-\ii \tau_{n+m}} \h O(0)\right)}_{T,c}\, .
\end{split}
\end{equation}
All the thermodynamic coefficients are linear combination of correlators~(\ref{mycorrrest})
with an appropriate $\wK$, $\wJ$ and $\h O$ components choice, and such expressions can be considered
as ``Kubo formulae'' for the corresponding coefficients. From the Lorentz generators definition~(\ref{eq:LorenzGenerators}),
it follows that the~(\ref{mycorrrest}) are given by means of correlators of the symmetric stress-energy tensor and the operator $\h O$.
In this work, we work out the (\ref{mycorrrest}) by using the imaginary time formalism.
The shifted boost and angular momentum generators, here obtained by definition~(\ref{eq:boostrot}) in the rest frame, are given by Eq.~(\ref{jshift})
\begin{equation*}
\wJ^{\mu\nu}_{-\I\tau}= \h {\sf T}((-\I\tau,\vec{0})) \wJ^{\mu\nu} \h {\sf T}^{-1}((-\I\tau,\vec{0}))
\end{equation*}
and, according to definition~(\ref{eq:LorenzGenerators}), arise from the spatial integration of the time evolved stress-energy tensor operator
\begin{equation}
\label{imagtime}
\wT^{\mu\nu}(\tau,\vec{x})=\h {\sf T}((-\I\tau,\vec{0})) \wT^{\mu\nu}(0,\vec{x}) \h {\sf T}^{-1}((-\I\tau,\vec{0})).
\end{equation}
Thus, the correlators~(\ref{mycorrrest}) are expressed as a linear combination of the following basic structure with suitable indices:
\begin{equation}
\begin{split}\label{basicstrut}
C_{\mu_1\nu_1|\cdots|\mu_n\nu_n| i_1\cdots i_n}\equiv& \int_0^{|\beta|} \frac{\di\tau_1\cdots\di\tau_n}{|\beta|^n}
\int \D^3x_1\cdots\D^3x_n\times\\
&\times  \mean{{\rm T}_\tau\left(\h{T}_{\mu_1\nu_1}(X_1)\cdots\h{T}_{\mu_n\nu_n}(X_n) \h O(0)\right)}\ped{$T$,c}\, x_i\cdots x_n,
\end{split}
\end{equation}
where $X_n=(\tau_n,\vec{x}_n)$.

We are now in a position to determine the expansion in thermal vorticity of conserved currents
including the axial chemical potential $\mu_A = \zeta_A T$. But before that, it should be pointed out
that any scalar or a pseudo-scalar operator, denoted as $\h\phi$ and $\h\phi\ped{A}$
in Table~\ref{tab:TransfProp}, does not have chiral corrections up to second order. Indeed, the correlator 
between $\h\phi$ or $\h\phi\ped{A}$ and $\wJ$ or $\wK$ simply vanish at first-order because the 
latter are both vectors under rotation. Besides, it can be readily seen again from
Table~\ref{tab:TransfProp}, that no second-order correlator can be formed with $\h K$ and $\h J$ 
that is simultaneously even under time reversal and odd under parity. Hence, there 
are no chiral-vortical corrections at second order for scalar and pseudo-scalar operators.
\newpage

\subsection{Stress-energy tensor}
The expansion up to second order in thermal vorticity of the stress-energy tensor with axial
chemical potential features additional terms with respect to the case $\zeta_A =0$ ~\cite{Buzzegoli:2017cqy,BecGro}.
Indeed, three new chiral coefficients appear at the first order in thermal vorticity,
$\mathbb{A},\mathbb{W}_1$ and $\mathbb{W}_2$:
\begin{equation}
\begin{split}\label{setdecomp}
\mean{\wT^{\mu\nu}}=&\,\mathbb{A}\,\epsilon^{\mu\nu\kappa\lambda}\alpha_\kappa u_\lambda+\mathbb{W}_1 w^\mu u^\nu +\mathbb{W}_2 w^\nu u^\mu\\
&+(\rho-\alpha^2 U_\alpha -w^2 U_w)u^\mu u^\nu -(p-\alpha^2D_\alpha-w^2D_w)\Delta^{\mu\nu}\\
&+A\,\alpha^\mu\alpha^\nu+Ww^\mu w^\nu+G_1 u^\mu\gamma^\nu+G_2 u^\nu\gamma^\mu+\mathcal{O}(\varpi^3)
\end{split}
\end{equation}
which can be obtained as:
\begin{equation}\label{eq:chiralsetcoeff}
\begin{split}
\mathbb{A}=\mycorr{\,\wK^{3}\,\frac{\wT^{12}-\wT^{21}}{2}}_T,\quad
\mathbb{W}_1=\mycorr{\,\wJ^{3}\,\wT^{30}}_T,\quad
\mathbb{W}_2=\mycorr{\,\wJ^{3}\,\wT^{03}}_T,
\end{split}
\end{equation}
while for non chiral coefficients we found~\cite{Buzzegoli:2017cqy}\footnote{Differently
	from~\cite{Buzzegoli:2017cqy} here the stress-energy tensor is taken to be the canonical
	Dirac stress-energy tensor, which is non-symmetric. The symmetric case is recovered
	with $\mathbb{A}=0$, $\mathbb{W}_1=\mathbb{W}_1$ and $G_1=G_2$.}
\begin{equation}\label{eq:setcoeff}
\begin{split}
U_\alpha&=\frac{1}{2}\mycorr{\,\wK^3\,\wK^3\,\wT^{00}}_T,\qquad\hphantom{G_2=-} U_w=\frac{1}{2}\mycorr{\,\wJ^3\,\wJ^3\,\wT^{00}}_T,\\
D_\alpha&=\frac{1}{2}\mycorr{\,\wK^3\,\wK^3\,\wT^{11}}_T-\frac{1}{3}\mycorr{\,\wK^1\,\wK^2\,\frac{\wT^{12}+\wT^{21}}{2}}_T,\\
D_w&=\frac{1}{2}\mycorr{\,\wJ^3\,\wJ^3\,\wT^{11}}_T-\frac{1}{3}\mycorr{\,\wJ^1\,\wJ^2\,\frac{\wT^{12}+\wT^{21}}{2}}_T,\\
A&=\mycorr{\,\wK^1\,\wK^2\,\frac{\wT^{12}+\wT^{21}}{2}}_T, \qquad  W=\mycorr{\,\wJ^1\,\wJ^2\,\frac{\wT^{12}+\wT^{21}}{2}}_T,\\
G_1&=-\frac{1}{2}\mycorr{\,\{\wK^1,\,\wJ^2\}\,\wT^{03}}_T,\, \qquad G_2=-\frac{1}{2}\mycorr{\,\{\wK^1,\,\wJ^2\}\,\wT^{30}}_T.
\end{split}
\end{equation}
The most evident consequence of parity breaking induced by $\zeta_A$ is an energy flux $q^\mu$ along the vorticity
$w^\mu$, see Eq.~(\ref{setdecomp}):
\begin{equation*}
q^\mu=\Delta^\mu_{\,\,\rho}\,u_\sigma\,\mean{\wT^{\sigma\rho}}=\mathbb{W}_2\,w^\mu+G_1\,\gamma^\mu.
\end{equation*}
The energy flux along $\gamma$ is not chiral but it is second order in thermal vorticity. The chiral
coefficients $\mathbb{W}_1$ and $\mathbb{W}_1$ were already addressed in
literature~\cite{Vilenkin:1979ui,Landsteiner:2011iq,Landsteiner:2012kd,Chen:2015gta,Hidaka:2017auj,Abbasi:2017tea,Chowdhury:2015pba}.
The coefficient $\mathbb{A}$ has never been reported because usually we are interested in symmetric stress-energy tensor, where
it is vanishing. Second order thermodynamic coefficients were classified in~\cite{Baier:2007ix,Romatschke,Moore:2010bu} using the
Landau frame and contained $U_\alpha,\,D_\alpha,\,U_w,\,D_w,A,\,W$. For a comparison of the two notations see~\cite{Buzzegoli:2017cqy};
the hydrodynamic frame change is discussed later in Sec.~\ref{sec:LandauFrame}.
Using the relation in Eq.~(\ref{eq:KuboFormula}) we verified that the coefficients given in~(\ref{eq:chiralsetcoeff})
coincide with Kubo Formulae in~\cite{Landsteiner:2011iq,Landsteiner:2012kd,Chowdhury:2015pba}.

The above coefficients are not all independent, in fact there are relations between them stemming
from the conservation equation of stress-energy tensor. In the case of global equilibrium, which we are
discussing, the statistical operator is actually stationary. As a consequence the conservation equations for
operators, and in general all their relations, translate directly to mean values of the corresponding operators.
For example, the conservation equation for the stress-energy tensor becomes:
\begin{equation}
\label{eq:SETConMean}
\de_\mu\mean{\wT^{\mu\nu}}=\de_\mu\tr\left[\wrho\,\,\wT^{\mu\nu}\right]
	=\tr\left[\wrho\,\de_\mu\wT^{\mu\nu}\right]=\mean{\de_\mu\wT^{\mu\nu}}=0.
\end{equation}
These conservation equations for mean values can be used together with decomposition in Eq.~(\ref{setdecomp}) to
find relations between thermal coefficients. All thermodynamic coefficients are scalars quantity and are obtained
with a statistical average and integrating out all the microscopic details. As a result those coefficients can
only depend on scalars built with thermodynamic fields and, if any, on the scales of the systems, such as the
mass of constituents, some characteristic length of the system, scales from interactions and so on. We are now
considering a system without relevant scales other than the thermodynamic fields $\beta,\,\zeta$ and $\zeta\ped{A}$.
Since only $\beta$ is not constant, the coordinate dependence of those coefficients is contained only in $\beta^2(x)$.
The value of a generic thermal coefficient can be then expressed as a function $f(|\beta|(x),\zeta,\zeta\ped{A})$
and its derivative is
\begin{equation*}
\de_\mu f(|\beta|,\zeta,\zeta\ped{A})=\de_\mu|\beta|\frac{\de f(|\beta|,\zeta,\zeta\ped{A})}{\de|\beta|}
	=-\alpha_\mu\frac{\de f(|\beta|,\zeta,\zeta\ped{A})}{\de|\beta|},
\end{equation*}
where the derivative on $|\beta|$ is to be taken with fixed $\zeta$ and $\zeta\ped{A}$. This holds for what concern the
thermodynamic coefficients itself, but the conservations equation are also written in terms of the four-vectors
composing the tetrad $\{u,\,\alpha,\,w,\,\gamma\}$. It must be mentioned that coordinate dependence of these
four-vectors at global equilibrium is fixed once the form of the inverse four-temperature is given. Therefore,
in writing the conservative equations for mean values, those dependence can be carried out explicitly.
Many properties of the derivatives of those four-vectors and some of their combination, which are used to derive
the following results, are reported in Appendix~\ref{sec:betaframeidentities}. Furthermore, we can show that
thanks to linear independence of the tetrad, all conservation equations contains several independent pieces each
one pertaining to only one particular order on thermal vorticity. Therefore, the perturbative expansions used
to derive mean values does not hinder the validity of the following relations between thermal coefficients.

General case being settled, we now move to the specific equations. If we replace the decomposition~(\ref{setdecomp}) 
in Eq.~(\ref{eq:SETConMean}) and we carry out the derivatives we find that five relations between thermal coefficients
must be satisfied. The first one relates zeroth order coefficients:
\begin{equation*}
\de_\mu\left[\left(\rho+p\right)u^\mu u^\nu\right]=\de^\nu p,
\end{equation*}
and is nothing than the ideal hydro-dynamic equation for energy conservation. In the case of global equilibrium with
fluid velocity $u=\beta/\sqrt{\beta^2}$, that equation can also be written simply as
\begin{equation*}
\rho+p+|\beta|\frac{\de  p}{\de|\beta|}=0.
\end{equation*}
For second order non-chiral coefficients conservation equation also gives the following relations~\cite{Buzzegoli:2017cqy}
\begin{equation}
\label{eq:setrel}
\begin{split}
U_\alpha&=-|\beta|\frac{\partial}{\partial|\beta|}\big(D_\alpha+A\big)-\big(D_\alpha+A\big),\\
U_w&=-|\beta|\frac{\partial}{\partial|\beta|}\big(D_w+W\big)-D_w+2A-3W,\\
G_1+G_2&=2\big(D_\alpha+D_w\big)+A+|\beta|\frac{\partial}{\partial|\beta|}W+3W,
\end{split}
\end{equation}
and for the first-order chiral coefficients it requires that
\begin{equation}\label{setchiralrel}
-2\mathbb{A}=|\beta|\frac{\de \mathbb{W}_1}{\de|\beta|}+3\mathbb{W}_1+\mathbb{W}_2.
\end{equation}
From the previous relations we see that out of the 13 (10 for symmetric stress-energy tensor) thermodynamic
coefficient, at thermal global equilibrium only 8 of them (5 for symmetric stress-energy tensor) are independent.
In this work, however, we have evaluated all the thermal coefficients for a massless Dirac field by explicitly
computing the correlators and afterward we used the previous relations as a consistency check.
In the case of massless fields, when no other physical scales are relevant, dimensional analysis requires that
all the coefficients related to stress-energy tensor must be proportional to $\beta(x)^{-4}$. All the previous
relations then simplify in the following linear equations:
\begin{equation}
\label{eq:confrelset}
\begin{split}
\rho=&\,3p,\quad \mathbb{A}=\frac{\mathbb{W}_1-\mathbb{W}_2}{2},\quad U_\alpha=\,3\left(D_\alpha+A\right),\\
U_w=&\,3D_w+W+2A,\quad G_1+G_2=\,2\left(D_\alpha+D_w\right)+A-W.
\end{split}
\end{equation}
For massless field we also have an additional constraints that must be satisfied. Indeed, the trace of stress-energy
tensor is proportional to the mass of the fields, and so is vanishing in the conformal case. Also the vanishing trace
translate directly from operators to mean values and using decomposition~(\ref{setdecomp}) we readily obtain
\begin{equation*}
\mean{\h{T}^\mu_{\hphantom{\mu}\mu}}=\rho-\alpha^2 U_\alpha-w^2 U_w-3\left(p-\alpha^2 D_\alpha-w^2 D_w\right)+\alpha^2 A +w^2 W=0;
\end{equation*}
this is realized only when the following relations are satisfied:
\begin{equation}
\label{eq:tracerel}
\rho=3p,\quad U_\alpha=3 D_\alpha+A,\quad U_w=3 D_w+W.
\end{equation}
Comparing the two relations we found for $U_\alpha$, one in~(\ref{eq:confrelset}) and the other in~(\ref{eq:tracerel}),
we realize that thermal coefficient $A$ for a massless field must be vanishing, as also advocated in~\cite{Moore:2012tc}.

In order to compute their values, it is useful to express the previous first order chiral coefficients~(\ref{eq:chiralsetcoeff})
in terms of the quantity derived from Eq.~(\ref{basicstrut}),
\begin{equation*}
C_{\mu\nu|\alpha\beta| i}\apic{se}\equiv\int_0^{|\beta|} \frac{\D\tau}{|\beta|}  \int \D^3x
\mean{{\rm T}_\tau\left(\h{T}_{\mu\nu}(X)\,\h{T}_{\alpha\beta}(0)\right)}\ped{$T$,c}\, x_i.
\end{equation*}
Reminding the definition for Lorentz generators~(\ref{eq:LorenzGenerators}) the chiral coefficients~(\ref{eq:chiralsetcoeff})
are written as
\begin{equation}\label{eq:chiralsetcoeff2}
\mathbb{A}=\frac{1}{2}\left(C_{00|21|3}\apic{se}-C_{00|12|3}\apic{se}\right),\quad
\mathbb{W}_1=C_{02|30|1}\apic{se}-C_{01|30|2}\apic{se},\quad
\mathbb{W}_2=C_{02|03|1}\apic{se}-C_{01|03|2}\apic{se}.
\end{equation}
In the same way second order non chiral coefficients~(\ref{eq:setcoeff}) are obtained with the auxiliary quantity
\begin{equation*}
C\apic{se}_{\mu\nu|\gamma\delta|\alpha\beta| ij}=\!\int_0^{|\beta|}\! \frac{\D\tau_1}{|\beta|}\! \int_0^{|\beta|}\! \frac{\D\tau_2}{|\beta|}\! \int\D^3x\!  \int\!\D^3y\,
	\mean{{\rm T}_\tau \left(\h{T}_{\mu\nu}(X)\,\h{T}_{\gamma\delta}(Y)\,\h{T}_{\alpha\beta}(0)\right)}\ped{$T$,c}\, x_i y_j,
\end{equation*}
through
\begin{equation}\label{coefficientiC}
\begin{split}
U_\alpha=&\frac{1}{2}C\apic{se}_{00|00|00|33},\quad 
U_w=\frac{1}{2}\left( C\apic{se}_{01|01|00|22}-C\apic{se}_{01|02|00|21}-C\apic{se}_{02|01|00|12}+C\apic{se}_{02|02|00|11}\right),\\
D_\alpha=&\frac{1}{2}C\apic{se}_{00|00|11|33}-\frac{1}{6} C\apic{se}_{00|00|12|12}-\frac{1}{6} C\apic{se}_{00|00|21|12},\\
D_w=& \frac{1}{2} \left(C\apic{se}_{01|01|11|22}-C\apic{se}_{01|02|11|21}-C\apic{se}_{02|01|11|12}+C\apic{se}_{02|02|11|11}\right)\\
	&-\frac{1}{6} \left(C\apic{se}_{02|03|12|31}-C\apic{se}_{03|03|12|21}-C\apic{se}_{02|01|12|33}+C\apic{se}_{03|01|12|23}\right),\\
	&-\frac{1}{6} \left(C\apic{se}_{02|03|21|31}-C\apic{se}_{03|03|21|21}-C\apic{se}_{02|01|21|33}+C\apic{se}_{03|01|21|23}\right),\\
A=&\,\frac{1}{2} C\apic{se}_{00|00|12|12}+\frac{1}{2} C\apic{se}_{00|00|21|12},\\
W=&\frac{1}{2}\left(C\apic{se}_{02|03|12|31}-C\apic{se}_{03|03|12|21}-C\apic{se}_{02|01|12|33}+C\apic{se}_{03|01|12|23}\right)\\
	&+\frac{1}{2}\left(C\apic{se}_{02|03|21|31}-C\apic{se}_{03|03|21|21}-C\apic{se}_{02|01|21|33}+C\apic{se}_{03|01|21|23}\right),\\
G_1=&-\frac{1}{2}\left(C\apic{se}_{00|03|03|11}-C\apic{se}_{00|01|03|13}+C\apic{se}_{03|00|03|11}-C\apic{se}_{01|00|03|31}\right),\\
G_2=&-\frac{1}{2}\left(C\apic{se}_{00|03|30|11}-C\apic{se}_{00|01|30|13}+C\apic{se}_{03|00|30|11}-C\apic{se}_{01|00|30|31}\right).
\end{split}
\end{equation}
The decompositions in Eq.~(\ref{eq:chiralsetcoeff2}) and Eq.~(\ref{coefficientiC}) are used in Section~\ref{sec:CVFree} to compute
the thermodynamic coefficients for the massless Dirac field.

\subsection{Electric and axial currents}
\label{sec:CurrentsDec}
Consider now the conserved current $\wj\ped{V}^\mu$ related to the charge $\wQ$ (\ref{eq:ConsCharges}) 
and its transformation properties in Table~\ref{tab:TransfProp}. The only non vanishing terms 
up to the second order in thermal vorticity expansion turn out to be:
\begin{equation}\label{vcurrdecomp}
\mean{\wj\ped{V}^\mu}=n\ped{V}\,u^\mu+\left(\alpha^2 N\apic{V}_\alpha+w^2 N\apic{V}_\omega\right)u^\mu+W\apic{V}w^\mu+G\apic{V}\gamma^\mu
+\mathcal{O}(\varpi^3),
\end{equation}
where~\cite{Buzzegoli:2017cqy}
\begin{equation}
\label{vcurrevencoeff}
N\apic{V}_\alpha=\frac{\mycorr{\,\wK^3\,\wK^3\,\wj^0\ped{V}}_T}{2},\quad
N\apic{V}_w=\frac{\mycorr{\,\wJ^3\,\wJ^3\,\wj^0\ped{V}}_T}{2},\quad
G\apic{V}=\frac{\mycorr{\,\{\wK^1,\,\wJ^2\}\,\wj^3\ped{V}}_T}{2},
\end{equation}
and the chiral coefficient $W^V$ reads:
\begin{equation}\label{CVEcoeff}
W\apic{V}=\mycorr{\,\wJ^{3}\,\wj^3\ped{V}\,}_T\, ,
\end{equation}
which is the conductivity of the so-called Chiral Vortical Effect (CVE). Second order
corrections of electric current were first identified and evaluated in~\cite{Buzzegoli:2017cqy}.

Taking the divergence of~(\ref{vcurrdecomp}), one realizes that the mean vector current is conserved
$\de_\mu\mean{\wj\ped{V}^\mu}=0$ if the $W\apic{V}$ coefficient fulfills the relation:
\begin{equation}\label{CVERel}
|\beta|\frac{\de W\apic{V}}{\de |\beta|}+3W\apic{V}=0.
\end{equation}
There are no constraints for the the other thermal coefficients. Solution of Eq.~(\ref{CVERel}) imposes that the
coefficient must be proportional to the cube of temperature:
\begin{equation*}
W\apic{V}=\frac{f\left(\zeta,\zeta\ped{A}\right)}{|\beta|^3},
\end{equation*}
where $f$ is an arbitrary smooth function. For a massless field this is the form we expect just by dimensional analysis
argument and the constraint does not add any additional information. In the massive case we can not define a conserved
axial current and global equilibrium of chiral matter, hence CVE, can not be addressed. However, for massless fields
even with interactions, and consequently anomalies, we can use Chern-Simon current to define axial charge and global
equilibrium is well defined. Since electric current is not anomalous, the previous relation~(\ref{CVERel}) is unmodified.
It must be noted that conservation equation does not constraints the conductivity of CVE to be any particular value,
it just need to be proportional to $|\beta|^{-3}$. In the case of global equilibrium with external electromagnetic field,
the same analysis reveals that the coefficient $W\apic{V}$ is related with the chiral magnetic effect, this is discussed
in Sec.~\ref{subsec:Feffect}.

When we want to evaluate $W\apic{V}$, we write it by means of the general correlator:
\begin{equation*}
C_{\mu\nu|\alpha| i}\apic{V}\equiv\int_0^{|\beta|} \frac{\D\tau}{|\beta|}  \int \D^3x 
\mean{{\rm T}_\tau\left(\h{T}_{\mu\nu}(X)\,\wj_{\alpha}\apic{V}(0)\right)}\ped{$T$,c}\, x_i,
\end{equation*}
an it reads:
\begin{equation}\label{CVEcoeff2}
W\apic{V}=C_{02|3|1}\apic{V}-C_{01|3|2}\apic{V}.
\end{equation}
Instead, second oder coefficients are expressed as
\begin{equation}\label{eq:vcurrcoeff2}
\begin{split}
N_\alpha\apic{V}&=\frac{1}{2}C_{00|00|0|33}\apic{V}, \\
N_w\apic{V}&=\frac{1}{2}\left(C_{01|01|0|22}\apic{v}-C_{01|02|0|21}\apic{V}-C_{02|01|0|12}\apic{V}+C_{02|02|0|11}\apic{V}\right),\\
G\apic{v}&=\frac{1}{2}\left(C_{00|03|3|11}\apic{V}-C_{00|01|3|13}\apic{V}+C_{03|00|3|11}\apic{V}-C_{01|00|3|31}\apic{V}\right).
\end{split}
\end{equation}
where we denoted
\begin{equation*}
C_{\mu\nu|\gamma\delta|\alpha| ij}\apic{V}=\int_0^{|\beta|}\frac{\D\tau_1}{|\beta|} \int_0^{|\beta|} \frac{\D\tau_2}{|\beta|} \int  \D^3x  \int  \D^3y
	\mean{{\rm T}_\tau\left(\h{T}_{\mu\nu}(X)\,\h{T}_{\gamma\delta}(Y)\,\wj_{\alpha}(0)\right)}\ped{$T$,c}\, x_i\, y_j .
\end{equation*}

Consider now the axial current. Similarly to electric current, taking into account its transformation properties,
the conserved axial current $\wj\ped{A}^\mu$ has the following thermal vorticity expansion:
\begin{equation}\label{acurrdecomp}
\mean{\wj\ped{A}^\mu}=n\ped{A}\,u^\mu+\left(\alpha^2 N_\alpha\apic{A}+w^2 N_\omega\apic{A}\right)u^\mu+W\apic{A}w^\mu+G\apic{A}\gamma^\mu
+\mathcal{O}(\varpi^3).
\end{equation}
The so-called Axial Vortical Effect (AVE) shows up in the non chiral coefficient $W\apic{A}$
\begin{equation}
\label{AVEcoeff}
W\apic{A}=\mycorr{\,\wJ^{3}\,\wj^3\ped{A}\,}_T.
\end{equation}
On the other hand, the properly chiral terms are given by the coefficients:
\begin{equation}\label{acurrcoeff}
\begin{aligned}
n\ped{A}&= \mean{\,\wj^{0}\ped{A}\,}_T, & N_\alpha\apic{A}&=\frac{1}{2} \mycorr{\,\wK^{3}\, \wK^{3}\,\wj^{0}\ped{A}\,}_T,\\
N_\omega\apic{A}&=\frac{1}{2} \mycorr{\,\wJ^{3}\, \wJ^{3}\,\wj^{0}\ped{A}\,}_T, & G\apic{A}&=\frac{1}{2} \mycorr{\{\wK^{1},\,\wJ^{2}\,\}\,\wj^{3}\ped{A}\,}_T,
\end{aligned}
\end{equation}
The second order coefficients $N_\alpha\apic{A},\,N_w\apic{A}$ and $G\apic{A}$, see 
eqs~(\ref{acurrdecomp}) and~(\ref{acurrcoeff}) are indeed newly obtained and they appear 
as quantum corrections of the axial current in presence of acceleration and rotation. They 
are the axial counterpart of the second-order equilibrium corrections of the vector current discussed in~\cite{Buzzegoli:2017cqy}.  
$N_\alpha\apic{A},\,N_w\apic{A}$ are corrections along the fluid velocity and hence modify 
the axial charge density, while $G\apic{A}$ yields a chiral flow along the four-vector 
$\gamma^\mu$ defined in Eq.~(\ref{eq:transversedir}).

Since axial current has opposite parity of electric current, despite their decomposition are identical
chirality in the coefficients unfolds in the opposite way: the response to rotation is not chiral for axial current, while
the other coefficients are. Conservation equation of the mean axial current only entails a condition for $W\apic{A}$
\begin{equation}\label{AVERel}
|\beta|\frac{\de W\apic{A}}{\de |\beta|}+3W\apic{A}=0.
\end{equation}
The same considerations we made for electric current coefficient $W\ped{V}$ applies to $W\ped{A}$ in the case of massless free
fields. In the framework we use to evaluate all the thermal coefficient in Sec.~\ref{sec:CVFree}, anomalies are not considered
simply because we choose to not include them and consequently they could not play any role here. There are no interactions
with gauge fields, the space-time is flat and axial current is conserved both at classical and quantum level. Moreover, all
the coefficients reported here are pertaining the global thermal equilibrium and are automatically satisfying a vanishing
entropy production. This means that other additional constraints could not arise from entropy conservation, which is instead
a consequence of global thermal equilibrium and conservation of stress-energy tensor and currents. Despite this onset, the
AVE is still present.

To understand the constraint~(\ref{AVERel}) and the relation between axial vortical effect and anomalies, we also consider
the case of free massive field. In that case, axial current is not conserved, see Eq.~(\ref{eq:consAxialcurr}), but it is given by 
\begin{equation*}
\de_\mu\h{j}\ped{A}^\mu=2m\I \bar\psi\gamma^5\psi.
\end{equation*}
It follows that global equilibrium with a conserved axial charge can not be reached. Then, to still use the global
equilibrium analysis to massive field, we simply set the axial chemical potential to zero and we consider global
equilibrium with thermal vorticity and finite electric charge. As a consequence, the symmetries impose that all chiral coefficients must
be vanishing. However, the term in $W\apic{A}$ of axial current decomposition is not chiral and could be non-vanishing.
Since the conservation equation is changed we expect that also the condition~(\ref{AVERel}) will be modified.
We then have to consider the pseudo-scalar operator $\I\bar\psi\gamma^5\psi$ that appears on divergence of axial current.
Pseudoscalar thermal expectation value can be decomposed at second order in thermal vorticity in the same way as other
local operators and we find that it is given by a single term:
\begin{equation*}
\mean{\I\bar\psi\gamma^5\psi}=(\alpha\cdot w)L^{\alpha\cdot w}
\end{equation*}
where the non chiral thermal coefficient is given by
\begin{equation*}
L^{\alpha\cdot w}=\frac{1}{2}\mycorr{\{\h{K}_3,\h{J}_3\}\I\bar\psi\gamma^5\psi}.
\end{equation*}
With this definition we find that the condition on axial vortical effect conductivity becomes:
\begin{equation}
\label{AVERelMass}
|\beta|\frac{\de W\apic{A}}{\de |\beta|}+3W\apic{A}=-2mL^{\alpha\cdot w}.
\end{equation}
It is now no more constrained to be proportional to the third power of temperature and could acquire terms which
depends on the mass of the fields. The value and the behavior of this coefficients for free fields is discussed
in Sec.~\ref{sec:Results}.

For the evaluation of $W\apic{A}$ we introduce the correlator:
\begin{equation*}
C_{\mu\nu|\alpha| i}\apic{A}\equiv\int_0^{|\beta|} \frac{\D\tau}{|\beta|}  \int \D^3x 
\mean{{\rm T}_\tau\left(\h{T}_{\mu\nu}(X)\,\wj_{\alpha}\apic{A}(0)\right)}\ped{$T$,c}\, x_i,
\end{equation*}
an we write it as
\begin{equation*}
W\apic{A}=C_{02|3|1}\apic{A}-C_{01|3|2}\apic{A}.
\end{equation*}
The others coefficients are written in terms of the auxiliary correlators
\begin{equation*}
C_{\mu\nu|\gamma\delta|\alpha| ij}\apic{A}=\int_0^{|\beta|}\frac{\D\tau_1}{|\beta|} \int_0^{|\beta|} \frac{\D\tau_2}{|\beta|} \int  \D^3x  \int  \D^3y
\mean{{\rm T}_\tau\left(\h{T}_{\mu\nu}(X)\,\h{T}_{\gamma\delta}(Y)\,\wj_{\alpha}\apic{A}(0)\right)}\ped{$T$,c}\, x_i y_j
\end{equation*}
as:
\begin{equation}\label{acurrcoeff2}
\begin{split}
N_\alpha\apic{A}&=\frac{1}{2}C_{00|00|0|33}\apic{A}, \\
N_w\apic{A}&=\frac{1}{2}\left(C_{01|01|0|22}\apic{A}-C_{01|02|0|21}\apic{A}-C_{02|01|0|12}\apic{A}+C_{02|02|0|11}\apic{A}\right),\\
G\apic{A}&=\frac{1}{2}\left(C_{00|03|3|11}\apic{A}-C_{00|01|3|13}\apic{A}+C_{03|00|3|11}\apic{A}-C_{01|00|3|31}\apic{A}\right).
\end{split}
\end{equation}
By turning the correlators of CVE~(\ref{CVEcoeff}) and of AVE~(\ref{AVEcoeff}) into Kubo
Formulae via the relation in Eq.~(\ref{eq:KuboFormula}) we also verified that they
coincide with those commonly used in literature.

\subsection{Spin tensor}
\label{sec:SpinTens}
An interesting link between spin, chirality and vorticity can be established if we
consider the canonical spin tensor for the Dirac field, defined as
\begin{equation}
\label{eq:defSpinTens}
\h{\mathcal{S}}^{\lambda,\mu\nu}\equiv\frac{\I}{8}\bar{\psi}\,\{\gamma^\lambda,\,[\gamma^\mu,\,\gamma^\nu]\,\}\,\psi.
\end{equation}
This part is related to internal (spin) symmetries of the Dirac field undergoing a Lorentz transformation and it is
the super-potential gauge we had to choose to symmetrize the stress-energy tensor using the Belinfante procedure.
It has also been recently suggested that spin tensor may have a relevant role in out of equilibrium dynamics for
quantum relativistic fluid with thermal vorticity~\cite{Becattini:2018duy}. As it is evident from Eq.~(\ref{eq:defSpinTens}),
the spin tenor is a bilinear local observable and as such its mean value can be determined with the same methods used
for stress-energy tensor and for the currents.

As a first step, we provide its decomposition in thermal vorticity. Notice that the spin tensor is antisymmetric in
the last two indices, then also its thermal expectation value $\mathcal{S}^{\lambda,\mu\nu}$ is a rank three tensor
antisymmetric in the last two indices. The decomposition in the tetrad $\{u,\alpha,w,\gamma\}$ up to the second
order in thermal vorticity of $\mathcal{S}^{\lambda,\mu\nu}$ is therefore given by:
\begin{equation*}
\begin{split}
\mathcal{S}^{\lambda,\mu\nu}=&\mathbb{S}\,\epsilon^{\lambda\mu\nu\rho}u_\rho+F_\alpha\, u^\lambda(\alpha^\mu u^\nu-\alpha^\nu u^\mu) +F_w\, u^\lambda(w^\mu u^\nu-w^\nu u^\mu)
		+F_\gamma\, u^\lambda(\gamma^\mu u^\nu-\gamma^\nu u^\mu) \\
&+ \Gamma_\alpha\, \epsilon^{\mu\nu\rho\sigma}u^\lambda\alpha_\rho u_\sigma+ \Gamma_w\, \epsilon^{\mu\nu\rho\sigma}u^\lambda w_\rho u_\sigma
		+ \Gamma_\gamma\, \epsilon^{\mu\nu\rho\sigma}u^\lambda\gamma_\rho u_\sigma+ \Gamma_{\alpha w}\, u^\lambda(\alpha^\mu w^\nu-\alpha^\nu w^\mu)\\
&+S_\alpha\, (\epsilon^{\lambda\mu\rho\sigma}\alpha_\rho u_\sigma u^\nu - \epsilon^{\lambda\nu\rho\sigma}\alpha_\rho u_\sigma u^\mu)
	+S_w\, (\epsilon^{\lambda\mu\rho\sigma}w_\rho u_\sigma u^\nu - \epsilon^{\lambda\nu\rho\sigma}w_\rho u_\sigma u^\mu)\\
&+S_\gamma\, (\epsilon^{\lambda\mu\rho\sigma}\gamma_\rho u_\sigma u^\nu - \epsilon^{\lambda\nu\rho\sigma}\gamma_\rho u_\sigma u^\mu)
+\Sigma_{\alpha}\,(\alpha^\lambda \alpha^\mu u^\nu - \alpha^\lambda \alpha^\nu u^\mu)\\
&+\Sigma_w\,(w^\lambda w^\mu u^\nu - w^\lambda w^\nu u^\mu)
+\Sigma\ped{A}\,[(\alpha^\lambda w^\mu -\alpha^\mu w^\lambda)u^\nu - (\alpha^\lambda w^\nu - \alpha^\nu w^\lambda)u^\mu]\\
&+\Sigma\ped{S}\,[(\alpha^\lambda w^\mu + \alpha^\mu w^\lambda ) u^\nu - (\alpha^\lambda w^\nu  +\alpha^\nu w^\lambda ) u^\mu)
+\mathcal{O}(\varpi^3).
\end{split} 
\end{equation*}
Each coefficient in front of the four-vectors combination could be a thermodynamic function. First, noticing that
\begin{equation*}
\h S^{0,12}=\h S^{1,02},\quad \h S^{0,i0}=0,
\end{equation*}
we conclude that
\begin{equation*}
\Gamma_w=S_w,\quad F_\alpha=F_w=F_\gamma=0.
\end{equation*}
Comparing this decomposition with linear response theory expansion~(\ref{meanvalueoper}), we can write each thermodynamic
function as a combination of the correlators in the form of Eq.~(\ref{mycorr_repeat}) in the local rest frame of inverse
four-temperature. Moreover, using the spin tensor properties under discrete transformation (that can be found in
Table~\ref{tab:ptcObservables} of Appendix~\ref{sec:Transf}) we can select only the non-vanishing contributions.

Consider first an even parity thermal equilibrium, i.e. with vanishing axial chemical potential. In this case, chiral
coefficients are not allowed and from Table~\ref{tab:ptcObservables} we obtain the first order decomposition of
spin tensor expectation value:
\begin{equation*}
\mathcal{S}^{\lambda,\mu\nu}=\Gamma_w \left(\epsilon^{\mu\nu\rho\sigma}u^\lambda w_\rho u_\sigma
	+\epsilon^{\lambda\mu\rho\sigma}w_\rho u_\sigma u^\nu - \epsilon^{\lambda\nu\rho\sigma}w_\rho u_\sigma u^\mu\right)+\mc{O}(\varpi^2).
\end{equation*}
Moreover, after some manipulation we can show that
\begin{equation*}
u^\lambda\varpi^{\mu\nu}+u^\nu\varpi^{\lambda\mu}+u^\mu\varpi^{\nu\lambda}=
\epsilon^{\mu\nu\rho\sigma}u^\lambda w_\rho u_\sigma+\epsilon^{\lambda\mu\rho\sigma}w_\rho u_\sigma u^\nu - \epsilon^{\lambda\nu\rho\sigma}w_\rho u_\sigma u^\mu
\end{equation*}
and the spin tensor can be cast into
\begin{equation*}
\mathcal{S}^{\lambda,\mu\nu}=\Gamma_w\left( u^\lambda\varpi^{\mu\nu}+u^\nu\varpi^{\lambda\mu}+u^\mu\varpi^{\nu\lambda}\right)
	+\mc{O}(\varpi^2).
\end{equation*}
This form was also obtained in~\cite{Becattini:2013fla} starting from an ansatz for the Wigner function of fermions with thermal vorticity.

Consider now the case of a parity violating fluid. In this case, an additional zero order thermodynamic function is allowed:
\begin{equation}
\label{eq:SpinTensDec}
\mathcal{S}^{\lambda,\mu\nu}=\mathbb{S}\,\epsilon^{\lambda\mu\nu\rho}u_\rho+\Gamma_w\, \left( u^\lambda\varpi^{\mu\nu}+u^\nu\varpi^{\lambda\mu}+u^\mu\varpi^{\nu\lambda}\right)
+\mc{O}(\varpi^2),
\end{equation}
where
\begin{equation*}
\mathbb{S}=\mean{\mean{\h S^{3,21}}}\ped{T},\quad \Gamma_w=\mean{\mean{\h J^3 \h S^{0,12}}}\ped{T}=C_{01|012|2}\apic{st}-C_{02|012|1}\apic{st}
\end{equation*}
and we have defined  the correlators
\begin{equation*}
C_{\mu\nu|\lambda\alpha\beta| i}\apic{st}\equiv\int_0^{|\beta|} \frac{\D\tau}{|\beta|}  \int \D^3x 
	\mean{{\rm T}_\tau\left(\h{T}_{\mu\nu}(X)\,\h{\mathcal{S}}_{\lambda,\alpha\beta}(0)\right)}\ped{$T$,c}\, x_i.
\end{equation*}

Consider now the operatorial relations related to spin tensor. This case is particularly interesting because it tight
together axial current thermal coefficients, i.e. chirality, and stress-energy tensor coefficients.
The key observation to link spin tensor with chirality is to notice that the axial current is dual to the spin tensor:
\begin{equation*}
\h{j}\apic{A}_\tau=-\frac{1}{3}\epsilon_{\tau\lambda\mu\nu}\h{S}^{\lambda,\mu\nu}.
\end{equation*}
The same relation is transported without modification to expectation values
\begin{equation*}
\mean{\h{j}\apic{A}_\tau}=-\frac{1}{3}\epsilon_{\tau\lambda\mu\nu}\mean{\h{S}^{\lambda,\mu\nu}}.
\end{equation*}
We can then replace the decomposition (\ref{eq:SpinTensDec}) and (\ref{acurrdecomp}) at first order in thermal vorticity
in the previous identity to establish relations between thermal coefficients of spin tensor and axial current.
Taking advantage of the identities
\begin{gather*}
\gamma^{\mu}\gamma^{\nu}\gamma^{\rho}=\eta^{\mu\nu}\gamma^{\rho}+\eta^{\nu\rho}\gamma^{\mu}-\eta^{\mu\rho}\gamma^{\nu}
-\I\epsilon^{\sigma\mu\nu\rho}\gamma_{\sigma}\gamma^{5},\\
\epsilon_{\tau\lambda\mu\nu}\epsilon^{\lambda\mu\nu\rho}=6\delta_\tau^{\,\rho},\quad
w_\mu=-\frac{1}{2}\epsilon_{\mu\nu\rho\sigma}\varpi^{\nu\rho}u^\sigma,
\end{gather*}
we can show that
\begin{equation*}
\epsilon_{\tau\lambda\mu\nu}\gamma^\lambda\gamma^\mu\gamma^\nu=6\I \gamma_\tau\gamma^5,\quad
\epsilon_{\tau\lambda\mu\nu}\epsilon^{\lambda\mu\nu\rho}u_\rho=6 u_\tau,\quad
\epsilon_{\tau\lambda\mu\nu} u^\lambda\varpi^{\mu\nu}=-2w_\tau,
\end{equation*}
from which follows that
\begin{equation*}
\epsilon_{\rho\lambda\mu\nu}\mathcal{S}^{\lambda,\mu\nu}=-3\mean{\h{j}_\rho\apic{A}}=-6\mathbb{S} u_\rho-6\,\Gamma_w\, w_\rho
\end{equation*}
or equivalently
\begin{equation*}
\mean{\h{j}\apic{A}_\mu}=2\mathbb{S} u_\mu+2\Gamma_w\, w_\mu=n\ped{A}\,u_\mu+W\apic{A}\, w_\mu.
\end{equation*}
We have then found that the chiral thermodynamic coefficient of spin tensor is proportional to axial charge density and
that the thermal expectation related to the response of spin tensor to rotation is proportional to the axial vortical
effect conductivity:
\begin{equation}
\label{eq:SpinTensAndAxialRel}
\mathbb{S}=\frac{1}{2}n\ped{A},\quad \Gamma_w=\frac{1}{2}W\apic{A}.
\end{equation}

We can also relate those coefficients to those of stress-energy tensor. Spin tensor divergence is equivalent to the
antisymmetric part of canonical stress-energy tensor
\begin{equation*}
\de_\lambda\h{\mc{S}}^{\lambda,\mu\nu}=\h{T}\ped{Can}^{\mu\nu}-\h{T}\ped{Can}^{\nu\mu}.
\end{equation*}
As we now show these thermal coefficients can also be made proportional to the coefficient of canonical stress-energy tensor.
The spin tensor satisfy the conservation equation
\begin{equation*}
\de_\lambda \mathcal{S}^{\lambda,\mu\nu}=\mean{\h T^{\mu\nu}\ped{Can}}-\mean{\h T^{\nu\mu}\ped{Can}},
\end{equation*}
where if we choose the spin tensor to be defined by Eq.~(\ref{eq:defSpinTens}) the stress-energy tensor $\h T^{\mu\nu}\ped{Can}$ 
in the previous equation can only be the \emph{canonical} stress-energy tensor. From their decomposition we can show that
\begin{gather*}
\de_\lambda\left(\mathbb{S}\epsilon^{\lambda\mu\nu\rho}u_\rho\right)=-\left(\frac{\mathbb{S}}{|\beta|}+\frac{\de \mathbb{S}}{\de|\beta|} \right)\epsilon^{\mu\nu\kappa\lambda}\alpha_k u_\lambda
	+2\frac{\mathbb{S}}{|\beta|}\left(u^\mu\gamma^\nu-u^\nu\gamma^\mu \right),\\
\de_\lambda\left(\Gamma_\alpha(\epsilon^{\mu\nu\rho\sigma}u^\lambda \alpha_\rho u_\sigma
	+\epsilon^{\lambda\mu\rho\sigma}\alpha_\rho u_\sigma u^\nu - \epsilon^{\lambda\nu\rho\sigma}\alpha_\rho u_\sigma u^\mu) \right)=0,\\
\de_\lambda\left(\Gamma_w\left( u^\lambda\varpi^{\mu\nu}+u^\nu\varpi^{\lambda\mu}+u^\mu\varpi^{\nu\lambda}\right)\right)
	=\left(\frac{\Gamma_w}{|\beta|}-\frac{\de\Gamma_w}{\de|\beta|} \right)\left(u^\mu\gamma^\nu-u^\nu\gamma^\mu \right),
\end{gather*}
and hence
\begin{equation*}
\de_\lambda \mathcal{S}^{\lambda,\mu\nu}=-\left(\frac{\mathbb{S}}{|\beta|}+\frac{\de \mathbb{S}}{\de|\beta|} \right)\epsilon^{\mu\nu\kappa\lambda}\alpha_k u_\lambda
+2\frac{\mathbb{S}}{|\beta|}\left(u^\mu\gamma^\nu-u^\nu\gamma^\mu \right)
+\left(\frac{\Gamma_w}{|\beta|}-\frac{\de\Gamma_w}{\de|\beta|} \right)\left(u^\mu\gamma^\nu-u^\nu\gamma^\mu \right)
\end{equation*}
whereas
\begin{equation*}
\begin{split}
\mean{\h T^{\mu\nu}\ped{Can}}-\mean{\h T^{\nu\mu}\ped{Can}}=&2\mathbb{A}\apic{Can}\epsilon^{\mu\nu\kappa\lambda}\alpha_k u_\lambda
+(\mathbb{W}_1\apic{Can}-\mathbb{W}_2\apic{Can})\left(u^\mu\gamma^\nu-u^\nu\gamma^\mu \right)+\\
&+(G_1\apic{Can}-G_2\apic{Can})\left(u^\mu\gamma^\nu-u^\nu\gamma^\mu \right).
\end{split}
\end{equation*}
By comparison, it is straightforward to obtain the following relations
\begin{equation}
\label{eq:SpinTensAndSETRel}
\begin{split}
-\left(\frac{\mathbb{S}}{|\beta|}+\frac{\de \mathbb{S}}{\de|\beta|} \right)&= 2\mathbb{A}\apic{Can},\\
2\frac{\mathbb{S}}{|\beta|}&=\mathbb{W}_1\apic{Can}-\mathbb{W}_2\apic{Can},\\
\frac{\Gamma_w}{|\beta|}-\frac{\de\Gamma_w}{\de|\beta|} &=4\frac{\Gamma_w}{|\beta|} =G_1\apic{Can}-G_2\apic{Can}.
\end{split}
\end{equation}
Then combining Eq.s (\ref{eq:SpinTensAndAxialRel}) and (\ref{eq:SpinTensAndSETRel}), the coefficients of canonical
stress-energy tensor and axial current are related by:
\begin{equation*}
\begin{split}
\mathbb{A}\apic{Can}=&-\left(\frac{n\ped{A}}{|\beta|}+\frac{\de n\ped{A}}{\de|\beta|} \right),\\
\frac{n\ped{A}}{|\beta|}=&\mathbb{W}_1\apic{Can}-\mathbb{W}_2\apic{Can},\\
\frac{W\apic{A}}{|\beta|}=&\frac{G_1\apic{Can}-G_2\apic{Can}}{2}.
\end{split}
\end{equation*}
%

\section{Landau and \texorpdfstring{$\beta$}{beta} frames in chiral hydrodynamics}
\label{sec:LandauFrame}
Before proceeding to the explicit calculation of the thermodynamic coefficients for free fields, 
it is worth discussing the choice of the hydrodynamic frame. This choice is relevant for the hydrodynamics
of chiral fluids, which can be defined in general as fluids where the axial current is relevant.
We remind that hydrodynamic frame refers to the choice made on the definition of temperature,
fluid velocity and chemical potentials. When those quantities have non vanishing derivatives,
several definitions can be adopted as long as they all agree on the value of temperature, fluid
velocity and of chemical potentials at homogeneous thermal equilibrium. As has been introduced in
Chapter~\ref{chp:QRelStatMech}, the four-velocity of the fluid, appearing in our formulae as $u$,
has been defined starting from the four-temperature vector $\beta$ which is a Killing
vector in global equilibrium. This frame is called $\beta$ or thermodynamic frame~\cite{BecaBetaF,Kovtun:2016lfw}.
While this is a natural frame for the generalized equilibrium, this is not the most common choice
in literature and it is not the frame typically used in numerical codes. It is then of interest
to translate the constitutive equations of previous sections into the more used Landau frame.

Indeed, in general Landau frame and $\beta$-frame do not coincide. In Landau frame the fluid velocity
$u$ is by definition an eigenvector of the (symmetrized) stress-energy tensor. Instead, we clearly
seen from Eq.~(\ref{setdecomp}) that stress-energy tensor in $\beta$-frame is not diagonal in fluid
velocity. In the non-chiral case the difference between Landau frame velocity and $\beta$-frame
velocity shows up only at second order in the gradients of $u$ or $\beta$~\cite{Buzzegoli:2017cqy} owing
to the terms in $u^\nu \gamma^\mu$. However, in chiral fluids, as we are going to show, the difference
between  the two velocities shows up at first order in the gradients because of the chiral terms
in $u^\nu w^\mu$ in Eq.~(\ref{setdecomp}). 

This is especially important for chiral hydrodynamics because it implies that one cannot formulate
it at first order without taking these equilibrium non-dissipative terms in $u^\nu w^\mu$ into 
account. Even if one tried to eliminate them in the stress-energy tensor by going to the Landau frame,
they would reappear in the constitutive equations of the axial current at first order in the 
gradients; this is just what happens, see Eq.s~(\ref{eq:VecCurrLandau}) and~(\ref{eq:AxialCurrLandau})
below. For chiral-magneto-hydrodynamics, with dynamical electro-magnetic fields, the transition to
Landau frame would be even more complicated  because a change of the velocity definition implies
a change of definition in the comoving electric field.

To move from $\beta$ frame to Landau frame it suffices to establish the relation between the
$\beta$-frame velocity $u$ and the Landau frame velocity $u\ped{L}$.
By definition, the Landau velocity $u\ped{L}$ is the eigenvector of the stress-energy tensor:
\begin{equation*}
u_{{\rm L}\mu}T^{\mu\nu}=\rho\ped{L}u\ped{L}^{\nu}
\end{equation*}
where $\rho\ped{L}$ is both the eigenvalue and the energy density in the Landau frame.
Suppose we are in the rotational case (see Sec.~\ref{sec:ThermalVort}), then the Landau velocity
$u\ped{L}$ can be expressed in terms of the tetrad $\{u,\alpha,w,\gamma\}$, where
the four-vector $\alpha,\,w$ and $\gamma$ are defined in Sec.~\ref{sec:ThermalVort} by Eq.s~(\ref{eq:accvort})
and~(\ref{eq:transversedir}):
\begin{equation}
\label{Landauvel}
u\ped{L}^\mu=a\,u^\mu+b\,\frac{w^\mu}{|w|}+c\,\frac{\gamma^\mu}{|\gamma|}
+d\,\frac{\alpha^\mu}{|\alpha|}\, ,
\end{equation}
where $|w|=\sqrt{-w^2},\,|\gamma|=\sqrt{-\gamma^2},\,|\alpha|=\sqrt{-\alpha^2}$
and $a,b,c$ and $d$ are four unknown constants such that $u\ped{L}^\mu u_{{\rm L}\mu}=1$, i.e.
\begin{equation*}
a^2-b^2-2bd\frac{\alpha\cdot w}{|\alpha||w|}-c^2-d^2=1.
\end{equation*}
Furthermore, since for a vanishing thermal vorticity we expect that the two thermodynamic frames
coincide, for $\varpi=0$ we must have that $a=1$ and $\rho\ped{L}=\rho$.
This means that the leading term of $a$ and $\rho\ped{L}$ is zeroth order in thermal vorticity.

Now, taking the stress-energy tensor mean value at the second order expansion on thermal vorticity
\begin{equation*}
\begin{split}
T^{\mu\nu}=&\,\mathbb{A}\,\epsilon^{\mu\nu\kappa\lambda}\alpha_\kappa u_\lambda+\mathbb{W}_1 w^\mu u^\nu +\mathbb{W}_2 w^\nu u^\mu\\
&+(\rho-\alpha^2 U_\alpha -w^2 U_w)u^\mu u^\nu -(p-\alpha^2D_\alpha-w^2D_w)\Delta^{\mu\nu}\\
&+A\,\alpha^\mu\alpha^\nu+Ww^\mu w^\nu+G_1 u^\mu\gamma^\nu+G_2 u^\nu\gamma^\mu+\mathcal{O}(\varpi^3)
\end{split}
\end{equation*}
and contracting it with the landau velocity (\ref{Landauvel}), we obtain
\begin{equation*}
\begin{split}
u_{{\rm L}\mu}T^{\mu\nu}=&
\left(a\,\rho\ped{eff}-b\,\mathbb{W}_1|w|-c\,G_2|\gamma|+d\,\mathbb{W}_1\frac{\alpha\cdot w}{|\alpha|}\right)u^\nu\\
&+\left(a\,\mathbb{W}_2-b\,p\ped{eff}\frac{1}{|w|}-b\,W|w|
+c\,\mathbb{A}\frac{|\alpha|^2}{|\gamma|}+d\,W\frac{\alpha\cdot w}{|\alpha|}\right)w^\nu+\\
&+\left(a\,G_1-b\,\mathbb{A}\frac{1}{|w|}-c\,p\ped{eff}\frac{1}{|\gamma|}\right)\gamma^\nu+\\
&+\left(b\,A\frac{\alpha\cdot w}{|w|}+c\,\mathbb{A}\frac{\alpha\cdot w}{|\gamma|}
-d\,p\ped{eff}\frac{1}{|\alpha|}-d\,A|\alpha|\right)\alpha^\nu+\mathcal{O}(\varpi^3),
\end{split}
\end{equation*}
where for simplicity we defined $\rho\ped{eff}\equiv\rho-\alpha^2 U_\alpha -w^2 U_w$ and
$p\ped{eff}\equiv p-\alpha^2D_\alpha-w^2D_w$.
Equating this expression to $\rho\ped{L}u\ped{L}^{\nu}$, with $u\ped{L}$ given by (\ref{Landauvel}),
we obtain the five equations in five unknown variables $(a,b,c,d,\rho\ped{L})$ that
diagonalize the stress-energy tensor (at second order in thermal vorticity):
\begin{equation}
\label{eqslin}
\begin{split}
a\,\rho\ped{eff}-b\,\mathbb{W}_1|w|-c\,G_2|\gamma|+d\,\mathbb{W}_1\frac{\alpha\cdot w}{|\alpha|} &= a\,\rho\ped{L}\\
a\,\mathbb{W}_2-b\,p\ped{eff}\frac{1}{|w|}-b\,W|w| +c\,\mathbb{A}\frac{|\alpha|^2}{|\gamma|}
+d\,W\frac{\alpha\cdot w}{|\alpha|}&=b\,\rho\ped{L}\,\frac{1}{|w|}\\
a\,G_1-b\,\mathbb{A}\frac{1}{|w|}-c\,p\ped{eff}\frac{1}{|\gamma|}
&=c\,\rho\ped{L}\frac{1}{|\gamma|}\\
b\,A\frac{\alpha\cdot w}{|w|}+c\,\mathbb{A}\frac{\alpha\cdot w}{|\gamma|}
-d\,p\ped{eff}\frac{1}{|\alpha|}-d\,A|\alpha|&=d\,\rho\ped{L}\frac{1}{|\alpha|}\\
a^2-b^2-2bd\frac{\alpha\cdot w}{|\alpha||w|}-c^2-d^2 &=1.
\end{split}
\end{equation}

To solve it, we can simplify the equations furtherer.
Indeed, we can write the Eq. in the first line of (\ref{eqslin}) as
\begin{equation}
\frac{b}{a}=\frac{\rho\ped{eff}-\rho\ped{L}}{\mathbb{W}_1}\frac{1}{|w|}
	+\frac{d}{a}\frac{\alpha\cdot w}{|\alpha||w|}-\frac{c}{a}\frac{G_2}{\mathbb{W}_1}\frac{|\gamma|}{|w|}.
\end{equation}
In the same way, we can write the third Eq. of (\ref{eqslin}) isolating $b/a$:
\begin{equation}
\label{eqforb}
\frac{b}{a}=\frac{G_1}{\mathbb{A}}|w|
	-\frac{c}{a}\frac{\rho\ped{L}+p\ped{eff}}{\mathbb{A}}\frac{|w|}{|\gamma|};
\end{equation}
hence, equating the previous two, we obtain
\begin{equation*}
\frac{c}{a}\left(\frac{\rho\ped{L}+p\ped{eff}}{\mathbb{A}}-\frac{G_2}{\mathbb{W}_1}\frac{|\gamma|^2}{|w|^2} \right)=
	\frac{G_1}{\mathbb{A}}|\gamma|-\frac{\rho\ped{eff}-\rho\ped{L}}{\mathbb{W}_1}\frac{|\gamma|}{|w|^2}
		-\frac{d}{a}\frac{(\alpha\cdot w)|\gamma|}{|\alpha||w|^2},
\end{equation*}
and then at second order in thermal vorticity (notice that $\rho\ped{eff}-\rho\ped{L}$ is at least first order on thermal vorticity):
\begin{equation}
\label{eqforc}
\frac{c}{a}=\frac{\mathbb{A}}{\rho+p}\left[\frac{G_1}{\mathbb{A}}|\gamma|-
	\frac{\rho\ped{eff}-\rho\ped{L}}{\mathbb{W}_1}\frac{|\gamma|}{|w|^2}-\frac{d}{a}\frac{(\alpha\cdot w)|\gamma|}{|\alpha||w|^2}\right]+\mathcal{O}(\varpi^3).
\end{equation}
From Eq. (\ref{eqforc}) we conclude that also $c$ is at least first order in $\varpi$.
While, from the Eq. in the fourth line of (\ref{eqslin}) we have
\begin{equation}
\label{eqford}
\frac{d}{a}=\frac{1}{\rho\ped{L}+p\ped{eff}+A|\alpha|^2}\l(\frac{b}{a}\,A\,|\gamma|+\frac{c}{a}\,\mathbb{A}\,|w|\r)\frac{\alpha\cdot w}{|w||\gamma|}|\alpha|,
\end{equation}
showing that $d$ is at least second order in thermal vorticity. This means that the term in $d/a$ of Eq.~(\ref{eqforc}) can be removed and we find:
\begin{equation}
\label{eqcovera}
\frac{c}{a}=\frac{\mathbb{A}}{\mathbb{W}_1}\frac{\rho\ped{L}-\rho\ped{eff}}{\rho+p}\frac{|\gamma|}{|w|^2}
	+\frac{G_1}{\rho+p}|\gamma|+\mathcal{O}(\varpi^3).
\end{equation}
Replacing this solution in Eq.~(\ref{eqforb}) we obtain
\begin{equation}
\label{eqbovera}
\frac{b}{a}=\frac{\rho\ped{eff}-\rho\ped{L}}{\mathbb{W}_1}\frac{1}{|w|}+\frac{G_1}{\mathbb{A}}\l(1-\frac{\rho\ped{L}+p\ped{eff}}{\rho+p}\r)|w|,
\end{equation}
and plugging the previous two Eq.s in (\ref{eqford}) we find
\begin{equation}
\label{eqdovera}
\frac{d}{a}=\frac{A(\rho+p)-\mathbb{A}^2}{\mathbb{W}_1(\rho+p)^2}\l(\rho\ped{eff}-\rho\ped{L}\r)\frac{|\alpha|(\alpha\cdot w)}{|w|^2}.
\end{equation}

To obtain $\rho\ped{L}$ we replace the Eq.s (\ref{eqcovera}), (\ref{eqbovera}) and (\ref{eqdovera}) in the second equation of (\ref{eqslin}).
This becomes a second grade equation for $\rho\ped{L}$, the only solution at second order in vorticity reproducing $\rho\ped{L}=\rho$ for $\varpi=0$ is:
\begin{equation*}
\rho\ped{L}=\rho\ped{eff}-\frac{\mathbb{W}_1\mathbb{W}_2}{\rho+p}|w|^2+\mathcal{O}(\varpi^4).
\end{equation*}
Now, plugging this into the previous ratios (\ref{eqcovera}), (\ref{eqbovera}) and (\ref{eqdovera}) we find
\begin{equation*}
\begin{split}
\frac{b}{a}=&\frac{\mathbb{W}_2}{\rho+p}|w|+\mathcal{O}(\varpi^3),\\
\frac{c}{a}=&\frac{G_1(\rho+p)-\mathbb{A}\mathbb{W}_2}{(\rho+p)^2}|\gamma|+\mathcal{O}(\varpi^3),\\
\frac{d}{a}=&\mathcal{O}(\varpi^3).
\end{split}
\end{equation*}
From these we learn that $b$ is actually first order on thermal vorticity, $c$ is second order and $d$ is in fact third order
and so can be set to zero. Consequently, the last equation to solve is $a^2-b^2-c^2=1$, that gives
\begin{equation*}
a=\frac{1}{\sqrt{1-(b/a)^2-(c/a)^2}}
=1+\frac{1}{2}\frac{\mathbb{W}_2^2}{(\rho+p)^2}|w|^2+\mathcal{O}(\varpi^3).
\end{equation*}
Where we choose the positive solution because it reproduces $a=1$ for $\varpi=0$.

In conclusion, we found that the relation between the Landau fluid velocity $u\ped{L}$
and the $\beta$-frame fluid velocity $u$ is
\begin{equation*}
u\ped{L}=\left(1+\frac{1}{2}\frac{\mathbb{W}_2^2}{(\rho+p)^2}|w|^2\right)u+\frac{\mathbb{W}_2}{\rho+p} w
	+\frac{G_1(\rho+p)-\mathbb{A}\mathbb{W}_2}{(\rho+p)^2}\gamma+\mathcal{O}(\varpi^3)
\end{equation*}
or, reverting it, $u$ is given by 
\begin{equation}
\label{uLandau}
u=\left(1-\frac{1}{2}\frac{\mathbb{W}_2^2}{(\rho+p)^2}|w|^2\right)u\ped{L}
	-\frac{\mathbb{W}_2}{\rho+p}w-\frac{G_1(\rho+p)-\mathbb{A}\mathbb{W}_2}{(\rho+p)^2}\gamma+\mathcal{O}(\varpi^3).
\end{equation}
This transformation at first order in vorticity and the following relation between CVE conductivity
in Landau frame with $\mathbb{W}_2$ and $W\apic{V}$ were also pointed out in~\cite{Landsteiner:2012kd}.
At this point, taking advantage of Eq.~(\ref{uLandau}), we can express the stress-energy tensor
mean value at second order in thermal vorticity of Eq.~(\ref{setdecomp}) in the Landau frame:
\begin{equation*}
\begin{split}
T^{\mu\nu}=&\,\mathbb{A}\epsilon^{\mu\nu\kappa\lambda}\alpha_\kappa u_{{\rm L}\lambda}
	-\frac{\mathbb{A}\mathbb{W}_2}{\rho+p}\epsilon^{\mu\nu\kappa\lambda}\alpha_\kappa w_\lambda
		+(\mathbb{W}_1-\mathbb{W}_2)u\ped{L}^\nu w^\mu - p\ped{eff}\,\Delta\ped{L}^{\mu\nu}\\
&+\left(\rho\ped{eff}-\frac{\mathbb{W}_2^2}{\rho+p}|w|^2\right)u\ped{L}^\mu u\ped{L}^\nu
	+A\alpha^\mu \alpha^\nu+\frac{W(\rho+p)+\mathbb{W}_1\mathbb{W}_2}{\rho+p}w^\mu w^\nu\\
&+\frac{\mathbb{A}\mathbb{W}_2}{\rho+p}u\ped{L}^\mu\gamma^\nu
	+\frac{(G_2-G_1)(\rho+p)+\mathbb{A}\mathbb{W}_2}{\rho+p}u\ped{L}^\nu\gamma^\mu.
\end{split}
\end{equation*}
If the stress-energy tensor is symmetric then we have $\mathbb{A}=0,\,\mathbb{W}_1=\mathbb{W}_1=\mathbb{W},\,G_1=G_2$
and the previous expression simplifies in:
\begin{equation*}
\begin{split}
T^{\mu\nu}\ped{Sym}=&\,\left(\rho\ped{eff}-\frac{\mathbb{W}_2^2}{\rho+p}|w|^2\right)u\ped{L}^\mu u\ped{L}^\nu - p\ped{eff}\,\Delta\ped{L}^{\mu\nu}
	+A\alpha^\mu \alpha^\nu+\frac{W(\rho+p)+\mathbb{W}^2}{\rho+p}w^\mu w^\nu.
\end{split}
\end{equation*}
Furthermore, using (\ref{uLandau}), we can also write the vectorial current (\ref{vcurrdecomp}) in the Landau frame
\begin{equation}
\label{eq:VecCurrLandau}
\begin{split}
j\ped{V}^\mu=& \, n\ped{V}\left(1-\frac{1}{2}\frac{\mathbb{W}_2^2}{(\rho+p)^2}|w|^2\right) 
u\ped{L}^\mu+\left(\alpha^2\,N\apic{V}_\alpha+w^2\,N\apic{V}_w\right)u\ped{L}^\mu\\
&+\left(W\apic{V}-n\ped{V}\frac{\mathbb{W}_2}{\rho+p} \right)w^\mu
+\left(G\apic{V}-n\ped{V}\frac{G_1(\rho+p)-\mathbb{A}\mathbb{W}_2}{(\rho+p)^2}\right)\gamma^\mu.
\end{split}
\end{equation}
Instead, replacing (\ref{uLandau}) in (\ref{acurrdecomp}), we obtain the axial current in the Landau frame
\begin{equation}
\label{eq:AxialCurrLandau}
\begin{split}
j\ped{A}^\mu=& \, n\ped{A}\left(1-\frac{1}{2}\frac{\mathbb{W}_2^2}{(\rho+p)^2}|w|^2\right) 
u\ped{L}^\mu+\left(\alpha^2\,N\apic{A}_\alpha+w^2\,N\apic{A}_w\right)u\ped{L}^\mu\\
&+\left(W\apic{A}-n\ped{A}\frac{\mathbb{W}_2}{\rho+p}\right)w^\mu
+\left(G\apic{A}-n\ped{A}\frac{G_1(\rho+p)-\mathbb{A}\mathbb{W}_2}{(\rho+p)^2}\right)\gamma^\mu.
\end{split}
\end{equation}
Similarly, the spin tensor in Landau frame at first order in thermal vorticity is
\begin{equation*}
\mathcal{S}^{\lambda,\mu\nu}=\mathbb{S}\,\epsilon^{\lambda\mu\nu\rho}u_\rho
	+\l(\Gamma_w+\frac{\mathbb{S}\mathbb{W}_2}{\rho+p}\r) \left( u^\lambda\varpi^{\mu\nu}+u^\nu\varpi^{\lambda\mu}+u^\mu\varpi^{\nu\lambda}\right),
\end{equation*}

In conclusion, the $\beta$ frame appears as the most appropriate to write down hydrodynamic equations,
as advocated in~\cite{BecaBetaF,Kovtun:2016lfw}.

\section{Free massless Dirac field } 
\label{sec:CVFree}
Here we describe the procedure we used for the evaluation of thermal coefficients described above
for a sysytem consisting of free massless Dirac particles. Thermal coefficients are evaluated directly
through the correlators in their definition, i.e. Eq.~(\ref{basicstrut}). These correlators are to be
computed at local rest frame, therefore we can use usual thermal field theory techniques to perform
the calculations. We reviewed finite temperature and density filed theory  with the inclusion of axial
chemical potential in chapter~\ref{ch:ftft}, where we also derived the thermal propagator for free Dirac
field in Sec.~\ref{sec:FermiProp}.

Here for convenience, we report again the notation and convention we are adopting. We use imaginary
time formalism and we use Euclidean gamma matrices $\gamma_\mu$ fulfilling the relation  $\{\gamma_\mu ,\gamma_\nu \}=2 \delta_{\mu\nu}$.
We also adopt the following notation
\begin{equation}
\label{symbols}
P^\pm=(p_n\pm\ii\mu, {\bm p} ),\quad \quad X=(\tau ,{\bm x}), 
\quad \quad\sumint_{\{P\}}=\frac{1}{|\beta|}\sum_{n=-\infty}^\infty\int \frac{\di ^3 p }{(2\pi)^3},
\end{equation}
where $p_n=2\pi (n+1/2)/|\beta|$ with $n\in\mathbb{Z}$ are the Matsubara frequencies. We introduced
the right and left chiral chemical potentials and the corresponding projection operator
\begin{equation*}
\mu\ped{R}=\mu+\mu\ped{A},\quad\mu\ped{L}=\mu-\mu\ped{A},\quad P\ped{R/L}^\pm=(\omega_n\pm\I \mu\ped{R/L},\vec{p}),
\quad \mathbb{P}_\chi=\frac{1+\chi\gamma_5}{2},
\end{equation*}
where $\chi=$R,L is assigned to be respectively $+1,-1$.
Then, the massless Dirac propagator in the imaginary time reads~(\ref{eq:propFermCoord}):
\begin{equation}\label{eq:Prop}
\mean{{\rm T}_\tau \wPsi_a(X)\bar{\wPsi}_b(Y)}_T=
\sum_\chi \sumint_{\{P\}} \frac{\e^{\ii P^+_\chi\cdot(X-Y)}}{{P^+_\chi}^2}\left(\mathbb{P}_\chi G(P^+_\chi)\right)_{ab},
\quad G(P)\equiv-\I\slashed{P}\, ,
\end{equation}
where latin characters $a,b,\dots$ denote spinorial indices, $\slashed P=\gamma_\mu P_\mu$ 
is the standard contraction between the (Euclidean) Dirac matrices $\gamma_\mu$ and the 
(Euclidean) four-momenta $P$.
The Euclidean canonical stress-energy tensor, see Eq.~(\ref{imagtime}), equals
\begin{equation*}
\h T_{\mu\nu}(X)=
\frac{\I^{\delta_{0\mu}+\delta_{0\nu}}}{2}
\bar{\wPsi}(X)\left[\gamma_\mu\oraw{\de}_\nu-\gamma_\mu\olaw{\de}_\nu\right]\wPsi(X),
\end{equation*}
where the $\I^{\delta_{0\mu}}$ factor stems from Wick rotation. The Belinfante-symmetrized 
stress-energy tensor used to construct all Poincaré generators, is simply the symmetrization 
of the previous one:
\begin{equation*}
\h T_{\mu\nu}(X)=
\frac{\I^{\delta_{0\mu}+\delta_{0\nu}}}{4}
\bar{\wPsi}(X)\left[\gamma_\mu\oraw{\de}_\nu-\gamma_\mu\olaw{\de}_\nu+\gamma_\nu\oraw{\de}_\mu-\gamma_\nu\olaw{\de}_\mu\right]\wPsi(X),
\end{equation*}
which can be expressed according to the point-splitting procedure as:
\begin{equation}\label{pointsplitset}
\h T_{\mu\nu}(X)= \lim_{X_1,X_2\to X} \mathcal{D}_{\mu\nu}(\de_{X_1},\de_{X_2})_{ab}\bar\wPsi (X_1)_a\wPsi (X_2)_b,
\end{equation}
where:
\begin{equation*}
\mathcal{D}_{\mu\nu} (\partial_{X_1},\partial_{X_2})=
\frac{\I^{\delta_{0\mu}+\delta_{0\nu}}}{4}\left[ \gamma_\mu (\partial_{X_2}-\partial_{X_1})_\nu
+\gamma_\nu (\partial_{X_2}-\partial_{X_1})_\mu\right].
\end{equation*}
%

\subsection{First order correlators}\label{sec:firstordercoeff}
Here we outline the procedure to evaluate first-order coefficients of a
general operator $\h{O}^{\alpha\dots\beta}(x)$ for free massless Dirac field theory.
We can write a generic correlator~(\ref{basicstrut}) related to first order corrections in vorticity as
\begin{equation}
\label{eq:firstorderprototype}
C_{\mu\nu|\alpha\dots\beta| i}\equiv\int_0^{|\beta|}   \frac{\D\tau}{|\beta|}  \int  \D^3x\,
\mean{{\rm T}_\tau\left(\h{T}_{\mu\nu}(X)\,\h{O}_{\alpha\dots\beta}(0)\right)}\ped{$T$,c}\, x_i,
\end{equation}
where $\wT_{\mu\nu}$ is the symmetric stress-energy tensor coming from Lorentz generators~(\ref{eq:LorenzGenerators})
and the two operators connected correlator is
\begin{equation*}
\mean{\h{A}\h{B}}\ped{$T$,c}= \mean{\h{A}\,\h{B}}_T- \mean{\h{A}}_T\, \mean{\h{B}}_T.
\end{equation*}

We use the point-splitting procedure, Sec.~\ref{subsec:pointsplit}, to express the operator $\h O$ as
\begin{equation*}
\h O_{\alpha\dots\beta}(Y)= \lim_{Y_1,Y_2\to Y} \mathcal{O}_{\alpha\dots\beta} (\de_{Y_1},\de_{Y_2})_{cd}\bar\wPsi (Y_1)_c\wPsi (Y_2)_d.
\end{equation*}
Thus, to compute~(\ref{eq:firstorderprototype}), we first consider
\begin{equation*}
\begin{split}
C(X,Y) \equiv& \mean{{\rm T}_\tau\left(\h{T}_{\mu\nu}(X)\,\h{O}_{\alpha\dots\beta}(Y)\right)}\ped{$T$,c} 
 = \lim_{\substack{X_1 X_2\to X\\Y_1 Y_2\to Y}} {\Dpi_{\mu\nu}(\de_{X_1},\de_{X_2})}_{ab} {\oper(\de_{Y_1},\de_{Y_2})}_{cd}\times\\
&\times \mean{{\rm T}_\tau\!\!\left( \bar{\wPsi}(X_1)_a\wPsi(X_2)_b \bar{\wPsi}(Y_1)_c\wPsi(Y_2)_d\right)}\ped{$T$,c}.
\end{split}
\end{equation*}
In this form, the connected correlator only concerns Dirac fields and can be computed using Wick's theorem,
which for four point functions - in simplified notation - states that
\begin{equation*}
\begin{split}
\mean{\bar{\psi}_1\psi_2\bar{\psi}_3\psi_4}\ped{c} & = \mean{\bar{\psi}_1\psi_2\bar{\psi}_3\psi_4}- \mean{\bar{\psi}_1\psi_2} \mean{\bar{\psi}_3\psi_4}\\
&= \mean{\bar{\psi}_1\psi_4} \mean{\psi_2\bar{\psi}_3}+ \mean{\bar{\psi}_1\psi_2} \mean{\bar{\psi}_3\psi_4}- \mean{\bar{\psi}_1\psi_2} \mean{\bar{\psi}_3\psi_4}\\
& = \mean{\bar{\psi}_1\psi_4} \mean{\psi_2\bar{\psi}_3}.
\end{split}
\end{equation*}
With this result, and exchanging the order of the two anti-commuting fields $\bar{\wPsi}(X_1)_a\wPsi(Y_2)_d$,
we are able to recreate a trace operation:
\begin{equation*}
\begin{split}
C(X,Y) =&    \lim_{\substack{X_1 X_2\to X\\Y_1 Y_2\to Y}}  {\Dpi_{\mu\nu}(\de_{X_1},\de_{X_2})}_{ab} {\oper(\de_{Y_1},\de_{Y_2})}_{cd}\times\\
&\hphantom{\lim_{\substack{X_1 X_2\to X\\Y_1 Y_2\to Y}}}\times\mean{{\rm T}_\tau \bar{\wPsi}(X_1)_a\wPsi(Y_2)_d} \mean{{\rm T}_\tau\wPsi(X_2)_b \bar{\wPsi}(Y_1)_c} \\
=& -\!\! \lim_{\substack{X_1 X_2\to X\\Y_1 Y_2\to Y}}\!\!\tr \left[ {\Dpi_{\mu\nu}(\de_{X_1},\de_{X_2})}  \mean{{\rm T}_\tau \wPsi(X_2) \bar{\wPsi}(Y_1)} {\oper(\de_{Y_1},\de_{Y_2})}
\mean{{\rm T}_\tau \wPsi(Y_2) \bar{\wPsi}(X_1)} \right].
\end{split}
\end{equation*}
Now, we express fermionic propagator in Fourier space as in~(\ref{eq:Prop})
\begin{gather*}
\mean{{\rm T}_\tau \wPsi(X_2)_a\bar{\wPsi}(Y_1)_b}=\sum_\chi \sumint_{\{P\}}\frac{\E^{\I P^+_\chi\cdot (X_2-Y_1)}}{{P^+_\chi}^2}\left(\mathbb{P}_\chi G(P^+_\chi)\right)_{ab} ,\\
\mean{{\rm T}_\tau \wPsi(Y_2)_a\bar{\wPsi}(X_1)_b}=\sum_{\chi'}\sumint_{\{Q\}}\frac{\E^{\I Q^+_{\chi'}\cdot (Y_2-X_1)}}{{Q^+_{\chi'}}^2}\left(\mathbb{P}_{\chi'} G(Q^+_{\chi'})\right)_{ab},
\end{gather*}
where, as previously, $P^+\ped{R/L}= (p_n+\I\mu\ped{R/L},\vec{p})$ and $Q^+\ped{R/L}=(q_m+\I\mu\ped{R/L},\vec{q})$.
The derivatives, appearing in the point-splitted operators $\Dpi$ and $\mc{O}$, act on the propagators and are readily obtained:
\begin{align*}
\de_X  \mean{{\rm T}_\tau \wPsi(X)_a\bar{\wPsi}(Y)_b}&= \sum_\chi \sumint_{\{P\}}\I P^+_\chi\E^{\I P^+_\chi\cdot (X-Y)}\frac{\left(\mathbb{P}_\chi G(P^+_\chi)\right)_{ab}}{{P^+_\chi}^2};\\
\de_Y  \mean{{\rm T}_\tau \wPsi(X)_a\bar{\wPsi}(Y)_b}&=\sum_\chi \sumint_{\{P\}}(-\I P^+_\chi)\E^{\I P^+_\chi\cdot (X-Y)}\frac{\left(\mathbb{P}_\chi G(P^+_\chi)\right)_{ab}}{{P^+_\chi}^2}.
\end{align*}
At this point, we take these derivatives, we send $X_1,X_2\to X$ and $Y_1,Y_2\to Y$ and we rename $Q$ into $-Q$; note that
$Q^+_{\chi'}=(q_m+\I\mu_{\chi'},\vec{q})$ goes to $-Q^-_{\chi'}=(\I\mu_{\chi'}-q_m,-\vec{q})$.
At the end, the correlator $C(X,Y)$ becomes
\begin{equation*}
C(X,Y) =  - \sum_\chi \sumint_{\{P,Q\}} \frac{\E^{\I (P^+_\chi +Q^-_{\chi'})(X-Y)}}{{P^+_\chi}^2\, {Q^-_\chi}^2}\,\,F_{\chi}(P^+,Q^-),
\end{equation*}
%
where we defined
\begin{equation*}
\begin{split}
F_{\chi}(P,Q)\equiv
\tr \left[ \mathbb{P}_\chi G(-Q_{\chi}) \Dpi_{\mu\nu}(\I Q_{\chi},\I P_\chi) G(P_\chi) {\oper(-\I P_\chi,-\I Q_{\chi})} \right].
\end{split}
\end{equation*}
The trace is to be carried out over spinorial indices by using the Euclidean $\gamma$ 
matrices properties:
\begin{align*}
\tr\left( \gamma_\mu \gamma_\nu\right)&=4\,\delta_{\mu\nu}, \\
\tr\left( \gamma_{\mu_{1}} \dots \gamma_{\mu_{2n+1}}\right)&=0, \\
\tr\left(\gamma_k\gamma_\lambda\gamma_\mu\gamma_\nu\gamma_5\right)&=4\epsilon^{k\lambda\mu\nu},\\
\tr\left( \gamma_k \gamma_\lambda \gamma_\mu \gamma_\nu\right)&=4\, \delta_{k\lambda}\delta_{\mu\nu}-4\, \delta_{k\mu}\delta_{\lambda\nu}+4\, \delta_{k\nu}\delta_{\lambda\mu}.
\end{align*}
However, all the operators $\h{O}$ that will be considered are analytic functions in the four-momentum 
$Q$ and $P$; as a result, the function $F_{\chi}$ will also be analytic in those variables. This
enables us to analytically entail the sum over the Matsubara frequencies.

Before carrying out the thermal sum, we send $Y$ to $0$, as dictated by~(\ref{eq:firstorderprototype}), 
and separate the momentum integral from the frequency sum in $\sumint$~(\ref{symbols})
\begin{equation*}
C(\tau,\vec{x})=C(X,0)=(-1)\sum_{\chi} \int \frac{\D^3 p}{(2\pi)^3}\int \frac{\D^3q}{(2\pi)^3} \E^{\I(\vec{p}+\vec{q})\cdot \vec{x}} \,S_{\chi}(\vec{p},\vec{q},\tau),
\end{equation*}
where we have defined
\begin{equation*}
S_{\chi}(\vec{p},\vec{q},\tau)\equiv \frac{1}{|\beta|}\sum_{n=-\infty}^\infty \frac{1}{|\beta|}
\sum_{m=-\infty}^\infty\frac{\E^{\I (p_n+\I\mu_\chi)\tau}}{(p_n+\I\mu_\chi)^2 +p^2}
\frac{\E^{\I (q_m -\I\mu_{\chi})\tau}}{(q_m-\I\mu_{\chi})^2 +q^2} F_{\chi}(P^+,Q^-).
\end{equation*}
As stated previously, $F_\chi$ is an analytic function, hence the sum over frequencies can be carried 
out using the formula, see Appendix~\ref{sec:thermalsums}:
\begin{equation}
\label{freqsum}
\frac{1}{|\beta|}\sum_{\{\omega_n\}} \frac{( \omega_n\pm\I\mu)^k \E^{\I (\omega_n\pm\I\mu) \tau}}{(\omega_n\pm\I\mu)^2 +\omega^2} =
\frac{1}{2\omega}\left[ (\I \omega)^k \E^{-\omega \tau}(1-n\ped{F}(\omega\mp\mu))-(-\I \omega)^k\E^{\omega \tau}n\ped{F}(\omega\pm\mu)\right],
\end{equation}
where $\omega$ could be the modulus of $p^2$ or of $q^2$ and $n\ped{F}$ is the Fermi-Dirac distribution function:
\begin{equation*}
n\ped{F}(\omega)=\frac{1}{\E^{|\beta|\omega}+1}.
\end{equation*}
Introducing the notation
\begin{equation}
\label{Ptilde}
\tilde{P}(\pm)=(\pm\I p,\vec{p}),\quad p^\pm_\chi= p\pm\mu_\chi
\end{equation}
and similarly for $\tilde{Q}(\pm)$, and taking advantage of Eq.~(\ref{freqsum}), we have
\begin{equation*}
\begin{split}
S_{\chi}(\vec{p},\vec{q},\tau) =\frac{1}{4 p q} \Bigl\{& F_{\chi}\left(\tilde{P}(+),\tilde{Q}(+) \right) \E^{-(p^-_\chi +q^+_{\chi})\tau}\big[1-n\ped{F}(p-\mu_\chi)\big]\big[1-n\ped{F}(q+\mu_{\chi})\big]+ \\
&- F_{\chi}\left(\tilde{P}(+),\tilde{Q}(-) \right) \E^{(-p^-_\chi +q^-_{\chi})\tau}\big[1-n\ped{F}(p-\mu_\chi)\big]n\ped{F}(q-\mu_{\chi})+ \\
&- F_{\chi}\left(\tilde{P}(-),\tilde{Q}(+) \right)\E^{(p^+_\chi -q^+_{\chi})\tau} n\ped{F}(p+\mu_\chi) \big[1-n\ped{F}(q+\mu_{\chi})\big]+ \\
&+F_{\chi}\left(\tilde{P}(-),\tilde{Q}(-) \right) \E^{(p^+_\chi +q^-_{\chi})\tau} n\ped{F}(p+\mu_\chi)  n\ped{F}(q-\mu_{\chi}) \Bigr\}.
\end{split}
\end{equation*}
Note that we have added and subtracted the chemical potential in the exponential in such a way 
that the corresponding energy couples with the relating argument of the Fermi distribution function 
$n\ped{F}$. This feature will be used later on.

Finally, putting together and integrating by parts, we can write the coefficient~(\ref{eq:firstorderprototype}) as:
\begin{equation*}
\begin{split}
C_{\mu\nu\alpha\dots\beta i} & =\int_0^{|\beta|}\frac{\D\tau}{|\beta|} \int\D^3 x\, x_i\, C(\tau,\vec{x}) \\
&=-\I\sum_{\chi}\int_0^{|\beta|}\frac{\D\tau}{|\beta|} \int \frac{\D^3p}{(2\pi)^3} \left[\frac{\de S_{\chi}(\vec{p},\vec{q},\tau)}{\de q^i}\right]_{\vec{q}=-\vec{p}},
\end{split}
\end{equation*}
that is an integral over time and momenta of a known function.
We can simplify this expression further with the following steps:
\begin{itemize}
	\item after the evaluation of $F_{\chi}(P,Q)$, calculate $S_{\chi}$;
	\item then, derive $S_{\chi}(\vec{p},\vec{q},\tau)$ and replace $\vec{q}\to-\vec{p}$;
	\item integrate over solid angle $\Omega$: $\D^3 p=p^2\D p\,\D\Omega$ first and over $\tau$ thereafter;
	\item express all exponential factors with the Fermi distribution function using
	\begin{equation*}
	\E^{|\beta| E}=\frac{1}{n\ped{F}(E)}-1,\quad E=p\pm\mu\ped{R,L};
	\end{equation*}
	\item express the derivative of $n\ped{F}(E)$ in terms of its powers
	\begin{align*}
	n\ped{F}'(E)&=-|\beta| n\ped{F}^2(E) \E^{|\beta| E}=-|\beta| n\ped{F}(E)+|\beta| n\ped{F}^2(E),\\
	n\ped{F}''(E)&=|\beta|^2 n\ped{F}(E)-3|\beta|^2 n\ped{F}^2(E)+2 |\beta|^2 n\ped{F}^3(E),\\
	n\ped{F}'''(E)&=6|\beta|^3 n\ped{F}^4(E)-12|\beta|^3 n\ped{F}^3(E)+7|\beta|^3 n\ped{F}^2(E)-|\beta|^3 n\ped{F}(E);
	\end{align*}
	\item after the due simplifications, express powers of $n\ped{F}(E)$ in terms of its derivative:
	\begin{align*}
	n\ped{F}^4(E)&=\frac{1}{6|\beta|^3}n\ped{F}'''(E) +2 n\ped{F}^3(E)-\frac{7}{6}n\ped{F}^2(E)+\frac{1}{6}n\ped{F}(E),\\
	n\ped{F}^3(E)&=\frac{1}{2|\beta|^2}n\ped{F}''(E)+\frac{3}{2} n\ped{F}^2(E)-\frac{1}{2}n\ped{F}(E),\\
	n\ped{F}^2(E)&=\frac{1}{|\beta|}n\ped{F}'(E)+n\ped{F}(E),
	\end{align*}
	in this way a simple expression made only of $n\ped{F}(E)$ and its derivative is obtained;
	\item finally, integrate by part $n\ped{F}^{(k)}(E)$ when possible;
	\item sum up all the pieces with different $\chi$.
\end{itemize}
This procedure always leads to expressions of the form
\begin{equation}\label{GenCoeff}
\begin{split}
C(|\beta|,\mu\ped{R},\mu\ped{L})=K\int_0^\infty P_N(p)
\Big[ &n\ped{F}(p-\mu\ped{R})+\eta\,\eta\ped{A}\, n\ped{F}(p+\mu\ped{R})\\
&+\eta\ped{A}\, n\ped{F}(p-\mu\ped{L})+\eta\, n\ped{F}(p+\mu\ped{L})\Big]\D p,
\end{split}
\end{equation}
where $P_N$ is a polynomial of $N$ degree, $\eta$ is $+1$ if the correlator $C$ is even under
charge transformation, otherwise it is $-1$, likewise  $\eta\ped{A}=\pm$ reflect parity
transformation, and $K$ is a numerical factor. The correlators in the form~(\ref{GenCoeff})
can be easily integrated once we know the polynomial $P_N$ as described in Appendix~\ref{sec:masslessmomint}.

The coefficients evaluated in~\cite{Buzzegoli:2017cqy} have either $\eta=+,\eta\ped{A}=+$ or $\eta=-,\eta\ped{A}=+$ and $\mu\ped{A}=0$, that means $\mu\ped{R}=\mu\ped{L}=\mu$ and thus have the form
\begin{equation*}
C(|\beta|,\mu)=2K\int_0^\infty P_\nu(p)
\left[ n\ped{F}^{(k)}(p-\mu)\pm\, n\ped{F}^{(k)}(p+\mu)\right]\D p.
\end{equation*}
So we can immediately obtain the dependence on chiral chemical potential for non chiral coefficients putting them in the form~(\ref{GenCoeff}) and setting
$K=\tilde{K}/2$, where $\tilde{K}$ is the numerical factor  evaluated in~\cite{Buzzegoli:2017cqy} with this same method.
The results obtained using this procedure are written in Section~\ref{sec:Results}.

\subsection{Second order correlators}
A second-order correlator of the perturbative expansion~(\ref{basicstrut}) is the mean value 
of the local operator $\h O_{\alpha\dots\beta}$ with two Lorentz generators. We will follow
similar steps as those used for the first order correlators.
We start by writing the second order generic correlator~(\ref{basicstrut}) in Euclidean space-time as
\begin{equation}
\label{eq:secondorderprototype}
C_{\mu\nu|\gamma\delta|\alpha\dots\beta| ij}=\!\int_0^{|\beta|}\! \frac{\D\tau_1}{|\beta|}\! \int_0^{|\beta|}\! \frac{\D\tau_2}{|\beta|}\! \int\D^3x\!  \int\!\D^3y\,
\mean{{\rm T}_\tau \left(\h{T}_{\mu\nu}(X)\,\h{T}_{\gamma\delta}(Y)\,\h{O}_{\alpha\dots\beta}(0)\right)}\ped{$T$,c}\, x_i y_j,
\end{equation}
where $X=(\tau_1,\vec{x})$, $Y=(\tau_2,\vec{y})$ and the connected correlator is given by
\begin{equation*}
\mean{\h{A}\,\h{B}\,\h{C}}\ped{c}=\mean{\h{A}\,\h{B}\,\h{C}} -\mean{\h{A}}\mean{\h{B}\,\h{C}}
-\mean{\h{B}}\mean{\h{A}\,\h{C}} -\mean{\h{C}}\mean{\h{A}\,\h{B}} +2\mean{\h{A}}\mean{\h{B}}\mean{\h{C}}.
\end{equation*}
Again, using the point splitting procedure and taking advantage of the Wick's theorem, we can 
split the mean value of~(\ref{eq:secondorderprototype}) in two parts
\begin{equation}
\label{threepointfunc}
C(X,Y,Z)\equiv\mean{{\rm T}_\tau \left(\h{T}_{\mu\nu}(X)\,\h{T}_{\gamma\delta}(Y)\,\h{O}_{\alpha\dots\beta}(Z)\right)}\ped{$T$,c}=C_1+C_2,
\end{equation}
where $Z=(\tau_2,\vec{z})$ and
\begin{align*}
C_1=&\lim_{\substack{X_1 X_2\to X\\ Y_1 Y_2\to Y \\Z_1 Z_2\to Z}} (-1) \tr\left[\Dpi_{\mu\nu}(\de_{X_1},\de_{X_2})
\mean{{\rm T}_\tau \wPsi(X_2) \bar{\wPsi}(Z_1)}\oper(\de_{Z_1},\de_{Z_2})\times\right. \\
&\hphantom{\lim_{\substack{X_1 X_2\to X\\ Y_1 Y_2\to Y \\Z_1 Z_2\to Z}} (-1) \tr\Big[}\left.\times \mean{{\rm T}_\tau \wPsi(Z_2) \bar{\wPsi}(Y_1)} \Dpi_{\gamma\delta}(\de_{Y_1},\de_{Y_2})\mean{{\rm T}_\tau \wPsi(Y_2) \bar{\wPsi}(X_1)} \right],\\
C_2=&\lim_{\substack{X_1 X_2\to X\\ Y_1 Y_2\to Y \\Z_1 Z_2\to Z}}(-1) \tr\left[\Dpi_{\mu\nu}(\de_{X_1},\de_{X_2})
\mean{{\rm T}_\tau \wPsi(X_2) \bar{\wPsi}(Y_1)}\Dpi_{\gamma\delta}(\de_{Y_1},\de_{Y_2})\times\right. \\
&\hphantom{\lim_{\substack{X_1 X_2\to X\\ Y_1 Y_2\to Y \\Z_1 Z_2\to Z}} (-1) \tr\Big[}\left.\times \mean{{\rm T}_\tau \wPsi(Y_2) \bar{\wPsi}(Z_1)}\oper(\de_{Z_1},\de_{Z_2})\mean{{\rm T}_\tau \wPsi(Z_2) \bar{\wPsi}(X_1)} \right].
\end{align*}
Following similar procedure of first order correlators, we express the two point function in Fourier space~(\ref{eq:Prop})
and we can show that $C_1$ and $C_2$ are equal to 
\begin{equation*}
\begin{split}
C_1&= (-1)\sum_{\chi}\sumint_{\{P,Q,K\}}
\frac{\E^{\I (P^+_{\chi} +K^-_{\chi})X} \E^{\I (Q^-_{\chi} -K^-_{\chi})Y} \E^{-\I (P^+_{\chi} +Q^-_{\chi})Z}}{{P^+_{\chi}}^2 {Q^-_{\chi}}^2 {K^-_{\chi}}^2}\,
F_{1\,\chi}(P,Q,K),\\
C_2&=(-1)\sum_{\chi}\sumint_{\{P,Q,K\}}
\frac{\E^{\I (P^-_{\chi} +K^+_{\chi})X} \E^{\I (Q^+_{\chi} -K^+_{\chi})Y} \E^{-\I (P^-_{\chi} +Q^+_{\chi})Z}}{{P^-_{\chi}}^2{Q^+_{\chi}}^2{K^+_{\chi}}^2}
F_{2\,\chi}(P,Q,K),
\end{split}
\end{equation*}
where we denoted the momenta as $P^+_\chi= (p_n+\I\mu_\chi,\vec{p})$, $Q^+_{\chi}=(q_m+\I\mu_{\chi},\vec{q})$,\\ \mbox{$K^+_{\chi}=(k_l+\I\mu_{\chi},\vec{k})$},
and we defined the trace functions as:
\begin{equation*}
\begin{split}
F_{1\,\chi}(P,Q,K) =&\tr \Big[ \mathbb{P}_{\chi} G(-Q^-_{\chi})\Dpi_{\gamma\delta}(\I Q^-_{\chi} ,-\I K^-_{\chi})
G(-K^-_{\chi})\Dpi_{\mu\nu}(\I K^-_{\chi} ,\I P^+_{\chi})\times \\
&\hphantom{\tr \Big[}\times G(P^+_{\chi}) \oper(-\I P^+_{\chi},-\I Q^-_{\chi}) \Big],\\
F_{2\,\chi}(P,Q,K) =&\tr \Big[ \mathbb{P}_{\chi} G(-P^-_{\chi}) \Dpi_{\mu\nu}(\I P^-_{\chi},\I K^+_{\chi})
G(K^+_{\chi}) \Dpi_{\gamma\delta}(-\I K^+_{\chi},\I Q^+_{\chi})\times \\
&\hphantom{\tr \Big[}\times  G(Q^+_{\chi}) \oper(-\I Q^+_{\chi},-\I P^-_{\chi}) \Big].
\end{split}
\end{equation*}
Once the form of $\mc{O}$ is explicitly given, these two trace functions are computed thanks to the well known gamma matrix traces.
In general, we can say that as far as $\mc{O}$ is an analytic function of the derivative operator $\de_X$, the two
trace functions $F_1$ and $F_2$ are analytic functions on the four-momenta $P$, $Q$ and $K$.

Now, we send $Z$ to $0$ in Eq.~(\ref{threepointfunc}), as required by~(\ref{eq:secondorderprototype}),
and we separate the momenta integrals from the frequencies sums
\begin{equation}
\label{eq:C}
C(\tau_1,\tau_2,\vec{x},\vec{y})\equiv C(X,Y,0)=  -\sum_{\chi}\int \frac{\D^3p\, \D^3q\, \D^3k}{(2\pi)^9}\,
\E^{\I(\vec{p}+\vec{k})\cdot \vec{x}} \E^{\I(\vec{q}-\vec{k})\cdot \vec{y}} \, S_{\chi}(\vec{p},\vec{q},\vec{k},\tau_1,\tau_2),
\end{equation}
where we have defined
\begin{equation*}
\begin{split}
S_{\chi}(\vec{p},\vec{q},\vec{k},\tau_1,\tau_2)\! \equiv& \frac{1}{|\beta|^3}\!\!\!\sum_{n,m,l=-\infty}^\infty\!\!\Bigg[
\frac{\E^{\I (p_n+\I\mu_\chi) \tau_1+\I (q_m-\I\mu_{\chi}) \tau_2+\I (k_l-\I\mu_{\chi}) (\tau_1 -\tau_2)} F_{1\,\chi}(P^+,Q^-,K^-)}{[(p_n+\I\mu_\chi)^2 +p^2][(q_m-\I\mu_{\chi})^2 +q^2][(k_l-\I\mu_{\chi})^2 +k^2]}\\
&+\frac{\E^{\I (p_n-\I\mu_\chi) \tau_1+\I (q_m+\I\mu_{\chi}) \tau_2+\I (k_l+\I\mu_{\chi}) (\tau_1 -\tau_2)}F_{2\,\chi}(P^-,Q^+,K^+)}{[(p_n-\I\mu_\chi)^2 +p^2][(q_m+\I\mu_{\chi})^2 +q^2][(k_l+\I\mu_{\chi})^2 +k^2]}\Bigg].
\end{split}
\end{equation*}
Since $\tau_1$ and $\tau_2$ are always larger than zero in the integration~(\ref{eq:secondorderprototype}), the sums over $p_n$ and $q_m$
are made using~(\ref{freqsum}). For the $k_l$ sum, the result is still~(\ref{freqsum}) when $\tau=\tau_1-\tau_2>0$,
instead when $\tau=\tau_1-\tau_2<0$ the sum yields
\begin{equation*}
\frac{1}{|\beta|}\sum_{\{\omega_n\}} \frac{( \omega_n\pm\I\mu)^k \E^{\I (\omega_n\pm\I\mu) \tau}}{(\omega_n\pm\I\mu)^2 +\omega^2} =
\frac{1}{2\omega}\left[ (-\I \omega)^k\E^{\omega \tau}(1-n\ped{F}(\omega\pm\mu)) - (\I \omega)^k \E^{-\omega \tau}n\ped{F}(\omega\mp\mu)\right].
\end{equation*}
Reminding the definition~(\ref{Ptilde}) for $\tilde{P}(\pm)$, $\tilde{Q}(\pm)$, $\tilde{K}(\pm)$ and for $p^\pm_\chi$, $q^\pm_\chi$ and $k^\pm_\chi$,
after all the sums, we find
\begin{equation*}
\begin{split}
S_{\chi}(\vec{p},\vec{q},\vec{k},\tau_1,\tau_2) =\frac{1}{8\, p\, q\, k} \sum_{s_1,s_2,s_3=\pm}
\Bigl[& F_{1\,\chi}\left(\tilde{P}(s_1),\tilde{Q}(s_2),\tilde{K}(s_3)\right)S_{1,\chi}\\
&+ F_{2\,\chi}\left(\tilde{P}(s_1),\tilde{Q}(s_2),\tilde{K}(s_3)\right)\,S_{2\,\chi}\Bigr]
\end{split}
\end{equation*}
with
\begin{align*}
S_{1,\chi}\equiv& \exp\big[\!-s_1 p^{-s_1}_\chi\tau_1 \!-s_2 q^{s_2}_\chi\tau_2 \!-s_3 k^{s_3}_\chi(\tau_1\!-\tau_2)\big]
\left[\theta(s_1)\!-n\ped{F}(p^{-s_1}_\chi) \right]\left[\theta(s_2)\!-n\ped{F}(q^{s_2}_\chi) \right]\times \\
&\times \left[\left(\theta(s_3)\!-n\ped{F}(k^{s_3}_\chi)\right)\theta(\tau_1 \!- \tau_2)+\left(\theta(-s_3)\!-n\ped{F}(k^{s_3}_\chi)\right)\theta(\tau_2 \!- \tau_1)\right], \\
S_{2,\chi}\equiv&  \exp\big[\!-s_1 p^{s_1}_\chi\tau_1 \!-s_2 q^{-s_2}_\chi\tau_2 \!-s_3 k^{-s_3}_\chi(\tau_1\!-\tau_2)\big]
\left[\theta(s_1)\!-n\ped{F}(p^{s_1}_\chi) \right]\left[\theta(s_2)\!-n\ped{F}(q^{-s_2}_\chi) \right]\times \\
&\times \left[\left(\theta(s_3)\!-n\ped{F}(k^{-s_3}_\chi)\right)\theta(\tau_1 \!- \tau_2)+\left(\theta(-s_3)\!-n\ped{F}(k^{-s_3}_\chi)\right)\theta(\tau_2 \!- \tau_1)\right],
\end{align*}
where $\theta$ is the Heaviside theta function.
Now, we can take advantage of the formula
\begin{equation*}
\begin{split}
\int \D^3 x \int\D^3 y \,\, \E^{\I ({\bm p}+{\bm k} )\cdot {\bm x}}\E^{\I({\bm q}-{\bm k} )\cdot {\bm x} } 
x^i y^j=-(2\pi)^6 \frac{\de^2}{\partial k_i\de q_j}\delta^{(3)}(\bm p +\bm k)\delta^{(3)}(\bm q -\bm k)
\end{split}
\end{equation*}
to integrate over the coordinates $x$ and $y$ in the Eq.~(\ref{eq:secondorderprototype}) and, using Eq.~(\ref{eq:C}), we obtain:
\begin{equation*}
C_{\mu\nu|\gamma\delta|\alpha\dots\beta| ij}= \sum_{\chi}
\int_0^{|\beta|}\frac{\D\tau_1}{|\beta|} \int_0^{|\beta|}\frac{\D\tau_2}{|\beta|} \int \frac{\D^3p}{(2\pi)^3} \left[ \frac{\de^2 S_{\chi}(\vec{p},\vec{q},\vec{k},\tau_1,\tau_2)}{\de p^i \de q^j} \right]_{\vec{q}=-\vec{p},\vec{k}=-\vec{p}}.
\end{equation*}
From this point, we can adopt the same algorithm listed in the previous section, with an additional time integration,
to obtain simple expressions of the form~(\ref{GenCoeff}).

\subsection{Results}\label{sec:Results}
Here we report the values for thermodynamic coefficients of the stress-energy tensor, axial and electric
current and spin tensor for a free Dirac field obtained with the methods described in the previous sections.
First, we consider both the canonical and symmetric stress-energy tensor. In the Euclidean space-time they are given by
\begin{equation*}
\begin{split}
\h T_{\mu\nu}\apic{Can}(X)&=\frac{\I^{\delta_{0\mu}+\delta_{0\nu}}}{2}
\bar{\wPsi}(X)\left[\gamma_\mu\oraw{\de}_\nu-\gamma_\mu\olaw{\de}_\nu\right]\wPsi(X),\\
\h T_{\mu\nu}\apic{Sym}(X)&=\frac{\I^{\delta_{0\mu}+\delta_{0\nu}}}{4}
\bar{\wPsi}(X)\left[\gamma_\mu\oraw{\de}_\nu-\gamma_\mu\olaw{\de}_\nu+\gamma_\nu\oraw{\de}_\mu-\gamma_\nu\olaw{\de}_\mu\right]\wPsi(X)
\end{split}
\end{equation*}
and they give different results for $\mathbb{A},\,\mathbb{W}_1,\,\mathbb{W}_2,\,G_1,\,G_2$, i.e. the coefficients related to the non symmetric part of the expansion
\begin{equation*}
\begin{split}
\mean{\wT^{\mu\nu}}=&\,\mathbb{A}\,\epsilon^{\mu\nu\kappa\lambda}\alpha_\kappa u_\lambda+\mathbb{W}_1 w^\mu u^\nu +\mathbb{W}_2 w^\nu u^\mu\\
&+(\rho-\alpha^2 U_\alpha -w^2 U_w)u^\mu u^\nu -(p-\alpha^2D_\alpha-w^2D_w)\Delta^{\mu\nu}\\
&+A\,\alpha^\mu\alpha^\nu+Ww^\mu w^\nu+G_1 u^\mu\gamma^\nu+G_2 u^\nu\gamma^\mu+\mathcal{O}(\varpi^3).
\end{split}
\end{equation*}
The non chiral coefficients values are obtained from~(\ref{coefficientiC}) and are
\begin{equation*}
\begin{split}
\rho&= 3 p=\frac{30 \pi^2 \left(\zeta^2+\zeta\ped{A}^2\right)+15 \left(\zeta^4+6 \zeta^2 \zeta\ped{A}^2+\zeta\ped{A}^4\right)+7 \pi^4}{60 \pi^2 |\beta|^4},\\
U_\alpha&= 3 D_\alpha=\frac{3 \left(\zeta^2+\zeta\ped{A}^2\right)+\pi^2}{24 \pi^2 |\beta|^4},\quad
U_w= 3 D_w=\frac{3 \left(\zeta^2+\zeta\ped{A}^2\right)+\pi^2}{8 \pi^2 |\beta|^4},\\
A&=W=0,\quad G_1\apic{Sym}= G_2\apic{Sym}=\frac{3 \left(\zeta^2+\zeta\ped{A}^2\right)+\pi^2}{18 \pi^2 |\beta|^4},\\
G_1\apic{Can} &=\frac{2 \left[3 \left(\zeta^2+\zeta\ped{A}^2\right)+\pi^2\right]}{9 \pi^2 |\beta|^4},\qquad
G_2\apic{Can} = -\frac{3 \left(\zeta^2+\zeta\ped{A}^2\right)+\pi^2}{9 \pi^2 |\beta|^4}.
\end{split}
\end{equation*}
These values indeed satisfy the relations~(\ref{eq:setrel}) required by stress-energy conservation.
All non-chiral coefficients reported in this section are also listed in Table~\ref{tab:EvenParity}.
%
\begin{table}[h!bt]
	\small
	\caption{The non-chiral thermodynamic coefficients up to second order in thermal vorticity
		of the stress-energy tensor, electric and axial current for a free massless Dirac field (see 
		Eqs.~(\ref{setdecomp}),(\ref{vcurrdecomp}),(\ref{acurrdecomp}) for definitions).
		Here we use $T=1/|\beta|$, $\mu=\zeta\, T$ and $\mu\ped{A}=\zeta\ped{A}\, T$. }
	\label{tab:EvenParity}
	\newcommand\bstrut{\vphantom{\displaystyle\sum^{\Lambda}}}
	\[
	\begin{array}{||c|c||c|c||}
	\hline
	\rho=3p & \bstrut\frac{7 \pi^2 T^4}{60}+\frac{(\mu^2+\mu\ped{A}^2) T^2}{2}+\frac{3 \mu^2 \mu\ped{A}^2}{2 \pi^2}+\frac{\mu^4+\mu\ped{A}^4}{4 \pi^2} &    A,\,W & 0\\
	U_\alpha= 3 D_\alpha & \frac{T^4}{24}+\frac{(\mu^2+\mu\ped{A}^2) T^2}{8 \pi^2} &    n\ped{c} &     \frac{\mu^3}{3 \pi^2}+\frac{\mu\,  \mu\ped{A}^2}{\pi^2}+\frac{\mu\,  T^2}{3}\\
	U_w = 3 D_w & \bstrut\frac{T^4}{8}+\frac{3 (\mu^2+\mu\ped{A}^2) T^2}{8 \pi^2} & N\apic{V}_\alpha & \frac{\mu\,  T^2}{4 \pi^2}\\
	G_1\apic{Sym}= G_2\apic{Sym} & \frac{T^4}{18}+\frac{(\mu^2+\mu\ped{A}^2) T^2}{6 \pi^2} &    N\apic{V}_w & \bstrut\frac{\mu\,  T^2}{4 \pi^2}\\
	G_1\apic{Can} & \bstrut\frac{2 T^4}{9}+\frac{2 (\mu^2+\mu\ped{A}^2) T^2}{3 \pi^2} & G\apic{V} & \frac{\mu\,  T^2}{6 \pi^2}\\
	G_2\apic{Can} & -\frac{T^4}{9}-\frac{(\mu^2+\mu\ped{A}^2) T^2}{3 \pi^2} & W\apic{A}=2\Gamma_w & \bstrut\frac{T^3}{6}+\frac{(\mu^2+\mu\ped{A}^2) T}{2 \pi^2}\\
	\hline
	\end{array}
	\]
\end{table}
%
Instead, the chiral coefficients are given by Eq.s~(\ref{eq:chiralsetcoeff}). They can be calculated
using the methods described above starting from Eq.s~(\ref{eq:chiralsetcoeff2}); we obtain
\begin{equation}
\begin{split}\label{setchiralresults}
\mathbb{A}\apic{Sym} &= 0,\qquad\hphantom{+\zeta\ped{A}^2,\,\quad\mathbb{W}_2\apic{Can}= } \mathbb{A}\apic{Can}=\frac{\zeta\ped{A}\left(\pi^2+3\zeta^2+\zeta\ped{A}^2\right)}{6\pi^2|\beta|^4},\\
\mathbb{W}_1\apic{Sym}&=\mathbb{W}_2\apic{Sym}=\frac{\zeta\ped{A}\left(\pi^2+3\zeta^2+\zeta\ped{A}^2\right)}{3\pi^2|\beta|^4},\\
\mathbb{W}_1\apic{Can} &= \frac{\zeta\ped{A}\left(\pi^2+3\zeta^2+\zeta\ped{A}^2\right)}{2\pi^2|\beta|^4},\quad
\mathbb{W}_2\apic{Can}=  \frac{\zeta\ped{A}\left(\pi^2+3\zeta^2+\zeta\ped{A}^2\right)}{6\pi^2|\beta|^4}.\\
\end{split}
\end{equation}
We can readily check that the coefficients~(\ref{setchiralresults}) satisfy the relation~(\ref{setchiralrel}).
All chiral coefficients reported in this section are also listed in Table~\ref{tab:OddParity}.
%
\begin{table}[tb]
	\small
	\caption{The chiral thermodynamic coefficients up to second order in thermal vorticity
		of the stress-energy tensor, electric and axial current for a free massless Dirac field 
		(see Eqs.~(\ref{setdecomp}),(\ref{vcurrdecomp}),(\ref{acurrdecomp}) for definitions).
		Here we use $T=1/|\beta|$, $\mu=\zeta\, T$ and $\mu\ped{A}=\zeta\ped{A}\, T$.}
	\label{tab:OddParity}
	\newcommand\bstrut{\vphantom{\displaystyle\sum^{\Lambda}}}
	\[
	\begin{array}{||c|c||c|c||}
	\hline
	\mathbb{A}\apic{Sym} & \bstrut 0 & W\apic{V} & \frac{\mu\, \mu\,\ped{A} T}{\pi^2 }\\
	\mathbb{A}\apic{Can} & \bstrut\frac{\mu\ped{A}\,T^3}{6\pi^2}+\frac{\mu\ped{A}\,\mu^2\, T}{2\pi^2}+\frac{\mu\ped{A}^3\,T}{6\pi^2} &    \,\,n\ped{A}=2\mathbb{S}\,\, & \frac{\mu\ped{A}\,T^2}{3\pi^2}+\frac{\mu\ped{A}\,\mu^2}{\pi^2}+\frac{\mu\ped{A}^3}{3\pi^2}\\
	\mathbb{W}_1\apic{Sym}=\mathbb{W}_2\apic{Sym} &    \bstrut\frac{\mu\ped{A}\,T^3}{3\pi^2}+\frac{\mu\ped{A}\,\mu^2\, T}{\pi^2}+\frac{\mu\ped{A}^3\,T}{3\pi^2} &    N\apic{A}_\alpha & \frac{\mu\ped{A} T^2}{4 \pi^2 }\\
	\mathbb{W}_1\apic{Can} & \bstrut\frac{\mu\ped{A}\,T^3}{2\pi^2}+\frac{3\mu\ped{A}\,\mu^2\, T}{2\pi^2}+\frac{\mu\ped{A}^3\,T}{2\pi^2} & N\apic{A}_w & \frac{\mu\ped{A} T^2}{4 \pi^2 }\\
	\mathbb{W}_2\apic{Can} & \bstrut\frac{\mu\ped{A}\,T^3}{6\pi^2}+\frac{\mu\ped{A}\,\mu^2\, T}{2\pi^2}+\frac{\mu\ped{A}^3\,T}{6\pi^2} & G\apic{A} & \frac{\mu\ped{A} T^2 }{6 \pi^2}\\
	\hline
	\end{array}
	\]
\end{table}
To our knowledge, the coefficient $\mathbb{W}_2$ was obtained first by~\cite{Vilenkin:1979ui} as 
an energy flux resulting from neutrinos emitted by a rotating black hole; indeed, our result 
of $\mathbb{W}_2\apic{Can}$ for a free Dirac field~(\ref{setchiralresults}) perfectly agrees 
with the one in~\cite{Vilenkin:1979ui}(there is a factor $\frac{1}{2}$ of a difference
because Vilenkin considered only left-handed neutrinos). Lately, the same result for this 
coefficient was obtained by~\cite{Landsteiner:2011iq,Landsteiner:2012kd} using holographic 
techniques, by~\cite{Chen:2015gta,Hidaka:2017auj,Abbasi:2017tea} in chiral kinetic theory
and in~\cite{Landsteiner:2012kd,Chowdhury:2015pba} evaluating Kubo formulae in finite temperature 
field theory.

It is also worth noting that thermodynamic equilibria with vorticity imply different mean
values for the canonical and symmetric stress-energy tensor~\cite{BecTin2011,BecTin2012}.
This is seen here for the coefficients $\mathbb{A},\,\mathbb{W}_1,\,\mathbb{W}_2,\,G_1$ and $G_2$.
Particularly, the coefficient $\mathbb{A}$ vanishes if the stress-energy tensor is symmetric
but not for the canonical one. 

The Euclidean version of the Dirac field electric vector current $\wj\ped{V}^\mu=\bar\wPsi\gamma^\mu\wPsi$ is
\begin{equation*}
\wj\ped{V}^\mu=(-\I)^{1-\delta_{0\mu}}\bar\wPsi \gamma_\mu\wPsi.
\end{equation*}
The non chiral coefficients of the decomposition
\begin{equation*}
\mean{\wj\ped{V}^\mu}=n\ped{V}\,u^\mu+\left(\alpha^2 N\apic{V}_\alpha+w^2 N\apic{V}_\omega\right)u^\mu+W\apic{V}w^\mu+G\apic{V}\gamma^\mu
+\mathcal{O}(\varpi^3)
\end{equation*}
are obtained using the Eq.s~(\ref{eq:vcurrcoeff2})
\begin{equation*}
\begin{aligned}
n\ped{V} &= \frac{\zeta  \left(\zeta^2+3 \zeta\ped{A}^2+\pi^2\right)}{3 \pi^2 |\beta|^3},&
N\apic{V}_\alpha &= \frac{\zeta }{4 \pi^2 |\beta|^3},\\
N\apic{V}_w &= \frac{\zeta }{4 \pi^2 |\beta|^3},&
G\apic{V} &= \frac{\zeta }{6 \pi^2 |\beta|^3}.
\end{aligned}
\end{equation*}
While for the CVE conductivity $W\apic{V}$~(\ref{CVEcoeff}), evaluating the correlators in Eq.~(\ref{CVEcoeff2}), we obtain the well-known result
\begin{equation}\label{CVE}
W\apic{V} = \frac{\zeta \zeta\ped{A}}{\pi^2 |\beta|^3}.
\end{equation}
The same is done for the axial current
\begin{equation*}
\wj\ped{A}^\mu=(-\I)^{1-\delta_{0\mu}}\bar\wPsi \gamma_\mu\gamma_5\wPsi,
\end{equation*}
which has similar expansion on thermal vorticity
\begin{equation*}
\mean{\wj\ped{A}^\mu}=n\ped{A}\,u^\mu+\left(\alpha^2 N_\alpha\apic{A}+w^2 N_\omega\apic{A}\right)u^\mu+W\apic{A}w^\mu+G\apic{A}\gamma^\mu
+\mathcal{O}(\varpi^3).
\end{equation*}
The only non chiral coefficient is the axial vortical effect conductivity $W\apic{A}$, for which we found~\cite{Buzzegoli:2017cqy}
\begin{equation}\label{AVE}
W\apic{A}=\frac{3 \left(\zeta^2+\zeta\ped{A}^2\right)+\pi^2}{6 \pi^2 |\beta|^3}.
\end{equation}
The chiral coefficients are again evaluated accounting for the relations~(\ref{acurrcoeff}) and~(\ref{acurrcoeff2}) 
\begin{equation*}
\begin{aligned}
n\ped{A}&= \frac{\zeta\ped{A}(\pi^2+3\zeta^2+\zeta\ped{A}^2)}{3\pi^2|\beta|^3}, &
N_\alpha\apic{A}&=\frac{\zeta\ped{A} }{4 \pi^2 |\beta|^3},\\
N_\omega\apic{A}&=\frac{\zeta\ped{A} }{4 \pi^2 |\beta|^3}, &
G\apic{A}&=\frac{\zeta\ped{A} }{6 \pi^2 |\beta|^3}.
\end{aligned}
\end{equation*}
As expected for dimensional analysis, the coefficients $W\apic{V}$~(\ref{CVE}) and $W\apic{A}$~(\ref{AVE}) respectively
fulfill the relations~(\ref{CVERel}) and~(\ref{AVERel}). In the case of massive Dirac fields, as discussed in Sec.~\ref{sec:CurrentsDec},
global thermal equilibrium with thermal vorticity and vanishing axial chemical potential is well defined and the axial currents
mean value can be directed along the rotation of the fluid. In that situation the AVE conductivity for a free massive Dirac field
is~\cite{Buzzegoli:2017cqy}
\begin{equation}\label{coeffw2}
W^A=\frac{1}{2 \pi^2 |\beta| }\int_0^\infty \,\di p \left[n_F(E_p-\mu)+n_F(E_p+\mu)\right](2p^2+m^2),
\end{equation}
where $E_p=\sqrt{p^2+m^2}$. This coefficients is related to pseudo-scalar thermal coefficient $L^{\alpha\cdot w}$ via Eq.~(\ref{AVERelMass})
and indeed is given by
\begin{equation}
\label{eq:PseudoL}
L^{\alpha\cdot w}=-\frac{m}{4 \pi^2 \beta^2 } \int_0^\infty \frac{\di p}{E_p} \left(n^{\prime}_F(E_p-\mu)+n^\prime_F(E_p+\mu)\right),
\end{equation}
where the prime on distribution functions stands for derivative respect to $E_p$.
We can give approximate results for integral in Eq.~(\ref{coeffw2}). For high temperature regime $T\gg m$, if the
gas in non-degenerate $|\mu|<m$, we extract the AVE conductivity behavior using the Mellin transformation technique~\cite{LandWeer}.
The result is
\begin{equation}
\label{eq:WAmassHighT}
\frac{W\apic{A}}{T}\simeq\frac{T^2}{6}+\frac{\mu^2}{2\pi^2}-\frac{m^2}{4\pi^2}-\frac{7\zeta'(-2)T^2}{8\pi^2}\left(\frac{m}{T}\right)^4+\mc{O}\left(\frac{m^6}{T^6}\right).
\end{equation}
The first term in mass was also obtained in~\cite{Flachi:2017vlp} where they evaluated axial vortical effect with
our same statistical operator in curved space-time. Low temperature behavior can also be extracted from~(\ref{coeffw2}),
see~\cite{Buzzegoli:2017cqy}. For a degenerate gas ($|\mu|>m$) at zero temperature we obtain\footnote{Notice that $W\apic{A}w^\mu\to(W\apic{A}/T)\vec{\omega}$,
	so there is no divergence for $T\to 0$.}:
\begin{equation*}
\frac{W\apic{A}}{T}=\frac{\mu^2}{2\pi^2}\frac{\sqrt{\mu^2-m^2}}{\mu}.
\end{equation*}
Instead for non degenerate gas ($|\mu|<m$) at low temperature $T<<m$ we have
\begin{equation}
\label{eq:WAmassLowT}
\frac{W\apic{A}}{T}\approx\frac{m\sqrt{m T}}{12\sqrt{2}\,\pi^{3/2}}\E^{|\beta|(\mu-m)}.
\end{equation}
Axial current corrections for rotating and accelerating fluids is also discussed in~\cite{Prokhorov:2017atp,Prokhorov:2018qhq}
for both massive and massless fields using an ansatz for Wigner function with thermal vorticity.

For the spin tensor decomposition
\begin{equation*}
\mathcal{S}^{\lambda,\mu\nu}=\mathbb{S}\,\epsilon^{\lambda\mu\nu\rho}u_\rho+\Gamma_w\, \left( u^\lambda\varpi^{\mu\nu}+u^\nu\varpi^{\lambda\mu}+u^\mu\varpi^{\nu\lambda}\right),
\end{equation*}
after calculation we find  for a massless field
\begin{equation*}
\mathbb{S}=\frac{\zeta\ped{A}(\pi^2+3\zeta^2+\zeta\ped{A}^2)}{6\pi^2|\beta|^3},\quad
\Gamma_w=\frac{3 \left(\zeta^2+\zeta\ped{A}^2\right)+\pi^2}{12 \pi^2 |\beta|^3},
\end{equation*}
which satisfy the relations in Eq.s~(\ref{eq:SpinTensAndAxialRel}) and~(\ref{eq:SpinTensAndSETRel}).

It is worth noting that the exact thermal expectation values of stress-energy tensor, electric and axial
current for a non-chiral fluid of free massless fermions under the effect of rotation has been found using finite
temperature field theory~\cite{Ambrus:2014uqa}. All the thermal coefficients related to rotation reported
above in this section agree with the corresponding term evaluated in~\cite{Ambrus:2014uqa}.

\section{Axial vortical effect origin}
\begin{figure}[htbp]
	\centering
	\includegraphics[width=\textwidth]{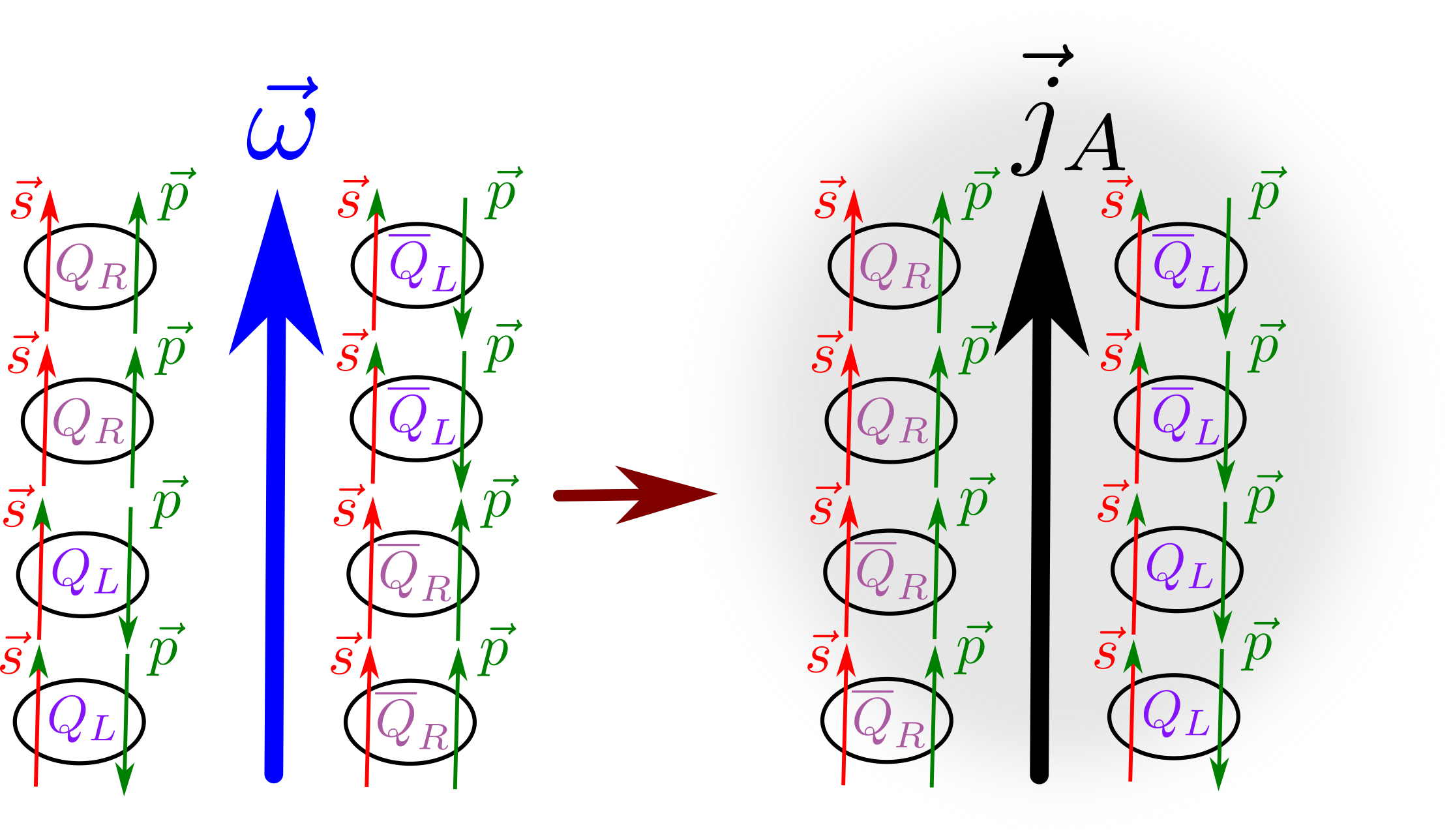}
	\caption{ Illustration of the Axial Vortical Effect. To be specific, the illustration is for
		just one kind of massless quarks with electric charge $Q$ and for the case of
		$\mu\ped{A}=\mu=0$. Even in the case of axial or electric imbalance, the axial current is
		aligned along the vorticity no matter what the signs of chemical potentials are.}
	\label{fig:AVECartoon}
\end{figure}
The Axial Vortical Effect (AVE) states that in rotating fermion gas, an axial current is induced
along rotation and consequently, the right and left chiral fermions get separated. This effect can be
illustrated with a simple argument, which is strictly valid only for massless particles.
In a rotating system, the fermion spin tends to align with the direction of rotation $\vec\omega$
independently of their charge~\cite{Becattini:2007nd}, in Fig.~\ref{fig:AVECartoon} for the sake of clarity
all spin are aligned along rotation. The right-handed particles must have their momentum aligned with the spin, consequently will
move in the direction of the spin, i.e. as shown in the right panel of Fig.~\ref{fig:AVECartoon} we get a net
right-handed particles flow in the direction of $\vec\omega$. On the other hand, the left-handed particles
will move in the opposite direction, giving a net left-handed particles flow opposite to $\vec\omega$.
Together these flows give an axial current $\vec{j}\ped{A}=n\ped{R}\vec{v}\ped{R}-n\ped{L}\vec{v}\ped{L}\propto \vec{\omega}$.
A chiral or electric imbalance is not needed to reproduce the effect.

The very reason of the non-vanishing axial current is simply that it is allowed by the symmetry of the
statistical operator, as recognized also in~\cite{Son:2009tf}. Indeed the current $\h{\vec{j}}\ped{A}$
has the same properties of the angular momentum operator $\h{\vec{J}}$ under reflection, time reversal
and charge conjugation; therefore its mean value is non vanishing
\begin{equation*}
\vec{j}\ped{A}=\frac{1}{\parz}\tr\left[\vec{\h{j}}\ped{A}\,\exp\left\{-\h{H}/T_0+\vec{\omega}\cdot\vec{\h{J}}/T_0 \right\} \right]\neq 0.
\end{equation*}
Notice that this symmetry is classical; rotational symmetry is broken by fluid rotation not by quantum processes.
For a free field, this procedure leads to the values of axial current reported in the previous section.

The AVE was originally addressed for neutrinos emitted by rotating black hole in
ref.~\cite{Vilenkin:1979ui}. Lately, this effect was addressed in the context of Chiral Vortical Effect (CVE)
and quantum anomalies~\cite{Son:2004tq,Son:2009tf}. Several terms were found to contribute to the proportionality
coefficient between $\vec{j}\ped{A}$ and $\vec\omega$, in the massless limit: a term proportional to the
chiral potential $\mu\ped{A}^2$ , also confirmed in holographic models~\cite{Torabian:2009qk,Erdmenger:2008rm,Banerjee:2008th,Amado:2011zx};
a term proportional to $T^2$~\cite{Neiman:2010zi,Landsteiner:2011iq,Gao:2012ix} whose existence was
attributed to the gravitational anomaly ~\cite{Landsteiner:2011cp} and to the modular anomaly~\cite{Golkar:2015oxw}.
It is also worth pointing out that in fact, all calculations of the $T^2$ term in Eq.~(\ref{AVE}) in
refs.~\cite{Vilenkin:1979ui,Landsteiner:2011iq,Gao:2012ix} give the same result, i.e. $T^2 /6$ like in Eq.~(\ref{AVE}).
We would also like to point out that in the derivation of Vilenkin~\cite{Vilenkin:1979ui} it is
clear that the effect is caused by the modified density operator in the presence of rotation,
exactly like in our case, and indeed we recover the same result in Eq.~(\ref{AVE}).

We conclude this chapter motivating why the Axial Vortical Effect (AVE) is not to be ascribed to quantum anomalies,
specifically chiral anomaly and gravitational anomaly. We understand that this statement goes against the conventional
interpretation in literature, and yet some other authors have lately cast doubts about the anomalous origin of
the AVE~\cite{Kalaydzhyan:2014bfa,Flachi:2017vlp}.

Regarding chirality, we have called the AVE coefficient non-chiral because the corresponding correlator has even parity.
As a consequence AVE does not vanish for $\zeta\ped{A}=\mu\ped{A}/T = 0$, as it is apparent from Eq.~(\ref{AVE}).
This means that the AVE does not need chiral imbalance to show up and persists for a perfectly chirally balanced system.
This observation suggests that the AVE is not entirely dictated by the chiral anomaly.
Furthermore, the connection with the chiral anomaly, namely the numerical factors in front of the chemical potentials in Eq.~(\ref{AVE}),
is completely lost for a massive field. Indeed, AVE conductivity for massive filed was first found in~\cite{Buzzegoli:2017cqy,Flachi:2017vlp}
and then in~\cite{Prokhorov:2018qhq,Lin:2018aon,Hattori:2019ahi}. The chiral anomaly factor is recovered at high temperature,
see Eq.~(\ref{eq:WAmassHighT}), but it is lost at low temperature, see Eq.~(\ref{eq:WAmassLowT}). Mass dependence also occurs
similarly for the Chiral Separation Effect (CSE) coefficient~\cite{Metlitski:2005pr}, which is axial current induced
by a magnetic field.

The temperature part of AVE is attributed to the gravitational anomaly or to be more specific to the mixed axial-gravitational
anomaly, which is the term depending on curvature in chiral anomaly, see Eq.~(\ref{eq:ChiralAnomaly}).
Again, this connection, at least in the form presented in~\cite{Landsteiner:2011cp}, is lost when we consider the AVE
for a massive field, see for instance Eq.~(\ref{eq:WAmassLowT}). Even in the case of a massless field, another evidence
against the anomalous origin of AVE temperature term is that it receives radiative corrections by gauge interactions such
as those of QED and QCD, as showed in~\cite{Golkar:2012kb,Hou:2012xg,Jensen:2013vta,Braguta:2014gea}.
We also show in Sec.~\ref{sec:AVENon-Univ}, that these radiative corrections are recovered and well explained
in the global equilibrium model we are presenting. Thus, radiative corrections do not hinder the cause we are suggesting
for AVE, i.e. the thermal states modified by rotation. On the contrary, quantum anomalies are exact results and should
not receive correction above 1-loop diagrams. Gravitational anomaly is not exceptional to this feature. The fact that
AVE does indeed receive such corrections is a clear sign that AVE could not be explained with (only) chiral and gravitational anomaly.
Similar radiative corrections are found for CSE~\cite{Gorbar:2013upa} but not for Chiral Magnetic Effect (CME)~\cite{Feng:2018tpb},
whose relationship with the chiral anomaly is more consolidated.

Addressing the proper cause of the effect surely helps to understand the phenomena and therefore, to develop
an accurate model to describe it. Knowing if it can be enhanced or diminished by mass or by the interactions
is crucial when designing experiments aimed to present experimental evidence of such effects. The experimental manifestation of AVE,
except the obvious axial-number separation, is linked to polarization. Indeed we showed that at global equilibrium,
spin tensor alignment along rotation coincides with AVE, see Eq.~(\ref{eq:SpinTensAndAxialRel}). Since polarization
is essentially the manifestation of spin-tensor anisotropy, corrections to AVE are directly transferred to polarization.
\chapter{Chiral fermions under electromagnetic field}
\label{ch:ExternalB}
In this chapter, we consider the effects of an external electromagnetic field on the thermodynamics
of chiral vorticous fluids. The effects of electromagnetic fields on relativistic quantum fluids
is already a consolidated subject, see for instance~\cite{HakimBook} and reference therein and
specifically for the effects of constant magnetic field on fermion gases see~\cite{Canuto:1969cs}.
More recently, motivated by relativistic heavy-ion collisions, this topic has been addressed
in~\cite{Huang:2011dc} using the Zubarev's non-equilibrium statistical operator that we are
also using, and in~\cite{Kovtun:2016lfw} using generating functional and also with Wigner
function using kinetic theory in~\cite{Weickgenannt:2019dks,Gao:2019znl,Chen:2012ca}.

In this chapter (and in a paper in preparation~\cite{Buzzegoli:magnetico}), we want to study
magnetized plasmas, highlighting the modifications caused by chirality and by thermal vorticity.
First, we review the relativistic quantum theory of fermions under the effect of an external
electromagnetic field. In the particular case of a constant homogeneous electromagnetic field,
we discuss the symmetries of the theory and we provide the generators of those symmetries,
which are the translations and a subgroup of Lorentz transformations.
Then, we give exact solutions for thermal states of a system consisting of chiral fermions
under an external constant magnetic field, which were not given before in literature using
Zubarev's density matrix formalism. We discuss the general properties of a system at thermal
equilibrium with vorticity and electromagnetic field in Sec.~\ref{sec:GlobEqEM}.
Since we derived exact solutions, we treat the electromagnetic field without approximation
(at thermal equilibrium) and we consider the effect of thermal vorticity with linear response
theory as done in the previous chapter. The consequences of electromagnetic field on chiral
vortical effect and on axial vortical effect are analyzed. At last, we consider instead
a dynamic gauge field and we compute the first-order radiative correction to the axial
vortical effect conductivity.

\section{Dirac Field in external electromagnetic field}
\label{sec:DiracInB}
Consider a Dirac field in external electromagnetic field. The Lagrangian of the theory is given by
Eq.~(\ref{eq:DiracLcurved}), which in flat space-time reads:
\begin{equation*}
\mathcal{L}= \frac{\I}{2}\left[ \bar{\psi}\gamma^\mu \oraw{\de}_\mu \psi -\bar{\psi}\gamma^\mu \olaw{\de}_\mu \psi\right]-m\bar{\psi}\psi-j^\mu A_\mu,
\end{equation*}
with $\h j^\mu=q\bar{\psi}\gamma^\mu\psi$ and where $q$ is the elementary electric charge of the field and where the gauge potential
$A^\mu$ is an external non dynamic field. This Lagrangian is the same as the one obtained from the free Dirac Lagrangian
via the  minimal coupling  substitution $\de_\mu\to\de_\mu+\I qA_\mu$,  which grants gauge invariance to the theory.
From Euler Eq.s we obtain the equations of motion (EOM) for the Dirac field:
\begin{equation*}
\slashed{\de}\psi=-\I(q\slashed{A}+m)\psi,\quad \slashed{\de}\bar{\psi}=\bar{\psi}\I(q\slashed{A}+m).
\end{equation*}
From the EOM it is straightforward to verify that the electric current is conserved $\de_\mu \h j^\mu=0$.
We can also use the minimal substitution
\begin{equation*}
\oraw{\de}\to\oraw{\nabla}_\mu=\oraw{\de}_\mu+\I qA_\mu,\quad \olaw{\de}\to\olaw{\nabla}_\mu=\olaw{\de}_\mu-\I qA_\mu
\end{equation*}
in the canonical stress-energy tensor of the free theory to obtain the stress-energy tensor of fermions in
external electromagnetic field:
\begin{equation*}
\h T^{\mu\nu}=\frac{\I}{2}\left[\bar{\psi}\gamma^\mu\oraw{\de}^\nu\psi-\bar{\psi}\gamma^\mu\olaw{\de}^\nu\psi\right]-\h j^\mu A^\nu.
\end{equation*}
Again, with the EOM (see Appendix~\ref{ch:SeTInB} for second order equations) we can verify that the stress-energy
tensor is not conserved but it is driven by Lorentz force:
\begin{equation}
\label{eq:sym_cons_set_em}
\de^\mu\h T_{\mu\nu}=\h j^\lambda F_{\nu\lambda}.
\end{equation}

We can show (see Appendix~\ref{ch:SeTInB}) that the correct symmetrization of this stress-energy tensor that leaves
the previous operator relations unmodified is
\begin{equation*}
\h T^{\mu\nu}=\frac{\I}{4}\left[\bar{\psi}\gamma^\mu\oraw{\de}^\nu\psi-\bar{\psi}\gamma^\mu\olaw{\de}^\nu\psi+\bar{\psi}\gamma^\nu\oraw{\de}^\mu\psi-\bar{\psi}\gamma^\nu\olaw{\de}^\mu\psi\right]
-\frac{1}{2} \left(\h j^\mu A^\nu+\h j^\nu A^\mu\right);
\end{equation*}
note that this is the stress-energy tensor we would obtain using the minimal substitution to the symmetric stress-energy 
tensor of free Dirac field.

\subsection{Symmetries in constant homogeneous electromagnetic field}
\label{subsec:SymInFconst}
It is worth noticing that the symmetries of the theory of fermions in external electromagnetic field
are different from those of free fermions or those of quantum electrodynamics. Indeed while a system
without external forces is symmetric for the full Poincaré group, some of the symmetries are lost
when external fields are considered, because those do not transform together with the rest of the system.
Here we discuss the symmetries of the system in the particular case of a constant homogeneous electromagnetic
field. We examine the transformations that are still symmetries of the theory, the consequent
conserved quantities of the system and the form of the generators of such transformations.

If the Lagrangian of our theory is invariant under translations, by Noether's theorem we can identify
and we can construct four operators that share three properties: they are conserved quantities, they are
the generators of translations and they constitute the four-momentum of the system. However, translational
invariance by itself does not guarantees that only a single quantity must have all the three above properties
altogether. Consider again a system under an external electromagnetic field. In this situation, Poincaré symmetry
of space-time is broken. Let the electromagnetic field be constant and homogeneous, then translational invariance
is still present. However, the Lagrangian is not invariant under space-time translation, but it acquires
a term that is a four-divergence. This term, under appropriate boundary conditions, does not affect the action
of the system and the overall invariance is preserved. Nevertheless, the consequence of the additional term is
that we can distinguish three different operators, each of them having one of the three properties stated above.
This is understood with the Noether-Tassie-Buchdahl theorem~\cite{MR0180181,MR0180182,Bacry:1970ye}:
given a Lagrangian $\mathcal{L}(\psi(x),\de_\mu\psi(x),x)$ and the infinitesimal transformation
$x'^\mu=x^\mu+\delta x^\mu$, $\psi'=\psi+\delta\psi$ such that  $\de_\mu\delta x^\mu=0$,
which transforms the Lagrangian in
\begin{equation*}
\mathcal{L}(\psi'(x'),\de'_\mu\psi'(x'),x')=\mathcal{L}(\psi(x),\de_\mu\psi(x),x)+\de_\mu X^\mu,
\end{equation*}
where $X^\mu$ is a functional depending exclusively on $\psi(x)$ and $x$, then the quantity
\begin{equation*}
\Gamma^\mu=\frac{\delta \mathcal{L}}{\delta \de_\mu \psi}\delta \psi
-\left(\frac{\delta \mathcal{L}}{\delta \de_\mu \psi}\de_\nu\psi-\mathcal{L}\,g^\mu_{\,\nu} \right)\delta x^\nu-X^\mu
\end{equation*}
is conserved, i.e. divergence-less.

Consider the Dirac Lagrangian in constant homogeneous electromagnetic field
\begin{equation*}
\mathcal{L}(\psi(x),\de_\mu\psi(x),x)=\bar\psi(\I\slashed{\de}-m)\psi-\h j^\mu A_\mu.
\end{equation*}
The translation transformation ($\delta\psi=0$, $\delta x^\mu=\epsilon^\mu$) acts on the Dirac field but does not
act directly to the external gauge field. Therefore, the Lagrangian variation by translation is
\begin{equation*}
\begin{split}
\delta\mathcal{L}=&\mathcal{L}(\psi'(x'),\de'_\mu\psi'(x'),x')-\mathcal{L}(\psi(x),\de_\mu\psi(x),x)\\
=&\h j^\mu \de_\nu A_\mu \epsilon^\nu
=\h j^\mu (F_{\nu\mu}+\de_\mu A_\nu)\epsilon^\nu=-\h j^\mu (F_{\mu\nu}-\de_\mu A_\nu)\epsilon^\nu.
\end{split}
\end{equation*}
We can then identify the quantity $X$ of Noether-Tassie-Buchdahl theorem 
\begin{equation*}
X^{\mu\nu}\equiv \h j^\mu( A^\nu + F^{\nu\lambda}x_\lambda).
\end{equation*}
Indeed its divergence is the variation of the Lagrangian
\begin{equation*}
\epsilon_\nu\de_\mu X^{\mu\nu}=(\de_\mu\h j^\mu)( A^\nu + F^{\nu\lambda}x_\lambda)\epsilon_\nu +\h j^\mu (\de_\mu A_\nu + F_{\nu\mu})\epsilon^\nu=
-\h j^\mu (F_{\mu\nu}-\de_\mu A_\nu)\epsilon^\nu=\delta\mathcal{L},
\end{equation*}
therefore, the theorem implies that the system has a canonical conserved tensor given by:
\begin{equation*}
\h\pi^{\mu\nu}\ped{can}=\h T^{\mu\nu}_0-\h j^\mu A^\nu-\h j^\mu F^{\nu\lambda} x_\lambda,
\end{equation*}
where $\h T^{\mu\nu}_0$ is the free canonical Dirac stress-energy tensor.
Using Belinfante procedure (see Appendix~\ref{ch:SeTInB}) we can transform $\h T^{\mu\nu}_0-\h j^\mu A^\nu$ into the symmetric
stress-energy tensor of Dirac field in external magnetic field $\h T^{\mu\nu}_S$ and the above conserved
tensor becomes:
\begin{equation}
\label{eq:GenTranInF}
\h\pi^{\mu\nu}\equiv\h T^{\mu\nu}_S-\h j^\mu F^{\nu\lambda} x_\lambda.
\end{equation}
From the above equation, we can simply verify that $\de_\mu \h\pi^{\mu\nu}=0$ form Eq.~(\ref{eq:sym_cons_set_em}).
However, note that $\h{\pi}^{\mu\nu}$ is not symmetric and that it is gauge invariant.
The conserved quantities related to this quantities are the four vectors
\begin{equation*}
\h\pi^{\mu}=\int \D^3 x\, \h{\pi}^{0\mu} 
\end{equation*}
and they also constitute the generator of the translations~\cite{Bacry:1970ye}.
The momentum of the system is still given by
\begin{equation*}
\h{P}^\mu=\int \D^3 x\, \h{T}^{0\mu} 
\end{equation*}
but it is not a conserved quantity and it is not the generator of translations.
Another difference from four-momenta is that different components of this
vectors do not commute, instead they satisfy~\cite{Bacry:1970ye}:
\begin{equation*}
[\h\pi^\mu,\h\pi^\nu]=\I \h Q F^{\mu\nu},
\end{equation*}
where $\h{Q}$ is the electric charge operator.

As for Lorentz's transformations, we expect that the variation of the Lagrangian is a full divergence
only for specific forms of transformations. For example, with a vanishing electric field and a constant
magnetic field, only the rotation along the direction of the magnetic field and the boost along the
magnetic field are symmetries of the theory. Therefore, only in these cases, the Lagrangian variation
could be vanishing or a full divergence.

Returning to the general case, we can repeat the previous argument of translations to the Lorentz transformation:
$\delta x^\mu=\omega^{\mu\nu}x_\nu,\, \delta\psi=-\frac{\I}{2}\omega_{\mu\nu}\sigma^{\mu\nu}\psi$;
we find that the transformed Lagrangian is
\begin{equation*}
\begin{split}
\delta\mathcal{L}=&\omega^{\mu\nu}\h j^\lambda (g_{\mu\lambda}A_\nu-g_{\nu\lambda}A_\mu+x_\mu \de_\nu A_\lambda-x_\nu \de_\mu A_\lambda)\\
=&\omega^{\mu\nu}\h j^\lambda\l[\de_\lambda\l(x_\mu A_\nu-x_\nu A_\mu\r)-x_\mu\de_\lambda A_\nu+x_\mu\de_\nu A_\lambda
+x_\nu\de_\lambda A_\mu-x_\nu\de_\mu A_\lambda \r]\\
=&\omega^{\mu\nu}\h j^\lambda\l[\de_\lambda\l(x_\mu A_\nu-x_\nu A_\mu\r)+x_\mu F_{\nu\lambda}-x_\nu F_{\mu\lambda} \r].
\end{split}
\end{equation*}
It is straightforward to check the identities:
\begin{gather*}
\omega^{\mu\nu}x_\mu F_{\nu\lambda}=x^\rho\omega_{\rho\sigma}F^\sigma_{\,\,\lambda},\\
\omega^{\mu\nu}x_\mu F_{\nu\lambda}=-\omega^{\mu\nu}\de_\lambda(x_\mu x^\sigma F_{\sigma\nu})-x^\rho \omega_{\lambda\sigma}F^\sigma_{\,\,\rho},
\end{gather*}
from which it follows
\begin{equation*}
\begin{split}
\omega^{\mu\nu}x_\mu F_{\nu\lambda}=&\frac{1}{2}\omega^{\mu\nu}x_\mu F_{\nu\lambda}+\frac{1}{2}\omega^{\mu\nu}x_\mu F_{\nu\lambda}\\
=&-\frac{1}{2}\omega^{\mu\nu}\de_\lambda(x_\mu x^\sigma F_{\sigma\nu})-\frac{1}{2}x^\rho (\omega_{\lambda\sigma}F^\sigma_{\,\,\rho}-\omega_{\rho\sigma}F^\sigma_{\,\,\lambda}),
\end{split}
\end{equation*}
and similarly
\begin{equation*}
\omega^{\mu\nu}x_\nu F_{\mu\lambda}=\frac{1}{2}\omega^{\mu\nu}\de_\lambda(x_\nu x^\sigma F_{\sigma\mu})
-\frac{1}{2}x^\rho (\omega_{\lambda\sigma}F^\sigma_{\,\,\rho}-\omega_{\rho\sigma}F^\sigma_{\,\,\lambda}).
\end{equation*}
Therefore, the Lagrangian variation becomes
\begin{equation*}
\begin{split}
\delta\mathcal{L}=&\frac{1}{2}\omega^{\mu\nu}\de_\lambda\l[\h j^\lambda x_\mu \l(A_\nu-\frac{1}{2}x^\sigma F_{\sigma\nu}\r)
-\h j^\lambda x_\nu \l(A_\mu-\frac{1}{2}x^\sigma F_{\sigma\mu}\r)\r]\\
&-\frac{1}{2}x^\rho \h j^\lambda (\omega_{\lambda\sigma}F^\sigma_{\,\,\rho}-\omega_{\rho\sigma}F^\sigma_{\,\,\lambda}).
\end{split}
\end{equation*}
The first part of r.h.s. is written as a four-divergence.  The remaining part can not be cast into
a four-divergence but  it is proportional to the following product
\begin{equation*}
(\omega\wedge F)_{\lambda\rho}=\omega_{\lambda\sigma}F^\sigma_{\,\,\rho}-\omega_{\rho\sigma}F^\sigma_{\,\,\lambda}.
\end{equation*}
The product of two non-vanishing anti-symmetric tensor of rank two, $\omega\wedge F$, is zero if and
only if $\omega$ is a linear combination of $F$ and its dual $F^*$ (or viceversa):
\begin{equation}
\label{eq:lemmaAntisymProd}
(\omega\wedge F)_{\lambda\rho}=0\quad\text{iff}\quad \omega_{\mu\nu}=k\,F_{\mu\nu}+k'\,F^*_{\mu\nu},\quad k,k'\in\mathbb{R}.
\end{equation}
The previous statement can be proven by denoting the tensor with the ``electric'' and ``magnetic'' vectors
\begin{gather*}
F=(\vec{E},\vec{B}),\quad\quad\text{meaning}\quad E^i=F^{0i},\quad B^i=\frac{1}{2}\epsilon^{ijk}F_{jk},\\
\omega=(\vec{\alpha},\vec{w}),\quad\quad\text{meaning}\quad \alpha^i=\omega^{0i},\quad w^i=\frac{1}{2}\epsilon^{ijk}\omega_{jk}.
\end{gather*}
Then, with this notation, the wedge product can be written as
\begin{equation*}
\omega\wedge F=(\vec{\alpha}\times\vec{B}+\vec{w}\times\vec{E},\,\vec{w}\times\vec{B}-\vec{\alpha}\times\vec{E}).
\end{equation*}
Then at each tensor one can associate a complex three-vector
\begin{equation*}
F\to\vec{F}=\vec{B}+\I\vec{E},\quad\quad \omega\to\vec{\omega}=\vec{w}+\I\vec{\alpha}
\end{equation*}
and the wedge product becomes
\begin{equation*}
\omega\wedge F\to \vec{\omega}\times\vec{F}=(\vec{w}+\I\vec{\alpha})\times(\vec{B}+\I\vec{E}).
\end{equation*}
It is now clear that the product is vanishing when $\vec{\omega}$ is parallel to $\vec{F}$, this
occur if they are proportional:
\begin{equation*}
(\vec{w}+\I\vec{\alpha})=(a+\I\,b)(\vec{B}+\I\vec{E}),\quad a,b\in \mathbb{R}.
\end{equation*}
This is equivalent to say that $\vec{\omega}$ is a real combination of $\vec{B}+\I\vec{E}$ and $\vec{B}-\I\vec{E}$,
or that we have two independent solutions of Eq.~(\ref{eq:LieDerivFconst}): $\omega=k(\vec{E},\vec{B})=k F$ and $\omega=k'(-\vec{B},\vec{E})=k' F^*$,
with $k$ and $k'$ real numbers. Therefore, we proved that the part of Lagrangian variation which is not
a divergence is vanishing when $\omega^{\mu\nu}$ is a linear combination of electromagnetic stress-energy
tensor and its dual:
\begin{equation}
\label{eq:Lorentz_param_inv}
\omega^{\mu\nu}=a\,F^{\mu\nu}+\frac{b}{2}\epsilon^{\mu\nu\rho\sigma}F_{\rho\sigma}.
\end{equation}
This means, as expected, that the action and the theory is invariant only under certain type of Lorentz transformations: the ones
generated with a parameter of the form (\ref{eq:Lorentz_param_inv}). For example, in the case of constant magnetic field, we recover
that the system is invariant only for rotation and boost along the magnetic field. Set then $\omega$ either as
$\omega_{\mu\nu}\propto F_{\mu\nu}$ or $\omega_{\mu\nu}\propto F^*_{\mu\nu}$, so that the Lagrangian variation is a four-divergence,
from Noether-Tassie-Buchdahl theorem, the two following quantities are divergence-less:
\begin{gather*}
\h{\Gamma}^\lambda=\frac{1}{2}F^{\mu\nu}\l[ x_\mu \l(\h{T}^\lambda_{0\,\,\nu}-\h{j}^\lambda A_\nu+\frac{1}{2}\h{j}^\lambda x^\rho F_{\rho\nu}\r)
-x_\nu \l(\h{T}^\lambda_{0\,\,\mu}-\h{j}^\lambda A_\mu+\frac{1}{2}\h{j}^\lambda x^\rho F_{\rho\mu}\r)+\h{S}^\lambda_{\,\mu\nu}\r],\\
\h{\Gamma}^{*\lambda}=\frac{1}{2}F^{*\mu\nu}\l[ x_\mu \l(\h{T}^\lambda_{0\,\,\nu}-\h{j}^\lambda A_\nu+\frac{1}{2}\h{j}^\lambda x^\rho F_{\rho\nu}\r)
-x_\nu \l(\h{T}^\lambda_{0\,\,\mu}-\h{j}^\lambda A_\mu+\frac{1}{2}\h{j}^\lambda x^\rho F_{\rho\mu}\r)+\h{S}^\lambda_{\,\mu\nu}\r],
\end{gather*}
with $\h{S}^\lambda_{\mu\nu}$ the canonical spin tensor of free Dirac field. After Belinfante transformation, the quantities become
\begin{equation}
\label{eq:LorentzInF}
\begin{split}
\h{\Gamma}^\lambda=&\frac{1}{2}F^{\mu\nu}\h{M}^\lambda_{\mu\nu},\quad
\h{\Gamma}^{*\lambda}=\frac{1}{2}F^{*\mu\nu}\h{M}^\lambda_{\mu\nu},\\
\h{M}^\lambda_{\mu\nu}\equiv&x_\mu \l(\h{T}^\lambda_{S\,\nu}+\frac{1}{2}\h{j}^\lambda x^\rho F_{\rho\nu}\r)
-x_\nu \l(\h{T}^\lambda_{S\,\mu}+\frac{1}{2}\h{j}^\lambda x^\rho F_{\rho\mu}\r)\\
=& x_\mu \l(\h{\pi}^\lambda_{\,\,\nu}-\frac{1}{2}\h{j}^\lambda x^\rho F_{\rho\nu}\r)
-x_\nu \l(\h{\pi}^\lambda_{\,\,\mu}-\frac{1}{2}\h{j}^\lambda x^\rho F_{\rho\mu}\r).
\end{split}
\end{equation}
We can define the integrals:
\begin{equation*}
\h{M}_{\mu\nu}=\int \D^3 x\,\h{M}^0_{\mu\nu},
\end{equation*}
which are conserved quantities only if projected with $F^{\mu\nu}$ and $F^{*\mu\nu}$. The operators $\h{M}_{\mu\nu}$ are the generators of
Lorentz transformations if they are also a symmetry for the theory, otherwise the Wigner's theorem does not applies and
we can not say that such transformations admit an unitary and linear (or anti-unitary and anti-linear) representation.
For those operators, the following Algebra holds~\cite{Bacry:1970ye}:
\begin{equation}\label{eq:MaxwellAlgebra}
\begin{split}
[\h{\pi}^\mu,\h{\pi}^\nu]=&\I F^{\mu\nu}\h{Q},\\
\frac{1}{2}F_{\rho\sigma}[\h{\pi}^\mu,\h{M}^{\rho\sigma}]=&\frac{\I}{2} F_{\rho\sigma}\l(\eta^{\mu\rho}\h\pi^\sigma-\eta^{\mu\sigma}\h\pi^\rho\r),\\
\frac{1}{2}F^*_{\rho\sigma}[\h{\pi}^\mu,\h{M}^{\rho\sigma}]=&\frac{\I}{2} F^*_{\rho\sigma}\l(\eta^{\mu\rho}\h\pi^\sigma-\eta^{\mu\sigma}\h\pi^\rho\r),
\end{split}
\end{equation}
where $\h{Q}$ is the electric charge operator.
In the particular case of vanishing electric field and constant magnetic field along the $z$ axis, the Algebra becomes:
\begin{gather*}
[\h\pi_x,\,\h\pi_y]=\I|\vec{B}|\h{Q},\\
[\h{J}_z,\h\pi_x]=\I\h\pi_y,\quad [\h{J}_z,\h\pi_y]=-\I\h\pi_x,\\
[\h{K}_z,\h\pi_t]=-\I\h\pi_z,\quad [\h{K}_z,\h\pi_z]=-\I\h\pi_t.
\end{gather*}
\newpage

\section{Chiral fermions in constant magnetic field}
\label{sec:ThermoInB}
Consider now a system consisting of chiral fermions in an external homogeneous constant magnetic field $\vec{B}$
at thermal equilibrium with vanishing thermal vorticity. The system Lagrangian and EOM were discussed
in the previous section. The local thermal equilibrium density operator is given by Eq.~(\ref{eq:LEDO}) with
the addition of an axial current:
\begin{equation*}
\h{\rho}\ped{LTE} = \frac{1}{Z} \exp \left[ -\int_{\Sigma} \di \Sigma_\mu \left( \h T^{\mu\nu}(x) \beta_\nu(x) - \zeta(x)\, \h j^\mu(x)
- \zeta\ped{A}(x)\, \h j\ped{A}^\mu(x)\right) \right],
\end{equation*}
where the operators are defined (see previous Section) as 
\begin{equation*}
\begin{split}
\h j^{\,\mu} =&\, q\bar{\psi} \gamma^\mu \psi, \quad \h j^{\,\mu}\ped{A} =\, \bar{\psi} \gamma^\mu\gamma^5 \psi,\\
\h T^{\mu\nu}= &\frac{\I}{4}\left[\bar{\psi}\gamma^\mu\oraw{\de}^\nu\psi-\bar{\psi}\gamma^\mu\olaw{\de}^\nu\psi+\bar{\psi}\gamma^\nu\oraw{\de}^\mu\psi
-\bar{\psi}\gamma^\nu\olaw{\de}^\mu\psi\right] - \frac{1}{2}\left(\h j^\mu A^\nu+\h j^\nu A^\mu\right),
\end{split}
\end{equation*}
and satisfy the following conservation equation:
\begin{equation*}
\de_\mu \h j^{\,\mu}=0,\quad \de_\mu\h j^{\,\mu}\ped{A}=0,\quad \de_\mu\h T^{\mu\nu}=\h j_\lambda F^{\nu\lambda}.
\end{equation*}
There is no chiral anomaly because there is no electric field.
As discussed in Sec~\ref{sec:GenGlobEquil}, the global equilibrium is reached when the divergence of the integrand inside the statistical
operator is vanishing. This occurs if the thermodynamic fields $\beta$, $\zeta$ and $\zeta\ped{A}$ satisfy:
\begin{equation*}
\de_{\mu}\beta_{\nu}(x)+\de_{\nu}\beta_{\mu}(x)=0,\qquad \de^\mu \zeta(x)=F^{\nu\mu}\beta_\nu(x)
,\qquad \de^\mu \zeta\ped{A}(x)=0.
\end{equation*}
In this section, we consider the case without thermal vorticity, therefore for the inverse four temperature $\beta$ we chose the
solution which is independent on the coordinate: $\de_\mu\beta_\nu=0$. In this case the condition for the electric chemical potential reads
\begin{equation*}
\de^\mu \zeta(x)=F^{\nu\mu}\beta_\nu=\sqrt{\beta^2}F^{\nu\mu}u_\nu=\sqrt{\beta^2}E^\mu
\end{equation*}
where $u$ is the fluid velocity and $E$ is the comoving electric field. Since we have chosen the case without electric field
the global equilibrium condition for the thermodynamic fields are simply:
\begin{equation*}
\beta_\mu=\text{cnst.},\quad \zeta=\text{cnst.},\quad\zeta\ped{A}=\text{cnst.}.
\end{equation*}
The global equilibrium statistical operator then becomes
\begin{equation*}
\h{\rho} =  \frac{1}{Z} \exp \left[-\h P^\mu \beta_\mu +\zeta \h Q+\zeta\ped{A}\h Q\ped{A}  \right].
\end{equation*}
Notice that the operators $\h{P}^\mu$ are not the generators of translation $\h\pi^\mu$ obtained by integrating the conserved current
in Eq.~(\ref{eq:GenTranInF}). However, in the case of vanishing comoving electric field the projection along the inverse temperature
of the four-momentum or of the generators of translations are equivalent:
\begin{equation*}
\h\pi^\mu \beta_\mu=\int_{\Sigma} \di \Sigma_\lambda \left(\h T^{\lambda\nu}-\h j^\lambda F^{\nu\sigma} x_\sigma\right)\beta_\nu
=\h P^\mu\beta_\mu-\sqrt{\beta^2}E^\sigma\int_{\Sigma} \di \Sigma_\lambda \h j^\lambda x_\sigma  =\h P^\mu\beta_\mu.
\end{equation*}
The statistical operator can now be written as
\begin{equation*}
\h{\rho} =  \frac{1}{Z} \exp \left[-\h \pi^\mu \beta_\mu +\zeta \h Q+\zeta\ped{A}\h Q\ped{A}  \right].
\end{equation*}
In this form, it is straightforward to use the algebra in Eq.~(\ref{eq:MaxwellAlgebra}) and translate the statistical operator
of a quantity $a$. We find
\begin{equation*}
\h {\sf T}(a)\, \h\rho\, \h {\sf T}^{-1}(a) =\E^{\I a\cdot\h\pi}\h\rho\,\E^{-\I a\cdot\h\pi}
=\frac{1}{Z} \exp \left[-\h \pi^\mu \beta_\mu +\zeta \h Q+\zeta\ped{A}\h Q\ped{A}+a_\mu F^{\mu\nu}\beta_\nu\h{Q}  \right]
\end{equation*}
but again $F^{\mu\nu}\beta_\nu$ is the comoving electric field which is vanishing. The statistical operator is therefore
homogeneous:
\begin{equation*}
\h {\sf T}(a)\, \h\rho\, \h {\sf T}^{-1}(a) =\h\rho .
\end{equation*}

Now that we have established the basic properties of thermal equilibrium with a constant and homogeneous magnetic field,
we adopt the techniques of thermal field theory (in Chapter~\ref{ch:ftft}) to find the exact solutions for thermal states
in a magnetic field. As we learned, the starting point is to give a path integral description of the partition function.
Since the partition function is a Lorentz invariant, we can choose the local rest frame where $u=(1,\vec{0})$.
In this frame, without loss in generality, the magnetic field is chosen along the $z$ axis and we adopt the Landau
gauge $A^\mu=(0,0,B x_1,0)$. The path integral formulation of the partition function in local rest frame
\begin{equation*}
\parz(T,\mu,\mu_5)=\tr\left[ \e^{-\beta (\h{H}-\mu \h{Q}-\mu\ped{A} \h{Q}\ped{A})}\right]
\end{equation*}
is given by
\begin{equation*}
\parz= C \int_{\Psi(\beta,\vec{x})=-\Psi(0,\vec{x})} \mathcal{D} \bar{\Psi}\,\mathcal{D} \Psi \,\, \exp \left( - S\ped{E}(\Psi,\bar{\Psi},\mu\ped{A}) \right)
\end{equation*}
where\footnote{We add a mass term for generalization, although with mass we cannot have a conserved axial current.},
in the notations of Chapter~\ref{ch:ftft}, the Euclidean action of Dirac fermions in external electromagnetic field is
\begin{equation*}
S\ped{E}(\Psi,\bar{\Psi},\mu_5)=\int_0^\beta\di\tau\int_{\vec{x}}
\bar{\Psi}(X)\left[\I(\gamma\cdot\pi^+)+m-\gamma_0 \gamma^5\mu\ped{A}  \right]\Psi(X)
\end{equation*}
and $\pi^+_\mu\equiv P^+_\mu-q A_\mu$, which is not to be confused with the generator of translations.

We are now interested in the exact solution, but instead of solving the Dirac equation directly, we use the
Ritus method~\cite{Ritus:1972ky}, see~\cite{Leung:2005yq} for a brief recap of the method. The core concept
of Ritus method is that we can construct a complete set of orthonormal function, called $E_p$ Ritus functions,
such that the Euclidean action is rendered formally identical to the Euclidean action of a free Dirac field
in absence of external fields. The $E_p$ functions are constructed such that they are the matrix of the
simultaneous eigenfunctions (eigenvectors) of the maximal set of mutually commuting operators
$\{(\gamma\cdot\pi)^2,\I\gamma_1\gamma_2,\gamma^5 \}$. From gamma algebra, it is straightforward to check that
\begin{equation*}
\I\gamma_1\gamma_2\delta(\sigma)=\sigma\Delta(\sigma),\quad
\frac{1+\chi\gamma^5}{2}\gamma^5=\chi\frac{1+\chi\gamma^5}{2},
\end{equation*}
with $\sigma=\pm$ and $\chi=\pm$ and we defined
\begin{equation*}
\Delta(\sigma)\equiv\frac{1+\I\sigma\gamma_1\gamma_2}{2}.
\end{equation*}
We can then show that~\cite{Ritus:1972ky,Leung:2005yq}
\begin{equation*}
(\gamma\cdot\pi^+)^2 E_{\h p \sigma}(X)={\underline{P}^+}^2 E_{\h p \sigma}(X)
\end{equation*}
where $\h p$ is a label for the quantum numbers $\{l,\omega_n,p_2,p_3\}$, the eigenvalues $\underline{P}^+$ are given by
\begin{equation*}
\underline{P}^+\equiv (\omega_n+\I\mu,0,-\bar\sigma\sqrt{2|qB|l},p_3),\quad \bar\sigma\equiv\text{sgn}(qB)
\end{equation*}
and the form of eigenfunction is
\begin{equation}
\label{eq:defEpsigma}
E_{\h p \sigma}(X)=N(n)\E^{\I(P_0 \tau + P_2 X_2+P_3 X_3)}D_n(\rho)
\end{equation}
where $N(n)=(4\pi |qB|)^{1/4}/\sqrt{n!}$ is a normalization factor, and $D_n(\rho)$ denotes the parabolic cylinder functions with argument
$\rho=\sqrt{2|qB|}(X_1 - p_2 /qB)$ and non-negative integer index $n = 0, 1, 2, \dots$ given by
\begin{equation*}
n=l+\frac{\sigma}{2}\text{sgn}(qB)-\frac{1}{2}.
\end{equation*}
Note that the form of the functions~(\ref{eq:defEpsigma}) strongly depends on the gauge chosen, in our case the Landau gauge.
Since the eigenfunction $E_{\h p \sigma}(X)$ does not depend on chirality, the maximal eigenfunctions of the operators
$\{(\gamma\cdot\pi)^2,\I\gamma_1\gamma_2,\gamma^5 \}$ are given by
\begin{equation}\label{eq:defE}
E_{\h p}(X)= \sideset{}{'}\sum_{\sigma=\pm} E_{\h p \sigma}(X)\Delta(\sigma),\quad
\bar{E}_{\h p}(X)=\gamma_0 E^\dagger_{\h p}(X)\gamma_0=\sideset{}{'}\sum_{\sigma'=\pm} E^*_{\h p \sigma'}(X)\Delta(\sigma')
\end{equation}
where a prime on the summation symbol means that it is subject to the constraint
\begin{equation*}
\sigma=\begin{cases} \text{sgn}(qB) & l=0\\ \pm & l>0 \end{cases}.
\end{equation*}
Some important properties can be derived from this definitions. First that the functions $E_p$ commute with $\gamma_0$ and with $\gamma^5$.
Moreover, they satisfy the orthogonality relation
\begin{equation*}
\int_X \bar{E}_{\h q}(X) E_{\h p}(X)=(2\pi)^4 \h\delta^{(4)}(\h p-\h q)\Pi(l)
\end{equation*}
where
\begin{gather*}
\delta^{(4)}(\h p-\h p')\equiv\delta_{l,l'}\beta\delta{\omega_n,\omega_{n'}}\delta(p_2-p_2')\delta(p_3-p_3')\\
\Pi(l)\equiv\begin{cases} \frac{1+\I\bar\sigma\gamma_1\gamma_2}{2} & l=0\\ 1 & l>0 \end{cases}.
\end{gather*}
And lastly, the action of the operator $(\gamma\cdot\pi^+)$ on these function is
\begin{equation*}
(\gamma\cdot\pi^+)E_{\h p}(X)=E_{\h p}(X)\gamma\cdot \underline{P}.
\end{equation*}

Since we showed that $E_p$ Ritus functions are a complete orthonormal functions, we can then expand
the Dirac fields in these functions:
\begin{equation*}
\begin{split}
\Psi(X)=& T\sum_{\{\omega_n\}}\sum_{l=0}^\infty\int\D p_2\int \frac{\D p_3}{(2\pi)^3}E_{\h p}(X) \Psi(\underline{P})
\equiv \sumint_{\h P} E_{\h p}(X) \Psi(\underline{P}),\\
\bar{\Psi}(X)=&\sumint_{\h Q} \bar{\Psi}(\underline{Q}) \bar{E}_{\h q}(X).
\end{split}
\end{equation*}
Replacing this expansion on the Euclidean action, we find
\begin{equation*}
S\ped{E}(\Psi,\bar{\Psi},\mu_5)=\sumint_{\h P}\sumint_{\h Q}\int_{X}
\bar{\Psi}(\underline{Q})\bar{E}_{\h q}(X)\left[\I(\gamma\cdot\pi^+)+m-\gamma_0 \gamma^5\mu\ped{A}  \right]E_{\h p}(X)\Psi(\underline{P})
\end{equation*}
and, using the above mentioned properties of $E_p$ functions, we obtain
\begin{equation*}
S\ped{E}(\Psi,\bar{\Psi},\mu_5)=\sumint_{\h P}\bar{\Psi}(\underline{P})\Pi(l)\left[\I\gamma\cdot\underline{P}^+
+m-\gamma_0 \gamma^5\mu\ped{A}  \right]\Psi(\underline{P}).
\end{equation*}
Notice that this is formally identical to the Euclidean action of free Dirac field obtained in Sec.~\ref{sec:PathIntFermion}.

We can then proceed to evaluate the partition function as in the free case. We change the  integration variables in the partition
function to the modes of $E_p$ functions. The partition function is then a Gaussian integral of Grassmann variables, whose result
is the exponent determinant. We then have
\begin{equation}\label{eq_parzmagnetic}
\begin{split}
\parz=& \tilde{C} \int_{\Psi(\beta,\vec{x})=-\Psi(0,\vec{x})} \mathcal{D} \bar{\Psi}(\underline{P})\,\mathcal{D} \Psi(\underline{P})\times \\
&\times\exp \left\{ - \sumint_{\h P}\bar{\Psi}(\underline{P})\Pi(l)\left[\I\gamma\cdot\underline{P}^+ +m-\gamma_0 \gamma^5\mu\ped{A}  \right]\Psi(\underline{P}) \right\}\\
=&\tilde{C}\det\left[ \Pi(l)\left(\I\gamma\cdot\underline{P}^+ +m-\gamma_0 \gamma^5\mu\ped{A}  \right)\right].
\end{split}
\end{equation}
For the sake of clarity, let's neglect $\Pi(l)$ for now on:
\begin{equation*}
\begin{split}
\parz=&\tilde{C}\det\left(\begin{array}{cc} m\idmat_{2\times 2}  & [\I(\omega_n+\I\mu)-\mu\ped{A}]\idmat_{2\times 2}+\sigma_i \underline{P}_i\\
(\I(\omega_n+\I\mu)+\mu\ped{A})\idmat_{2\times 2}-\sigma_i \underline{P}_i & m\idmat_{2\times 2}\end{array}\right).
\end{split}
\end{equation*}
The determinant is evaluated using the standard formula for block matrices
\begin{equation*}
\det\left(\begin{array}{cc} A & B\\ C & D\end{array}\right)= \det\left( AD-BD^{-1}CD\right);
\end{equation*}
replacing that into the partition function, we have
\begin{equation*}
\begin{split}
\parz& =\tilde{C}\det\left[({\underline{P}^+}^2+m^2+\mu\ped{A}^2)\mathbb{1}_{2\times 2}-2\sigma_i \underline{P}_i\mu\ped{A} \right] \\
&= \tilde{C}\det\left[({\underline{P}^+}^2+m^2+\mu\ped{A}^2)^2-4|\vec{\underline{P}}|^2\mu\ped{A}^2 \right]\\
&= \prod_{\omega_n,l,p_3}\tilde{C}\left[({\underline{P}^+}^2+m^2+\mu\ped{A}^2)^2-4|\vec{\underline{P}}|^2\mu\ped{A}^2 \right],
\end{split}
\end{equation*}
where we evaluated the determinant as the product of the eigenvalues of the matrix.
To connect this quantity to the thermodynamics of the system, we are actually interested in its logarithm:
\begin{equation}
\label{eq:logparzinB}
\log \parz=\sum_{\omega_n,l,p_3}\log\left[({\underline{P}^+}^2+m^2+\mu\ped{A}^2)^2-4|\vec{\underline{P}}|^2\mu\ped{A}^2 \right]+\text{cnst}.
\end{equation}
From the logarithm of partition function we obtain the thermodynamic potential of the system and then, by simple derivation, we could
obtain other thermodynamic quantities. In the next subsection, we evaluate the thermodynamic potential of the system. Then, from
partition function in Eq.~(\ref{eq_parzmagnetic}), we obtain the thermal propagator of a chiral fermion in a magnetic field. Once we
have the propagator, we can use the point-splitting procedure to evaluate other quantities that are not related to the thermodynamic
potential. We use this method to evaluate the mean value of electric current and axial current in the following subsections.

\subsection{Thermodynamic potential}
The thermodynamic potential is derived from the partition function as following
\begin{equation*}
\Omega=\lim_{V\to\infty}-\frac{T}{V}\log \parz,
\end{equation*}
where the logarithm of partition function is given by Eq.~(\ref{eq:logparzinB}).
We can follow the steps of Sec.~\ref{subsec:thermpot} to evaluate the thermodynamic potential and to sum the Matsubara frequencies.
However, we must first consider that in this case the Landau levels generated by the magnetic field have different degeneracy factors
ad must be properly taken into account when performing the infinite volume limit. Let be $S$ the area in the $x-y$ plane and
$p_{\perp 1}$ and $p_{\perp 2}$ the momenta in that plane. In the infinity area limit, the sum on modes becomes the following integrals:
\begin{equation*}
\lim_{S\to\infty}\frac{1}{S}\sum_{p_{\perp 1}}\sum_{p_{\perp 2}}=\int_{-\infty}^\infty\frac{\D p_{\perp 1}}{2\pi} \int_{-\infty}^\infty\frac{\D p_{\perp 2}}{2\pi}.
\end{equation*}
Each Landau level has degeneracy associated with some quantum numbers; this degeneracy is gauge independent and it is given by:
\begin{equation*}
d_l=\left\lfloor \frac{|qB| S}{2\pi}\right\rfloor.
\end{equation*}
To obtain this degeneracy, we just have to evaluate the quantity
\begin{equation*}
\frac{\D p_{\perp 1}}{2\pi}\frac{\D p_{\perp 2}}{2\pi}
\end{equation*}
between two consecutive energy levels:
\begin{equation*}
d_l=\left\lfloor \frac{|qB| S}{2\pi}\right\rfloor=\int_l^{l+1}\frac{\D p_{\perp 1}}{2\pi}\frac{\D p_{\perp 2}}{2\pi}.
\end{equation*}
Therefore, removing the floor function, the infinite volume limit of the sum on the states of the system gives:
\begin{equation*}
\lim_{V\to\infty}\frac{1}{V}\sum_{\omega_n,l,p_3}=\frac{|qB|}{2\pi}\sum_{l=0}^\infty\int_{-\infty}^\infty\frac{\D p_3}{2\pi} \sum_{\{\omega_n\}}.
\end{equation*}
Consequently the thermodynamic potential reads:
\begin{equation*}
\begin{split}
\Omega=&\lim_{V\to\infty}-\frac{T}{V}\log \parz \\
=&-\frac{|qB|}{2\pi}\sum_{l=0}^\infty\int_{-\infty}^\infty\frac{\D p_3}{2\pi} 
T\sum_{\{\omega_n\}}\log\left[({\underline{P}^+}^2+m^2+\mu\ped{A}^2)^2-4|\vec{\underline{P}}|^2\mu\ped{A}^2 \right]+\text{cnst}.
\end{split}
\end{equation*}

The Matsubara sum can be performed with the help of the auxiliary function
\begin{equation*}
j(y)\equiv T\sum_{\{\omega_n\}}\log\left[({\underline{P}^+}^2+m^2+\mu\ped{A}^2)^2-y^2 \right];
\end{equation*}
in terms of which the thermodynamic potential is
\begin{equation*}
\Omega=-\frac{|qB|}{2\pi}\sum_{l=0}^\infty\int_{-\infty}^\infty\frac{\D p_3}{2\pi} j(y_0)+\text{cnst}
\end{equation*}
with $y_0=2\mu\ped{A}$. Instead of summing inside the function $j(y)$, we can sum inside the following function
\begin{equation*}
i(y)\equiv\frac{1}{2y}\frac{\D}{\D y}j(y)=T\sum_{\{\omega_n\}}\left[({\underline{P}^+}^2+m^2+\mu\ped{A}^2)^2-y^2 \right]^{-1},
\end{equation*}
and recover the $j(y)$ function using 
\begin{equation*}
j(y_0)=\int_c^{y_0}2y\, i(y)\D y+\text{constant}.
\end{equation*}
The final result for thermodynamic potential is
\begin{equation*}
\Omega=-\frac{|qB|}{2\pi}\sum_{l=0}^\infty\sideset{}{'}\sum_{s=\pm}\int_{-\infty}^\infty\frac{\D p_3}{2\pi}
\left[E_s+T\sum_\pm\log\left(1+\E^{-\beta(E_s\pm\mu)}\right) \right]+\text{cnst},
\end{equation*}
where $E_s^2=[(p_3^2+2qBl)^{1/2}+s\mu\ped{A}]^2+m^2$ and the constraint of $s=\bar{\sigma}$ for $l=0$ is caused
by the projector $\Pi(l)$. This same thermodynamic potential for chiral fermions in external magnetic
field was used in~\cite{Fukushima:2008xe} to derive the chiral magnetic effect (CME). This expression can be used to obtain
the electric and axial charge density, but instead we are using the point-splitting procedure because the latter can also be used to
evaluate other thermodynamic functions related to currents. To do that, we first need the thermal propagator of chiral fermions.

\subsection{Chiral fermion propagator in magnetic field}
Since we have used the Ritus method, the form of Euclidean action is formally identical to those of a free Dirac field.
It then comes with no surprise that the fermionic propagator is obtained identically to the free case as explained
in Sec.~\ref{sec:FermiProp}. The propagator in Fourier modes is obtained in path integral formulation by
\begin{equation*}
\mean{\tilde{\psi}_a(P)\bar{\tilde\psi}_b(Q)}_T=\frac{\int \Dpi \tilde\Psi \,\Dpi \tilde{\bar{\Psi}} \; \exp \l( - S\ped{E}\r) \tilde{\psi}_a(P)\bar{\tilde\psi}_b(Q)}{\int\Dpi \tilde\Psi \,\Dpi \tilde{\bar{\Psi}}\; \exp \l( - S\ped{E}\r)}-
\end{equation*}
It is convinient to use the partition function as given in the first line of Eq.~(\ref{eq_parzmagnetic}):
\begin{equation*}
\parz= \tilde{C} \int_{\Psi(\beta,\vec{x})=-\Psi(0,\vec{x})}\hspace{-1cm} \mathcal{D} \bar{\Psi}(\underline{P})\,\mathcal{D} \Psi(\underline{P}) \,\,
\exp \left\{ - \sumint_{\h P}\bar{\Psi}(\underline{P})\Pi(l)\left[\I\gamma\cdot\underline{P}^+ +m-\gamma_0 \gamma^5\mu\ped{A}  \right]\Psi(\underline{P}) \right\}.
\end{equation*}
The Grassmann integrals are straightforward and gives
\begin{equation*}
\mean{\bar{\Psi}(\underline{Q})_a \Psi(\underline{P})_b}=\delta^{(4)}(\underline{P}-\underline{Q})\mathcal{M}^{-1}_{ab}
\end{equation*}
where $a,b$ denotes spinorial indices and
\begin{equation*}
\mathcal{M}=\left[ \I\gamma\cdot \underline{P}^+ -\gamma_0\gamma^5\mu\ped{A}\right].
\end{equation*}
The inverse of $\mathcal{M}$ was obtained in Section \ref{sec:FermiProp}, therefore we have
\begin{equation*}
\mean{\bar{\Psi}(\underline{Q})_a \Psi(\underline{P})_b}=\delta^{(4)}(\underline{P}-\underline{Q})\sum_\chi
\left(\mathbb{P}_\chi\frac{-\I\slashed{\underline{P}}_\chi^+}{{\underline{P}^+_\chi}^2} \right)_{ab}.
\end{equation*}
This is the generalization in Euclidean spacetime with chemical potentials of the propagator in \cite{Ritus:1972ky,Leung:2005yq}.
In the configuration space, the two-point function is
\begin{equation*}
\begin{split}
\mean{\bar{\Psi}(X)_a \Psi(Y)_b}&
=\sumint_{\h P} \sumint_{\h Q}\bar{E}_{\h p}(X)_{a'a} E_{\h q}(Y)_{bb'}\mean{\bar{\Psi}(\underline{P})_{a'} \Psi(\underline{Q})_{b'}}\\
&=-\sumint_{\h P} \sumint_{\h Q}\bar{E}_{\h p}(X)_{a'a} E_{\h q}(Y)_{bb'}\mean{\Psi(\underline{Q})_{b'} \bar{\Psi}(\underline{P})_{a'}}\\
&=-\sumint_{\h P} \sumint_{\h Q}\bar{E}_{\h p}(X)_{a'a} E_{\h q}(Y)_{bb'}\delta^{(4)}(\underline{P}-\underline{Q})\sum_\chi
\left(\mathbb{P}_\chi\frac{-\I\slashed{\underline{P}}_\chi^+}{{\underline{P}^+_\chi}^2} \right)_{b'a'}
\end{split}
\end{equation*}
where to go to second line we used fermion anti-commutation; finally, integrating the delta, we have
\begin{equation}
\label{eq:PropinB}
\mean{\bar{\Psi}(X)_a \Psi(Y)_b}=-\sumint_{\h P}\sum_\chi E_{\h p}(Y)_{bb'}
\left(\mathbb{P}_\chi\frac{-\I\slashed{\underline{P}}_\chi^+}{{\underline{P}^+_\chi}^2} \right)_{b'a'} \bar{E}_{\h p}(X)_{a'a}.
\end{equation}
\newpage

\subsection{Electric current mean value}
\label{sec:CME}
Now we use the propagator to evaluate the mean value of electric current.
This method is similar to the one used in~\cite{Fukushima:2009ft}, as it is also using the Ritus method.
As we learned, the mean value can be obtained with point-splitting procedure (see Sec.~\ref{subsec:pointsplit} and~\ref{subsec:homogcurrents}).
Following that prescription, we write the current in Euclidean space-time, we recover the thermal propagator
of fermionic field and then we reconstruct the trace on spinorial indices; those steps give:
\begin{equation*}
\begin{split}
\mean{\h{j}_\mu(X)}&=(-\ii)^{1-n_\mu}q\mean{\h{\bar{\psi}}(X)\gamma_\mu \h{\psi}(X)}
=\lim_{X_1,X_2\to X} (-\ii)^{1-n_\mu}q\left(\gamma_\mu\right)_{ab}\mean{\h{\bar{\psi}}_a(X_1)\h{\psi}_b(X_2)}\\
&=\lim_{X_1,X_2\to X} -(-\ii)^{1-n_\mu} q\sumint_{\h P}\sum_\chi\left(\gamma_\mu\right)_{ab}E_{\h p}(X_2)_{bb'}
\left(\mathbb{P}_\chi\frac{-\I\slashed{\underline{P}}_\chi^+}{{\underline{P}^+_\chi}^2} \right)_{b'a'} \bar{E}_{\h p}(X_1)_{a'a}\\
&=-(-\ii)^{1-n_\mu} q\sumint_{\h P}\sum_\chi \tr\left[\gamma_\mu E_{\h p}(X)\mathbb{P}_\chi
\frac{-\I\slashed{\underline{P}}_\chi^+}{{\underline{P}^+_\chi}^2} \bar{E}_{\h p}(X)\right]\\
&=-(-\ii)^{1-n_\mu} q\sumint_{\h P}\sum_\chi \tr\left[\bar{E}_{\h p}(X) \gamma_\mu E_{\h p}(X)\mathbb{P}_\chi
\frac{-\I\slashed{\underline{P}}_\chi^+}{{\underline{P}^+_\chi}^2} \right].
\end{split}
\end{equation*}
It is convenient to indicate with the parallel symbol ``$\parallel$'', the components parallel to the magnetic field,
which are the time component and the $z$ component. For those components it holds:
\begin{equation*}
\left[ \gamma^\parallel_\mu, E_{\h p}(X)\right]=0.
\end{equation*}
Therefore, reminding the definitions of the Ritus $E_{\h p}(X)$ functions~(\ref{eq:defE}), we obtain for the parallel component
of electric current
\begin{equation*}
\begin{split}
\mean{\h{j}^\parallel_\mu(X)}=&-(-\ii)^{1-n_\mu} q\sum_{l=0}^\infty\int\frac{\D p_2}{2\pi}\int\frac{\D p_3}{(2\pi)^2}\sum_\chi T\sum_{\{\omega_n\}}
\tr\left[\bar{E}_{\h p}(X) E_{\h p}(X)\gamma_\mu^\parallel \mathbb{P}_\chi\frac{-\I\slashed{\underline{P}}_\chi^+}{{\underline{P}^+_\chi}^2} \right]\\
=&-(-\ii)^{1-n_\mu} q\sum_{l=0}^\infty\int\frac{\D p_2}{2\pi}\int\frac{\D p_3}{(2\pi)^2}\sum_\chi T\sum_{\{\omega_n\}}
\sideset{}{'}\sum_{\sigma=\pm}\sideset{}{'}\sum_{\sigma'=\pm} E^*_{\h p\sigma'}(X) E_{\h p\sigma}(X)\times\\
&\times\tr\left[\Delta(\sigma')\Delta(\sigma)\gamma_\mu^\parallel \mathbb{P}_\chi\frac{-\I\slashed{\underline{P}}_\chi^+}{{\underline{P}^+_\chi}^2} \right].
\end{split}
\end{equation*}
We can simplify the expression by taking advantage of the following identity
\begin{equation*}
\Delta(\sigma)\Delta(\sigma')=\frac{1+\sigma\sigma'+\I(\sigma+\sigma')\gamma_1\gamma_2}{4}=\delta_{\sigma,\sigma'}\Delta(\sigma).
\end{equation*}
Notice that the only dependence on $p_2$ is inside $E^*_{\h p\sigma'}(X) E_{\h p\sigma}(X)$. We can then show that
the integration over $p_2$ gives
\begin{equation*}
\int_{-\infty}^\infty \frac{\D p_2}{2\pi} E^*_{\h p\sigma'}(X) E_{\h p\sigma}(X)=|qB|\delta_{n,n'}.
\end{equation*}
%
%
It is furthermore convenient to split between lowest Landau level (LLL) $l=0$ and higher Landau level (HLL) $l>1$. For $l=0$ the sums on $\sigma$ are
constrained to be equal to $\sigma=\sigma'=\bar\sigma=$sgn$(eB)$, and the momenta are given by $\underline{P}^+_\chi=(\omega_n+\I\mu_\chi,0,0,p_3)$.
At the lowest Landau level we have
\begin{equation*}
\mean{\h{j}^\parallel_\mu(X)}\ped{LLL}=-(-\ii)^{1-n_\mu} q|qB|\int_{-\infty}^\infty\frac{\D p_3}{(2\pi)^2}\sum_\chi T\sum_{\{\omega_n\}}
\tr\left[\frac{1+\I\bar\sigma\gamma_1\gamma_2}{2}\gamma_\mu^\parallel \mathbb{P}_\chi\frac{-\I\slashed{\underline{P}}_\chi^+}{{\underline{P}^+_\chi}^2} \right].
\end{equation*}
After the trace, we find that the zero component is
\begin{equation*}
\begin{split}
\mean{\h{j}_0(X)}\ped{LLL}=&-\int_{-\infty}^\infty\frac{q|qB|\D p_3}{(2\pi)^2}\sum_\chi T\sum_{\{\omega_n\}}
\frac{-\I \left[(\omega_n+\I\mu_\chi)-\I p_3 \bar\sigma  \chi \right]}{(\omega_n+\I\mu_\chi)^2+p_3^2}\\
=&\int_{-\infty}^\infty\frac{q|qB|\D p_3}{(2\pi)^2}\sum_\chi T\sum_{\{\omega_n\}}
\frac{\I \left[(\omega_n+\I\mu_\chi)\right]}{(\omega_n+\I\mu_\chi)^2+p_3^2},
\end{split}
\end{equation*}
while the $z$ component is
\begin{equation*}
\begin{split}
\mean{\h{j}_3(X)}\ped{LLL}=&\ii \int_{-\infty}^\infty\frac{q|qB|\D p_3}{(2\pi)^2}\sum_\chi T\sum_{\{\omega_n\}}
\frac{-\I \left[p_3+\I (\omega_n+\I\mu_\chi) \bar\sigma  \chi \right]}{(\omega_n+\I\mu_\chi)^2+p_3^2}\\
=& \int_{-\infty}^\infty\frac{\bar\sigma q|qB|\D p_3}{(2\pi)^2}\sum_\chi T\sum_{\{\omega_n\}}
\frac{\chi\I (\omega_n+\I\mu_\chi)}{(\omega_n+\I\mu_\chi)^2+p_3^2},
\end{split}
\end{equation*}
where the part with $p_3$ on the numerator was dropped because it is odd on $p_3$ and as such vanish when integrated.
After the Matsubara sum, we have
\begin{gather*}
\mean{\h{j}_0(X)}\ped{LLL}=q|qB|\sum_\chi\int_{-\infty}^\infty\frac{\D p_3}{(2\pi)^2}\frac{1}{2}\left[n\ped{F}(p_3-\mu_\chi)-n\ped{F}(p_3+\mu_\chi)\right],\\
\mean{\h{j}_3(X)}\ped{LLL}=q^2 B \sum_\chi\int_{-\infty}^\infty\frac{\D p_3}{(2\pi)^2}\frac{\chi}{2}\left[n\ped{F}(p_3-\mu_\chi)-n\ped{F}(p_3+\mu_\chi)\right].
\end{gather*}
Then the integral over $p_3$ gives
\begin{equation*}
\int_{-\infty}^\infty \frac{\D p_3 }{2}\left[n\ped{F}(p_3-\mu_\chi)-n\ped{F}(p_3+\mu_\chi)\right]=\mu_\chi,
\end{equation*}
therefore, after the sum on chirality we finally obtain
\begin{gather*}
\mean{\h{j}_0(X)}\ped{LLL}=\frac{q|qB|}{(2\pi)^2}(\mu\ped{R}+\mu\ped{L})=\frac{\mu q|qB|}{2\pi^2},\\
\mean{\h{j}_3(X)}\ped{LLL}=\frac{q^2 B}{(2\pi)^2}(\mu\ped{R}-\mu\ped{L})=\frac{q^2\mu\ped{A}}{2\pi^2}B.
\end{gather*}

We now consider the higher Lanndau levels:
\begin{equation*}
\begin{split}
\mean{\h{j}^\parallel_\mu(X)}\ped{HLL}=&-(-\ii)^{1-n_\mu} q|qB|\sum_{l=1}^\infty\int_{-\infty}^\infty\!\!\frac{\D p_3}{(2\pi)^2}\sum_\chi T\sum_{\{\omega_n\}}
\sum_{\sigma,\sigma'=\pm}\!\delta_{n,n'}\times\\
&\times\tr\left[\Delta(\sigma')\Delta(\sigma)\gamma_\mu^\parallel \mathbb{P}_\chi\frac{-\I\slashed{\underline{P}}_\chi^+}{{\underline{P}^+_\chi}^2} \right].
\end{split}
\end{equation*}
When $l$ is fixed we can replace the $\delta_{n,n'}$ with the $\delta_{\sigma,\sigma'}$ and we immediately sum on $\sigma'$. The expression is similar
to the LLL case, we just have to replace $\bar\sigma$ with $\sigma$ and sum over $\sigma=\pm$ and remind that now $\underline{P}$ has also the $y$ component.
Indeed, after the trace and removing $p_3$ odd parts, we find
\begin{equation*}
\begin{split}
\mean{\h{j}_0(X)}\ped{HLL}=&\sum_{l=1}^\infty\int_{-\infty}^\infty\frac{q|qB|\D p_3}{(2\pi)^2}\sum_\chi T\sum_{\{\omega_n\}}\sum_{\sigma=\pm}
\frac{\I \left[(\omega_n+\I\mu_\chi)\right]+p_3\sigma\chi}{{\underline{P}^+_\chi}^2}\\
=&\sum_{l=1}^\infty\int_{-\infty}^\infty\frac{q|qB|\D p_3}{(2\pi)^2}\sum_\chi T\sum_{\{\omega_n\}}
\frac{2\I \left[(\omega_n+\I\mu_\chi)\right]}{{\underline{P}^+_\chi}^2},\\
\mean{\h{j}_3(X)}\ped{HLL}=&\sum_{l=1}^\infty\int_{-\infty}^\infty\frac{q|qB|\D p_3}{(2\pi)^2}\sum_\chi T\sum_{\{\omega_n\}}
\sum_{\sigma=\pm}\sigma\frac{\chi\I (\omega_n+\I\mu_\chi)+p_3}{{\underline{P}^+_\chi}^2}=0.
\end{split}
\end{equation*}
We find out that the third component does not get corrections from HLL, instead for the time component after the frequency sum we obtain
\begin{equation*}
\begin{split}
\mean{\h{j}_0(X)}\ped{HLL}&=\sum_{l=1}^\infty\int_{-\infty}^\infty\frac{q|qB|\D p_3}{(2\pi)^2}\sum_\chi \left[n\ped{F}(E_{p_3,l}-\mu_\chi)-n\ped{F}(E_{p_3,l}+\mu_\chi)\right],
\end{split}
\end{equation*}
where $E_{p_3,l}\equiv\sqrt{p_3^2+2|qB|l}$. For the perpendicular components we expect a vanishing result because is not allowed by the symmetries of the system.
Anyhow, we want to explicitly show that. We start from the point-splitting expression of the mean value
\begin{equation*}
\mean{\h{j}^\perp_\mu(X)}=-(-\ii)^{1-n_\mu} \frac{|qB|}{2\pi}\int_{-\infty}^\infty\frac{\D p_3}{2\pi}\sum_\chi T\sum_{\{\omega_n\}}\sideset{}{'}\sum_{\sigma=\pm}
\tr\left[\Delta(\sigma)\gamma_\mu^\perp\Delta(\sigma) \mathbb{P}_\chi\frac{-\I\slashed{\underline{P}}_\chi^+}{{\underline{P}^+_\chi}^2} \right].
\end{equation*}
Inside the trace we commute the Gamma matrix with the $\Delta$ matrix and we obtain
\begin{equation*}
\gamma_\mu^\perp \Delta(\sigma)=\Delta(-\sigma)\gamma_\mu^\perp.
\end{equation*}
Since the product $\Delta(\sigma)\Delta(-\sigma)$ is vanishing, also $\mean{\h{j}^\perp_\mu(X)}=0$.
Summarizing the results and returning to covariant formalism, we found that the electric current has two thermodynamic function:
the electric charge density $n\ped{c}$ and the chiral magnetic effect (CME) conductivity $\sigma\ped{B}$:
\begin{equation*}
\mean{\h{j}_\mu(X)}=n\ped{c}\, u_\mu + \sigma\ped{B}\,B_\mu.
\end{equation*}
The electric charge density is the mean value in the local rest frame evaluated above: $\mean{\h{j}_0(X)}$, i.e.
\begin{equation*}
\begin{split}
n\ped{c}=\frac{q|qB|}{2\pi^2}\Bigg\{\mu +\sum_{l=1}^\infty\int_{-\infty}^\infty\frac{\D p_3}{2}&
\Big[n\ped{F}(E_{p_3,l}-\mu\ped{R})-n\ped{F}(E_{p_3,l}+\mu\ped{R})+\\
&\hphantom{\Big[}+n\ped{F}(E_{p_3,l}-\mu\ped{L})-n\ped{F}(E_{p_3,l}+\mu\ped{L})\Big]\Bigg\},
\end{split}
\end{equation*}
while the CME conductivity is $\mean{\h{j}_3(X)}/B$:
\begin{equation}
\label{eq:CME_exact}
\sigma\ped{B}=\frac{q^2\mu\ped{A}}{2\pi^2}.
\end{equation}
To our knowledge the equation for the electric charge density of an electron gas in a magnetic medium was first given in~\cite{Canuto:1969cs} and coincides
with the expression above. The CME effect evaluated here coincides with the one obtained with many other derivations~\cite{Kharzeev:2015znc}.

\subsection{No magnetic field limit}
It is useful to know how to perform the $|qB|\to 0$ limit in expressions like the one for electric charge density.
For the sake of clarity, consider the case with vanishing axial chemical potential, i.e. $\mu\ped{A}=0$ (hence $\mu\ped{R}=\mu\ped{L}=\mu$).
The limit is done noticing that the energy of the $l$-th Landau level must go from $\sqrt{p_z^2+2|qB|l}$ to $\sqrt{p_z^2+p_\perp^2}$.
To move the discrete sum on $l$ to the continuum integrals over the perpendicular momentum, we replace $\sum_{l=0}^\infty$
with $\int_0^\infty\D l$. We then want to change the integration variable from $l$ to $p_\perp=\sqrt{p_x^2+p_y^2}$.
This is achieved setting $l=p_\perp^2/2|qB|$ and using $2|qB|\D l=2 p_\perp\D p_\perp$. If we use this replacement on the
electric charge density we find
\begin{align*}
n\ped{c}&=q\int_{0}^\infty\D l\int_{-\infty}^\infty\frac{2|qB|\D p_3}{(2\pi)^2} \left[n\ped{F}(E_{p_3,l}-\mu)-n\ped{F}(E_{p_3,l}+\mu)\right]\\
&=q\int_0^\infty \frac{2p_\perp\D p_\perp}{(2\pi)^2} \int_{-\infty}^\infty \D p_3 \left[n\ped{F}(E_p-\mu)-n\ped{F}(E_p+\mu)\right]\\
&=\frac{2q}{(2\pi)^3}\int_{-\infty}^\infty \D p_x \int_{-\infty}^\infty \D p_y \int_{-\infty}^\infty \D p_z \left[n\ped{F}(E_p-\mu)-n\ped{F}(E_p+\mu)\right]\\
&=\frac{q}{\pi^2}\int_0^\infty |\vec{p}|^2 \D |\vec{p}|\left[n\ped{F}(E_p-\mu)-n\ped{F}(E_p+\mu)\right],
\end{align*}
with $E_p=\sqrt{p_x^2+p_y^2+p_z^2}$. This is exactly the electric charge density for a free Dirac field, compare with Eq.~(\ref{eq:ncfreefield}).

\subsection{Axial current mean value}
In the same way, as done for electric current, we can compute the axial current mean value.
Since $\Delta(\sigma)$ commute with $\gamma^5$, at the lowest Landau level the mean value of the parallel components
of axial current are
\begin{equation*}
\mean{\h{j}^\parallel\ped{A$\mu$}(X)}\ped{LLL}=-(-\ii)^{1-n_\mu} |qB|\int_{-\infty}^\infty\frac{\D p_3}{(2\pi)^2}\sum_\chi T\sum_{\{\omega_n\}}
\tr\left[\frac{1+\I\bar\sigma\gamma_1\gamma_2}{2}\gamma_\mu^\parallel\gamma^5 \mathbb{P}_\chi\frac{-\I\slashed{\underline{P}}_\chi^+}{{\underline{P}^+_\chi}^2} \right].
\end{equation*}
After we perform the trace on spinorial index and we remove $p_3$ odd parts, we find
\begin{equation*}
\begin{split}
\mean{\h{j}\ped{A0}(X)}\ped{LLL}=&-\int_{-\infty}^\infty\frac{|qB|\D p_3}{(2\pi)^2}\sum_\chi T\sum_{\{\omega_n\}}
\frac{-\I \left[(\omega_n+\I\mu_\chi)\chi-\I p_3 \bar\sigma \right]}{(\omega_n+\I\mu_\chi)^2+p_3^2}\\
=&\int_{-\infty}^\infty\frac{|qB|\D p_3}{(2\pi)^2}\sum_\chi T\sum_{\{\omega_n\}}
\frac{\chi\I \left[(\omega_n+\I\mu_\chi)\right]}{(\omega_n+\I\mu_\chi)^2+p_3^2},\\
\mean{\h{j}\ped{A3}(X)}\ped{LLL}=&\ii \int_{-\infty}^\infty\frac{|qB|\D p_3}{(2\pi)^2}\sum_\chi T\sum_{\{\omega_n\}}
\frac{-\I \left[p_3 \chi+\I (\omega_n+\I\mu_\chi) \bar\sigma  \right]}{(\omega_n+\I\mu_\chi)^2+p_3^2}\\
=& \int_{-\infty}^\infty\frac{\bar\sigma|qB|\D p_3}{(2\pi)^2}\sum_\chi T\sum_{\{\omega_n\}}
\frac{\I (\omega_n+\I\mu_\chi)}{(\omega_n+\I\mu_\chi)^2+p_3^2}.
\end{split}
\end{equation*}
After the Matsubara sum we obtain
\begin{gather*}
\mean{\h{j}\ped{A0}(X)}\ped{LLL}=|qB|\sum_\chi\int_{-\infty}^\infty\frac{\D p_3}{(2\pi)^2}\frac{\chi}{2}\left[n\ped{F}(p_3-\mu_\chi)-n\ped{F}(p_3+\mu_\chi)\right],\\
\mean{\h{j}\ped{A3}(X)}\ped{LLL}=qB \sum_\chi\int_{-\infty}^\infty\frac{\D p_3}{(2\pi)^2}\frac{1}{2}\left[n\ped{F}(p_3-\mu_\chi)-n\ped{F}(p_3+\mu_\chi)\right].
\end{gather*}
The integration over $p_3$ is made reminding the result
\begin{equation*}
\int_{-\infty}^\infty \frac{\D p_3 }{2}\left[n\ped{F}(p_3-\mu_\chi)-n\ped{F}(p_3+\mu_\chi)\right]=\mu_\chi.
\end{equation*}
After the sum on chirality, we obtain
\begin{gather*}
\mean{\h{j}\ped{A0}(X)}\ped{LLL}=\frac{|qB|}{(2\pi)^2}(\mu\ped{R}+\mu\ped{L})=\frac{\mu\ped{A}|qB|}{2\pi^2},\\
\mean{\h{j}\ped{A3}(X)}\ped{LLL}=\frac{qB}{(2\pi)^2}(\mu\ped{R}-\mu\ped{L})=\frac{q\mu}{2\pi^2}B.
\end{gather*}

Consider now the higher Landau levels for the parallel components
\begin{equation*}
\begin{split}
\mean{\h{j}^\parallel\ped{A$\mu$}(X)}\ped{HLL}=&-(-\ii)^{1-n_\mu} |qB|\sum_{l=1}^\infty\int_{-\infty}^\infty
	\frac{\D p_3}{(2\pi)^2}\sum_\chi T\sum_{\{\omega_n\}}\sum_{\sigma,\sigma'=\pm}\times\\
&\times\tr\left[\Delta(\sigma')\Delta(\sigma)\gamma_\mu^\parallel\gamma^5 \mathbb{P}_\chi\frac{-\I\slashed{\underline{P}}_\chi^+}{{\underline{P}^+_\chi}^2} \right];
\end{split}
\end{equation*}
when $l$ is fixed the $\delta_{n,n'}$ is just a $\delta_{\sigma,\sigma'}$, therefore we can drop the sum on $\sigma'$.
We then have to trace and sum over $\sigma$ which is now not constrained. For the time component, we find:
\begin{equation*}
\begin{split}
\mean{\h{j}\ped{A0}(X)}\ped{HLL}=&\sum_{l=1}^\infty\int_{-\infty}^\infty\frac{|qB|\D p_3}{(2\pi)^2}\sum_\chi T\sum_{\{\omega_n\}}\sum_{\sigma=\pm}
\frac{\I \left[(\omega_n+\I\mu_\chi)\right]\chi+p_3\sigma}{{\underline{P}^+_\chi}^2}\\
=&\sum_{l=1}^\infty\int_{-\infty}^\infty\frac{|qB|\D p_3}{(2\pi)^2}\sum_\chi T\sum_{\{\omega_n\}}\frac{2\I\chi \left[(\omega_n+\I\mu_\chi)\right]}{{\underline{P}^+_\chi}^2},
\end{split}
\end{equation*}
while the $z$ component is vanishing
\begin{equation*}
\mean{\h{j}\ped{A3}(X)}\ped{HLL}=\sum_{l=1}^\infty\int_{-\infty}^\infty\frac{|qB|\D p_3}{(2\pi)^2}\sum_\chi T\sum_{\{\omega_n\}}
\sum_{\sigma=\pm}\sigma\frac{\I (\omega_n+\I\mu_\chi)+p_3\chi}{{\underline{P}^+_\chi}^2}=0.
\end{equation*}
At last, the frequencies sum for the time component gives
\begin{equation*}
\begin{split}
\mean{\h{j}\ped{A0}(X)}\ped{HLL}&=\sum_{l=1}^\infty\int_{-\infty}^\infty\frac{|qB|\D p_3}{(2\pi)^2}
\sum_\chi \chi\left[n\ped{F}(E_{p_3,l}-\mu_\chi)-n\ped{F}(E_{p_3,l}+\mu_\chi)\right]
\end{split}
\end{equation*}
with $E_{p_3,l}=\sqrt{p_3^2+2|qB|l}$. Also for the axial current we can define an axial charge density $n\ped{A}$ and
a Chiral Separation Effect (CSE) conductivity $\sigma\ped{s}$:
\begin{equation*}
\mean{\h{j}_{A\mu}}=n\ped{A}\,u_\mu+\sigma\ped{s} B_\mu.
\end{equation*}
For what we already evaluated, we have
\begin{align*}
n\ped{A}&=\frac{|qB|}{2\pi^2}\Bigg\{\mu\ped{A}+\\     &+\sum_{l=1}^\infty\int_{-\infty}^\infty\frac{\D p_3}{2} 
\left[n\ped{F}(E_{p_3,l}-\mu\ped{R})-n\ped{F}(E_{p_3,l}+\mu\ped{R})-n\ped{F}(E_{p_3,l}-\mu\ped{L})+n\ped{F}(E_{p_3,l}+\mu\ped{L})\right]\Bigg\},\\
\sigma\ped{s}&=\frac{q\mu}{2\pi^2}.
\end{align*}
The last thermal coefficient is exactly the Chiral Separation Effect (CSE) conductivity. The same steps can be followed to evaluate the axial charge density and
the CSE conductivity of massive fermions with vanishing axial chemical potential $\mu\ped{A}=0$. In that case, the thermal equilibrium
can be reached and all the quantity discussed in this section are still well defined; the results for the thermodynamic functions
related to axial current are:
\begin{align*}
n\ped{A}=&0,\\
\sigma\ped{s}=&qB\int_{-\infty}^\infty \frac{\D p_3}{(2\pi)^2}\left[n\ped{F}(E_{p_3}-\mu)-n\ped{F}(E_{p_3}+\mu)\right],
\end{align*}
with $E_{p_3}^2=p_3^2+m^2$. The CSE induces an axial current even if the system is not chiral.
We have reported this result here to show that it exhibits an explicit mass dependence, as is known that it should~\cite{Metlitski:2005pr}.

\section{General global equilibrium with electromagnetic field}
\label{sec:GlobEqEM}
Now that we recovered the main properties of chiral fermions in magnetic field, we move to discuss
the effects of thermal vorticity in such systems. At the beginning of this section, we provide the general properties
of general global equilibrium (defined in Sec.~\ref{sec:GenGlobEquil}) with external electromagnetic field.
Then, we restrict the discussion to the case of a constant and homogeneous electromagnetic.

We remind that for a system consisting of chiral fermions in external electromagnetic field
the local thermal equilibrium statistical operator is given by:
\begin{equation}
\label{eq:rhoLTEF}
\h{\rho}\ped{LTE} = \frac{1}{Z} \exp \left[ -\int_{\Sigma} \di \Sigma_\mu \left( \h T^{\mu\nu}(x) \beta_\nu(x) - \zeta(x)\, \h j^\mu(x)
- \zeta\ped{A}(x)\, \h j\ped{A}^\mu(x)\right) \right].
\end{equation}
The operators above are those related to Dirac fermions interacting with an external gauge fields 
\begin{equation*}
\begin{split}
\h j^{\,\mu} =&\, q\bar{\psi} \gamma^\mu \psi, \quad \h j^{\,\mu}\ped{A} =\, \bar{\psi} \gamma^\mu\gamma^5 \psi,\\
\h T^{\mu\nu}= &\frac{\I}{4}\left[\bar{\psi}\gamma^\mu\oraw{\de}^\nu\psi-\bar{\psi}\gamma^\mu\olaw{\de}^\nu\psi+\bar{\psi}\gamma^\nu\oraw{\de}^\mu\psi
-\bar{\psi}\gamma^\nu\olaw{\de}^\mu\psi\right] - \frac{1}{2}\left(\h j^\mu A^\nu+\h j^\nu A^\mu\right),
\end{split}
\end{equation*}
and they satisfy the operatorial relations:
\begin{equation*}
\de_\mu \h j^{\,\mu}=0,\quad \de_\mu\h T^{\mu\nu}=\h j_\lambda F^{\nu\lambda}.
\end{equation*}
Regarding the axial current $\h{j}\ped{A}$, we also have to take into account the chiral anomaly.
The chiral anomaly affects the axial current divergence as follows
\begin{equation*}
\de_\mu \h{j}^\mu\ped{A}=-\frac{1}{8}\epsilon^{\mu\nu\rho\lambda}\frac{q^2}{2\pi^2}F_{\mu\nu} F_{\rho\lambda}
=-\frac{q^2}{2\pi^2}(E\cdot B),
\end{equation*}
where $q$ is the electric charge of the fermion. We will show that even in the general
global thermal equilibrium case, the product $E\cdot B$ is non-vanishing and hence
the axial current is not conserved. As discussed in the first chapter (Sec.~\ref{sec:Symm}),
we can still define a new conserved ``axial'' current by means of the Chern-Simons
current $K$, whose divergence gives the chiral anomaly:
\begin{equation*}
K^\mu=\epsilon^{\mu\nu\rho\sigma}A_\nu F_{\rho\sigma},\quad
\frac{q^2}{8\pi^2}\de_\mu K^\mu=\frac{q^2}{2\pi^2}(E\cdot B).
\end{equation*}
The new axial conserved current $\h{j}\ped{CS}$ is then defined as
\begin{equation*}
\h{j}^\mu\ped{CS}=\h{j}^\mu\ped{A}+\frac{q^2}{8\pi^2}K^\mu,\quad
\de_\mu \h{j}^\mu\ped{CS}=0,
\end{equation*}
and the axial chemical potential $\mu\ped{A}$ is to be associated to this current.
Since the additional current $K$ depends only on external fields, it is not
a quantum operator and it does not contribute to thermal averages. Therefore, all the
results we discuss in the absence of chiral anomaly are still valid and unchanged.
Because there is no difference in the results, we continue to denote the conserved
chiral current with $\h{j}\ped{A}$, even when the chiral anomaly is non-vanishing.

With that considered, the system reach a global thermal equilibrium if the
thermodynamic fields satisfy:
\begin{equation*}
\de_{\mu}\beta_{\nu}(x)+\de_{\nu}\beta_{\mu}(x)=0,\qquad \de^\mu \zeta(x)=F^{\nu\mu}\beta_\nu(x)
,\qquad \de^\mu \zeta\ped{A}(x)=0.
\end{equation*}
These equations has been solved in Sec.~\ref{sec:ThermoInB} for the case of vanishing thermal vorticity and constant
homogeneous magnetic field. Here we provide the solution for the general case of non-vanishing thermal vorticity
and any electromagnetic field. The inverse four-temperature and the axial chemical potential solves the previous
conditions in the general case if they are given by (see Sec.~\ref{sec:GenGlobEquil}):
\begin{equation*}
\beta_\mu(x)=b_\mu+\varpi_{\mu\rho}x^\rho,\quad \zeta\ped{A}=\text{const}.
\end{equation*}
where $b$ is a constant time-like four-vector and $\varpi$ is a constant anti-symmetric tensor.
The equation we are left to solve is the one for the electric chemical potential:
\begin{equation}
\label{eq_ZetaGeneral}
\de^\mu \zeta(x)= F^{\sigma\mu}\beta_\sigma.
\end{equation}
To find the solution, we first derive it respect to $\de^\nu$:
\begin{equation*}
\de^\nu\de^\mu \zeta=\de^\nu( F^{\sigma\mu}\beta_\sigma).
\end{equation*}
Since we can exchange the derivatives $\de^\nu\de^\mu$ on the l.h.s., the antisymmetric part respect
to $\mu$ and $\nu$ of this equation must be vanishing:
\begin{equation*}
\begin{split}
\de^\nu \de^\mu \zeta -\de^\mu \de^\nu \zeta& =0=\left[\de^\nu( F^{\sigma\mu}\beta_\sigma)-\de^\mu( F^{\sigma\nu}\beta_\sigma)\right]\\
&= \left[\beta_\sigma\de^\nu F^{\sigma\mu}+ F^{\sigma\mu}\de^\nu\beta_\sigma-\beta_\sigma\de^\mu F^{\sigma\nu}- F^{\sigma\nu}\de^\mu\beta_\sigma\right]\\
&=\left[\beta_\sigma(\de^\nu F^{\sigma\mu}-\de^\mu F^{\sigma\nu})
+(\de^\nu\beta_\sigma) F^{\sigma\mu}+(\de^\mu\beta_\sigma) F^{\nu\sigma}\right].
\end{split}
\end{equation*}
Using the first Bianchi identity $\de^\nu F^{\sigma\mu}+\de^\mu F^{\nu\sigma}+\de^\sigma F^{\mu\nu}=0$, we obtain
\begin{equation*}
\beta_\sigma \de^\sigma F^{\mu\nu}+(\de^\nu\beta_\sigma) F^{\mu\sigma}+(\de^\mu\beta_\sigma) F^{\sigma\nu}=0.
\end{equation*}
We may recognize the Lie derivative of $F$ along $\beta$ in the previous equation. This constitutes a first
condition for global equilibrium, the system can reach global equilibrium only if
\begin{equation}
\label{eq_EqConstraintFmunu}
\mc{L}_\beta (F)=0,\quad\leftrightarrow\quad
\beta_\sigma(x)\de^\sigma F^{\mu\nu}(x)=\varpi^\mu_{\hphantom{\mu}\sigma} F^{\sigma\nu}(x)-\varpi^\nu_{\hphantom{\nu}\sigma} F^{\sigma\mu}(x),
\end{equation}
that it is to say if the electromagnetic field follows the fields lines of inverse-four temperature.

To actually solve Eq.~(\ref{eq_ZetaGeneral}), we translate the global equilibrium condition of the strength tensor~(\ref{eq_EqConstraintFmunu})
to the four-vector potential $A^\mu$; the constraint (\ref{eq_EqConstraintFmunu}) is satisfied if $A$ solve
\begin{equation}
\label{eq_EqConstraintAmu}
\beta_\sigma(x)\de^\sigma A^\mu(x)=\varpi^\mu_{\hphantom{\mu}\sigma}A^\sigma(x)+\de^\mu\Phi(x),
\end{equation}
where $\Phi$ is a smooth function of $x$. We can directly check that Eq.~(\ref{eq_EqConstraintFmunu}) follows from Eq.~(\ref{eq_EqConstraintAmu}) 
by evaluating the derivative of $F$ along $\beta$ and taking advantage of Eq.~(\ref{eq_EqConstraintAmu}):
\begin{equation*}
\begin{split}
\beta_\sigma\de^\sigma F^{\mu\nu}&=\beta_\sigma \de^\sigma(\de^\mu A^\nu-\de^\nu A^\mu)
=\beta_\sigma \de^\sigma(\de^\mu A^\nu)-\beta_\sigma \de^\sigma(\de^\nu A^\mu)\\
&= \de^\mu (\beta_\sigma \de^\sigma A^\nu)-\de^\nu (\beta_\sigma \de^\sigma A^\mu)-(\de^\mu\beta_\sigma)\de^\sigma A^\nu+(\de^\nu\beta_\sigma)\de^\sigma A^\mu\\
&= \de^\mu (\varpi^\nu_{\,\sigma} A^\sigma+\de^\nu\Phi)-\de^\nu (\varpi^\mu_{\,\sigma} A^\sigma+\de^\mu\Phi)
-\varpi_\sigma^{\,\,\mu}\de^\sigma A^\nu+\varpi_\sigma^{\,\,\nu}\de^\sigma A^\mu\\
&= \varpi^\nu_{\,\sigma}\de^\mu A^\sigma-\varpi^\mu_{\,\sigma}\de^\nu A^\sigma+(\de^\mu\de^\nu-\de^\nu\de^\mu)\Phi
+\varpi_{\,\sigma}^\mu \de^\sigma A^\nu-\varpi_{\,\sigma}^\nu \de^\sigma A^\mu\\
&=\varpi^\mu_{\,\sigma}(\de^\sigma A^\nu-\de^\nu A^\sigma)-\varpi^\nu_{\,\sigma}(\de^\sigma A^\mu-\de^\mu A^\sigma)\\
&=\varpi^\mu_{\,\sigma} F^{\sigma\nu}-\varpi^\nu_{\,\sigma} F^{\sigma\mu}.
\end{split}
\end{equation*}
It is important to stress out that after a gauge transformation, condition~(\ref{eq_EqConstraintAmu}) still holds for the new
gauge potential because the function $\Phi$ is also affected by the gauge transformation. Indeed, let $A^\mu$ satisfies Eq.~(\ref{eq_EqConstraintAmu}),
after the gauge transformation $A^{\prime\mu}=A^\mu+\de^\mu\Lambda$ we find:
\begin{equation*}
\begin{split}
\beta_\sigma\de^\sigma A^{\prime\mu}=&\beta_\sigma \de^\sigma A^\mu+\beta_\sigma\de^\sigma \de^\mu\Lambda
=\omega^\mu_{\,\,\sigma}A^\sigma+\de^\mu\Phi+\de^\mu(\beta_\sigma\de^\sigma\Lambda)-(\de^\mu\beta_\sigma)\de^\sigma\Lambda\\
=&\varpi^\mu_{\hphantom{\mu}\sigma}(A^\sigma+\de^\sigma\Lambda)+\de^\mu(\Phi+\beta_\sigma\de^\sigma\Lambda)
=\varpi^\mu_{\hphantom{\mu}\sigma}A^{\prime\sigma}+\de^\mu \Phi',
\end{split}
\end{equation*}
that is exactly condition~(\ref{eq_EqConstraintAmu}) for $A^{\prime\mu}$ but with a transformed function $\Phi'$, 
which is $\Phi$  shifted by the transport of $\Lambda$ along $\beta$.

We can now write Eq.~(\ref{eq_ZetaGeneral}) by taking advantage of Eq.~(\ref{eq_EqConstraintAmu}):
\begin{equation*}
\begin{split}
\de^\mu \zeta &= F^{\sigma\mu}\beta_\sigma=\beta_\sigma (\de^\sigma A^\mu-\de^\mu A^\sigma)\\
&=\beta_\sigma \de^\sigma A^\mu -\de^\mu (\beta_\sigma A^\sigma)+(\de^\mu \beta_\sigma)A^\sigma\\
&= \varpi^\mu_{\,\sigma}A^\sigma +\de^\mu\Phi- \varpi^\mu_{\,\sigma}A^\sigma-\de^\mu(\beta_\sigma A^\sigma).
\end{split}
\end{equation*}
We can then collect all the derivatives together into the equation
\begin{equation*}
\de^\mu\left(\zeta-\Phi+\beta_\sigma A^\sigma\right)=0,
\end{equation*}
from which we immediately get the solution:
\begin{equation}
\label{eq_ZetaEmEq}
\zeta(x)=\zeta_0-\beta_\sigma(x)A^\sigma(x)+\Phi(x),
\end{equation}
where $\zeta_0$ is a constant. The parameter $\Phi$ is analogous to the parameter which grants gauge invariance to
chemical potential in~\cite{Jensen:2013kka}. Even though Eq.~(\ref{eq_ZetaEmEq}) is given in terms of the gauge potential it
is still gauge invariant. Indeed, we showed that $A^\mu$ and $\Phi$ transform under a gauge transformation as
\begin{equation*}
A^{\prime\mu}=A^\mu+\de^\mu\Lambda,\quad \Phi'=\Phi+\beta_\sigma\de^\sigma\Lambda;
\end{equation*}
therefore, the chemical potential $\zeta$ is overall unaffected by gauge transformations:
\begin{equation*}
\zeta(x)'=\zeta_0-\beta_\sigma(x)A^{\prime\sigma}+\Phi'
=\zeta_0-\beta_\sigma(x)A^{\sigma}+\Phi-\beta_\sigma\de^\sigma\Lambda+\beta_\sigma\de^\sigma\Lambda
=\zeta(x).
\end{equation*}
The global equilibrium statistical operator is then obtained from the local one in Eq.~(\ref{eq:rhoLTEF})
replacing the thermal equilibrium form of the thermodynamic field $\beta,\zeta,\zeta\ped{A}$:
\begin{equation}
\label{eq:GEDO_EM}
\h{\rho} = \frac{1}{Z} \exp \left[ -\int_{\Sigma} \di \Sigma_\mu \left[ \left(\h T^{\mu\nu}(x) + \h j^\mu(x) A^\nu(x)\right)\beta_\nu(x)
- \left(\zeta_0+\Phi(x)\right) \h j^\mu(x) - \zeta\ped{A}\, \h j\ped{A}^\mu(x)\right] \right].
\end{equation}

The general form of electromagnetic fields required for the global equilibrium can be given in terms
of the Lie transported basis of Sec.~\ref{sec:ThermalVort}. Indeed, it is easy to realize that the
electromagnetic strength tensor $F$ has vanishing Lie derivative along the $\beta$ field if and only if
both the comoving and electromagnetic field $E^\mu$ and $B^\mu$ have vanishing Lie derivative
along the $\beta$ field. Being $E$ and $B$ four-vectors, they can be decomposed in the basis built
from the thermal vorticity $\varpi$. We have seen that we can distinguish two cases. In the rotational
case, a four-vector $V$ is given by
\begin{equation*}
V^\mu=c_1(x)\,u^\mu+c_2(x)\,\alpha^\mu+c_3(x)\,w^\mu+c_4(x)\,\gamma^\mu
\end{equation*}
and it is Lie transported along $\beta$ if
$$\mathcal{L}_\beta(c_1(x))=\mathcal{L}_\beta(c_2(x))=\mathcal{L}_\beta(c_3(x))=\mathcal{L}_\beta(c_4(x))= 0.$$
Equivalently, in the irrotational case, is given by
\begin{equation*}
V^\mu=c_1(x)\,u^\mu+c_2(x)\,\alpha^\mu+c_3(x)\,V\ped{Irr}^\mu+c_4(x)\,\gamma\ped{Irr}^\mu
\end{equation*}
and it is Lie transported along $\beta$ under the same condition. When $E$ and $B$ are given in the
previous Lie transported form, we are insured that the system can reach the global thermal equilibrium.

We acquire a better understanding and insight of the previous statistical operator~(\ref{eq:GEDO_EM}), if we give the explicit solutions in the case of 
constant homogeneous electromagnetic field $F^{\mu\nu}=$constant for which we already studied the symmetries (Sec.~\ref{subsec:SymInFconst}).
In this case the Lie derivative of electromagnetic strength tensor becomes vanishing when
\begin{equation}
\label{eq:LieDerivFconst}
\mc{L}_\beta(F^{\mu\nu})=\varpi^\mu_{\,\,\sigma}F^{\sigma\nu}-\varpi^\nu_{\,\,\sigma}F^{\sigma\mu}\equiv(\varpi\wedge F)^{\mu\nu}=0.
\end{equation}
We already encountered this wedge product and we proved, see Eq.~(\ref{eq:lemmaAntisymProd}), that  Eq.~(\ref{eq:LieDerivFconst})
has two independent solutions: $F=k\varpi$ and $F=k'\varpi^*$, with $k$ and $k'$ real numbers. In either cases we can then choose
the covariant gauge $A^\mu=\frac{1}{2}F^{\rho\mu}x_\rho$ and taking advantage of Eq.~(\ref{eq:LieDerivFconst}) we find that 
the condition~(\ref{eq_EqConstraintAmu}) is satisfied with $\Phi=\frac{1}{2}b_\sigma F^{\sigma\lambda}x_\lambda$.
The equilibrium chemical potential~(\ref{eq_ZetaEmEq}) is then written as
\begin{equation}
\label{eq:zetaconstF}
\zeta(x)=\zeta_0-\beta_\sigma(x) F^{\lambda\sigma}x_\lambda+\frac{1}{2}\varpi_{\sigma\rho}x^\rho F^{\lambda\sigma}x_\lambda.
\end{equation}
The same solution can be found searching the direct solution of Eq.~(\ref{eq_ZetaGeneral}) with the aid of Eq.~(\ref{eq:LieDerivFconst}).
This last method to obtain the solution proves that the chemical potential in Eq.~(\ref{eq:zetaconstF}) is not gauge dependent.
For constant magnetic field and vanishing thermal vorticity the solution (\ref{eq:zetaconstF}) reduces to $\zeta=$constant,
as it was correctly used in Sec.~\ref{sec:ThermoInB}.

Plugging the form (\ref{eq:zetaconstF}) on the operator Eq.~(\ref{eq:rhoLTEF}), we find:
\begin{equation*}
\h\rho=\frac{1}{\parz}\exp\l\{-\int\D\Sigma_\lambda \l[\l(\h{T}^{\lambda\nu}-\h{j}^\lambda F^{\nu\rho}x_\rho\r)\beta_\nu
-\frac{1}{2}\varpi_{\sigma\rho}\h{j}^\lambda x^\rho F^{\tau\sigma}x_\tau-\zeta_0\h{j}^\lambda\r] \r\}.
\end{equation*}
Inside the round bracket we recognize the divergence-less operator $\h\pi^{\lambda\nu}$ of Eq.~(\ref{eq:GenTranInF}),
whose integrals are the generators of translations. Expressing the coordinate dependence of $\beta$, we can then write
\begin{equation*}
\begin{split}
\h\rho=&\frac{1}{\parz}\exp\l\{-\int\D\Sigma_\lambda \l[\h{\pi}^{\lambda\nu}b_\nu+\varpi_{\nu\tau}x^\tau\h{\pi}^{\lambda\nu}
-\frac{1}{2}\varpi_{\mu\nu}\h{j}^\lambda x^\nu F^{\tau\mu}x_\tau-\zeta_0\h{j}^\lambda\r] \r\}\\
=&\frac{1}{\parz}\exp\l\{-\int\D\Sigma_\lambda \l[\h{\pi}^{\lambda\nu}b_\nu
+\varpi_{\mu\nu}x^\nu\l(\h{\pi}^{\lambda\mu}-\frac{1}{2}\h{j}^\lambda F^{\rho\mu}x_\rho\r)-\zeta_0\h{j}^\lambda\r] \r\}\\
=&\frac{1}{\parz}\exp\l\{-\int\D\Sigma_\lambda \l[\h{\pi}^{\lambda\nu}b_\nu
-\frac{1}{2}\varpi_{\mu\nu}\l[x^\mu\l(\h{\pi}^{\lambda\nu}-\frac{1}{2}\h{j}^\lambda F^{\rho\nu}x_\rho\r)
-x^\nu\l(\h{\pi}^{\lambda\mu}-\frac{1}{2}\h{j}^\lambda F^{\rho\mu}x_\rho\r)\r]\r.\r.\\
&\l.\l.-\zeta_0\h{j}^\lambda\r] \r\};
\end{split}
\end{equation*}
this time we have recreated the divergence-less quantity $\varpi_{\mu\nu}\h{M}^{\lambda,\mu\nu}$ of Eq.~(\ref{eq:LorentzInF}) that generates
those Lorentz transformations which are symmetries of the system. We can then integrate over the coordinate and we find:
\begin{equation*}
\h\rho=\frac{1}{\parz}\exp\left\{-b\cdot\h{\pi}+\frac{1}{2}\varpi:\h{M}+\zeta_0\h{Q}\right\}.
\end{equation*}
In the above form, the analogy with statistical operator without electromagnetic field in Eq.~(\ref{eq:GEDO}) is evident.
In both cases the statistical operator is written with the sum of conserved operators, each one weighted with constant
Lagrange multiplier. We can also write the statistical operator around a specif point $x$. We just need to chose a
fixed point $x$, then we can write the thermodynamic constant vector $b$ and constant scalar $\zeta_0$ starting from that
as follows
\begin{equation*}
b_\mu=\beta(x)_\mu-\varpi_{\mu\nu}x^\nu,\quad
\zeta_0=\zeta(x)+\beta_\sigma(x) F^{\lambda\sigma}x_\lambda-\frac{1}{2}\varpi_{\sigma\rho}x^\rho F^{\lambda\sigma}x_\lambda,
\end{equation*}
from which the statistical operator becomes
\begin{equation*}
\begin{split}
\h\rho=&\frac{1}{\parz}\exp\Big\{-\beta(x)_\mu\left(\h{\pi}^\mu-F^{\lambda\mu}x_\lambda\h Q \right)+\\
&+\frac{1}{2}\varpi_{\mu\nu}\left(\h{M}^{\mu\nu}+x^\nu\h\pi^\mu-x^\mu\h\pi^\nu-x^\nu F^{\lambda\mu}x_\lambda\h{Q}\right)+\zeta(x)\h{Q}\Big\}.
\end{split}
\end{equation*}

It is important to stress-out that with an external magnetic field the Poincaré algebra is modified and becomes
the Algebra in Eq.~(\ref{eq:MaxwellAlgebra}), which we report here for convenience:
\begin{equation*}
\begin{split}
[\h{\pi}^\mu,\h{\pi}^\nu]=&\I F^{\mu\nu}\h{Q},\\
\frac{1}{2}F_{\rho\sigma}[\h{\pi}^\mu,\h{M}^{\rho\sigma}]=&\frac{\I}{2} F_{\rho\sigma}\l(\eta^{\mu\rho}\h\pi^\sigma-\eta^{\mu\sigma}\h\pi^\rho\r),\\
\frac{1}{2}F^*_{\rho\sigma}[\h{\pi}^\mu,\h{M}^{\rho\sigma}]=&\frac{\I}{2} F^*_{\rho\sigma}\l(\eta^{\mu\rho}\h\pi^\sigma-\eta^{\mu\sigma}\h\pi^\rho\r).
\end{split}
\end{equation*}
Notice that because $F$ must have vanishing Lie derivative along $\beta$, if we replace $F$ with $\varpi$ and $F^*$ with $\varpi^*$
the last two algebra identities still hold.
Since the Algebra is known, we can evaluate the effect of a translation to the statistical
operator. Since the translation transformation is unitary, we have
\begin{equation*}
\begin{split}
\h {\sf T}(x)\, \h \rho\, \h {\sf T}^{-1}(x)
	=&\frac{1}{\parz}\exp\Big\{-\h {\sf T}(x)\l(b\cdot\h{\pi}\r) \h {\sf T}^{-1}(x)+\\
	&+\h {\sf T}(x)\l(\frac{\varpi:\h{M}}{2}\r) \h {\sf T}^{-1}(x)+\zeta_0\h {\sf T}(x)\,\h{Q}\, \h {\sf T}^{-1}(x)\Big\}.
\end{split}
\end{equation*}
Therefore we just need to evaluate how the operators transform under translations. 
In general for a unitary transformation we have the expansion:
\begin{equation*}
\E^{\I\h{A}}\h K \E^{-\I\h{A}}\simeq \h{K}-\I\left[\h K,\h A\right]-\frac{1}{2}\left[\left[\h{K},\h{A}\right] ,\h{A}\right]
+\frac{\I}{6}\left[\left[\left[\h{K},\h{A}\right] ,\h{A}\right],\h{A}\right]+\cdots.
\end{equation*}
Applying this formula to our operators, we get an exact results because after a certain order
all the commutators are vanishing. In particular, for Lorentz transformation operator we find
\begin{equation*}
\frac{1}{2}\varpi_{\mu\nu}\h M^{\mu\nu}_x\equiv
\frac{1}{2}\varpi_{\mu\nu}\h {\sf T}(x) \h M^{\mu\nu} \h {\sf T}^{-1}(x)=\frac{1}{2}\varpi_{\mu\nu}\left(\h{M}^{\mu\nu}
+x^\nu\h\pi^\mu-x^\mu\h\pi^\nu-x^\nu F^{\lambda\mu}x_\lambda\h{Q}\right).
\end{equation*}
Notice the difference with the free case in Eq.~(\ref{jshift}).
For the other operators instead we find:
\begin{equation*}
\h \pi^{\mu}_x\equiv \h {\sf T}(x)\, \h \pi^{\mu}\, \h {\sf T}^{-1}(x)=\h \pi^{\mu}-x_\rho F^{\rho\mu}\h{Q},\quad
\h {\sf T}(x)\, \h Q\, \h {\sf T}^{-1}(x)=\h Q.
\end{equation*}
With these definition of translated operators, the density matrix is written around a point $x$ as:
\begin{equation}
\label{eq:GEDObetaF}
\h\rho=\frac{1}{\parz}\exp\left\{-\beta(x)\cdot \h{\pi}_x+\frac{1}{2}\varpi:\h{M}_x+\zeta(x)\h{Q}\right\}.
\end{equation}
Moreover, for what we have learned, a translation transformation on the statistical operator
act as following:
\begin{equation*}
\begin{split}
\h {\sf T}(a)\,\h\rho\, \h {\sf T}^{-1}(a)=&\frac{1}{\parz}\exp\left\{-\beta(x)\cdot \h{\pi}_{x+a}+\frac{1}{2}\varpi:\h{M}_{x+a}+\zeta(x)\h{Q}\right\}\\
=&\frac{1}{\parz}\exp\left\{-\beta(x-a)\cdot \h{\pi}_x+\frac{1}{2}\varpi:\h{M}_x+\zeta(x-a)\h{Q}\right\}.
\end{split}
\end{equation*}
%

\subsection{Expansion on thermal vorticity}
\label{sec:VortExpanF}
Here we use linear response theory, as done in Section~\ref{sec:VortExpan}, to evaluate thermal expectation values in the case of constant electromagnetic field.
We are then looking for a thermal vorticity expansion for the mean value of an operator $\h O$  at the point $x$.
It is convenient to use the statistical operator cast in the form of Eq.~(\ref{eq:GEDObetaF}) with the same point $x$.
Using the properties of the trace, we can transfer the $x$ dependence from the operator $\h O$ to density matrix:
\begin{equation*}
\begin{split}
\mean{\h{O}(x)}=&\frac{1}{\parz}\tr\l[\exp\left\{-\beta(x)\cdot \h{\pi}_x+\frac{1}{2}\varpi:\h{M}_x+\zeta(x)\h{Q}\right\}\h{O}(x)\r]\\
=&\frac{1}{\parz}\tr\l[\h {\sf T}(-x)\exp\left\{-\beta(x)\cdot \h{\pi}_x+\frac{1}{2}\varpi:\h{M}_x+\zeta(x)\h{Q}\right\}\h {\sf T}^{-1}(-x) \h{O}(0)\r]\\
=&=\frac{1}{\parz}\tr\l[\exp\left\{-\beta(x)\cdot \h{\pi}+\frac{1}{2}\varpi:\h{M}+\zeta(x)\h{Q}\right\}\h{O}(0)\r].
\end{split}
\end{equation*}

Therefore, to evaluate the mean value, we expand the statistical operator in the form of the last equality of the previous equation.
To do that, we split the operator exponent in two parts defined as following
\begin{equation*}
\h{\rho}=\dfrac{1}{\parz}\exp\left[\h{A}+\h{B}\right],\quad
\h{A}\equiv-\beta_\mu(x)\h{\pi}^\mu+\zeta(x)\h{Q},\quad\h{B}\equiv\frac{1}{2}\varpi:\h{M},
\end{equation*}
and we expand on $\h{B}$, which is the part containing thermal vorticity.
Since $\h{B}$ and $\h{A}$ have the same commutator as in Section~\ref{sec:VortExpan}, the expansion will leads to the
same result, which is
\begin{equation}
\label{eq:VortExpEM}
\begin{split}
\mean {\h O(x)}=&\mean{\h O(0)}_{\beta(x)}-\alpha_\rho \mycorr{\,\h K^\rho \h O\,}-w_\rho \mycorr{\,\h J^\rho \h O\,}
+\frac{\alpha_\rho\alpha_\sigma}{2}\mycorr{\,\h K^\rho \h K^\sigma \h O\,}\\
&+\frac{w_\rho w_\sigma}{2} \mycorr{\,\h J^\rho \h J^\sigma \h O\,} +\frac{\alpha_\rho w_\sigma}{2}\mycorr{\,\{\h K^\rho,\h J^\sigma\}\h O\,}+\mathcal{O}(\varpi^3).
\end{split}
\end{equation}
where, similarly to Sec.~\ref{sec:VortExpan}, we defined
\begin{equation*}
\begin{split}
\mycorr{\h K^{\rho_1}\cdots \h K^{\rho_n} \h J^{\sigma_1}\cdots \h J^{\sigma_m} \h O} \equiv&
\int_0^{|\beta|} \frac{\di\tau_1\cdots\di\tau_{n+m}}{|\beta|^{n+m}}\times\\
&\times\mean{{\rm T}_\tau\left(\h K^{\rho_1}_{-\ii \tau_1 u}\cdots\h K^{\rho_n}_{-\ii \tau_n u}
	\h J^{\sigma_1}_{-\ii \tau_{n+1} u} \cdots\h J^{\sigma_m}_{-\ii \tau_{n+m} u} \h O(0)\right)}_{\beta(x),c}.
\end{split}
\end{equation*}
The difference in this case is that boost and rotation are defined starting from $\h{M}^{\mu\nu}$, which is
different from Lorentz generators in a system without external electromagnetic field,
\begin{equation*}
\h{K}^\mu=u_\lambda\h{M}^{\lambda\mu},\quad
\h{J}^\mu=\f{1}{2}\epsilon^{\alpha\beta\gamma\mu}u_\alpha\h{M}_{\beta\gamma},
\end{equation*}
and that the averages $\mean{\cdots}_{\beta(x)}$ are made with the statistical operator
\begin{equation*}
\h\rho_0=\frac{1}{\parz_0}\exp\l\{-\beta(x)\cdot\h{\pi}+\zeta(x)\h{Q}\r\}.
\end{equation*}
%

\subsection{Currents and chiral anomaly}
\label{subsec:Feffect}
Here, we determine constitutive equations for electric and axial current at first
order in thermal vorticity and we investigate the contributions from electric
and magnetic fields to thermal coefficients related to vorticity. The constitutive
equations are obtained from the previous expansion on thermal vorticity. For thermal
coefficients, instead of a direct evaluation, we use conservation equations to
show that indeed no additional corrections from electric and magnetic field occur
to first order vorticous coefficients, such as the conductivity of Chiral Vortical
Effect (CVE) and of the Axial Vortical Effect (AVE). Moreover, in this way we exploit
relations between those coefficients and how they are related to the chiral anomaly.

Consider the case of global thermal equilibrium with constant vorticity $\varpi_{\mu\nu}$
and an electromagnetic field with strength tensor $F_{\mu\nu}=k\,\varpi_{\mu\nu}$,
with $k$ a constant.  It then follows that the comoving magnetic and electric fields
are parallel respectively to thermal rotation and thermal acceleration;
to be specific, they are given by:
\begin{equation*}
B^\mu (x)=-k\,w^\mu(x),\quad E^\mu(x)=k\,\alpha^\mu(x).
\end{equation*}
For instance, in the case of constant thermal vorticity caused by a rigid rotation 
long the $z$ axis and a constant magnetic field along $z$, we have:
\begin{equation*}
\varpi_{\mu\nu}=\frac{\Omega}{T_0}\l(\eta_{\mu 1}\eta_{\nu 2}-\eta_{\nu 1}\eta_{\mu 2}\r),\quad
F_{\mu\nu}=B\l(\eta_{\mu 1}\eta_{\nu 2}-\eta_{\nu 1}\eta_{\mu 2}\r),\quad
k=\frac{B T_0}{\Omega},
\end{equation*}
where $\Omega,\,T_0$ and $B$ are constants. In this example, electric and magnetic
fields are orthogonal and there is no chiral anomaly. However, in the general case,
the product $E\cdot B$ is non-vanishing. In that case, we showed in Sec.~\ref{sec:GlobEqEM},
that we can still discuss global equilibrium with chiral imbalance by defining a conserved Chern-Simons
current. Although the statistical operator depends on this new current, we evaluate
the thermal expectation value of the ``true'' axial current $\h{j}\ped{A}^\mu=\bar{\psi}\gamma^\mu\gamma^5\psi$,
which in general is not conserved. Indeed, the form of the statistical operator does
not impose any restriction on the choice of the operator to average.

Now, we use the previous thermal vorticity expansion~(\ref{eq:VortExpEM}) to write the thermal expectation
value of electric current at first order in thermal vorticity. We want to stress-out
that in the expansion~(\ref{eq:VortExpEM}) no approximations are made on the effects of the external electric
and magnetic fields; the expansion~(\ref{eq:VortExpEM}) only approximates the effects of vorticity.
At first order on thermal vorticity, the only quantities that can contribute to the mean value of a current are the
four-vectors $w^\mu,\,\alpha^\mu$ and the scalars $E\cdot\alpha$,  $E\cdot w=B\cdot\alpha$ and $B\cdot w$.
We therefore write the thermal expansion in terms of these quantities, which will select different thermal coefficients.
Moreover, considering that the thermal equilibrium statistical operator can only connect time-even correlators,
the thermal vorticity expansion of electric current is
\begin{equation}
\label{eq:ElecCurrEMVort}
\begin{split}
\mean{\,\h j^{\,\mu}(x)}=&\left[n\ped{c}^0+n\ped{c}^{E\cdot \alpha}(E\cdot \alpha)+n\ped{c}^{B\cdot w}(B\cdot w) \right]u^\mu
+\left[\sigma_B^0+\sigma_B^{E\cdot \alpha}(E\cdot \alpha)+\sigma_B^{B\cdot w}(B\cdot w) \right] B^\mu\\
&+W\apic{V} w^\mu+\sigma_E^{B\cdot \alpha} (B\cdot \alpha)E^\mu+\mc{O}\left(\varpi^2\right).
\end{split}
\end{equation}
Since the thermal coefficients $n\ped{c}^0$ and $\sigma_B^0$ are not related to thermal
vorticity, they are exactly those computed in Sec.~\ref{sec:CME} (for vanishing electric field).
In particular, the Chiral Magnetic Effect (CME) conductivity at vanishing vorticity,
Eq.~(\ref{eq:CME_exact}), is
\begin{equation}
\label{eq:CME_repeat}
\sigma_B^0(x)=\frac{q\zeta_A}{2\pi^2|\beta(x)|}.
\end{equation}
All the other coefficients are related to thermal vorticity and they have the following
properties under parity, time-reversal and charge conjugation
\begin{equation}
\label{tab:ptcElectroMagnetic}
\begin{array}{lccc|cccc|ccc}
& E\cdot\alpha & E\cdot w=B\cdot\alpha & B\cdot w & n\ped{c}^0 & \sigma_B^0 & W\apic{V} & \sigma_E^0 & \sigma_B^{E\cdot\alpha} & \sigma_B^{B\cdot w} & \sigma_E^{B\cdot\alpha} \\ 
\hline
\group{P} & + & - & + & + & - & - & + & - & - & - \\
\group{T} & + & - & + & + & + & + & - & + & + & + \\
\group{C} & - & - & - & - & + & - & + & - & - & - \\
\end{array}
\end{equation}

Moreover, each thermal coefficients is a function depending only on
\begin{equation*}
|\beta|,\, \zeta,\, \zeta\ped{A},\, B^2,\, E^2,\, E\cdot B.
\end{equation*}
The coordinate dependence of thermal coefficients is completely contained inside
the previous thermodynamic scalars. Analogous arguments can be made for the axial
current, which have the following expansion
\begin{equation*}
\begin{split}
\mean{\,\h j^{\,\mu}\ped{A}(x)}=&\left[n\ped{A}^0+n\ped{A}^{E\cdot \alpha}(E\cdot \alpha)+n\ped{A}^{B\cdot w}(B\cdot w) \right]u^\mu
+\left[\sigma_s^0+\sigma_s^{E\cdot \alpha}(E\cdot \alpha)+\sigma_s^{B\cdot w}(B\cdot w) \right] B^\mu\\
&+W\apic{A} w^\mu+\sigma_{sE}^{B\cdot \alpha} (B\cdot \alpha)E^\mu+\mc{O}\left(\varpi^2\right).
\end{split}
\end{equation*}

We are now looking for relations and constraints between those thermodynamic
coefficients by imposing the conservation of electric current, which for mean
values implies
\begin{equation*}
\de_\mu\mean{\,\h j^{\,\mu}(x)}=\mean{\,\de_\mu\h j^{\,\mu}(x)}=0.
\end{equation*}
The coordinate derivative acts both on thermal coefficients and on thermodynamic fields.
We need to establish how the derivative acts on those quantities.
For thermodynamic fields, using the equilibrium conditions and the identities in
Appendix~\ref{sec:betaframeidentities}, we find
\begin{equation*}
\begin{split}
\de_\mu u^\mu=0,\quad
\de_\mu w^\mu=&-3\frac{w\cdot \alpha}{|\beta|},\quad
\de_\mu\alpha^\mu=\frac{2w^2-\alpha^2}{|\beta|},\\
\de_\mu B^\mu=-3\frac{B\cdot \alpha}{|\beta|},\quad
\de_\mu E^\mu=&-\frac{2(w\cdot B)+(\alpha\cdot E)}{|\beta|},\quad
\de_\mu(B\cdot \alpha)=0,\\
\de_\mu(E\cdot \alpha)=-\frac{2}{|\beta|}\left[(w\cdot B)\alpha_\mu+(E\cdot w)w_\mu\right],&\quad
\de_\mu(B\cdot w)=\frac{2}{|\beta|}\left[(w\cdot B)\alpha_\mu+(E\cdot w)w_\mu\right].
\end{split}
\end{equation*}
Moreover, we can also show that
\begin{equation*}
\begin{split}
\de_\mu |\beta|=-\alpha_\mu,\quad
\de_\mu\zeta=\beta^\nu F_{\nu\mu}=-|\beta| E_\mu,\quad
\de_\mu\zeta\ped{A}=0,\quad
\de_\mu(E\cdot B)=0,\\
\de_\mu|B|=-\frac{(E\cdot B)w_\mu+B^2\alpha_\mu}{|\beta||B|},\quad
\de_\mu|E|=-\frac{(E\cdot B)w_\mu+B^2\alpha_\mu}{|\beta||E|},
\end{split}
\end{equation*}
where 
\begin{equation*}
|\beta|=\sqrt{\beta^\sigma \beta_\sigma},\quad |B|=\sqrt{-B^\sigma B_\sigma},\quad
|E|=\sqrt{-E^\sigma E_\sigma}.
\end{equation*}
The derivative respect to coordinates of a thermodynamic function is
\begin{equation*}
\begin{split}
\de_\mu f(|\beta|,\zeta,\zeta_A,|B|,|E|,E\cdot B)=&\left(-\de_\mu|\beta| \frac{\de }{\de |\beta|}+\de_\mu\zeta \frac{\de }{\de \zeta}
+\de_\mu\zeta\ped{A} \frac{\de }{\de \zeta\ped{A}} +\de_\mu |B| \frac{\de }{\de |B|}\right.\\
&\left. +\de_\mu |E| \frac{\de }{\de |E|}+\de_\mu (E\cdot B) \frac{\de }{\de (E\cdot B)}\right) f.
\end{split}
\end{equation*}
Therefore, using the previous identities, the derivative of a thermodynamic function becomes
\begin{equation*}
\begin{split}
\de_\mu f=&\left[-\alpha_\mu\left( \frac{\de }{\de |\beta|}-\frac{|B|^2}{|\beta|}\frac{1}{|B|}\frac{\de }{\de |B|} -\frac{|B|^2}{|\beta|}\frac{1}{|E|}\frac{\de }{\de |E|} \right)\right.\\
&\left. -|\beta| E_\mu\frac{\de }{\de \zeta} -\frac{(E\cdot B)w_\mu}{|\beta|}\left( \frac{1}{|B|}\frac{\de }{\de |B|}+ \frac{1}{|E|} \frac{\de }{\de |E|}\right)\right] f.
\end{split}
\end{equation*}
We can also define the following short-hand notation:
\begin{equation*}
\de_{\tilde{\beta}}\equiv \frac{\de }{\de |\beta|}-\frac{|B|^2}{|\beta|}\de_{\tilde{B}},\quad
\de_{\tilde{B}}\equiv \frac{1}{|B|}\frac{\de }{\de |B|}+ \frac{1}{|E|} \frac{\de }{\de |E|},
\end{equation*}
from which the previous derivative is written as
\begin{equation*}
\de_\mu f=\left[-\alpha_\mu\de_{\tilde{\beta}}  -|\beta| E_\mu \de_\zeta -\frac{(E\cdot B)w_\mu}{|\beta|}\de_{\tilde{B}}\right] f.
\end{equation*}

We can now use the previous relations to impose electric current conservation by 
evaluating the divergence of the expansion in Eq.~(\ref{eq:ElecCurrEMVort}).
For thermal coefficients along the fluid velocity, we find that no additional constraints
are required:
\begin{equation*}
\de_\mu\left(n^0 u^\mu\right)=\de_\mu\left(n^{E\cdot\alpha} (E\cdot \alpha) u^\mu\right)=\de_\mu\left(n^{B\cdot w} (B\cdot w) u^\mu\right)=0.
\end{equation*}
For the terms along the magnetic field, we find:
\begin{equation*}
\begin{split}
\de_\mu\left(\sigma_B^0 B^\mu\right)=&-(B\cdot \alpha)\left[\frac{3}{|\beta|}+\de_{\tilde{\beta}} \right]\sigma_B^0
-(E\cdot B)|\beta|\de_\zeta \sigma_B^0-\frac{(E\cdot B)(B\cdot w)}{|\beta|}\de_{\tilde{B}}\sigma_B^0,\\
\de_\mu\left(\sigma_B^{E\cdot\alpha}(E\cdot\alpha) B^\mu\right)=&-(B\cdot \alpha)(E\cdot\alpha)\left[\frac{3}{|\beta|}+\de_{\tilde{\beta}} \right]\sigma_B^{E\cdot\alpha}
-(E\cdot\alpha)(E\cdot B)|\beta|\de_\zeta \sigma_B^{E\cdot\alpha}\\
&-\frac{(E\cdot B)(B\cdot w)(E\cdot\alpha)}{|\beta|}\de_{\tilde{B}}\sigma_B^{E\cdot\alpha}\\
&-\frac{2}{|\beta|}\left[(w\cdot B)(B\cdot\alpha)+(E\cdot w)(B\cdot w) \right]\sigma_B^{E\cdot\alpha},\\
\de_\mu\left(\sigma_B^{B\cdot w}(B\cdot w) B^\mu\right)=& -(B\cdot \alpha)(B\cdot w)\left[\frac{3}{|\beta|}+\de_{\tilde{\beta}} \right]\sigma_B^{B\cdot w}
-(B\cdot w)(E\cdot B)|\beta|\de_\zeta \sigma_B^{B\cdot w}\\
&-\frac{(E\cdot B)(B\cdot w)^2}{|\beta|}\de_{\tilde{B}}\sigma_B^{B\cdot w}\\
&+\frac{2}{|\beta|}\left[(w\cdot B)(B\cdot\alpha)+(E\cdot w)(B\cdot w) \right]\sigma_B^{B\cdot w}.
\end{split}
\end{equation*}
Along electric field, we have
\begin{equation*}
\begin{split}
\de_\mu\left(\sigma_E(B\cdot\alpha)E^\mu\right)=&-(B\cdot\alpha)(\alpha\cdot E)\left[\frac{1}{|\beta|}+\de_{\tilde{\beta}} \right]\sigma_E
-(B\cdot \alpha)E^2|\beta|\de_\zeta\sigma_E\\
&-\frac{(E\cdot B)(E\cdot w)}{|\beta|}\de_{\tilde{B}}\sigma_E.
\end{split}
\end{equation*}
Lastly, the divergence of the term along rotation is
\begin{equation*}
\de_\mu\left(W\apic{V} w^\mu\right)=-(w\cdot\alpha)\left[\frac{3}{|\beta|}+\de_{\tilde{\beta}} \right]W\apic{V}-(E\cdot w)|\beta|\de_\zeta W\apic{V}
-\frac{(E\cdot B)w^2}{|\beta|}\de_{\tilde{B}}W\apic{V}.
\end{equation*}

To require that $\de_\mu \mean{\,\h j^{\,\mu}(x)}=0$, we sum all the previous pieces and
we select the linear independent terms. Those terms must vanish independently of the
values of the electromagnetic field and of thermal vorticity. We then find that we must
impose several equalities. Among those identities, first consider the following:
\begin{equation*}
\begin{split}
\de_\zeta \sigma_B^0=&0,\quad
\de_{\tilde{B}} W\apic{V}=0,\quad
\de_\zeta \sigma_B^{E\cdot \alpha}=0,\quad
\de_{\tilde{B}} \sigma_B^{E\cdot \alpha}=0,\quad
\de_{\tilde{B}} \sigma_B^{B\cdot w}=0,\\
\de_\zeta \sigma_E^{B\cdot\alpha}=&0,\quad
\de_{\tilde{B}} \sigma_E^{B\cdot\alpha}=0.
\end{split}
\end{equation*}
Notice from the table in~(\ref{tab:ptcElectroMagnetic}) that $\sigma_B^{E\cdot \alpha}$
and $\sigma_E^{B\cdot\alpha}$ are related to $\group{C}$-odd correlator. Therefore, they
must be odd functions of the electric chemical potential $\zeta$. But the previous
constraints require that they do not depend on $\zeta$, therefore they must be vanishing
\begin{equation*}
\sigma_B^{E\cdot \alpha}=0,\quad \sigma_E^{B\cdot\alpha}=0.
\end{equation*}
The previous constraints also require that $W\apic{V}$ and $\sigma_B^{B\cdot w}$
do not depend on $|B|$ and $|E|$\footnote{Note that to reduce the numbers of relations,
	we have indicated electric field and magnetic field derivative together with one derivative $\de_{\tilde{B}}$. However, electric and magnetic fields are independent and each derivative must be considered independently.}.
Then the CVE conductivity $W\apic{V}$ is the same as evaluated without electric and magnetic
field. Moreover, electric current conservation also imposes that
\begin{equation*}
\begin{split}
\left(3+|\beta|\de_{\tilde{\beta}}\right)\sigma_B^0-|\beta|^2\de_\zeta W\apic{V}=&0,\\
\de_{\tilde{B}}\sigma_B^0+|\beta|^2\de_\zeta \sigma_B^{B\cdot w}=&0,\\
\left(3+|\beta|\de_{\tilde{\beta}}\right)\sigma_B^{B\cdot w}=&0,\\
\left(3+|\beta|\de_{\beta}\right)W\apic{V}=&0.
\end{split}
\end{equation*}
From the previous constraints we can replace the known result for the CME conductivity
$\sigma_B^0$ in Eq.~(\ref{eq:CME_repeat}), we then find
\begin{equation*}
\begin{split}
\de_\zeta W\apic{V}=&\frac{1}{|\beta|^2}\left(3+|\beta|\de_{\tilde{\beta}}\right)\sigma_B^0=\frac{q\zeta\ped{A}}{\pi^2|\beta|^3},\\
\de_\zeta \sigma_B^{B\cdot w}=&0,\\
\left(3+|\beta|\frac{\de}{\de\beta}\right)W\apic{V}=&0.
\end{split}
\end{equation*}
Again, since $\sigma_B^{B\cdot w}$ is $\group{C}$-odd it follows form second
equation that it must be vanishing
\begin{equation*}
\sigma_B^{B\cdot w}=0.
\end{equation*}

We want to stress-out that we have found a relation between CVE and CME conductivities:
\begin{equation}
\label{eq:RelCME_CVE}
\de_\zeta W\apic{V}=\frac{1}{|\beta|^2}\left(3+|\beta|\de_{\tilde{\beta}}\right)\sigma_B^0.
\end{equation}
The CVE conductivity $W\apic{V}$~(\ref{CVE}) evaluated before without electromagnetic
field satisfies this relation. We want to stress-out that with Eq.~(\ref{eq:RelCME_CVE})
the CVE conductivity is completely determined from the CME one. Indeed, since $W\apic{V}$
is odd under charge conjugation, fixing the $\zeta$ part of $W\apic{V}$ gives the entire coefficient.
Therefore, CVE acquires all the properties proved for CME. For instance, CME conductivity
is completely dictated by the chiral anomaly and it is therefore protected from corrections coming
from interactions~\cite{Feng:2018tpb}. Since relation~(\ref{eq:RelCME_CVE}) holds not only
for a free theory but for any microscopic interactions, as long as global thermal equilibrium
is concerned, then also CVE conductivity is dictated by the chiral anomaly and it is universal.

Consider now the axial current. Similar steps can be followed to derive constraint equations
between the thermal coefficients of axial current. However, for the axial current, we can consider
the general case in which the divergence of axial current is
\begin{equation*}
\de_\mu \mean{\,\h j^{\,\mu}\ped{A}(x)}= 2m\mean{\,\I\bar{\psi}\gamma^5\psi}-\frac{q^2(E\cdot B)}{2\pi^2}.
\end{equation*}
Remind that if we consider a massive field, we can not define a global equilibrium with axial chemical
potential and all the chiral coefficients are vanishing. Nevertheless, we can write all the equations
with both mass and chiral coefficients. If we consider the mass term we also have to provide the
constitutive equation for the pseudo-scalar. From symmetries, we obtain
\begin{equation*}
\begin{split}
\mean{\,\I\bar{\psi}\gamma^5\psi}=&L^{E\cdot B}(E\cdot B)+L^{\alpha\cdot w}(\alpha\cdot w)+L^{E\cdot w}(E\cdot w)\\
&+L^{(E\cdot w)(E\cdot \alpha)}(E\cdot w)(E\cdot\alpha)+L^{(E\cdot w)(B\cdot w)}(E\cdot w)(B\cdot w)+\mc{O}\left(\varpi^3\right).
\end{split}
\end{equation*}
The value of $L^{\alpha\cdot w}$ for free Dirac field has been reported in Eq.~(\ref{eq:PseudoL}), and
the other coefficients can be computed with Ritus method when both electric and magnetic field
are present.

These terms proportional to the mass and to the chiral anomaly change the constraint equations compared
to the previous case of the electric current. For instance, the relation related to the response to magnetic
field, i.e. the Chiral Separation Effect (CSE) conductivity $\sigma_s^0$, and to the AVE conductivity are
\begin{equation*}
\begin{split}
\de_\zeta \sigma_s^0=&\frac{q^2}{2\pi^2|\beta|}-2mL^{E\cdot B},\\
\left(3+|\beta|\frac{\de}{\de\beta}\right)W\apic{A}=&-2mL^{\alpha\cdot w},\\
\de_{\tilde{B}}\sigma^0_s=&-|\beta|^2\de_\zeta\sigma^{B\cdot w}_s,\\
\de_{\tilde{B}}W\apic{A}=&-2m|\beta|L^{(E\cdot B) w^2},\\
\de_\zeta W\apic{A}=&\frac{1}{|\beta|^2}\left(3+|\beta|\de_{\tilde{\beta}}\right)\sigma_s^0-\frac{2m}{|\beta|}L^{E\cdot w}
\end{split}
\end{equation*}
The second equation has been discussed in~\cite{Buzzegoli:2018wpy}.
The first equation is similar to the second: a term is coming from the chiral anomaly but we also have to
consider the term proportional to the mass. Therefore  for a massive field, as discussed for the AVE,
the CSE is not entirely dictated by the anomaly.
It is then not surprising to find a correction to the CSE~\cite{Gorbar:2013upa}.

On the other hand, for massless field, those constraints becomes
\begin{equation*}
\begin{split}
\de_\zeta \sigma_s^0=&\frac{q^2}{2\pi^2|\beta|},\\
\left(3+|\beta|\frac{\de}{\de\beta}\right)W\apic{A}=&0,\quad \de_{\tilde{B}}W\apic{A}=0, \\
\de_\zeta W\apic{A}=&\frac{1}{|\beta|^2}\left(3+|\beta|\de_{\beta}\right)\sigma_s^0.
\end{split}
\end{equation*}
In this case the CSE conductivity is completely fixed by the chiral anomaly because of the first
equation and by the fact that $\sigma_s^0$ must be an odd function of $\zeta$. Notice that, contrary to CME
and CVE, only in the massless case there is a relation with just CSE and AVE.
In the massive case, other than the mass correction we can also expect corrections from the external
electromagnetic field both in the AVE and in the CSE conductivities.
\newpage

\section{Axial vortical effect non-universality}
\label{sec:AVENon-Univ}
Here, we show that AVE conductivity, or to be precise the part of it which only depend on
temperature, is not protected by radiative corrections. This observation suggests
that the temperature part of AVE is not related to quantum anomalies. Indeed,
quantum anomalies are universal, i.e. they are protected against interactions~\cite{Adler:2004qt}.
We use a Coleman-Hill-like argument~\cite{Coleman:1985zi} to identify
a class of diagrams that indeed gives a contribution and we then evaluate the first-order correction
for a QED-like gauge field plasma. The same argument applied to electric
current shows that the chiral vortical effect is instead protected against radiative corrections.
The crucial difference between CVE and AVE is that the axial current is not conserved when
we consider dynamical gauge fields. If we apply the same argument for scalars interactions,
we find that both CVE and AVE are protected. That occurs because scalars interactions do not modify
the conservation equation of axial current.

We already saw that AVE does not require a chiral imbalance to occur. Then, for the
sake of clarity we consider the global thermal equilibrium with thermal vorticity but
without chiral imbalance. The Lagrangian density of a system consisting of charged massless
fermions interacting through a QED like theory is
\begin{equation*}
\mathcal{L}=-\frac{1}{4}F^{\mu\nu}F_{\mu\nu}-\I\bar{\psi}\gamma^\mu D_\mu\psi,
\end{equation*}
where $F_{\mu\nu}=\de_\mu A_\nu-\de_\nu A_\mu$ is the electromagnetic field strength tensor
associated with the gauge field $A_\mu$ and $D_\mu=\de_\mu-\I e A_\mu$ is the covariant derivative
and $e$ the elementary charge of the particle. The corresponding symmetric stress-energy tensor is
\begin{equation}
\label{eq:GaugeSET}
\h T_{\mu\nu}=F_\mu^{\,\rho}F_{\nu\rho}-\frac{1}{4}\eta_{\mu\nu}F^{\rho\lambda}F_{\rho\lambda}
+\frac{1}{2}\left(\bar{\psi}\gamma_{(\mu}D_{\nu)}\psi-D_{(\mu}\bar{\psi}\gamma_{\nu)}\psi\right).
\end{equation}
Notice that the first terms are related to the gauge bosons, while the remaining terms are
related to the fermionic fields.
The axial current $\h j\ped{A}^\mu=\bar{\psi}\gamma^\mu\gamma^5\psi$, because of the gauge interactions,
is not conserved but it is anomalous:
\begin{equation}
\label{axialanom}
\de_\mu \h j\ped{A}^\mu=-\frac{e^2}{16\pi^2}\epsilon^{\mu\nu\rho\sigma}F_{\mu\nu}F_{\rho\sigma}.
\end{equation}

In the density matrix formalism, when the system is at global thermal equilibrium with
a inverse four-temperature $\beta^\mu=u^\mu/T$ possessing a non-vanishing thermal vorticity
and a chemical potential $\mu$ related to the conserved electric charge of fermions,
the CVE conductivity $W\apic{A}$ is computed using the following formula in the imaginary
time formalism, see Eq.~(\ref{AVEcoeff}):
\begin{equation}
\label{eq:AVERadiative}
\begin{split}
W\apic{A} =&\mycorr{\,\wJ^{3}\,\wj^3\ped{A}\,}_T \\
=&\int_0^{|\beta|}\frac{\D \tau}{|\beta|}\int \D^3 x \left[\mean{\h T^{02}(\tau ,\vec x) \h j^{\,3}\ped{A}(0)}_{T,c} x^1
- \mean{\h T^{01}(\tau ,\vec x)\h j^{\,3}\ped{A}(0)}_{T,c}x^2\right]\\
\equiv & C\apic{A}_{02|3|1}-C\apic{A}_{01|3|2}
\end{split}
\end{equation}
where the subscript c in the correlators stands for connected quantities and we denote
the mean values in rest frame with a subscript $T$, that is:
\begin{equation*}
\mean{\cdots}_{T}=\frac{\tr\left[\exp\left(-\frac{\h H}{T}+\frac{\mu}{T}\h Q\right)\cdots\right]}{\tr\left[\exp\left(-\frac{\h H}{T}+\frac{\mu}{T}\h Q\right)\right]}.
\end{equation*}
The main ingredient in Eq.~(\ref{eq:AVERadiative}), as in the corresponding Kubo
formula~\cite{Amado:2011zx}, is the correlator between the axial current density
and energy flux density. The correlator for the CVE is analogous, we just need
to replace the axial current with the electric current.

We report here the Coleman-Hill like argument~\cite{Coleman:1985zi} used in~\cite{Golkar:2012kb,Hou:2012xg}
to select the class of diagrams which brings contribution to the AVE conductivity.
First, it is important to stress out that at finite temperature gauge theories produce
a non-perturbative mass gap~\cite{Gross:1980br}. Therefore the theory is free of infrared
singularities and the diagrams we take into account can be considered analytic functions.

Consider then a generic diagram contributing to the AVE conductivity. We can obtain
such diagrams employing effective vertices obtained integrating out fermionic fields.
Those vertices have gauge fields, stress-energy tensor and axial current as external legs.
A diagram is then recovered by contracting external gauge bosons and integrating
over momenta. 

\begin{figure}[thb]
	\centering
	\subfigure[A generic diagram]{\label{fig:GenDiag}\includegraphics[width=0.45\columnwidth]{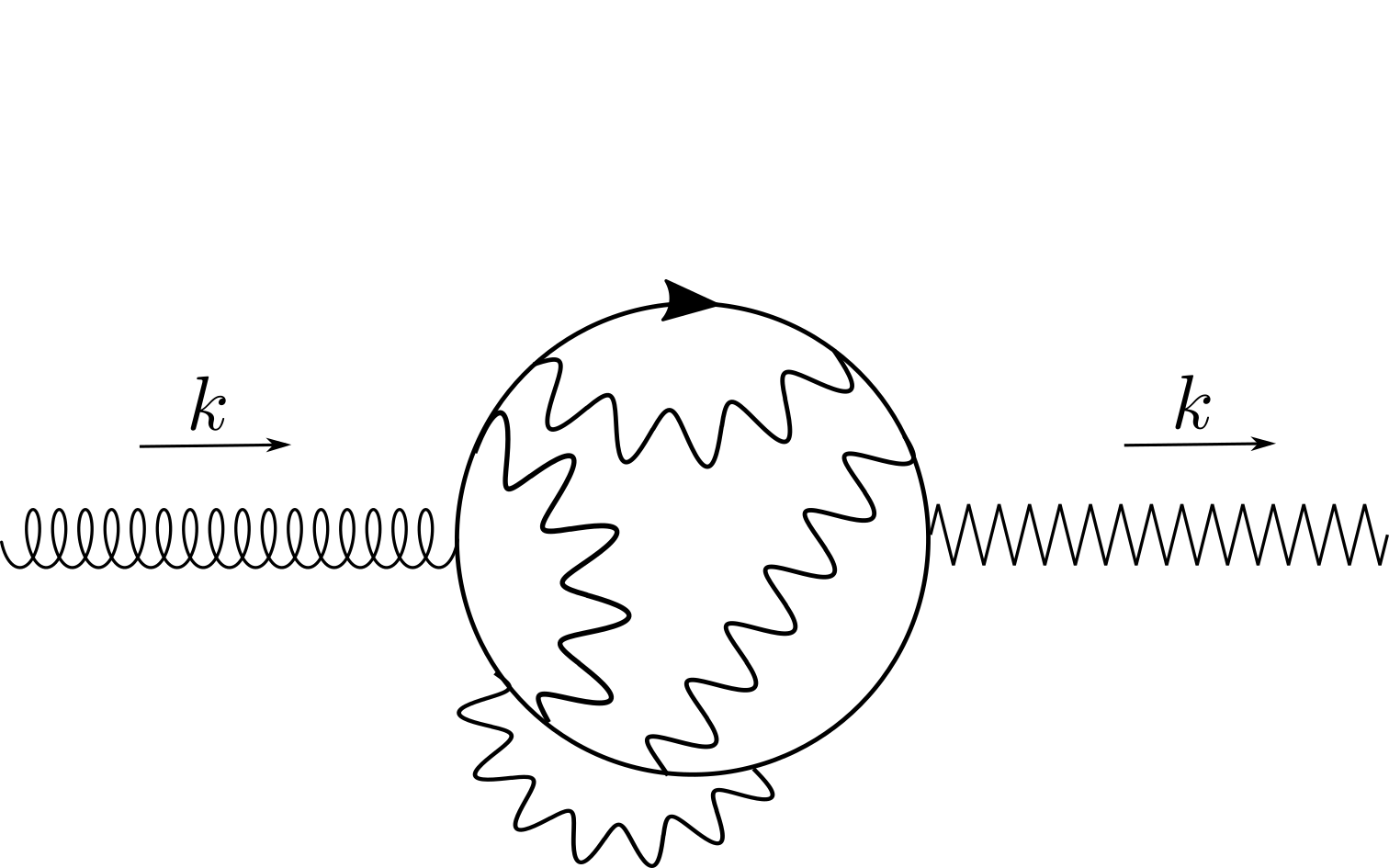}}
	\subfigure[$n$-gauge boson effective vertex]{\label{fig:NEffectiveDiag}\includegraphics[width=0.45\columnwidth]{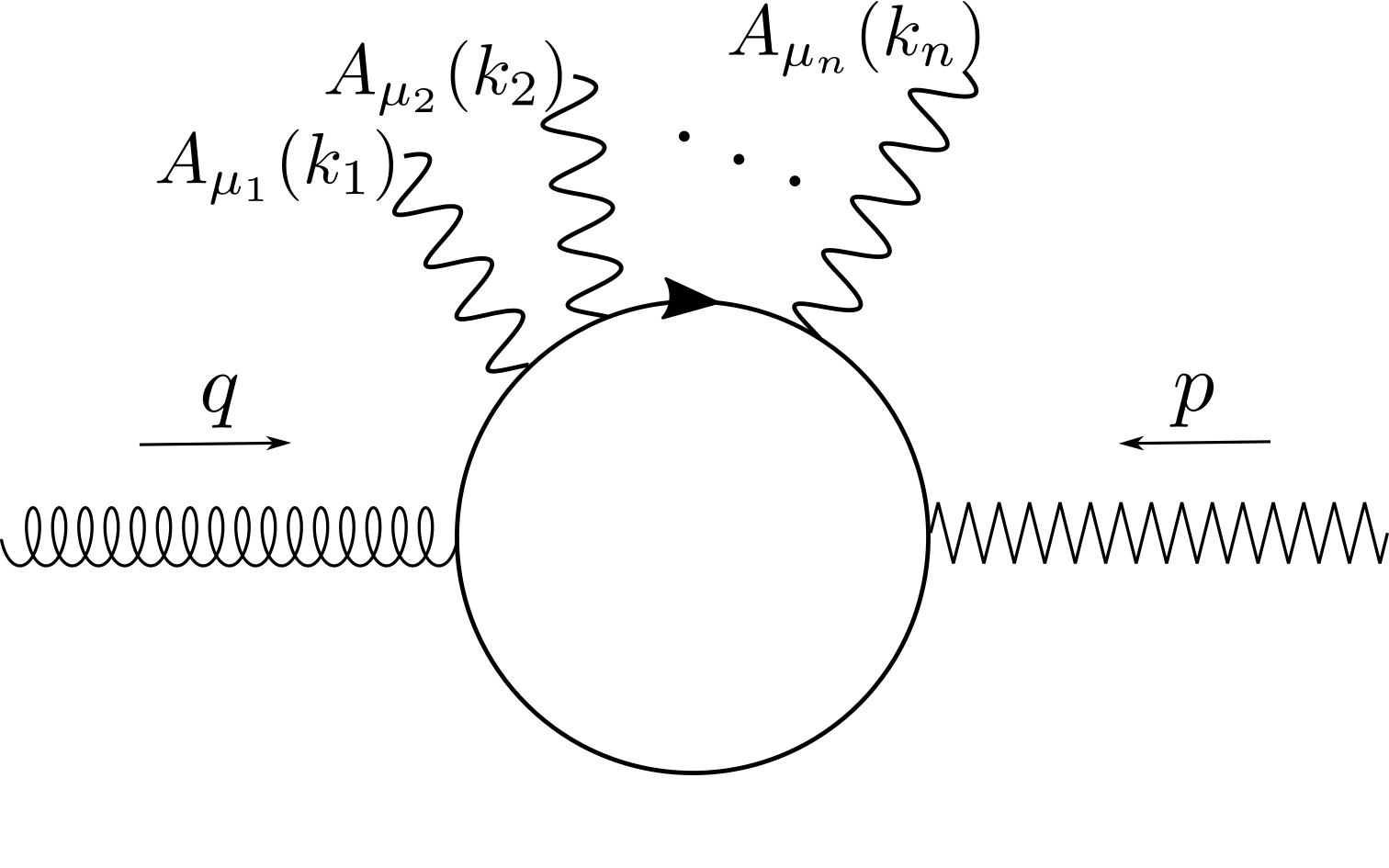}}
	\caption{(a) A generic diagram for gauge interactions. The internal lines are the gauge fields and the external lines
		are the stress-energy tensor and the axial current insertions. (b) An $n$ gauge boson effective vertex.}
	\label{fig:GenericDiag}
\end{figure}

Suppose for now that there is no chiral anomaly (hence the following argument holds for
the CVE but not for the AVE, which is fully addressed later). Then, a generic diagram is
depicted in Fig.~\ref{fig:GenDiag}. The $n$-gauge boson effective diagram
\begin{equation*}
\Gamma^{(n)}_{ij}(p,q,k_1,\dots,k_n)
\end{equation*}
is a one-loop graph with $n$ external dynamical gauge boson, see Fig.~\ref{fig:NEffectiveDiag}.
Gauge invariance imposes the following Ward identities:
\begin{equation}
\label{eq:WardId}
p^i \,\Gamma^{(n)}_{ij}(p,\, q,\, k_1,\dots,\, k_n)=0,\quad
q^i \,\Gamma^{(n)}_{ij}(p,\, q,\, k_1,\dots,\, k_n)=0.
\end{equation}
If we differentiate the Ward identities with respect to $p^r$ and $q^r$
and we then set $p$ and $q$ to zero, we obtain:
\begin{equation*}
\Gamma^{(n)}_{ij\mu_1 \dots \mu_n}(0,\, q,\, k_1,\dots,\, k_n)=0,\quad
\Gamma^{(n)}_{ij\mu_1 \dots \mu_n}(p,\, 0,\, k_1,\dots,\, k_n)=0;
\end{equation*}
therefore, since these functions are analytic, it follows that 
\begin{equation}
\label{eq:EffN}
\Gamma^{(n)}_{ij\mu_1 \dots \mu_n}(p,\, q,\, k_1,\dots,\, k_n)=\mc{O}(p),\quad
\Gamma^{(n)}_{ij\mu_1 \dots \mu_n}(p,\, k,\, k_1,\dots,\, k_n)=\mc{O}(k).
\end{equation}
When $n=0$, for momentum conservation, it must be $q=-p$ and hence
\begin{equation*}
\Gamma^{(0)}_{ij\mu_1 \dots \mu_n}(p,\, -p,\, k_1,\dots,\, k_n)=\mc{O}(p).
\end{equation*}
Instead, when $n>0$ the two momenta are independent and we conclude that
\begin{equation}
\label{eq:EffNl2}
\Gamma^{(n)}_{ij\mu_1 \dots \mu_n}(p,\, -p,\, k_1,\dots,\, k_n)=\mc{O}(pq).
\end{equation}
Then, if the stress-energy tensor and the axial current are attached to the same
fermion loop, the Eq.~(\ref{eq:EffNl2}) states that the total diagram is $\mc{O}(k^2)$.
If instead they are attached to two different loops, we see by using the Eq.~(\ref{eq:EffN})
for each loop, that also these diagrams are $\mc{O}(k^2)$. We would then conclude that only
the one-loop free diagram contributes to the AVE (CVE). This argument can be successfully be
applied to the chiral vortical effect because electric current is conserved. We conclude
that the chiral vortical effect is not affected by radiative corrections.

\begin{figure}[thb]
	\centering
	\subfigure[Fermionic stress tensor diagram]{\label{fig:FermionicDiag}\includegraphics[width=0.45\columnwidth]{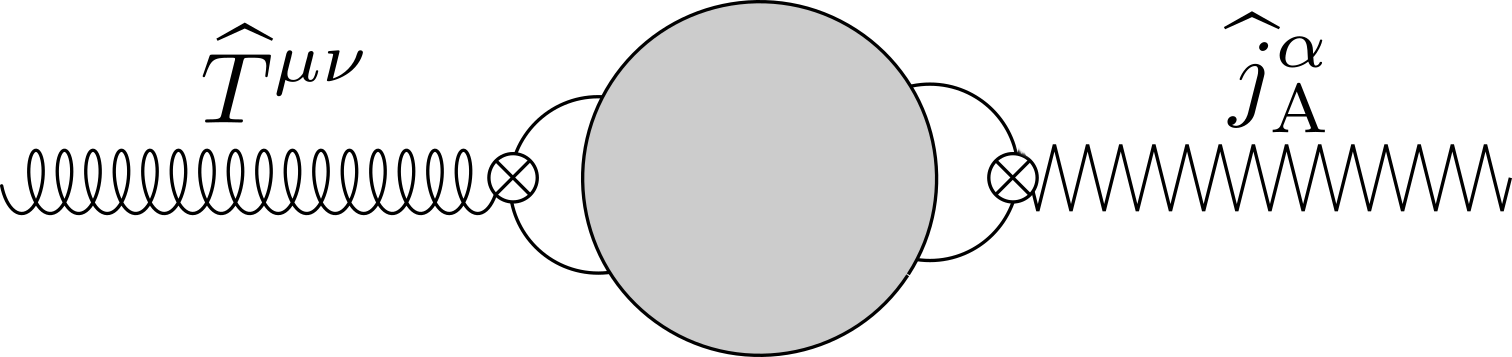}}
	\subfigure[Gauge stress tensor diagram]{\label{fig:GaugeDiag}\includegraphics[width=0.45\columnwidth]{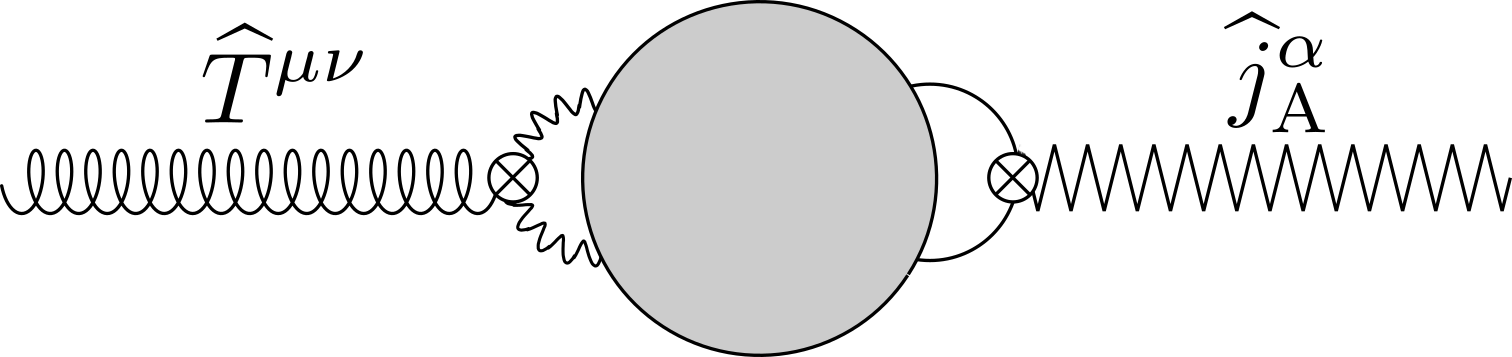}}
	\caption{The two diagram classes.}
	\label{fig:TwoDiagClass}
\end{figure}

However, this argument does not apply to AVE because the axial current is not conserved, and consequently,
we can not use the first Ward identity in Eq.~(\ref{eq:WardId}). To address the problem, we
then split the diagrams into two classes, see Fig.~\ref{fig:TwoDiagClass}.
In the first class, Fig.~\ref{fig:FermionicDiag}, the stress-energy tensor is
connected through the fermionic field. The axial anomaly is captured by the
Eq.~(\ref{axialanom}). Then, expanding in metric perturbation, the anomaly
couples only to the trace of the metric perturbation. Since these diagrams
involves the insertion of an off-diagonal component of stress-energy tensor
they do not give contribution.

\begin{figure}[thb]
	\centering
	\subfigure[Anomalous diagram]{\label{fig:AnomDiags}\includegraphics[width=0.45\columnwidth]{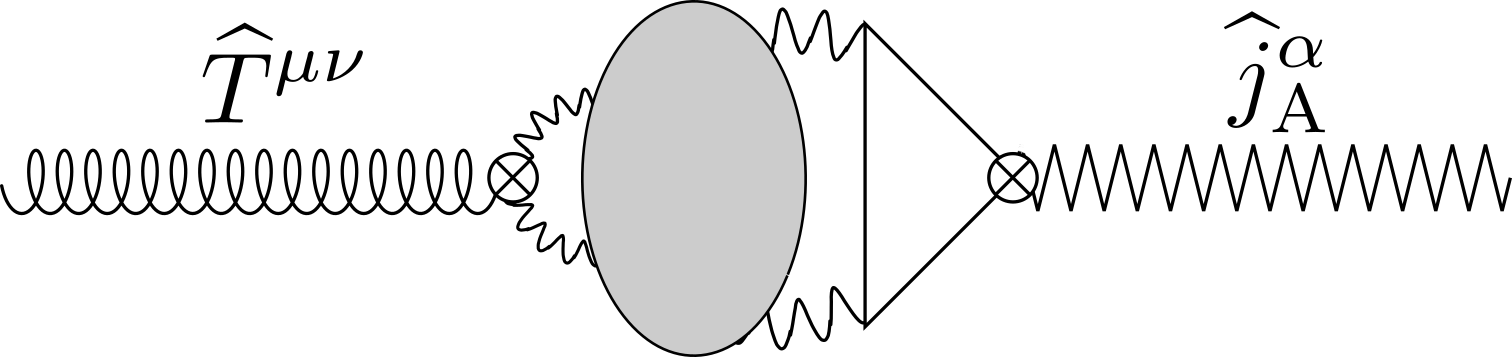}}
	\subfigure[Boxes blob]{\label{fig:Boxes}\includegraphics[width=0.45\columnwidth]{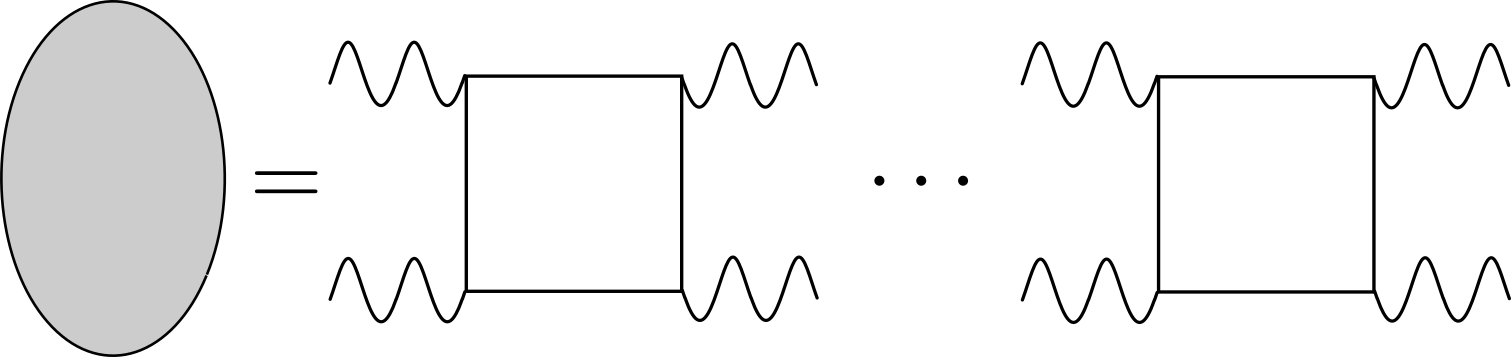}}
	\caption{(a) The diagrams that give contribution to the AVE. The blob in (a) is explicitly given in (b)
		as a ladder of box subdiagrams.}
	\label{fig:TriangleDiag}
\end{figure}

In the second class, Fig.~\ref{fig:GaugeDiag}, the stress-energy tensor is
connected through the gauge field. Since we proved that in the absence of
chiral anomaly no contribution arises for higher orders, the diagrams
of this class that actually give contribution are those containing the
triangle diagram, which is ``responsible'' for the chiral anomaly.
The generic form of these diagrams are given in Fig.~\ref{fig:TriangleDiag}.
The axial current is inserted with the triangle subdiagram and the
stress-energy tensor is connected to it through a ladder of box subdiagrams,
see Fig.~\ref{fig:Boxes}.

\begin{figure}[thb!]
	\centering
	\subfigure[Anomalous diagram]{\label{fig:FeynDiag2loop}\includegraphics[width=0.45\columnwidth]{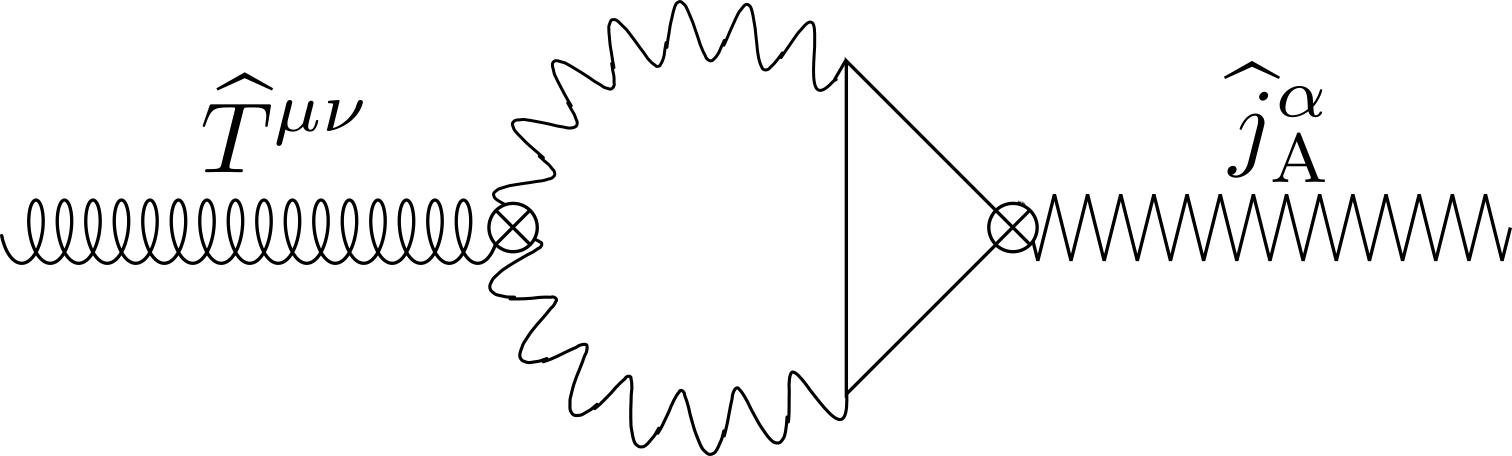}}
	\subfigure[Effective diagram]{\label{fig:FeynDiagEff}\includegraphics[width=0.45\columnwidth]{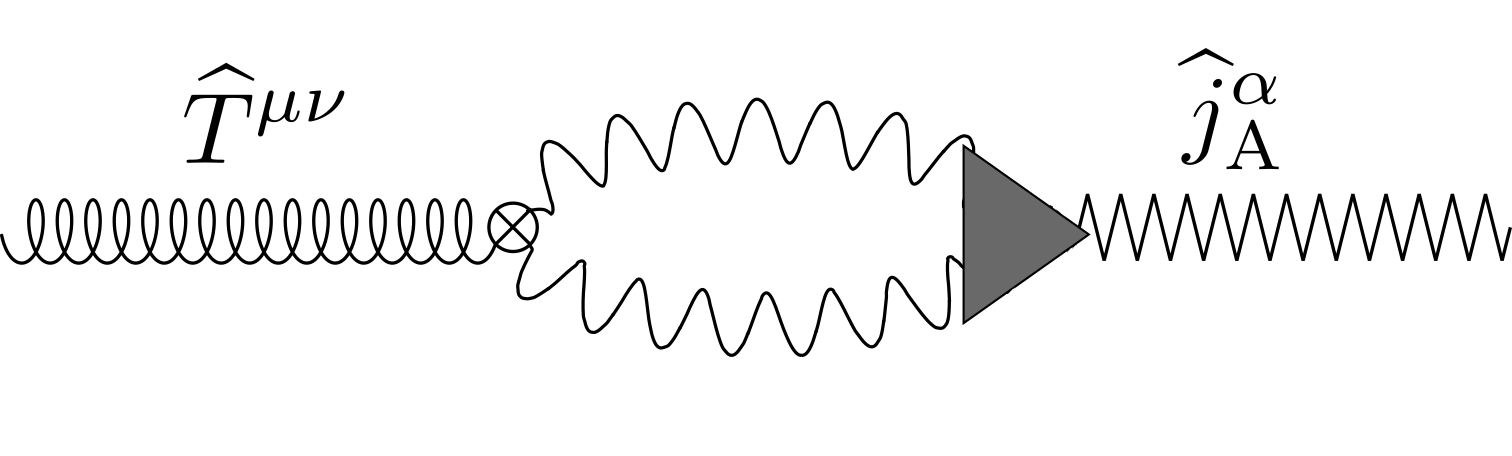}}
	\caption{(a) The leading order diagram and (b) the same diagram after replacing the effective vertex, representing the anomalous effect of the triangle diagram.}
	\label{fig:FeynDiag}
\end{figure}

We now proceed to evaluate the two-loop contribution to the AVE using
the density matrix formalism. As argued, the first radiative correction comes
from the diagram in Fig.~\ref{fig:FeynDiag2loop}. This means that only the
gauge part of the stress-energy tensor~(\ref{eq:GaugeSET}) gives corrections to the AVE conductivity.

Moreover, since the triangle part is constrained by the anomalous Ward identity~(\ref{axialanom})
we can effectively reduce to one-loop calculation by replacing the triangle part of the diagram
with an effective vertex, see Fig.~\ref{fig:FeynDiagEff}. This effective vertex is captured by
replacing the axial current in Eq.~(\ref{eq:AVERadiative}) with the following current
\begin{equation*}
\h j^\mu\ped{Anom}=-\frac{e^2}{4\pi^2}\epsilon^{\mu\nu\lambda\sigma}A_\nu\de_\lambda A_\sigma,
\end{equation*}
which reproduces the chiral anomaly.

To calculate the latter two terms in~(\ref{eq:AVERadiative}) we write the correlation functions in the
Euclidean form. Before doing that, we remind the notation:
\begin{equation*}
P=(p_n, {\vec p} ),\quad X=(\tau ,{\vec x}), 
\quad \sumint_{P}=\frac{1}{|\beta|}\sum_{p_n=-\infty}^\infty\int \frac{\D ^3 p }{(2\pi)^3}
\end{equation*}
and: 
\begin{equation*}
\tilde\delta(P)=\int_0^{|\beta|}\D \tau\int \D^3 x \,\E^{\I X\cdot P}=|\beta| (2\pi)^3 \delta_{p_n,0}\,\delta^{(3)}({\vec p})
\end{equation*}
where $X\cdot P=\tau\,p_n-{\vec x}{\vec p}$ and $p_n=2\pi n /|\beta|$ are the bosonic Matsubara frequencies.
The photon propagator, see Eq.~(\ref{propden}), in the imaginary time reads:
\begin{equation}\label{photprop}
\mean{ {\rm T}_\tau A_\mu(X)A_\nu(Y)}_T= \sumint_{P}\E^{\I P(X-Y)} \frac{G_{\mu\nu}(P)}{P^2}
\end{equation}
where:
\begin{equation*}
G_{\mu\nu}(P) = \delta_{\mu,\nu}+(1-\xi)\frac{P_\mu P_\nu}{P^2}
\end{equation*}
and $P^2$ is the Euclidean squared magnitude of the four vector $P$.

The Euclidean effective axial current using the point splitting procedure (see Sec.~\ref{subsec:pointsplit}) reads:
\begin{equation*}
\begin{split}
\h j&\ped{Anom}^\alpha(X)=-\frac{e^2}{4\pi^2}\epsilon^{\alpha\nu\lambda\sigma}A_\nu(X)\de_\lambda A_\sigma(X)\\
&=\lim_{X_1,X_2\to X} \frac{e^2}{8\pi^2}\epsilon^{\alpha\rho\lambda\tau}(\de_{X_1\lambda}-\de_{X_2\lambda}) A_\rho(X_1)A_\tau(X_2)\\
&\equiv\mathcal{J}\ped{Anom}^{\alpha,\,\rho\tau}(\de_{X_1},\de_{X_2})A_\rho(X_1)A_\tau(X_2).
\end{split}
\end{equation*}
Instead the electromagnetic stress-energy tensor is
\begin{equation*}
\h T\ped{em}^{\mu\nu}(X)=\lim_{X_1,X_2\to X} \mathcal{D}\ped{em}^{\mu\nu,\,\kappa\sigma}(\de_{X_1},\de_{X_2})A_\kappa(X_1)A_\sigma(X_2)
\end{equation*}
with 
\begin{equation*}
\begin{split}
\mathcal{D}\ped{em}^{\mu\nu,\,\kappa\sigma}(\de_{X_1},\de_{X_2})=\frac{1}{2}\Big[&\delta^{\kappa,\sigma}\!\left(\de^\mu_{X_1}\de^\nu_{X_2}\!+\!\de^\nu_{X_1}\de^\mu_{X_2}\right) 
-\!\left(\delta^{\sigma,\nu}\de^\mu_{X_1}\de^\kappa_{X_2}+\delta^{\kappa,\nu}\de^\sigma_{X_1}\de^\mu_{X_2}\right)\\
&\!-\!\left(\delta^{\kappa,\mu}\de^\sigma_{X_1}\de^\nu_{X_2} \!+\!\delta^{\sigma,\mu}\de^\nu_{X_1}\de^\kappa_{X_2}\right)
+\left(\delta^{\kappa,\mu}\delta^{\sigma,\nu}\!+\!\delta^{\kappa,\nu}\delta^{\sigma,\mu}\right)\de_{X_1}\!\!\!\cdot\!\de_{X_2}\\
&\!\!-\!\eta^{\mu\nu}\!\left(\delta^{\kappa,\sigma}\de_{X_1}\!\!\!\cdot\!\de_{X_2}\!-\!\de^\sigma_{X_1}\de^\kappa_{X_2}\right)\Big].
\end{split}
\end{equation*}

Now we are ready to evaluate the two-loop correction of the coefficient from Eq.~(\ref{eq:AVERadiative})
\begin{equation}\label{Archetype}
C_{\mu\nu\alpha i}\apic{A}=\int_0^{|\beta|}\D \tau\int \D^3 x \mean{\h T^{\mu\nu}(X) \h j^{\,\alpha}\ped{A}(0)}_{T,c} x^i.
\end{equation}
First, we consider:
\begin{equation*}
\begin{split}
\mean{{\rm T}_\tau \h T^{\mu\nu}(X)\h j\ped{Anom}^\alpha(Y)}_{T,c}=&\lim_{\substack{X_1,X_2\to X \\ Y_1,Y_2\to Y}}
\mathcal{D}\ped{em}^{\mu\nu,\,\kappa\sigma}(\partial_{X_1},\partial_{X_2})\times\\
&\times \mathcal{J}\ped{Anom}^{\alpha,\,\rho\tau}(\partial_{Y_1},\partial_{Y_2}) \mean{{\rm T}_\tau A_\kappa(X_1)A_\sigma(X_2)A_\rho(Y_1)A_\tau(Y_2)}_{T,c}.
\end{split}
\end{equation*}
The evaluation of the four gauge field correlator can be done by employing the standard Wick 
theorem and since only its connected part appears, only the following two terms survive:
\begin{equation*}
\begin{split}
\mean{A_1 A_2 A_3 A_4}_{c}&=\mean{A_1 A_2 A_3 A_4}-\mean{A_1 A_2}\mean{A_3 A_4}\\
&=\mean{A_1 A_3}\mean{A_2 A_4} + \mean{A_1 A_4}\mean{A_2 A_3}.
\end{split}
\end{equation*}
It can be shown that these two latters terms give the same contribution, therefore
we consider just the first one. The only correlators left are the already known
two point correlators of the photon; in the Fourier space we can choose the 
following notation to express them as in~(\ref{photprop})
\begin{gather*}
\mean{ {\rm T}_\tau A_\kappa(X_1)A_\rho(Y_1)}_T= \sumint_{P}\E^{\I P(X_1-Y_1)} \frac{G_{\kappa\rho}(P)}{P^2},\\
\mean{ {\rm T}_\tau A_\sigma(X_2)A_\tau(Y_2)}_T= \sumint_{Q}\E^{\I Q(X_2-Y_2)} \frac{G_{\sigma\tau}(Q)}{Q^2}.
\end{gather*}
Since the coordinate dependence is now only in the exponential factor, the derivative
action present in the operators $\mathcal{D}\ped{em}$ and $\mathcal{J}\ped{Anom}$
is easily found:
\begin{gather*}
\de_X \mean{ {\rm T}_\tau A_\mu(X)A_\nu(Y)}_T= \I P\mean{ {\rm T}_\tau A_\mu(X)A_\nu(Y)}_T;\\
\de_Y \mean{ {\rm T}_\tau A_\mu(X)A_\nu(Y)}_T= - \I P\mean{ {\rm T}_\tau A_\mu(X)A_\nu(Y)}_T.
\end{gather*}
With these substitutions and after the limit of $X_1,X_2\to X$ and $Y_1,Y_2\to Y$
the correlator in $Y=0$ becomes
\begin{equation*}
\begin{split}
\mean{{\rm T}_\tau &\h T^{\mu\nu}(X)\h j\ped{Anom}^\alpha(0)}_{T,c}=\sumint_{P,Q}\frac{F^{\mu\nu\alpha}(P,Q)}{P^2 Q^2}\E^{\I (P+Q)\cdot X},
\end{split}
\end{equation*}
where the function $F^{\mu\nu\alpha}$ results from the composite operator
in the correlation functions,  in this case:
\begin{equation}\label{Ffunz}
\begin{split}
F^{\mu\nu\alpha}(P, Q)=& 2\, \mathcal{D}\ped{em}^{\mu\nu,\,\kappa\sigma}(\I P,\I Q )\mathcal{J}\ped{Anom}^{\alpha,\,\rho\tau}(-\I Q,-\I P ) G_{\kappa\rho}(P)G_{\sigma\tau}(Q)\\
=&\frac{\I e^2}{8\pi^2}\left[\left(P_\mu-Q_\mu\right)\epsilon_{\alpha\nu\rho\sigma}P_\rho Q_\sigma+\left(P_\nu-Q_\nu\right)\epsilon_{\alpha\mu\rho\sigma}P_\rho Q_\sigma\right].
\end{split}
\end{equation}

So the coefficient (\ref{Archetype}) becomes
\begin{equation*}
C\apic{A}_{\mu\nu\alpha i}=\int_0^{|\beta|}\frac{\D \tau}{|\beta|}\int \D^3 x \sumint_{P,Q}\E^{\I(P+Q)X}x_i \frac{F_{\mu\nu\alpha}(P,Q)}{P^2 Q^2}
\end{equation*}
and rewriting the coordinate $x_i$ as derivative of the exponential
and integrating by part
\begin{equation*}
\begin{split}
C\apic{A}_{\mu\nu\alpha i} &= -\I\int_0^{|\beta|}\frac{\D \tau}{|\beta|}\int \D^3 x \sumint_{P,Q}\left(\frac{\de}{\de q^i}\E^{\I(P+Q)X}\right) \frac{F_{\mu\nu\alpha}(P,Q)}{P^2 Q^2}\\
&= \I\!\sumint_{P,Q}\!\int_0^{|\beta|}\frac{\D \tau}{|\beta|}\int \D^3 x \,\E^{\I(P+Q)X}\!\! \left(\frac{\de}{\de q^i}\frac{F_{\mu\nu\alpha}(P,Q)}{P^2 Q^2}\right).
\end{split}
\end{equation*}
At this point, the integral over the coordinates gives a $\tilde\delta(P+Q)$,
that we can use to sum over the momentum $Q$
\begin{equation*}
C\apic{A}_{\mu\nu\alpha i} = \frac{\I}{|\beta|}\sumint_{P}\left(\frac{\de}{\de q^i}\frac{F_{\mu\nu\alpha}(P,Q)}{P^2 Q^2}\right)_{Q=-P}.
\end{equation*}

The derivative of the quantity in the bracket is simplified by the property
$F_{\mu\nu\alpha}(P,-P)=0$, as we see from~(\ref{Ffunz}):
\begin{equation*}
\begin{split}
\left(\frac{\de}{\de q^i}\frac{F_{\mu\nu\alpha}(P,Q)}{P^2 Q^2}\right)_{Q=-P}&=\frac{1}{P^4}\left(\frac{\de}{\de q^i}F_{\mu\nu\alpha}(P,Q)\right)_{Q=-P}\\
&\equiv \frac{f_{\mu\nu\alpha i}(p_n,\vec p)}{(p_n^2+{\vec p}^2)^2},
\end{split}
\end{equation*}
where we have defined the function $f_{\mu\nu\alpha i}(p_n,\vec p)$, which
is a polynomial function. For instance we need
\begin{equation*}
f_{0132}=\frac{\I e^2}{4\pi^2}(p_n^2-{p^{1}}^2),\quad f_{0231}=-\frac{\I e^2}{4\pi^2}(p_n^2-{p^{2}}^2).
\end{equation*}

We are then reduced to a single momentum integral that can be made explicitly
with standard technique. Consider first
\begin{equation*}
\begin{split}
C\apic{A}_{0132}&=-\frac{e^2}{4\pi^2|\beta|}\int\D p \frac{1}{|\beta|}\sum_{p_n}\int_{4\pi} \frac{\D \Omega_p}{(2\pi)^3}\frac{p^2(p_n^2-p_{1}^2)}{(p_n^2+{\vec p}^2)^2}\\
& =-\frac{e^2}{8\pi^4|\beta|}\int\D p \frac{1}{|\beta|}\sum_{p_n}\frac{p^2(p_n^2-p^2/3)}{(p_n^2+{\vec p}^2)^2}\\
& =-\frac{e^2}{24\pi^4|\beta|}\int\D p \left(p\, n\ped{B}(p)+2p^2 n\ped{B}'(p)\right),
\end{split}
\end{equation*}
where $n\ped{B}$ is the Bose-Einstein distribution function
\begin{equation*}
n\ped{B}(p)=\frac{1}{\E^{|\beta| p}-1}
\end{equation*}
and $n\ped{B}'(p)$ its derivative that can be integrated by parts:
\begin{equation*}
\int\D p\, 2p^2\, n\ped{B}'(p)=-4 \int\D p \,p\, n\ped{B}(p).
\end{equation*}
At the end we are left with
\begin{equation*}
\begin{split}
C\apic{A}_{0132}&=\frac{e^2}{8\pi^4|\beta|}\int\D p \,p\, n\ped{B}(p)=\frac{e^2}{48\pi^2}T^3.
\end{split}
\end{equation*}
In the same way we can show that $C\apic{A}_{0231}=-C\apic{A}_{0132}$ so that the two-loop
correction of~(\ref{eq:AVERadiative}) is
\begin{equation}
W\apic{A}_{(2)}=\frac{e^2}{24\pi^2}T^3.
\end{equation}
Our result is twice the value of \cite{Hou:2012xg,Golkar:2012kb} because we have
considered both right and left fermions contribution instead that just one chirality.
We have then explicitly shown a radiative correction of AVE.
Notice that this term is compatible with the constraint we have found for this conductivity:
\begin{equation*}
\left(3+|\beta|\frac{\de}{\de\beta}\right)W\apic{A}=0.
\end{equation*}

This result is also easily generalized to a QCD like non-Abelian gauge theory with
$N_c$ colors and $N_f$ flavors, we obtain
\begin{equation*}
W\apic{A}=N_c N_f\left(\frac{\zeta\ped{A}^2 T^3}{2\pi^2}+\frac{T^3}{6}+\frac{N_c^2-1}{2N_c}\frac{g_0^2}{24\pi^2}T^3\right),
\end{equation*}
where $g_0$ is the Yang-Mills coupling. This kinds of corrections could play a significant
role in the phenomenology of the Quark-Gluon Plasma, where thanks to the strong coupling
this effect could be enhanced. It would be particularly important to address the case of
massive fields. Because if it turns out that this correction is also proportional to
the naive chiral anomaly, then the intensity of this correction can be estimated in the
order of the pion mass squared. Indeed, using the partial conservation of axial current
the naive chiral anomaly is
\begin{equation*}
\de_\mu \h{j}^\mu\ped{A}=F_\pi m_\pi^2 \pi^0,
\end{equation*}
where $F_\pi$ is the pion decay constant, $m_\pi$ the pion mass, and $\pi^0$ the pion field.

\chapter*{Conclusions}
\label{ch:conclusion}
\addcontentsline{toc}{chapter}{Conclusions}
The present thesis aimed to examine the effects of vorticity on the thermodynamics
of relativistic quantum systems. The main goal of the current study was to determine
a theoretical framework to address quantum effects induced by vorticity, keeping full
covariant and quantum properties of the system. In particular, since a complete quantum
regime could hinder a kinetic theory, the method proposed does not rely on it.

The second aim of this study was to investigate the effects of vorticity in the presence
of chiral matter and external electromagnetic field. To that objective, this work has been
focused on systems consisting of massless chiral fermions. For those systems, this thesis
had the purpose of determining and examining quantum thermodynamics phenomena induced by
chirality, vorticity and magnetic field.

This thesis has identified in Zubarev's non-equilibrium statistical operator the quantitative
framework for detecting the effects of thermal vorticity on relativistic quantum systems.
This study has also shown that Zubarev's method can be reliably used to study global thermal
equilibrium in the mutual presence of thermal vorticity, chirality and external electromagnetic
field. The analysis has shown that the usual ideal thermodynamics is to be changed in the
presence of thermal vorticity. This method also recovered the significant quantum phenomena
known in the literature, namely the chiral magnetic effect, the chiral vortical effect, the
axial vortical effect and the chiral separation effect. The investigation of constitutive
equations has also revealed the presence of additional effects at second-order on thermal
vorticity. This thesis has provided a deeper insight into the interconnection between spin,
chirality and vorticity; for instance, it showed the relations between expectation values
of spin tensor and axial current.

This study has also identified and presented the exact solutions of thermal states for
a system at global thermal equilibrium consisting of chiral massless fermions under the
action of an external constant homogeneous magnetic field. Taking advantage of these exact solutions
and conservation equations, the study also proved that the thermal coefficients related to
first-order effects on thermal vorticity do not receive corrections from the external
electromagnetic field. The same argument revealed existing relations between
those thermal coefficients, even connecting coefficients related to vorticity to other related
to electromagnetic field. For instance, this analysis has found that the chiral vortical effect
and the chiral magnetic effect conductivities are connected one to the other by a differential equation.
Therefore, this research provides insights into the relations between the effects and the interplay
of electromagnetic fields and vorticity.

One of the more significant findings to emerge from this study is that the axial vortical effect is
not entirely dictated by the chiral anomaly or by other types of quantum anomalies.
Indeed, the origination of the axial vortical effect is explained by just the symmetries of
the statistical operator. The results of this investigation support this view by showing that,
for massive fields, any relations with quantum anomalies are lost. As additional evidence,
a Coleman-Hill-like argument revealed that the axial vortical effect conductivity is not protected
against radiative corrections. This fact clearly indicated that quantum anomalies could not cause
the effects, because non-renormalization theorems should instead protect them. The explicit calculation
of the first-order corrections, made with Zubarev's methods, is in agreement with previous predictions.

\appendix
\chapter{Useful resources}
\section{Summary of \texorpdfstring{$\group{P}$}{P}, \texorpdfstring{$\group{T}$}{T} and \texorpdfstring{$\group{C}$}{C} transformations}
\label{sec:Transf}
We report transformation parities of various operators relevant for hydrodynamics and of operators built with Dirac algebra.
We indicate with $\group{P},\,\group{T},\,\group{C}$ respectively parity, time reversal and charge conjugation transformations.
While $\h{T},\h{j}\ped{V},\h{j}\ped{A},\h{S}$ are the stress-energy tensor, vector current, axial current and spin operators,
$\h\psi$ is Dirac field and $\h{\sigma}^{\mu\nu}=\h{\bar{\psi}}[\gamma^\mu,\gamma^\nu]\h\psi$. In the local rest frame of thermal
bath $u=(1,\vec{0})$ the previous quantities transform as follow:
\begin{equation}
\label{tab:ptcObservables}
\begin{array}{lccc|ccccccccccc}
& \h{T}_{00} & \h{T}_{0i} & \h{T}_{ij} & \h{j}^0\ped{V} & \h{\vec{j}}\ped{V} & \h{j}^0\ped{A} & \h{\vec{j}}\ped{A} & \h{\sigma}^{0i} & \h{\sigma}^{ij} & \h{\bar{\psi}}\h\psi
& \ii\h{\bar{\psi}}\gamma^5\h\psi & \h{S}^{0,ij} & \h{S}^{i,0j} & \h{S}^{i,jk} \\ 
\hline
\group{P} & + & - & + & + & - & - & + & - & + & + & - & + & + & - \\
\group{T} & + & - & + & + & - & + & - & + & - & + & - & \pm & \pm & \pm \\
\group{C} & + & + & + & - & - & + & + & - & - & + & + & + & + & + \\
\end{array}
\end{equation}
Consider then the inverse four-temperature $\beta$, the thermal vorticity tensor $\varpi$ and
for an electromagnetic field, the gauge potential $A$, the electric field $E$ and the magnetic field $B$.
We indicate with $\h{\vec{K}}$ and $\h{\vec{J}}$, respectively the boosts and rotations generators.
$\h{Q}$ is electric charge operator, $\h{Q}\ped{A}$ is axial charge operator and $\h\rho$ is the statistical
operator with both conserved vector and axial currents. Those quantities transform as follow
\begin{equation}
\label{tab:ptcParameters}  
\begin{array}{lcccc|cccc|cccc|ccccc}
& \beta_0 & \vec{\beta} & \varpi_{i0} & \varpi_{ij} & A_{0} & \vec{A} & \vec{E} & \vec{B} & \h{\vec{K}} & \h{\vec{J}} & (\h{K}\h{K},\,\h{J}\h{J}) & \h{K}\h{J} 
& \mu & \h{Q} & \mu_5 &  \h{Q}\ped{A} & \h\rho\\ 
\hline
\group{P} & + & - & - & + & + & - & - & + & - & + & + & - & + & + & - & - & \pm\\
\group{T} & + & - & + & - & + & - & + & - & + & - & + & - & + & + & + & + & +\\
\group{C} & + & + & + & + & - & - & - & - & + & + & + & + & - & - & + & + & \pm\\
\end{array}
\end{equation}
where with $\pm$ we indicate that it does not simply acquires a phase but transformation is more complex.
From the inverse four-temperature and thermal vorticity $\varpi$  we can construct a tetrad $\{\beta,\alpha,w,\gamma\}$,
which transform as following
\begin{equation}
\begin{array}{lcccccccc}
& \beta_0 & \vec{\beta} & \alpha_0 & \vec{\alpha} & w_0 & \vec{w} & \gamma_0 & \vec{\gamma} \\ 
\hline
\group{P} & + & - & + & - & - & + & + & - \\
\group{T} & + & - & - & + & + & - & + & - \\
\group{C} & + & + & + & + & + & + & + & + \\
\end{array}
\end{equation}
\newpage

\section{Thermodynamic relations in \texorpdfstring{$\beta$}{beta} frame}
\label{sec:betaframeidentities}
At global thermal equilibrium with vorticity, thermodynamic fields satisfy several
equilibrium relations which constraints their coordinate dependence. In the $\beta$-frame,
we can build several quantities from the four-vector $\beta$ and thermal vorticity $\varpi$:
\begin{gather*}
u_\mu=\f{\beta_\mu}{\sqrt{\beta^2}};\quad\Delta^{\mu\nu}=g^{\mu\nu}-u^\mu u^\nu;\quad
\varpi_{\mu\nu}=\de_\nu \beta_\mu = \epsilon_{\mu\nu\rho\sigma}w^\rho u^\sigma+\alpha_\mu u_\nu - \alpha_\nu u_\mu; \\
\alpha_\mu = \varpi_{\mu\nu} u^\nu; \quad w_\mu=-\frac{1}{2} \epsilon_{\mu\nu\rho\sigma}\varpi^{\nu\rho}u^\sigma; \quad
\gamma_\mu=(\alpha\cdot\varpi)^\lambda\Delta_{\lambda\mu}=\epsilon_{\mu\nu\rho\sigma} w^\nu \alpha^\rho u^\sigma.
\end{gather*}
Most of these quantities depend on coordinates and their derivatives are~\cite{Buzzegoli:2017cqy}:
\begin{gather*}
\de_\nu \beta_\mu = \varpi_{\mu\nu}; \quad
\de_\nu=-\alpha_\nu \frac{\de}{\de\sqrt{\beta^2}};\quad \varpi : \varpi=2\left(\alpha^2-w^2\right)\\
\de_\nu u_\mu=\frac{1}{\sqrt{\beta^2}}\big(\varpi_{\mu\nu}+\alpha_\nu u_\mu\big);\quad
\de^\alpha u_\alpha=0;\quad
u_\alpha\de^\alpha u_\mu=\frac{\alpha_\mu}{\sqrt{\beta^2}};\\
\de_\mu \alpha_\nu=\frac{1}{\sqrt{\beta^2}}\big( \varpi_{\nu\rho}\varpi^\rho_{\;\mu}+\alpha_\mu \alpha_\nu\big);\quad
\de^\alpha \alpha_\alpha=\f{1}{\sqrt{\beta^2}}\big( 2w^2-\alpha^2\big);\quad
u_\alpha\de^\alpha \alpha^2=0;\\
\de_\mu w_\nu=\frac{1}{\sqrt{\beta^2}}\big(\alpha_\mu w_\nu -\f{1}{2}\epsilon_{\nu\rho\sigma\lambda}\varpi^{\rho\sigma}\varpi^\lambda_{\;\mu}\big);\quad
\de^\alpha w_\alpha=-3\f{w\cdot\alpha}{\sqrt{\beta^2}}; \quad u_\alpha\de^\alpha w^2=0;\\
\alpha^\sigma\de_\mu \alpha_\sigma=w^\sigma\de_\mu w_\sigma=\frac{1}{\sqrt{\beta^2}}\big(w^2\alpha_\mu-(\alpha\cdot w)w_\mu\big);\quad \de_\mu(\alpha\cdot w)=0;\\
\de_\alpha\gamma^\alpha=0;\quad \de^\alpha \Delta_{\alpha\beta}=-\frac{\alpha_\beta}{\sqrt{\beta^2}}.
\end{gather*}
%
\section{Identities for massless momentum integrals}
\label{sec:masslessmomint}
\newcommand{\Lii}[2]{\text{Li}_{#1}\left( #2\right)}
The integrals in thermodynamics quantities related to massless fermions can be turn into the general integral:
\begin{equation*}
\begin{split}
\int_0^\infty \di p \,p^k \left(\frac{1}{\e^{|\beta| (p-\mu)}+1}+\frac{1}{\e^{|\beta| (p-\mu)}+1}\right)
	=&-|\beta|^{-k-1} k\, \Gamma(k)\times\\
&\times\left(\mathrm{Li}_{k+1}(-\e^{|\beta|\mu})+\mathrm{Li}_{k+1}(-\e^{-|\beta|\mu})\right).
\end{split}
\end{equation*}
Furthermore, the Lie function (or polylogarithmic function) $\Lii{n}{z}$ obtained from the previous integral can be further
simplified into polynomials in the potential $\zeta$ thanks to the following identities:
\begin{gather*}
-\left[\Lii{2}{-\E^\zeta}+\Lii{2}{-\E^{-\zeta}}\right]=\frac{\pi^2}{6}+\frac{\zeta^2}{2},\\
-2\left[\Lii{3}{-\E^\zeta}-\Lii{3}{-\E^{-\zeta}}\right]=\frac{\zeta}{3}\left(\pi^2+\zeta^2\right),\\
-\left[\Lii{4}{-\E^\zeta}+\Lii{4}{-\E^{-\zeta}}\right]=\frac{\zeta^4}{24}+\frac{\pi^2\zeta^2}{12}+\frac{7\pi^4}{360}.
\end{gather*}
\newpage

\section{Sums on Matsubara frequencies}
\label{sec:thermalsums}
Since we are using thermal field theory in imaginary time, in order to acquire physical quantities we
have to sum over Matsubara frequency, either bosonic or fermionic. This is accomplished first transforming
the sum into a complex integral and using the residue theorem. Consider the bosonic and fermionic thermal sums
\begin{equation*}
\sigma\ped{B}\equiv T\sum_{\omega_n}f(\omega_n),\quad\sigma\ped{F}\equiv T\sum_{\{\omega_n\}}f(\omega_n),
\end{equation*}
where $f(z)$ is a general function which is analytic in the complex plane (apart from singularities), and regular on the real axis.
Sums such as these can be turned into complex integral with the help of $\cosh$ function. Instead, it would be more relevant for thermodynamics
to consider the following auxiliary functions
\begin{equation*}
\I n\ped{B}(\I\, z)\equiv \f{\I}{\exp(\I\beta z)-1},\quad-\I n\ped{F}(\I\, z)\equiv \f{-\I}{\exp(\I\beta z)+1},
\end{equation*}
where $n\ped{B}$ is Bose-Einstein distribution function and $n\ped{F}$ is the Fermi-Dirac distribution.
The poles of these functions are respectively located exactly where the relative bosonic or fermion sum are evaluated, i.e. at $z = \omega_n$,
and the residue at each pole is $T$ for both functions. The sums in $\sigma$ can be then replaced by a complex integral:
\begin{equation*}
\begin{split}
\sigma\ped{B}=&\oint \f{\D z}{2\pi\I}f(z)\I n\ped{B}(\I\,z)\equiv \int_{-\infty-\I 0^+}^{+\infty-\I 0^+} \f{\D z}{2\pi}f(z)n\ped{B}(\I\, z)
	+\int_{+\infty+\I 0^+}^{-\infty+\I 0^+} \f{\D z}{2\pi}f(z)n\ped{B}(\I\, z),\\
\sigma\ped{F}=&\oint \f{\D z}{2\pi\I}f(z)(-\I) n\ped{F}(\I\,z)\equiv - \int_{-\infty-\I 0^+}^{+\infty-\I 0^+} \f{\D z}{2\pi}f(z)n\ped{F}(\I\, z)
	-\int_{+\infty+\I 0^+}^{-\infty+\I 0^+} \f{\D z}{2\pi}f(z)n\ped{F}(\I\, z),
\end{split}
\end{equation*}
where the integration contour runs anti-clockwise around the real axis of the complex $z$-plane.
We change integration variables into $z\to -z$ to the last terms of both equation and taking advantage of the realtion
\begin{equation*}
n\ped{B}(-\I\, z)=-1-n\ped{B}(\I z),\quad n\ped{F}(-\I\, z)=1-n\ped{F}(\I z),
\end{equation*}
we obtain
\begin{equation*}
\begin{split}
\sigma\ped{B}&= \int_{-\infty}^{+\infty} \f{\D z}{2\pi} f(z)+\int_{-\infty-\I 0^+}^{+\infty-\I 0^+} \f{\D z}{2\pi}[f(z)+f(-z)]n\ped{B}(\I\, z),\\
\sigma\ped{F}&= \int_{-\infty}^{+\infty} \f{\D z}{2\pi} f(z)-\int_{-\infty-\I 0^+}^{+\infty-\I 0^+} \f{\D z}{2\pi}[f(z)+f(-z)]n\ped{F}(\I\, z),
\end{split}
\end{equation*}
where in the first term we returned to the real axis (because there are no singularities), and replaced again $z\to -z$.
The general thermal sums on bosonic and fermionic frequencies are then expressed as a complex integral weighted with Bose-Einstein or Fermi-Dirac distribution function:
\begin{equation}
\label{eq:thermalsumrule}
\begin{split}
\f{1}{\beta} \sum_{\omega_n} f (\omega_n)=&\int_{-\infty}^{+\infty} \f{\D z}{2\pi} f(z)+\int_{-\infty-\I 0^+}^{+\infty-\I 0^+} \f{\D z}{2\pi} [ f(z) + f(-z) ] n\ped{B}(\I z),\\
\f{1}{\beta} \sum_{\{\omega_n\}} f (\omega_n)=&\int_{-\infty}^{+\infty} \f{\D z}{2\pi} f(z)-\int_{-\infty-\I 0^+}^{+\infty-\I 0^+} \f{\D z}{2\pi} [ f(z) + f(-z) ] n\ped{F}(\I z).
\end{split}
\end{equation}

Only the last terms of Eq.~(\ref{eq:thermalsumrule}) gives a results that is affected by the temperature, which is contained inside $n\ped{B}$ and $n\ped{F}$.
While the first term in equation~(\ref{eq:thermalsumrule}) is temperature-independent. It corresponds to the zero temperature vacuum contribution. This often
rise divergences but it cancels after a proper renormalization. As closing remark, note that in the lower half-plane
\begin{equation*}
\mod{n\ped{B/F}(\I\, z)}^{z=x-\I y}=\mod{\f{1}{\E^{\I\beta x}\E^{\beta y}\mp 1}} \overset{y\gg T}{\approx} \E^{-\beta y}\overset{y\gg x}{\approx}\E^{-\beta\mod{z}}.
\end{equation*}
This means that if the function $f(z)$ grows slower than $\E^{\beta\mod{z}}$ at large $\mod{z}$, then the integration contour can be closed in the lower half-plane,
and the result is determined by the poles and residues of the function $f(z)+f(-z)$.

As an application of Eq.~(\ref{eq:thermalsumrule}), let $f$ be an analytic function, then for either $\tau>0$ or $\tau<0$ it holds:
\begin{equation*}
\frac{1}{|\beta|}\sum_{\omega_n}\frac{f(\omega_n\pm \ii \mu) \e^{\ii (\omega_n\pm \ii \mu) \tau}}{(\omega_n\pm \ii \mu)^2+E^2}
=\frac{1}{2E}\sum_{s=\pm 1} f(-\ii s E)\e^{\tau  s E}[\theta(- s \tau)+n_B(E\pm s \mu)].
\end{equation*}
Equivalently for fermionic sums we find:
\begin{equation*}
\frac{1}{\beta}\sum_{\{\omega_n\}}\frac{f(\omega_n\pm \ii \mu) \e^{\ii (\omega_n\pm \ii \mu) \tau}}{(\omega_n\pm \ii \mu)^2+E^2}
=\frac{1}{2E}\sum_{s=\pm 1} f(-\ii s E)\e^{\tau  s E}[\theta(- s \tau)-n_{F}(E\pm s\mu)];
\end{equation*}
Instead, if the poles are of second order we obtain:
\begin{equation*}
\begin{split}
\frac{1}{\beta}\!\sum_{\{\omega_n\}}\frac{f(\omega_n \pm \ii \mu)\e^{\ii (\omega_n\pm \ii \mu) \tau} }{\left[(\omega_n\pm \ii \mu)^2+E^2\right]^2}
=\! \sum_{s=\pm 1}\! \E^{\tau  s E}& \left\{\!\frac{(1-\tau E)f(-\ii sE)+\ii E f'(-\ii sE) }{4 E^3}[\theta(- s \tau)+\right.\\
&\left.- n\ped{F}(E\pm s\mu)]+\frac{f(-\ii s E)}{4E^2}n'\ped{F}(E\pm s\mu)\right\}.
\end{split}
\end{equation*}

Other useful Matsubara sum are
\begin{equation*}
\begin{split}
T\sum_{\{\omega_n\}}&\frac{(\omega_n+\I\mu)^2}{\l[(\omega_n+\I\mu)^2+E_1^2\r]\l[(\omega_n+\I\mu)^2+E_2^2\r]}=\\
&=\frac{1}{2(E_1+E_2)}+\frac{1}{2(E_2^2-E_1^2)}\l[E_1\tilde{n}\ped{F}(E_1,\mu)-E_2\tilde{n}\ped{F}(E_2,\mu)\r],
\end{split}
\end{equation*}
where
\begin{equation*}
\tilde{n}\ped{F}(E,\mu)=n\ped{F}(E-\mu)+n\ped{F}(E+\mu),
\end{equation*}
and similarly we have
\begin{equation*}
\begin{split}
T\sum_{\omega_n}&\frac{(\omega_n+\I\mu)^2}{\l[(\omega_n+\I\mu)^2+E_1^2\r]\l[(\omega_n+\I\mu)^2+E_2^2\r]}=\\
&=\frac{1}{2(E_1+E_2)}+\frac{1}{2(E_2^2-E_1^2)}\l[E_1\tilde{n}\ped{B}(E_1,\mu)-E_2\tilde{n}\ped{B}(E_2,\mu)\r].
\end{split}
\end{equation*}
%

\chapter{Stress-energy tensor in external electromagnetic field}
\label{ch:SeTInB}
\addtocontents{toc}{\protect\setcounter{tocdepth}{0}}
The Lagrangian of Dirac field in external electromagnetic field is
\begin{equation*}
\mathcal{L}= \frac{\I}{2}\left[ \bar{\psi}\gamma^\mu \oraw{\de}_\mu \psi -\bar{\psi}\gamma^\mu \olaw{\de}_\mu \psi\right]-m\bar{\psi}\psi-j^\mu A_\mu.
\end{equation*}
From Lagrangian we derive the equations of motion (EOM)
\begin{equation*}
\slashed{\de}\psi=-\I(q\slashed{A}+m)\psi,\quad \slashed{\de}\bar{\psi}=\bar{\psi}\I(q\slashed{A}+m),
\end{equation*}
and the canonical stress-energy tensor to obtain a stress-energy tensor
\begin{equation*}
\h T^{\mu\nu}\ped{can}=\frac{\I}{2}\left[\bar{\psi}\gamma^\mu\oraw{\de}^\nu\psi-\bar{\psi}\gamma^\mu\olaw{\de}^\nu\psi\right]-\h j^\mu A^\nu.
\end{equation*}
The canonical stress-energy tensor satisfies the conserved equation
\begin{equation}
\label{eq:SeTConsEqInF}
\de^\mu\h T_{\mu\nu}=\h j^\lambda F_{\nu\lambda}.
\end{equation}

Here we prove that the symmetric stress-energy tensor
\begin{equation*}
\begin{split}
\h T^{\mu\nu}\ped{S}=&\frac{1}{2}\h T^{\mu\nu}\ped{can}+\frac{1}{2}\h T^{\nu\mu}\ped{can}\\
=&\frac{\I}{4}\left[\bar{\psi}\gamma^\mu\oraw{\de}^\nu\psi-\bar{\psi}\gamma^\mu\olaw{\de}^\nu\psi+\bar{\psi}\gamma^\nu\oraw{\de}^\mu\psi-\bar{\psi}\gamma^\nu\olaw{\de}^\mu\psi\right]
-\frac{1}{2} \left(\h j^\mu A^\nu+\h j^\nu A^\mu\right),
\end{split}
\end{equation*}
also satisfy the same conservation equation~(\ref{eq:SeTConsEqInF}).

To verify this we first need the relativistic Pauli equations. One can find them applying the EOM two times.
Since it holds
\begin{equation*}
\left[\I\gamma^\mu(\oraw{\de}_\mu+\I qA_\mu)-m\right]\psi=0
\end{equation*}
we can write
\begin{equation*}
\left[\I\gamma^\mu(\oraw{\de}_\mu+\I qA_\mu)+m\right]\left[\I\gamma^\nu(\oraw{\de}_\nu+\I qA_\nu)-m\right]\psi=0,
\end{equation*}
and hence 
\begin{equation*}
\left[-\gamma^\mu\gamma^\nu(\oraw{\de}_\mu+\I qA_\mu)(\oraw{\de}_\nu+\I qA_\nu)-m^2\right]\psi=0.
\end{equation*}
Then, we separate the $\gamma^\mu\gamma^\nu$ product in a symmetric and antisymmetric part and we take advantage of their algebra:
\begin{equation*}
\gamma^\mu\gamma^\nu=g^{\mu\nu}+\frac{1}{2}\sigma^{\mu\nu},\quad\sigma^{\mu\nu}=\left[\gamma^\mu,\,\gamma^\nu\right].
\end{equation*}
The symmetric part gives
\begin{equation*}
\begin{split}
g^{\mu\nu}(\oraw{\de}_\mu+\I qA_\mu)(\oraw{\de}_\nu+\I qA_\nu) & =g^{\mu\nu}\left[\oraw{\de}_\mu\oraw{\de}_\nu+\I q\left(A_\mu\oraw{\de}_\nu+A_\nu\oraw{\de}_\mu+\de_\mu A_\nu\right)-q^2 A_\mu A_\nu\right]\\
&=\oraw{\square}+2\I qA_\rho\oraw{\de}^\rho+\I q\de_\rho A^\rho-q^2 A^2
\end{split}
\end{equation*}
instead the antisymmetric part gives
\begin{equation*}
\begin{split}
\frac{1}{2}\sigma^{\mu\nu}(\oraw{\de}_\mu+\I qA_\mu)(\oraw{\de}_\nu+\I qA_\nu) =&\frac{1}{4}\sigma^{\mu\nu}\left[(\oraw{\de}_\mu+\I qA_\mu)(\oraw{\de}_\nu+\I qA_\nu)+\right.\\
&\left.-(\oraw{\de}_\nu+\I qA_\nu)(\oraw{\de}_\mu+\I qA_\mu)\right]\\
=&\frac{\I q}{4}\sigma^{\mu\nu}(\de_\mu A_\nu-\de_\nu A_\mu)=\frac{\I q}{4}\sigma^{\mu\nu}F_{\mu\nu}.
\end{split}
\end{equation*}
Putting the two pieces together we obtain the relativistic Pauli equation for the $\psi$ field:
\begin{equation*}
\square\psi=\left( q^2 A^2 -2\I qA_\rho\oraw{\de}^\rho-\I q(\de_\rho A^\rho)-\frac{\I q}{4}\sigma^{\mu\nu}F_{\mu\nu}-m^2\right)\psi.
\end{equation*}
For the $\bar{\psi}$ we can follow the same procedure. We start from EOM
\begin{equation*}
\bar{\psi}\left[\I\gamma^\mu(\olaw{\de}_\mu-\I qA_\mu)+m\right]=0
\end{equation*}
and we apply it an other time with the mass sign flipped
\begin{equation*}
\bar{\psi}\left[\I\gamma^\mu(\olaw{\de}_\mu-\I qA_\mu)+m\right]\left[\I\gamma^\nu(\olaw{\de}_\nu-\I qA_\nu)-m\right]=0,
\end{equation*}
and hence 
\begin{equation*}
\bar{\psi}\left[-\gamma^\mu\gamma^\nu(\olaw{\de}_\mu-\I qA_\mu)(\olaw{\de}_\nu-\I qA_\nu)-m^2\right]=0.
\end{equation*}
Again we split the gamma matrix product into a symmetric and into an asymmetric part. The symmetric part gives
\begin{equation*}
\begin{split}
g^{\mu\nu}(\olaw{\de}_\mu-\I qA_\mu)(\olaw{\de}_\nu-\I qA_\nu) & =g^{\mu\nu}\left[\olaw{\de}_\mu\olaw{\de}_\nu-\I q\left(\olaw{\de}_\nu A_\mu+\olaw{\de}_\mu A_\nu+\de_\nu A_\mu\right)-q^2 A_\mu A_\nu\right]\\
&=\olaw{\square}-2\I q\olaw{\de}^\rho A_\rho-\I q\de_\rho A^\rho-q^2 A^2
\end{split}
\end{equation*}
instead the antisymmetric part gives
\begin{equation*}
\begin{split}
\frac{1}{2}\sigma^{\mu\nu}(\olaw{\de}_\mu-\I qA_\mu)(\olaw{\de}_\nu-\I qA_\nu) =&\frac{1}{4}\sigma^{\mu\nu}\left[(\olaw{\de}_\mu-\I qA_\mu)(\olaw{\de}_\nu-\I qA_\nu)+\right.\\
&\left.-(\olaw{\de}_\nu-\I qA_\nu)(\olaw{\de}_\mu-\I qA_\mu)\right]\\
=&-\frac{\I q}{4}\sigma^{\mu\nu}(\de_\nu A_\mu-\de_\mu A_\nu)=\frac{\I q}{4}\sigma^{\mu\nu}F_{\mu\nu}.
\end{split}
\end{equation*}
The relativistic Pauli equation for the field $\bar{\psi}$ is then
\begin{equation*}
\square\bar{\psi}=\bar{\psi}\left( q^2 A^2 +2\I q\olaw{\de}^\rho A_\rho+\I q(\de_\rho A^\rho)-\frac{\I q}{4}\sigma^{\mu\nu}F_{\mu\nu}-m^2\right).
\end{equation*}

Using the relativistic Pauli equations we can evaluate the divergence of the canonical stress-energy tensor with inverted indices:
\begin{equation*}
\begin{split}
\de_\mu T^{\nu\mu}\ped{can}
=&\de_\mu\left\{ \frac{\I}{2}\left[\bar{\psi}\gamma^\nu\oraw{\de}^\mu\psi-\bar{\psi}\gamma^\nu\olaw{\de}^\mu\psi\right]-q \h j^\nu A^\mu\right\}\\
=& \frac{\I}{2}\left[(\de_\mu\bar{\psi})\gamma^\nu(\de^\mu\psi)+\bar{\psi}\gamma^\nu(\square\psi)-(\square\bar{\psi})\gamma^\nu\psi-(\de_\mu\bar{\psi})\gamma^\nu(\de^\mu\psi)\right]\\
	&-q \h (\de_\mu j^\nu) A^\mu-q \h j^\nu (\de_\mu A^\mu)\\
=& \frac{\I}{2}\bar{\psi}\gamma^\nu \left( q^2 A^2 -2\I qA_\rho\oraw{\de}^\rho-\I q(\de_\rho A^\rho)-\frac{\I q}{4}\sigma^{\mu\nu}F_{\mu\nu}-m^2\right)\psi\\
&-\frac{\I}{2}\bar{\psi}\left( q^2 A^2 +2\I q\olaw{\de}^\rho A_\rho+\I q(\de_\rho A^\rho)-\frac{\I q}{4}\sigma^{\mu\nu}F_{\mu\nu}-m^2\right)\gamma^\nu\psi\\
&-q \bar{\psi}\left( \gamma^\nu A_\rho\oraw{\de}^\rho+\olaw{\de}^\rho A_\rho\gamma^\nu\right) \psi -q \h j^\nu (\de_\rho A^\rho)
\end{split}
\end{equation*}
and after the simplifications it becomes
\begin{equation*}
\begin{split}
\de_\mu T^{\nu\mu}\ped{can}=& q \bar{\psi}\left( \gamma^\nu A_\rho\oraw{\de}^\rho+\olaw{\de}^\rho A_\rho\gamma^\nu\right) \psi
-q \bar{\psi}\left( \gamma^\nu A_\rho\oraw{\de}^\rho+\olaw{\de}^\rho A_\rho\gamma^\nu\right) \psi
+\frac{q}{8}\bar{\psi}[\gamma^\nu,\sigma^{\rho\lambda}] F_{\rho\lambda}\psi\\
&+ \frac{q}{2}(\de_\rho A^\rho)\left(\bar{\psi}\gamma^\nu\psi+\bar{\psi}\gamma^\nu\psi\right) - \h j^\nu (\de_\rho A^\rho)
=\h j_\lambda F^{\nu\lambda},
\end{split}
\end{equation*}
where we used $[\gamma^\nu,\sigma^{\rho\lambda}]=4(g^{\nu\rho} \gamma^\lambda-g^{\nu\lambda} \gamma^\rho)$.
Therefore, thanks to the previous result the four-divergence of symmetric stress-energy tensor is:
\begin{equation}
\de_\mu T^{\mu\nu}\ped{S}=\de_\mu\left( \frac{1}{2}T^{\mu\nu}\ped{can}+\frac{1}{2}T^{\nu\mu}\ped{can}\right)=\h j_\lambda F^{\nu\lambda},
\end{equation}
which is the same for the canonical one. We would have obtained the same result with the Belinfante procedure:
\begin{equation*}
\begin{split}
T^{\mu\nu}\ped{S}=&T^{\mu\nu}\ped{can}+\frac{1}{2}\de_\lambda\left(\Phi^{\lambda,\mu\nu}-\Phi^{\mu,\lambda\nu}-\Phi^{\nu,\lambda\mu}\right)
\end{split}
\end{equation*}
and choosing
\begin{equation*}
\begin{split}
\Phi^{\mu,\lambda\nu}=\frac{\I}{8}\bar{\psi}\{\gamma^\mu,[\gamma^\lambda,\gamma^\nu]\}\psi.
\end{split}
\end{equation*}

\backmatter
\cleardoublepage
\chapter{Acknowledgments}

Foremost, I would like to express my very great appreciation to my advisor Professor Francesco Becattini for
the continuous support of my Ph.D. study and research, for his guidance, motivation, suggestion and immense expertise.
His advice on my career was precious and has been very much valued.
I could not have imagined having a better advisor and mentor for my Ph.D. study.

I would like to express my deep gratitude to Professor Dmitri Kharzeev for hosting me one year at Stony Brook University.
Discussions with him are always enlightening, insightful and helped me to deepen my understanding of Physics.

My sincere thanks also go to Dr. Eduardo Grossi for leading my first steps on research and for sharing his vast knowledge.
He is still a reference point for me.

Furthermore, I am indebted to all other members of the research team I met in these three years: Dr. Claudio Bonati,
Dr. Iurii Karpenko and my fellow doctoral student Davide Rindori for stimulating discussions and knowledgeable comments.

Finally, although this is not the appropriate place, I am glad to express my very profound gratitude to the people
close to my heart: my parents, my brother and my friends for providing me with unfailing support and continuous
encouragement throughout my years of study and through the process of researching and writing this thesis.
This accomplishment would not have been possible without them. Thank you.
\clearpage
\printbibliography
\addcontentsline{toc}{chapter}{\bibname}

\end{document}